%% file: thesis.tex
\input harvmac2.tex
\input labeldefs.tex
\input amssym.def
\input amssym
\input epsf.tex
\magnification\magstep1
\baselineskip 18pt
\font\bigcmsy=cmsy10 scaled 1500
\parskip 6pt
\newdimen\itemindent \itemindent=32pt
\def\textindent#1{\parindent=\itemindent\let\par=\resetpar%
\indent\llap{#1\enspace}\ignorespaces}

\let\oldpar=\par
\def\resetpar{\oldpar\parindent=20pt\let\par=\oldpar}

\font\ninerm=cmr9 \font\ninesy=cmsy9
\font\eightrm=cmr8 \font\sixrm=cmr6
\font\eighti=cmmi8 \font\sixi=cmmi6
\font\eightsy=cmsy8 \font\sixsy=cmsy6
\font\eightbf=cmbx8 \font\sixbf=cmbx6
\font\eightit=cmti8
\def\eightpoint{\def\rm{\fam0\eightrm}
  \textfont0=\eightrm \scriptfont0=\sixrm \scriptscriptfont0=\fiverm
  \textfont1=\eighti  \scriptfont1=\sixi  \scriptscriptfont1=\fivei
  \textfont2=\eightsy \scriptfont2=\sixsy \scriptscriptfont2=\fivesy
  \textfont3=\tenex   \scriptfont3=\tenex \scriptscriptfont3=\tenex
  \textfont\itfam=\eightit  \def\it{\fam\itfam\eightit}%
  \textfont\bffam=\eightbf  \scriptfont\bffam=\sixbf
  \scriptscriptfont\bffam=\fivebf  \def\bf{\fam\bffam\eightbf}%
  \normalbaselineskip=9pt
  \setbox\strutbox=\hbox{\vrule height7pt depth2pt width0pt}%
  \let\big=\eightbig  \normalbaselines\rm}
\catcode`@=11 %
\def\eightbig#1{{\hbox{$\textfont0=\ninerm\textfont2=\ninesy
  \left#1\vbox to6.5pt{}\right.\n@@space$}}}
\def\vfootnote#1{\insert\footins\bgroup\eightpoint
  \interlinepenalty=\interfootnotelinepenalty
  \splittopskip=\ht\strutbox %
  \splitmaxdepth=\dp\strutbox %
  \leftskip=0pt \rightskip=0pt \spaceskip=0pt \xspaceskip=0pt
  \textindent{#1}\footstrut\futurelet\next\fo@t}
\catcode`@=12 %
\def \de{\delta}

\def \si{\sigma}

\def \ga{\gamma}

\def \pr{\partial}
\def \tr{{\rm tr }}

\def \ha{{\hat a}}

\def \hf{{\hat f}}

\def \J{{\rm J}}
\def \T{{\rm T}}
\def \X{{\rm X}}
\def \hD{{\hat D}}
\def \cD{{\hat {\cal D}}}
\def \hG{{\hat {\cal G}}}
\def \hV{{\hat V}}
\def \zz{{\bar x}}
\def \yz{{\bar y}}
\def \bz{{\bar z}}
\def \bta{{\bar \eta}}

\def \l{\big \langle}
\def \r{\big \rangle}
\def \ep{\epsilon}
\def \vep{\varepsilon}
\def \hhalf{{1\over 2}}
\def \half{{\textstyle {1 \over 2}}}
\def \thir{{\textstyle {1 \over 3}}}
\def \quar{{\textstyle {1 \over 4}}}
\def \ts{\textstyle}

\def \d{{\rm d}}
\def \v{{\rm v}}
\def \x{{\rm x}}
\def \y{{\rm y}}

\def \tx{{\tilde {\rm x}}}

\def \A{{\cal A}}
\def \B{{\cal B}}
\def \C{{\cal C}}
\def \D{{\cal D}}
\def \E{{\cal E}}
\def \F{{\cal F}}
\def \G{{\cal G}}
\def \H{{\cal H}}
\def \I{{\cal I}}
\def \J{{\cal J}}
\def \K{{\cal K}}
\def \L{{\cal L}}
\def \N{{\cal N}}
\def \O{{\cal O}}
\def \P{{\cal P}}
\def \Q{{\cal Q}}
\def \S{{\cal S}}
\def \U{{\cal U}}
\def \V{{\cal V}}
\def \W{{\cal W}}
\def \Y{{\cal Y}}

\def \hD{{\hat D}}
\def \tQ{{\tilde Q}{}}
\def \dal{{\dot \alpha}}
\def \dbe{{\dot \beta}}

\def \bQ{{\bar Q}}
\def \bS{{\bar S}}
\def \bsi{\bar \sigma}
\def \bj{{\bar \jmath}}
\def \bJ{{\bar J}}
\def \hF{{\hat \F}}
\def \hU{{\hat \U}}

\def \tx{{\tilde {\rm x}}}

\def \tF{{\tilde \F}}
\def \tsi{{\tilde \sigma}}

\def \x{{\rm x}}
\def \y{{\rm y}}

\def \vphi{{\varphi}}
\def \bpsi{{\overline \psi}}
\def \bep{{\bar \epsilon}}
\def \hep{{\hat {\epsilon}}}
\def \hbep{{\hat {\bep}}}
\def \bet{{\bar \alpha}}

\def \dal{{\dot \alpha}}
\def \dbe{{\dot \beta}}

\def \bga{{\bar \gamma}}
\def \tx{{\tilde {\rm x}}}
\def \bsi{\bar \sigma}
\def \lam{\sigma}
\def \mun{\tau}
\def \tu{{\tilde u}}
\def \oD{{\overline D}}

\def \scs{\scriptstyle}

\def \Bsw{\!\mathrel{\hbox{\bigcmsy\char'056}}\!}
\def \Bse{\!\mathrel{\hbox{\bigcmsy\char'046}}\!}
\def \bchi{{\overline \chi}}
\def \bpsi{{\overline \psi}}

\def \xb{\overline{x}}
\font \bigbf=cmbx10 scaled \magstep1
\font \hgrm=cmr10 scaled \magstep2
\font \hgbf=cmbx10 scaled \magstep3

\font \largebf=cmbx10
\font \largerm=cmr10
\font \rms=cmr9
\def\toinf#1{\mathrel{\mathop{\longrightarrow}\limits_{\scriptstyle{#1}}}}

\writedefs

\lref\adscft{J.M. Maldacena, The Large N Limit of Superconformal Field Theories and Supergravity, 
Adv. Theor. Math. Phys. 2 (1998) 231; Int. J. Theor. Phys. 38 (1999) 1113, hep-th/9711200}
\lref\DZ{E. D'Hoker and D.Z. Freedman, Supersymmetric Gauge Theories
and the AdS/CFT Correspondence, Lectures given at Theoretical Advanced 
Study Institute in Elementary Particle Physics (TASI 2001), hep-th/0201253.}
\lref\hughtwo{J. Erdmenger and H. Osborn, {Conserved Currents and the 
Energy Momentum Tensor in Conformally Invariant Theories for General 
Dimensions}, Nucl. Phys. B483 (1997) 431, hep-th/9605009.}
\lref\hughone{H. Osborn and A. Petkou, {Implications of Conformal 
Invariance for Quantum Field Theories in $d>2$}
Ann. Phys. (N.Y.) {231} (1994) 311, hep-th/9307010.}
\lref\HO{H. Osborn, {$\N=1$ Superconformal Symmetry in Four-Dimensional
Quantum Field Theory}, Ann. Phys. (N.Y.) 272 (1999) 243, hep-th/9808041.}
\lref\Sei{S. Lee, S. Minwalla, M. Rangamani and N. Seiberg, {Three-Point
Functions of Chiral Operators in $D=4$, $\N=4$ SYM at Large $N$},
Adv. Theor. Math. Phys.  2 (1998) 697, hep-th/9806074.}
\lref\HFS{E. D'Hoker, D.Z. Freedman and W. Skiba, Field Theory Tests for Correlators 
in the AdS/CFT Correspondence, Phys. Rev. D59 (1999) 045008, hep-th/9807098.}
\lref\HSW{P.S. Howe, E. Sokatchev and P.C. West, {3-Point Functions in
$N=4$ Yang-Mills}, Phys. Lett. B444 (1998) 341, hep-th/9808162.}
\lref\OST{K.~Okuyama and L.~S.~Tseng, Three-point functions in N = 4 SYM theory at one-loop,
  JHEP 0408 (2004) 055, hep-th/0404190.}

\lref\LW{K. Lang and W. R\"uhl, Nucl. Phys. {B402} (1993) 573.}
\lref\Pet{A.C. Petkou, Ann. Phys. (N.Y.) 249 (1996) 180, hep-th/9410093.}
\lref\Fone{S. Ferrara, A.F. Grillo, R. Gatto and G. Parisi, Nucl. Phys. 
B49 (1972) 77\semi
S. Ferrara, A.F. Grillo, R. Gatto and G. Parisi, Nuovo Cimento 19A
(1974) 667.}
\lref\Ftwo{S. Ferrara, A.F. Grillo and R. Gatto, Ann. Phys. 76 (1973) 161.}
\lref\Dob{V.K. Dobrev, V.B. Petkova, S.G. Petrova and I.T. Todorov,
Phys. Rev. D13 (1976) 887.}
\lref\Ext{E. D'Hoker, D.Z. Freedman, S.D. Mathur, A. Matusis and
L. Rastelli, Extremal Correlators in the AdS/CFT correspondence,
{\it in} The Many Faces of the Superworld, ed. M.A. Shifman,
hep-th/9908160\semi
B. Eden, P.S. Howe, C. Schubert, E. Sokatchev and P.C. West,
Extremal Correlators in Four-dimensional SCFT, Phys. Lett. B472 (2000) 323,
hep-th/9910150\semi
B. Eden, P.S. Howe, E. Sokatchev and P.C. West, Extremal and
Next-to-Extremal N Point Correlators in Four-dimensional SCFT,
Phys. Lett. B494 (2000), hep-th/0004102\semi
M. Bianchi and S. Kovacs, Nonrenormalization of Extremal Correlators in
N=4 SYM Theory, Phys. Lett. B468 (1999) 102, hep-th/9910016\semi
J. Erdmenger and M. P\'erez-Victoria, Nonrenormalization of 
Next-to-Extremal Correlators in N=4 SYM and the AdS/CFT correspondence,
Phys. Rev. D62 (2000) 045008, hep-th/9912250\semi
E. D'Hoker, J. Erdmenger, D.Z. Freedman and M. P\'erez-Victoria,
Near Extremal Correlators and Vanishing Supergravity Couplings in 
AdS/CFT, Nucl. Phys. B589 (2000) 3, hep-th/0003218.}
\lref\Free{D.Z. Freedman, S.D. Mathur, A. Matusis and L. Rastelli, 
Nucl. Phys. B546 (1999) 96, hep-th/9804058\semi
D.Z. Freedman, S.D. Mathur, A. Matusis and L. Rastelli, 
Phys. Lett. B452 (1999) 61, hep-th/9808006\semi
E. D'Hoker and D.Z. Freedman, Nucl. Phys. B550 (1999) 261,
hep-th/9811257\semi
E. D'Hoker and D.Z. Freedman, Nucl. Phys. B544 (1999) 612, hep-th/9809179.}
\lref\FreeI{E. D'Hoker, D.Z. Freedman and L. Rastelli, 
Nucl. Phys. B562 (1999) 395, hep-th/9905049.}
\lref\FreeD{E. D'Hoker, D.Z. Freedman, S.D. Mathur, A. Matusis and 
L. Rastelli, Nucl. Phys. B562 (1999) 353, hep-th/9903196.}
\lref\Hokone{E. D'Hoker, S.D. Mathur, A. Matusis and L. Rastelli, {The 
Operator Product Expansion of $N=4$ SYM and the 4-point Functions of 
Supergravity}, Nucl. Phys. B589 (2000) 38, hep-th/9911222.}
\lref\AF{L. Andrianopoli and S. Ferrara, {On short and long $SU(2,2/4)$
multiplets in the AdS/CFT correspondence}, Lett. Math. Phys. 48 (1999) 145, 
hep-th/9812067.}
\lref\Short{S. Ferrara and A. Zaffaroni, {Superconformal Field Theories,
Multiplet Shortening, and the AdS${}_5$/SCFT${}_4$ Correspondence},
Proceedings of the Conf\'erence Mosh\'e Flato 1999, vol. 1, ed. G. Dito and 
D. Sternheimer, Kluwer Academic Publishers (2000), hep-th/9908163.}
\lref\short{F.A. Dolan and H. Osborn, {On short and semi-short
representations for four-dimensional superconformal symmetry}, Ann. Phys.
307 (2003) 41, hep-th/0209056.} 
\lref\pert{F. Gonzalez-Rey, I. Park and K. Schalm, {A note on four-point
functions of conformal operators in $N=4$ Super-Yang Mills},
Phys. Lett. B448 (1999) 37, hep-th/9811155\semi
B. Eden, P.S. Howe, C. Schubert, E. Sokatchev and P.C. West,
{Four-point functions in $N=4$ supersymmetric Yang-Mills theory at
two loops}, Nucl. Phys. B557 (1999) 355, hep-th/9811172; {Simplifications
of four-point functions in $N=4$ supersymmetric Yang-Mills theory at
two loops}, Phys. Lett. B466 (1999) 20, hep-th/9906051\semi
M. Bianchi, S. Kovacs, G. Rossi and Y.S. Stanev, {On the
logarithmic behaviour in $\N=4$ SYM theory}, JHEP 9908 (1999) 020, 
hep-th/9906188\semi
M. Bianchi, S. Kovacs, G. Rossi and Y.S. Stanev, Anomalous dimensions
in $\N$=4 SYM theory at order $g^4$, Nucl. Phys. B584 (2000) 216, 
hep-th/0003203\semi
B. Eden, C. Schubert and E. Sokatchev, Three loop four point
correlator in N=4 SYM, Phys. Lett. B482 (2000) 309, hep-th/0003096.}
\lref\hpert{G. Arutyunov, S. Penati, A. Santambrogio and E.Sokatchev,
Four-point correlators of BPS operators in $\N=4$ SYM at order $g^4$,
Nucl. Phys. B670 (2003) 103, hep-th/0305060.}
\lref\Arut{G. Arutyunov and S. Frolov, {Four-point Functions of Lowest
Weight CPOs in $\N=4$ SYM${}_4$ in Supergravity Approximation},
Phys. Rev. D62 (2000) 064016, hep-th/0002170.}
\lref\Three{G. Arutyunov and S. Frolov, {Three-point function of the 
stress-tensor in the AdS/CFT correspondence}, Phys. Rev. D60 (1999) 026004,
hep-th/9901121.}
\lref\ADHS{G. Arutyunov, F.A. Dolan, H. Osborn and E. Sokatchev,
Correlation Functions and Massive Kaluza-Klein Modes in the AdS/CFT
Correspondence, Nucl. Phys. B665 (2003) 273, hep-th/0212116.}
\lref\Degen{G. Arutyunov and E. Sokatchev, On a Large N Degeneracy
in $\N=4$ SYM and the AdS/CFT Correspondence, Nucl. Phys. B663 (2003) 163,
hep-th/0301058.}
\lref\ASok{F.~A.~Dolan, L.~Gallot and E.~Sokatchev, On four-point functions of 1/2-BPS operators in general dimensions,
  JHEP 0409 (2004) 056, hep-th/0405180.}
\lref\OPEN{G. Arutyunov, S. Frolov and A.C. Petkou, {Operator Product
Expansion of the Lowest Weight CPOs in $\N=4$ SYM${}_4$ at Strong Coupling},
Nucl. Phys. B586 (2000) 547, hep-th/0005182; 
(E) Nucl. Phys. B609 (2001) 539.}
\lref\OPEW{G. Arutyunov, S. Frolov and A.C. Petkou, {Perturbative and
instanton corrections to the OPE of CPOs in $\N=4$ SYM${}_4$}, Nucl. 
Phys. B602 (2001) 238, hep-th/0010137; (E) Nucl. Phys. B609 (2001) 540.}
\lref\Edent{B. Eden, A.C. Petkou, C. Schubert and E. Sokatchev, {Partial
non-renormalisation of the stress-tensor four-point function in $N=4$
SYM and AdS/CFT}, Nucl. Phys. B607 (2001) 191, hep-th/0009106.}
\lref\Except{G. Arutyunov, B. Eden, A.C. Petkou and E. Sokatchev, 
{Exceptional non-renorm-alization properties and OPE analysis of
chiral four-point functions in $\N=4$ SYM${}_4$}, 
Nucl. Phys. B620 (2002) 380, hep-th/0103230.}
\lref\BKRS{M. Bianchi, S. Kovacs, G. Rossi and Y.S. Stanev, {Properties
of the Konishi multiplet in  $\N=4$ SYM theory}, JHEP 0105 (2001) 042,
hep-th/0104016.}
\lref\HMR{L. Hoffmann, L. Mesref and W. R\"uhl, {Conformal partial wave
analysis of AdS amplitudes for dilaton-axion four-point functions},
Nucl. Phys. B608 (2001) 177, hep-th/0012153\semi
T. Leonhardt, A. Meziane and W. R\"uhl, Fractional BPS Multi-Trace
Fields of $\N=4$ SYM${}_4$ from AdS/CFT, Phys. Lett. B552 (2003) 87, 
hep-th/0209184.}

\lref\one{F.A. Dolan and H. Osborn, {Implications of $\N=1$ 
Superconformal Symmetry for Chiral Fields}, Nucl. Phys. B593 (2001) 599, 
hep-th/0006098.}
\lref\Dos{F.A. Dolan and H. Osborn, {Conformal four point functions
and the operator product expansion}, Nucl. Phys. B599 (2001) 459, 
hep-th/0011040.}
\lref\scft{F.A. Dolan and H. Osborn, {Superconformal symmetry,
correlation functions and the operator product expansion}, Nucl. Phys. B629
(2002) 3, hep-th/0112251.}

\lref\Hoff{L.C. Hoffmann, A.C. Petkou and W. R\"uhl, 
Phys. Lett. B478 (2000) 320, hep-th/0002025\semi
L.C. Hoffmann, A.C. Petkou and W. R\"uhl, hep-th/0002154.}
\lref\HoffR{L. Hoffmann, L. Mesref and W. R\"uhl, AdS box graphs, unitarity
and operator product expansions,
Nucl. Phys. B589 (2000) 337,  hep-th/0006165.}
\lref\Witt{E. Witten, Adv. Theor. Math. Phys. 2 (1998) 253, hep-th/9802150.}
\lref\Eden{B. Eden, P.S. Howe, A. Pickering, E. Sokatchev and P.C. West,
{Four-point functions in $N=2$ superconformal field theories},
Nucl. Phys. B581 (2000) 523, hep-th/0001138.}
\lref\Intr{K. Intriligator, {Bonus Symmetries of ${\cal N} =4$ 
Super-Yang-Mills Correlation Functions via AdS Duality}, 
Nucl. Phys. B551 (1999) 575, hep-th/9811047.}
\lref\Bon{K. Intriligator and W. Skiba, {Bonus Symmetry and the Operator
Product Expansion of ${\cal N} =4$ Super-Yang-Mills},
Nucl. Phys. B559 (1999) 165, hep-th/9905020.}
\lref\Non{G. Arutyunov, B. Eden and E. Sokatchev, {On 
Non-renormalization and OPE in Superconformal Field Theories}, Nucl. Phys. 
B619 (2001) 359, hep-th/0105254.}
\lref\bpsN{B. Eden and E. Sokatchev, {On the OPE of 1/2 BPS Short
Operators in $N=4$ SCFT${}_4$}, Nucl. Phys. B618 (2001) 259, hep-th/0106249.}
\lref\Hes{P.J. Heslop and P.S. Howe, {OPEs and 3-point correlators of
protected operators in $N=4$ SYM}, Nucl. Phys. B626 (2002) 265, 
hep-th/0107212.}
\lref\Pen{S. Penati and A. Santambrogio, {Superspace approach to 
anomalous dimensions in ${\N}=4$ SYM}, Nucl. Phys. B614 (2001) 367, 
hep-th/0107071.}
\lref\Hok{A.V. Ryzhov, {Quarter BPS Operators in $\N=4$ SYM}, 
JHEP 0111 (2001) 046, hep-th/0109064\semi
E. D'Hoker and A.V. Ryzhov, {Three Point Functions of Quarter BPS 
Operators in $\N=4$ SYM}, JHEP 0202 (2002) 047, hep-th/0109065.}
\lref\HH{P.J. Heslop and P.S. Howe, {A note on composite operators in
$N=4$ SYM}, Phys. Lett. 516B (2001) 367, hep-th/0106238.}
\lref\HesHo{P.J. Heslop and P.S. Howe, Aspects of $\N=4$ SYM, JHEP {\bf 0401}, 058 (2004), hep-th/0307210.}
\lref\Howe{P.J. Heslop and P.S. Howe, Four-point functions in $N=4$ SYM,
JHEP 0301 (2003) 043, hep-th/0211252.}

\lref\Vretare{L. Vretare, Formulas for Elementary Spherical Functions and
Generalized Jacobi Polynomials, Siam J. on Mathematical Analysis, 15 (1984)
805.}
\lref\Koo{T. Koornwinder, Two-variable Analogues of the Classical Orthogonal
Polynomials, in Theory and Applications of Special Functions, ed. R. Askey,
Academic Press, New York (1975).}
\lref\ortho{C.F. Dunkl and Y. Xu, Orthogonal Polynomials of Several Variables,
Cambridge University Press, Cambridge (2001).}
\lref\Class{M. G\"unaydin and N. Marcus, {The spectrum of the $S^5$
compactification of the chiral $N=2$, $D=10$ supergravity and the unitary
supermultiplets of $U(2,2/4)$}, Class. and Quantum Gravity, 2 (1985) L11.}
\lref\phd{F.A. Dolan, Aspects of Superconformal Quantum Field Theory,
University of Cambridge PhD thesis (2003).}

\lref\Dirac{P.A.M. Dirac, Wave Equations in Conformal Space,
Annals of Math. 37 (1936) 823; {\it in} ``The
Collected Works of P.A.M. Dirac 1924-1948'', edited by R.H. Dalitz, CUP
(Cambridge) 1995.}
\lref\CFT{F.A. Dolan and H. Osborn, {Conformal Partial Waves and the Operator 
Product Expansion}, Nucl. Phys. B678 (2004) 491, hep-th/0309180.}

\lref\Barg{V. Bargmann and I.T. Todorov, Spaces of analytic functions on a complex
cone as carriers for the symmetric tensor representations of SO(n), J. of Math.
Phys. 18 (1977) 1141.}
\lref\Dobb{V.K. Dobrev, G. Mack, V.B. Petkova, S.G. Petrova and I.T. Todorov,
Harmonic Analysis on the n-dimensional Lorentz Group and its Application to
Conformal Quantum Field Theory, Springer Lecture Notes in Physics, 63, 
Springer-Verlag (Berlin - Heidelberg) 1977.}

\lref\nilpo{B. Eden, P.S. Howe and P.C. West, {Nilpotent invariants in $N=4$ SYM}, Phys.Lett. B463 (1999) 19-26, hep-th/9905085}
\lref\od{F.A. Dolan, H. Osborn, {Conformal Partial Wave Expansions for N=4 Chiral Four Point Functions}, hep-th/0412335}
\lref\nh{M. Nirschl and H. Osborn, {Superconformal Ward Identities and their Solution}, Nucl.\ Phys.\ B711 (2005) 409, hep-th/0407060}

{\nopagenumbers
{\ }\vskip 2 true cm
\centerline {\hgbf Superconformal Symmetry}
\vskip 12 pt
\centerline {\hgbf and}
\vskip 12 pt
\centerline {\hgbf Correlation Functions}
\vskip 5 true cm
\centerline {\hgrm Michael Nirschl}
\vskip .5 true cm
\centerline {\hgrm Darwin College}
\vskip 3 true cm
\centerline{\hbox{\epsffile{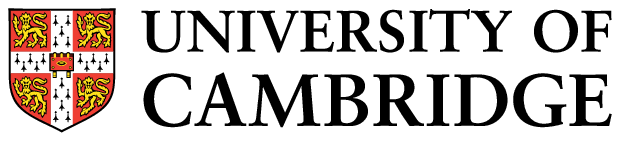}}}
\vskip .5 true cm
\centerline {Dissertation submitted for the degree of}
\centerline {Doctor of Philosophy} 
\vskip .5 true cm
\centerline {University of Cambridge}
\centerline {May 2005}
\vfill
\eject
{\ }
\eject
\centerline{\bigbf Declaration}

This dissertation is my own original work, except where explicitly stated in the text or the acknowledgements. Chapters 2-4, parts of chapter 5 and appendices A-D are based on 
published work \nh\ with my supervisor, Prof. Hugh Osborn. Parts of chapter 5 and appendices E-G are unpublished work in progress with my supervisor.

No part of this dissertation has been submitted for a degree, diploma, or other qualification at any other university, 
or is concurrently submitted for any degree, diploma or qualification.

{\ }
\eject
{\ }
\eject
{\bigbf Acknowledgements}

First of all I would like to thank my PhD Supervisor, Prof. Hugh Osborn, for constant support, advise and guidance as well as patience during the whole of my PhD.
Also, I would like to thank Francis Dolan for many useful discussions and encouragement throughout. I am very grateful to my office mates Christian Stahn and Axel Kleinschmidt
for having been very nice company and teaching me most of what I know about using Mathematica.

Furthermore I am grateful to my mother, Monika Nirschl, who supported me in many ways wherever she could and without who I never would have reached this stage.
I would also like to thank my friends from my undergraduate time in Bonn, Moritz, Damian and Georg 
whose friendship has lasted despite the distance.

I have made many friends in Cambridge. I want to thank especially Patrick, David, Fady, Takis, Hrobjatur, Rainer, Adonai, Aldo, Wing, Jabed, Tigran, Uche, Amit 
for making my time in Cambridge unforgettable. I know I have not mentioned 
many others who can be assured that I have not forgotten them.
\vfill
\eject
{\ }

\eject

{
\baselineskip 16pt
\centerline{\largerm Michael Nirschl}
\vskip .1 true cm
\centerline{\largerm Superconformal Symmetry and Correlation Functions}
\vskip .1 true cm
\centerline{\largebf Summary}

{\rms The constraints that $\N=2,4$ superconformal symmetry imposes in $d=4$ for four point functions of chiral primary $\half$-BPS 
operators are derived. The operators are described by symmetric traceless tensors of the internal $R$-symmetry group. A substantial simplification compared to earlier work is achieved by introduction
of null vectors. They reduce complicated tensorial expressions to polynomials of one or two invariant cross ratios for $\N=2,4$ respectively, similar to the 
2 conformally invariant cross ratios of $4$ points. Two variable polynomials corresponding to the different $R$-symmetry representations are constructed.
The Ward identities for superconformal symmetry are then obtained as simple differential equations. 
The general solution of these identities is presented in terms of a constant, a single variable function and a two variable function. In the extremal case it is shown that the amplitude has to be a constant.
In the next-to-extremal case the amplitude contains a constant and a single variable function only.
An interpretation in terms of the operator product expansion is given for the case of fields of equal dimension and 
for the so called (next-to)extremal cases. The result is shown to accommodate long multiplets as well as 
semishort and short multiplets with protected dimension. Generically also non-unitary multiplets can appear. It is 
shown how to remove them using appropriate semishort and long multiplets to obtain a unitary theory. Where possible, 
positivity of OPE coefficients required by unitarity is confirmed. Implications of crossing symmetry for the 
four point functions studied are derived and discussed. It is shown that crossing 
symmetry fixes the single variable function in the general solution to be of free field form using singularity arguments. 
For a restricted set of next-to-extremal correlation functions with $\S_3$ symmetry amongst the first three fields
it is shown that the amplitude is fixed up to normalization to be of free field form. 
Starting from a known expression for the large $N$ amplitude of $[0,4,0]$ operators we simplify it further and present it in a manifestly crossing symmetric form. 
We compute the coefficients of the conformal partial wave expansion of all 
representations in this amplitude and use them to compute an averaged value of the anomalous dimensions for long
multiplets given spin and twist in each relevant representation at first
order in 1/$N$. Finally assuming the previously observed universal singularity structure in the large $N$ amplitude we derive the 
general large $N$ amplitude of four identical $\half$-BPS operators in the $[0,p,0]$ representation in terms of $\oD$ functions. Explicit expressions
for all coefficients are given.}
\vfill
}
\eject
{\ }
\eject}
\pageno=1
\centerline{\bf Contents}\nobreak\medskip{\baselineskip=12pt
 \parskip=0pt\catcode`\@=11 \input toc.tex \catcode`\@=12 \bigbreak\bigskip}\writetoc
\vfill
\eject

\newsec{Introduction}

\subsec{Motivation}

One of the most important and powerful concepts in the development of physics in the 20th century was
the concept of symmetry. Einstein discovered the Poincar\'e symmetry of spacetime which was the integral part of his special theory of relativity in 1905. Later in 1915 he developed the general theory of
relativity which forms our basis of understanding of gravity. The key concept is again considering the general symmetry 
of change of coordinates of an observer. 

Later in the 20th century gauge symmetry became the central tool to construct
quantum theories which describe all the forces we observe in nature apart from gravity. Using the same formalism
with different symmetries we can accurately describe the electromagnetic force as well
as the weak and the nuclear forces which forms the basis of the Standard Model. Therefore gauge field theories, also called Yang
Mills theories (YM), have been and are in the very centre of research of quantum field theories.

Generally, all particles we observe fall into two classes characterized by
their spin. Particles with integer spin are called bosons, and those are the particles
which transmit all the forces we observe in nature. Particles with half-integer spin are
called fermions and constitute all the matter in the universe. Bosons and fermions show very
different behaviour. Fermions obey Fermi-Dirac statistics, i.e. they can never be in the same state as another one, which is a
generalization of the Pauli principle and is responsible for the stability of atoms and
thus the world surrounding us. Bosons on the other hand obey Bose-Einstein statistics, i.e. they tend to clump together, an effect
which is used in lasers to generate light with very long coherence length as well as in 
Bose-Einstein condensates to generate a fourth phase of matter.

In 1971 a new symmetry was proposed, supersymmetry. It unifies the two different classes of particles, 
bosons and fermions, by allowing them to
transform into each other. This was first studied by Ramond, Neveu and Schwarz in the context of string theory. 
1974 Wess and Zumino wrote down the first example of a 4-dimensional QFT with supersymmetry.
Haag, Sohniues and Lopuszanski showed in 1975 that the unification of supersymmetry and conformal symmetry, called superconformal symmetry, 
is actually the maximal spacetime symmetry group possible for a 
QFT permitting a non-trivial $S$-matrix. This was an extension of the Coleman-Mandula theorem that had established that the maximal symmetry Lie group
for a non-trivial $S$-matrix is conformal symmetry.

It turns out that supersymmetry has
several very desirable effects, which is why many theories studied until today are
supersymmetric although we still wait for experimental confirmation of its existence. 
First of all, supersymmetry cures some of the divergences which appear in
quantum field theories and in string theory. This makes the theories look mathematically more consistent 
and in the case of superstring theory or the quantum field theory we will study they actually become finite.
Another technical advantage of supersymmetry is that it gives more control over computations and therefore
more quantities are accessible in supersymmetric theories, generically because of so
called nonrenormalization theorems. Most quantities in a QFT depend strongly on the coupling constant which measures the strength of the forces.
The two extreme regimes are on the one hand the perturbative regime where the interactions are very weak, and the strong coupling regime with very intense forces on the other hand.
Supersymmetry guarantees that certain quantities are not 
affected by the renormalization flow which describes the transition between the different regimes of the theory.
Thus they can be computed in the perturbative regime and then 
be extrapolated to the strong coupling regime.
This is why supersymmetric theories are also often used as models for their non-supersymmetric counterparts. 
Finally from a phenomenological point of view supersymmetry is very desirable if one wants to unify electroweak and color forces. 
Without supersymmetry the three coupling constants never quite meet. Supersymmetry achieves exactly that. Also the hierarchy problem, or the question why the Higgs mass
is so small compared to the Planck scale, can be solved with supersymmetry.

In 1997 Maldacena \adscft\ found that there exists a very deep connection between string theory
and quantum field theories. The AdS/CFT correspondence conjectures that string theory in the geometric
background of $AdS_5\times S^5$ with 5-form flux of $N$ units is equivalent to $\N=4$
Supersymmetric Yang Mills (SYM) on $S^4$, the boundary of $AdS_5$, with gauge group $SU(N)$. The isometry group of $S^5$ is $SO(6)\simeq SU(4)$ which is the internal $R$-symmetry group.
$\N=4$ SYM is the unique unitary quantum field
theory (QFT) in 4 dimensions with $\N=4$ superconformal symmetry apart from the choice of the gauge group. Remarkably the AdS/CFT correspondence relates the weak
coupling limit of one theory to the strong coupling region of the other. This allows
obtaining nonperturbative information about string and quantum field theory from
perturbative calculations in the respective other theory. For a review of this and a complete list of references check \DZ.

The maximal amount of symmetry in $d=4$ apart from internal symmetries is $\N=4$ extended superconformal symmetry
\footnote{${ }^1$}{More supersymmetry is excluded because the theory would include a graviton and since gravity is not
renormalizable this theory cannot easily make sense as a QFT.}.
This is the symmetry we will study in this thesis along with its sibling $\N=2$ superconformal symmetry, both in the context 
of SYM theories.

One interesting question to ask is what constraints superconformal symmetry imposes on correlation functions. That is
we will purely focus on the symmetry aspect of the theory. Therefore all our results are
independent of the interactions of the theory and apply to all possible dynamics.
Conformal symmetry fixes the two and three point functions up to normalization. Therefore it will be natural to study four-point functions
which from a conformal symmetry point of view may contain arbitrary functions of two conformal invariants that one can construct out of four points.
We will look at a special set of operators, $\half$-BPS operators which in the AdS/CFT correspondence correspond to fundamental fields
 and Kaluza Klein modes. They are described by shortening conditions, i.e. some supercharges annihilate them. From this it follows that
 their dimension is not renormalized and thus their anomalous dimensions vanish.
 
Also one can use four point functions to analyse the operator product expansion (OPE) of the
theory. By taking a limit where two fields approach the other two fields, the four point function reduces 
to a sum of two point functions multiplied by OPE coefficients.
Finally we use conformal partial wave expansions to obtain further information about the four point functions we study. Conformal partial waves 
correspond to intermediate states when interpreting the correlation function as describing the scattering of 2 incoming states into 2 outgoing states. An expansion
of the amplitude in a basis of partial waves describing different intermediate states can reveal information about the intermediate states appearing.

Also in the light of the AdS/CFT correspondence we will examine the four point functions of four identical operators in the strong coupling limit more closely. 
We will study the amplitudes obtained using supergravity techniques as well as from a more field theoretic point of view by 
considering their singularity structure. We will see how the singularity structure actually uniquely determines these amplitudes together
with crossing symmetry.

\subsec{Superconformal Symmetry in $d=4$}

The conformal group in $d=4$ is $SO(2,4)\simeq SU(2,2)$. It contains translations generated 
by $P_{\mu}$, Lorentz transformations generated by $M_{\mu\nu}$, dilations generated by
$D$ and special conformal transformations generated by $K_{\mu}$. The scale dimensions of the
generators are
\eqn\dims{[M]=[D]=0,\quad[P]=1,\quad[K]=-1.}
In any theory there might be internal symmetries generated by $T^A$.

The superconformal group in $d=4$ is given by $SU(2,2|\N)$.
The bosonic part of $SU(2,2|\N)$ consists of the conformal group and
the $R$-symmetry $U(\N)$ generated by $R^i_j$. The $R$-symmetry corresponds to
automorphisms of the SUSY generators. For the case of $\N=4$ supersymmetry one can read
off from the commutation relations \short\ of the $R$-symmetry that it is consistent to remove the
traces and thus restrict the $R$-symmetry to $SU(4)$.
The fermionic part contains Poincar\'e supersymmetries generated by $Q_{\alpha}^i$ and
$\bar{Q}_{i\dal}$, also called SUSY generators in the following. Since their action does not commute with the action of the
special conformal transformations, the algebra is closed by generators of conformal
supersymmetries $S_{i\alpha}$ and $\bar{S}^i_{\dal}$. The scale dimensions of these additional
generators are
\eqn\dims{[T^A]=0,\quad[Q]=[\bar{Q}]=\half,\quad[S]=[\bar{S}]=-\half.}
For the complete commutation relations check \short.

\subsec{Physical Fields in $\N=4$ SYM}

The $\N=4$ Super Yang Mills theory contains 6 scalars $X_r$, 4 gauginos $\lambda_i$ and
the gauge field $A_{\mu}$ with field strength $F_{\mu\nu}$. The scalars transform in the
adjoint representation of $SU(4)$, the gauginos in the fundamental representation.
The scalars and the gauge field have mass dimension 1, the gauginos $\textstyle{3\over 2}$.
The gauge group $\G$ can be arbitrary. The moduli of the theory are given by the gauge
coupling $g$, the instanton angle $\theta$ and the expectation values of the scalars
$\left<X_r\right>$.
We will only be interested in the superconformal phase of the theory when all the
expectations values of the scalars vanish $\left<X_r\right>=0$. Then the gauge group $\G$
is unbroken and the theory possesses full superconformal symmetry given by the
supergroup $PSU(2,2|4)$. This symmetry holds classically and also at the quantum level. The
operators we consider are gauge invariant combinations of the physical fields and transform under unitary representations of $PSU(2,2|4)$.

\subsec{Conformal Scalar $2,3,4$-point Functions in $d=4$}

As mentioned before, conformal symmetry restricts two and three point functions completely up to normalization. This is due to the fact
that the conformal group can be used to map any set of two or three points into any other set of two or three points. Or in other words, 
there are no invariants under conformal symmetry that can be constructed out of two or three points. The spacetime dependence is completely fixed by 
the scaling behaviour.

First we define for convenience
\eqn\defrij{x_{ij} = x_i - x_j,\quad r_{ij} = x_{ij}^2}
and
\eqn\defdij{ \Delta_{ij} = \Delta_i - \Delta_j.}
For a scalar field $\varphi_i(x)$ with scale dimension $\Delta_i$ the two point function reads
\eqn\stpf{\left<\varphi_i(x_1)\varphi_j(x_2)\right>=\delta_{ij}{N\over r_{12}^{\Delta_i}},}
where $N$ is an arbitrary normalization which we choose to be $N=1$ in the following.
The three point functions for scalar fields are also determined by conformal symmetry
\eqn\sthpf{\left<\varphi_i(x_1)\varphi_j(x_2)\varphi_k(x_3)\right>=C_{ijk}{1\over 
r_{12}^{\hhalf(\Delta_i+\Delta_j-\Delta_k)}
r_{13}^{\hhalf(\Delta_i+\Delta_k-\Delta_j)}
r_{23}^{\hhalf(\Delta_j+\Delta_k-\Delta_i)}}.}
The constants $C_{ijk}$ only depend on which fields are present in the correlation function and will find an 
interpretation in the next section in the context of the operator product expansion.

The four point function may be expressed as
\eqn\sfpf{\eqalign{
\left<\varphi_1(x_1)\varphi_2(x_2)\varphi_3(x_3)\varphi_4(x_4)\right>
={}&{r_{23}^{\Sigma-\Delta_2-\Delta_3}\,r_{34}^{\Sigma-\Delta_3-\Delta_4}\over r_{13}^{\Delta_1}\,r_{24}^{\Sigma-\Delta_3}}F(u,v)\cr
={}&{1\over r_{12}^{\hhalf(\Delta_1+\Delta_2)}r_{34}^{\hhalf(\Delta_3+\Delta_4)}}\left({r_{24}\over r_{14}}\right)^{\!\hhalf\Delta_{12}}
\left({r_{14}\over r_{13}}\right)^{\!\hhalf\Delta_{34}}G(u,v) \, ,}}
where $2\,\Sigma=\Delta_1+\Delta_2+\Delta_3+\Delta_4$ and $u,v$ are the two independent conformal invariants of four points
\eqn\defuv{
u= {r_{12} \, r_{34} \over r_{13} \, r_{24}} \, , \qquad \qquad
v= {r_{14} \, r_{23} \over r_{13} \, r_{24}} \, .
}
$F(u,v)$ or $G(u,v)$ are arbitrary functions of $u,v$ within the context of conformal symmetry and can only be constrained by the dynamics of a 
particular theory considered.
$F(u,v),\,G(u,v)$ are obviously related by
\eqn\fgrel{F(u,v)=u^{-\hhalf(\Delta_1+\Delta_2)}v^{-\hhalf(\Delta_{12}-\Delta_{34})}\, G(u,v).}
As was shown in \scft\ the superconformal Ward
identities are greatly simplified if they are expressed in terms of new
variables $x,\zz$ rather than the usual conformal invariants $u,v$. In
terms of the standard correspondence for the space-time coordinates
$x^\mu \to \x = x^\mu\si_\mu $\foot{Thus $4$-vectors
are identified with $2\times 2$ matrices using the hermitian $\si$-matrices
$\si_\mu, \, \tsi_\mu, \, \si_{(\mu} \tsi_{\nu)} = - \eta_{\mu\nu}1$,
$x^\mu \to \x_{\alpha\dal} =
x^\mu (\si_\mu)_{\alpha\dal}, \ \tx^{\dal\alpha} = x^\mu (\tsi_\mu)^{\dal\alpha}
= \ep^{\alpha\beta}\ep^{\smash {\dal \dbe}} \x_{\smash {\beta \dbe}}$,
with inverse $x^\mu = - {1\over 2}{\rm tr}(\si^\mu \tx)$. We have $x{\cdot y}
= x^\mu y_\mu = - {1\over 2}{\rm tr}(\tx \y)$, $\det \x = -x^2$, 
$\x^{-1} = -\tx/x^2$.} for four points $x_1,x_2,x_3,x_4$, 
$x,\zz$ may be defined, as shown in \Howe, as the 
eigenvalues of $\x_{12}\,\x_{42}{}^{\!\!-1} \x_{43} \, \x_{13}{}^{\!\! -1}$. 
By conformal transformations we may choose a frame such that
$\x_2=0, \, \x_3=\infty, \, \x_4 = 1$ and $\x_1 = \pmatrix{x & 0 \cr 0& \zz}$.
The two conformal invariants in \defuv\ are then given in terms of $x,\zz$ by\foot{Since
$1+u-v = x+ \zz$ and $1+u^2+v^2-2uv-2u-2v=(x-\zz)^2$ it is easy to invert 
these results to obtain  $x,\zz$ in terms of $u,v$ up to the
arbitrary sign of the square root $\sqrt{(x-\zz)^2}$.  For any $f(u,v)$
there is a corresponding symmetric function ${\hat f}(x,\zz)={\hat f}(\zz,x)$
such that ${\hat f}(x,\zz)=f(u,v)$.},
\eqn\eiguv{\eqalign{
u={}&\det( \x_{12}\,\x_{42}{}^{\!\!-1} \x_{43}\, \x_{13}{}^{\!\! -1})= x\zz \, , \cr
v= {}& \det(1- \x_{12}\,\x_{42}{}^{\!\!-1}\x_{43} \, \x_{13}{}^{\!\! -1}) = 
\det( \x_{14}\,\x_{24}{}^{\!\!-1} \x_{23} \, \x_{13}{}^{\!\! -1}) =
(1-x)(1-\zz) \, .\cr}
}
For a Euclidean metric on space-time $x,\zz$ are complex conjugates.
We will see how this change of variables strongly simplifies not only the identities but also the derivation of those. 
They are also essential in writing down expressions for conformal partial waves.

\subsec{Operator Product Expansion}

Four point functions are also very useful to study the properties of the operator product expansion OPE of a QFT. In a conformal field theory a complete set of operators locally satisfies an OPE
\eqn\OPE{\varphi_i(x)\varphi_j(0)\sim\sum_k C_{ij\O} {1\over (x^2)^{{1\over 2}(\Delta_i+\Delta_j-\Delta+\ell)}}{x_{\mu_1}\ldots x_{\mu_\ell}\O_{\mu_1\ldots\mu_\ell}(0)}+\ldots} for $x\sim 0$.
Here $\O_{\mu_1\ldots\mu_\ell}(0)$ is a symmetric traceless tensor operator of scale dimension $\Delta$ and spin $\ell$.

To derive a short distance limit for $x_1\rightarrow x_2$ of the scalar four point function \sfpf\ we first write down the following expression for the three point function
of a tensor operator of spin $\ell$ and two scalar fields
\eqn\opp{\eqalign{
\left<\O_{\mu_1\ldots\mu_\ell}(x_2)\varphi_3(x_3)\varphi_4(x_4)\right>={}&C_{34\O}
{1\over r_{23}^{\hhalf(\Delta+\Delta_{34}-\ell)}r_{24}^{\hhalf(\Delta+\Delta_{43}-\ell)}r_{34}^{\hhalf(\Delta_3+\Delta_4-\Delta-\ell)}}
X_{\{\mu_1}\ldots X_{\mu_\ell\}},
}}
where
\eqn\oppp{X_{\mu}={x_{24\mu}\over r_{24}}-{x_{23\mu}\over r_{23}},\quad X^2={r_{34}\over r_{23}r_{24}}}
Using \OPE\ and \opp\ we can derive the following short distance limit for $x_1\rightarrow x_2$ for the four point function of four scalar fields
\eqn\sdl{\eqalign{
&\left<\varphi_1(x_1)\varphi_2(x_2)\varphi_3(x_3)\varphi_4(x_4)\right>\cr
&\mathop{\sim}\limits_{x_{12}\rightarrow 0}{C_{12\O}\,C_{34\O}\over(r_{12})^{\hhalf(\Delta_1+\Delta_2)}(r_{34})^{\hhalf(\Delta_3+\Delta_4}}\left({r_{24}\over r_{23}}\right)^{\hhalf \Delta_{34}}\left({r_{12}\,r_{34}\over r_{23}\,r_{24}}\right)^{\hhalf\Delta}C^{\hhalf d-1}_{\ell}(t){\ell!\over 2^\ell(\hhalf d-1)_\ell},
}}
where
\eqn\deft{t={x_{12}\cdot X\over (r_{12}X^2)^{\hhalf}}\mathop{\sim}\limits_{x_{12}\rightarrow 0}-{1-v\over 2 \sqrt{u}}+O(\sqrt{u},1-v)=-{x+\xb\over 2 (x\xb)^\hhalf}+O(\sqrt{x\xb},x+\xb).}
The function $C^{\hhalf d-1}_{\ell}(t)$ is a Gegenbauer polynomial which arises from
\eqn\gegendef{a_{\mu_1}\ldots a_{\mu_\ell}b_{\{\mu_1}\ldots b_{\mu_\ell\}}=(a^2)^{\hhalf\ell}(b^2)^{\hhalf \ell}C^{\hhalf d-1}_{\ell}(t){\ell!\over 2^\ell(\hhalf d-1)_\ell},}
for $a, b$ $d$-dimensional vectors.
It follows that the amplitude $G(u,v)$ in the four point function \sfpf\ has the following form
\eqn\Gform{G(u,v)\mathop{\sim}\limits_{x_{12}\rightarrow 0}C_{12\O}\,C_{34\O}\,u^{\hhalf\Delta}C^{\hhalf d-1}_{\ell}(t){\ell!\over 2^\ell(\hhalf d-1)_\ell}.}
Now for $d=4$ the Gegenbauer polynomial reduces to the following form
\eqn\gegendf{C^1_\ell\left(-{x+\xb\over 2 (x\xb)^{\hhalf}}\right)=(-1)^\ell(x\xb)^{-\hhalf\ell}{x^{\ell+1}-\xb{}^{\ell+1}\over x-\xb},}
and thus $G(u,v)$ simplifies to
\eqn\Gsimp{G(u,v)\mathop{\sim}\limits_{x_{12}\rightarrow 0} C_{12\O}\,C_{34\O}\,(x\xb)^{\hhalf(\Delta-\ell)}(-\half)^\ell{x^{\ell+1}-\xb{}^{\ell+1}\over x-\xb},}
a simple expression for the leading term contribution in the short distance limit of the four point function for an operator of scale dimension $\Delta$ and spin $\ell$.

The full analysis of the OPE is facilitated by a simple expression for the contribution of the conformal 
block of a quasi-primary operator of dimension $\Delta$ and spin $\ell$ to the four point function. The conformal block of a quasi-primary operator
includes the operator itself, and all its descendants formed by the actions of derivatives. The explicit expression was obtained in \Dos\
and gives a concrete expression for the contribution in the function $G(u,v)$ in \sfpf
\eqn\ope{\eqalign{G(u,v)={}&\G^{(\ell)}_{\Delta}(u,v;\Delta_{21},\Delta_{43})
={}u^{\hhalf(\Delta-\ell)}G^{(\ell)}_{\Delta}(u,v)\,.}}
The functions $G^{(\ell)}_{\Delta}(u,v)$ are given in terms of hypergeometric functions of type $_{2}F_{1}(a,b;c;x)$ by
\eqn\GF{\eqalign{
G^{(\ell)}_{\Delta}(u,v)={}&G^{(\ell)}(\half(\Delta-\Delta_{12}-\ell),\half(\Delta+\Delta_{34}-\ell),\Delta;u,v)), \cr
G^{(\ell)}(a,b,c;u,v)={}&{{(-\half x)^{\ell} x F(a+\ell,b+\ell;c+\ell;x)F(a-1,b-1;c-2-\ell;\xb)-x\leftrightarrow \xb}\over x-\xb}.}}
Note that the leading terms in \GF\ agrees with \Gsimp\ as required.
The conformal partial wave functions satisfy the following relations
\eqn\Gsym{\eqalign{
\G_\Delta^{(\ell)}(u,v;\Delta_{21}, \Delta_{43}) = {}& (-1)^\ell v^{{1\over 2}
\Delta_{43}} \, \G_\Delta^{(\ell)}(u/v,1/v;-\Delta_{21}, \Delta_{43}) \cr
= {}&  v^{{1\over 2}( \Delta_{43} - \Delta_{21})} \,
\G_\Delta^{(\ell)}(u,v;-\Delta_{21}, - \Delta_{43}) \, .\cr}
}

Finally the OPE requires an expansion of $G(u,v)$ in terms of conformal partial waves
\eqn\Gexpansion{G(u,v)=\sum_{\Delta,\ell}a_{\Delta,\ell}\,\G_{\Delta}^{(\ell)}(u,v),}
which determines the spectrum of possible operators in the OPE. The conformal partial waves coefficients are given by the OPE coefficients as
\eqn\cpwope{a_{\Delta,\ell}=C_{12\O}C_{34\O}.}

\subsec{$\N=4$ Superconformal Multiplets}

The unitary irreducible representations of $SU(2,2|4)$ with positive energy are labelled
by the quantum numbers of the maximal bosonic subgroup which is
$SO(2,4)\times SU(4)\simeq SO(2)\times SO(4)\times SU(4)$. There is a mapping of
representations of this group to representations of $SO(1,1)\times SO(1,3)\times SU(4)$. The quantum number of $SO(1,1)$ is the
scale dimension $\Delta$ which is non-negative for positive energy representations. The quantum numbers of 
$SO(1,3)\simeq SU(2)\times SU(2)$ are the `spins' $(j_1,j_2)$ where $j_1,j_2$ are non-negative half-integers. The
representations of $SU(4)$ are labelled by their Dynkin labels $[p_1,p_2,p_3]$ where $p_i$
are non-negative integers. Complex conjugation of a representation $[p_1,p_2,p_3]$ yields
$[p_3,p_2,p_1]$. 

In any superconformal multiplet there is one unique operator $\O$, called the
superconformal primary operator, which commutes with the generators of the conformal
supersymmetries. Notice that because of $\{S,S\}\sim K$ this is a stricter condition than commuting with special
conformal generators, i.e. being a conformal primary operator. Looking at the dimensions of the generators, $[S,\O]=0$ implies that $\O$ is the
operator of lowest dimension in the multiplet. The rest of the multiplet is generated by the action of 
the Poincar\'e supersymmetries and the operators are therefore called descendants of $\O$.
Their dimension is related to the one of $\O$ since acting with Poincar\'e supersymmetries
raises the dimension by $\half$. 

There exists a complete classification of unitary representations for primary operators with vanishing `spin' $j_1,j_2$ of
$SU(2,2|4)$\Dob :
\eqn\sc{\eqalign{
&1.)\ \Delta={}2p_1+p_2 \hbox{\ where\ } p_1=p_3\cr
&2.)\ \Delta={}\textstyle{3\over 2}p_1+p_2+\half p_3 \hbox{\ where\ }p_1\geq p_3+2\cr
&3.)\ \Delta={}\half p_1+p_2+\textstyle{3\over 2} p_3\hbox{\ where\ }p_3\geq p_1+2\cr 
&4.)\ \Delta\geq \hbox{Max}[2+\textstyle{3\over 2}p_1+p_2+\half p_3;2+\half p_1+p_2+\textstyle{3\over 2}p_3].
}}

The first three classes are discrete. They correspond to BPS operators, also
called short or chiral since they are chiral under a \ $\N=1$ subalgebra of $SU(2,2|4)$. The
shortness implies that the operators do not receive perturbative corrections to their
dimension, i.e. their anomalous dimensions vanish. This originates from a shortening condition that states that some of the supercharges annihilate the chiral primary operator (CPO).

The fourth class has continuous scale dimension. Primary operators in this class do not satisfy any shortening condition and thus the multiplets are long multiplets with dimension proportional to $2^{16}$.

Primary operators in the fourth class, satisfying the threshold for which equality holds in \sc\ are called semi-short. They transform in the $[q,p,q]$ representation
and thus are self-conjugate. Their dimension is $\Delta=2q+p+2$.

Notice that there is a gap of dimension 2 between the discrete classes and the continuous one. 
Therefore a long multiplet cannot be made short by going to the free field limit. We will actually study in chapter 3 how a 
long multiplet at the threshhold decomposes into 2 semi-short multiplets. Conversely this also implies that semi-short multiplets
are protected against anomalous dimensions unless there exists a second multiplet to pair up with to form a long multiplet.

We will mostly be interested in $\half$-BPS operators belonging to the first
class of representations. They transform in the $[0,p,0]$, $p\geq 2$ representation and thus
have dimension $p$. These operators are self-conjugate. The simplest case is a
single-trace operator containing $p$ scalars. The single trace operators we consider are of the form, $r=1,\ldots,6$
\eqn\sto{\varphi_{r_1\ldots r_p}=\tr(X_{\{r_1}\cdots X_{r_p\}}).}
Curly brackets always denote the completely symmetric and traceless part. These $\half$-BPS trace
operators are local gauge invariant operators of the theory with protected dimension. The fact that they are superconformal primaries 
can be read off the schematic action of the Poincar\'e supersymmetries on the
fields
\eqn\coms{\eqalign{
\{Q,\lambda\}&=F^++[X,X],\cr
\{Q,\bar{\lambda}\}&=DX,\cr
[Q,X]&=\lambda,\cr
[Q,F]&=D\lambda.
}}
Trace operators containing only products of scalars in a totally symmetrized way do not
contain any of the terms on the right hand side. Thus they cannot be generated by the action
of $Q$ on another operator. Therefore they are superconformal primaries. Since we are
interested in operators in irreducible representations we will in addition impose that the
products of scalars are traceless in their $R$-symmetry indices. The simplest operators
are given by single-trace operators like tr($X_rX_r$), the Konishi operator, which is a
singlet, or tr($X_{\{r}X_{s\}}$) which transforms in the $[0,2,0]$ representation of the
$R$-symmetry and contains the energy-momentum tensor. There are of course multi-trace operators given by products of single trace
operators projected onto a $[0,p,0]$ $R$-symmetry representation. Since the $X_r$ are
traceless the multi trace operators have dimension $\Delta=p\geq 4$.

The first three levels of the short $[0,p,0]$\ $\N=4$ multiplets we will use are given by
\eqn\Nfour{\def\normalbaselines{\baselineskip16pt\lineskip3pt
 \lineskiplimit3pt}
\matrix{&&&{~~}&&{~~~}&&{~~}&&{~~~}&&{~~}&&{~~~}&&{~~}&&\cr
&&&{~~}&&{~~~}&&{~~}&&\vphi_{r_1\ldots r_{p}}&&{~~}&&{~~~}&&{~~}&&\cr
&&&&&&&&\Bsw&&\Bse&&&&&&&\cr
&&&&&&&\hidewidth\psi_{r_1\ldots r_{p-1}\alpha}\hidewidth&&&&
\hidewidth\bpsi{}^{r_1\ldots r_{p-1}}{}_{\dal}\hidewidth&&&&&&\cr
&&&&&&\Bsw&&\Bse&&\Bsw&&\Bse&&&&&\cr
&&&&&\hidewidth \ldots\hidewidth&&&&
\hidewidth J_{r_1\ldots r_{p-1}\alpha\dal}\hidewidth&&&&
\hidewidth\ldots\hidewidth&&&&\cr}.
}
All fields are traceless and symmetric in the $r$-indices of the $R$-symmetry $SU(4)$. The $\alpha, \dal$ indices are spacetime spinor indices.
In addition the fields at level 2 satisfy an irreducibility constraint
\eqn\irr{\ga\cdot\psi=\bpsi\cdot\bga=0.}
where $\ga,\bga$ are the $SU(4)$ gamma matrices, $\ga_r\bga_s+\ga_s\bga_r=-2\delta_{rs}1$.

Tensorial complications in the analysis of superconformal Ward identities 
and also in applying the operator product expansion are avoided here
by taking
\eqn\tphi{
\vphi_{r_1 \dots r_p}(x) \to
\vphi^{(p)}(x,t) = \vphi_{r_1\dots r_p}(x)\, t_{r_1} \dots t_{r_p} \, ,
}
where $t$ is an arbitrary complex vector satisfying\foot{Such null vectors may also
be motivated by considering the harmonic superspace approach and were used
similarly for instance in \refs{\ADHS,\hpert}. Our application is independent 
of the harmonic superspace formalism and is essentially
motivated just by the requirement of simplifying the treatment of arbitrary rank
symmetric traceless tensors, we do not anywhere consider the conjugate of $t$.}
\eqn\nul{
t^2=0 \, .
}
(For a more mathematical discussion of using such vectors for the treatment of
representations of $SO(n)$ see \Barg, see also appendix A in \Dobb).
Clearly $\vphi_{r_1\dots r_p}$ can be recovered from $\vphi^{(p)}$.
The four point function then becomes a homogeneous polynomial in 
$t_1,t_2,t_3,t_4$, of respective degree $p_1,p_2,p_3,p_4$, invariant under
simultaneous rotations on all $t_i$'s. 

The four point correlation functions of interest then have the form
\eqn\Fourp{\eqalign{
\l \vphi^{(p_1)}& (x_1,t_1) \, \vphi^{(p_2)}(x_2,t_2)\, 
\vphi^{(p_3)}(x_3,t_3)\, \vphi^{(p_4)}(x_4,t_4 )\r \cr
&{} = { r_{23}^{\, \, \Sigma - p_2 - p_3}  \,
r_{34}^{\, \, \Sigma - p_3 - p_4}\over
r_{13}^{\, \, p_1}\,  r_{24}^{\, \, \Sigma - p_3} }\, F(u,v;t) \, ,
\quad 2\Sigma = p_1+p_2+p_3+p_4 \, , \cr}
}

Due to the condition \nul\ for each 
$t_i$ the conformally covariant four point function is reducible to an
invariant function $\F(u,v;\lam,\mun)$ with $\lam,\mun$ the two independent
invariants, homogeneous of degree zero in each $t_i$, which are analogous to the 
conformal invariants $u,v$,
\eqn\deflu{
\lam = {t_1{\cdot t_3} \, t_2{\cdot t_4} \over
t_1{\cdot t_2} \, t_3{\cdot t_4}} \, , \qquad
\mun = {t_1{\cdot t_4} \, t_2{\cdot t_3} \over
t_1{\cdot t_2} \, t_3{\cdot t_4}} \, .
}

We make the following choice for the definition of $\F(u,v;\lam,\mun)$
\eqn\Fourpp{
F(u,v,t) = (t_1\cdot t_4)^{p_1-E}(t_2\cdot t_4)^{p_2-E}(t_1\cdot t_2)^E(t_3\cdot t_4)^{p_3} \F(u,v;\sigma,\tau) \, , 
}
where we ordered $p_1\leq p_2\leq p_3\leq p_4$ and
\eqn\SigE{\eqalign{
2E ={}& p_1+p_2+p_3-p_4\, .
}}

In general $\F(u,v;\lam,\mun)$ is a polynomial in $\lam,\mun$, with  degree 
determined by the $p_i$, where the number of independent terms match exactly 
the number of tensorial invariants necessary for the general decomposition 
of the four  point function for the corresponding symmetric traceless 
tensorial fields, for $p_i=p$ there are $\half (p+1)(p+2)$ terms.

\subsec{$\N=2$ Superconformal Multiplets}

Analogously to the $\N=4$ case the unitary irreducible representations of $SU(2,2|2)$ with positive energy are labelled
by the quantum numbers of $SO(2,4)\times U(2)=SO(2)\times SO(4)\times SU(2)_R\times U_R(1)$. 
There is a mapping of representations of
this group to representations of $SO(1,1)\times SO(1,3)\times SU(2)_R\times U_R(1)$. The quantum number of $SO(1,1)$ is 
the scale dimension $\Delta$ which is non-negative for positive energy representations. The quantum numbers 
of $SO(1,3)\sim SU(2)\times SU(2)$ are the `spins' $(j_1,j_2)$ where $j_1,j_2$ are non-negative
half-integers. The representations of $SU(2)_R$
are labelled by the $R$-symmetry spin $R$, the representations of $U_R(1)$ by the
$R$-symmetry charge $r$.
We will encounter short or $\half$-BPS, semi-short and long multiplets. Short multiplets are
characterized by the condition $\Delta=2R+r$. We will actually consider operators analogous to the $\half$-BPS 
CPO's in $\N=4$ which will have
$j_1=j_2=r=0$. Therefore the shortening condition in our case will always be $\Delta=2R$.
Long representations are characterized by the bound $\Delta\geq 2R+r+2$. If the dimension
actually satisfies the bound, the multiplet is called semi-short with $\Delta=2R+r+2$.
Notice the gap in dimension between short and semi-short multiplets with the same quantum
numbers.
The first three levels of the short $\N=2$ multiplets with $R$-spin $n$ are given by, $r_i=1,2,3$
\eqn\Ntwo{\def\normalbaselines{\baselineskip16pt\lineskip3pt
 \lineskiplimit3pt}
\matrix{&&&&\vphi_{r_1\ldots r_n}&&&&\cr
&&&\Bsw&&\Bse&&&\cr
&&\hidewidth\psi_{r_1\ldots r_{n-1}}{}_{\,\alpha}&&&&\bchi_{r_1\ldots r_{n-1}\dal}\hidewidth&&\cr
&\Bsw~~&&\Bse&&\Bsw&&~~\Bse&\cr
\ldots\hidewidth&&&&J_{r_1\ldots r_{n-1}\alpha\dal}&&&&\hidewidth\ldots\cr}.
}
All fields are symmetric and traceless in the $r$-indices and the fermions at level 2 satisfy the irreducibility constraint
\eqn\irrr{\tau\cdot\psi=\bpsi\cdot\tau=0.} where $\tau$ are the $SU(2)$ Pauli matrices with the usual algebra 
$\tau_r\tau_s+\tau_s\tau_r=2\delta_{rs}$.

We will make the same definitions for the vectors $t_i$ as for the $\N=4$ case. For $\N=2$ there will be a slight difference since in 3 dimensions
four vectors are not linearly independent. This implies a relation between the four $t_i$ of a four point function. It turns out that this relation
translates into the following relation for $\sigma, \tau$ in the $\N=2$ case $\sigma^2+\tau^2+1=2(\sigma+\tau+\sigma \tau)$.

In appendix E we present an alternative treatment of the $\N=2$ case. Basically we consider fields with symmetric fundamental $SU(2)$ indices instead of symmetric traceless adjoint $SU(2)$ indices. 
There is a one to one mapping between the two pictures for an even number of fundamental $SU(2)$ indices. The fact that there is no trace condition on the fundamental $SU(2)$ indices will
simplify the treatment considerably since this renders derivatives with respect to $u$ ordinary derivatives and all the complications discussed in appendix A will be unnecessary to consider.

\subsec{Large $N$ Amplitudes}

In the large $N$ limit four point functions of four identical operators have been computed using supergravity techniques and the AdS/CFT correspondence \refs{\Arut,\ADHS,\Degen}. They can be 
expressed in terms of $\oD$-functions $\oD_{\Delta_1\Delta_2\Delta_3\Delta_4}(u,v)$ which are defined by conformal 4-point integrals and only
depend on the two conformal cross-ratios of two points $u,v$. 
We will take the results available for the case of $[0,4,0]$ operators \Degen\ and perform a conformal
partial wave expansion to ultimately obtain averaged first order anomalous dimensions. 
A conformal partial wave expansion has an analogue from studying scattering amplitudes. Four point functions can be interpreted as two incoming fields scattering and thus producing two outgoing fields.
The amplitude can now be expanded in a basis of functions corresponding to intermediate fields. In our context these are called conformal partial waves characterized by the dimension $\Delta$ and the spin $\ell$.
The crucial ingredients will be the explicit expression for
the contribution of a conformal block \sfpf\ and \ope, and a power series expansion of the part of the $\oD$ functions 
containing a factor of $\log u$ \ADHS. The aim of the partial wave expansion is to find the coefficients of the 
expansion of the amplitude consisting of $\oD$ functions in terms of the $G^{(\ell)}_{\Delta}$ describing a conformal block.

The $\oD_{\Delta_1\Delta_2\Delta_3\Delta_4}(u,v)$ may be expressed in three parts. First we define
\eqn\dss{
\Sigma = \half ( \Delta_1 + \Delta_2 + \Delta_3 + \Delta_4) \, , \qquad
s = \half ( \Delta_1 + \Delta_2 - \Delta_3 - \Delta_4) \, ,
}
and, restricting $\Sigma, s$ to be integer, for $s=1,2,\dots$ we have \refs{\ADHS,\od}
\eqnn\Df
$$\eqalignno{
\oD_{\Delta_1\Delta_2\Delta_3\Delta_4}(u,v) = {}& \sum_{m=0}^{s-1} u^{-s+m} 
{(-1)^m \over m!} \, (s-m-1)! \,f_{\Delta_1-s+m\, \Delta_2-s+m\,\Delta_3+m\, \Delta_4+m} (v)\cr 
&{} + \log u \, a_{\Delta_1\Delta_2\Delta_3\Delta_4}(u,v) + b_{\Delta_1\Delta_2\Delta_3\Delta_4}(u,v) \, ,  & \Df\cr}
$$
where $a_{\Delta_1\Delta_2\Delta_3\Delta_4}(u,v), \, b_{\Delta_1\Delta_2\Delta_3\Delta_4}(u,v)$ are both given by power 
series in $u,1-v$. For $s=0$ the first term in \Df\ is omitted. If $s<0$ we can use \didst{b} to invert the sign of $s$.
In \Df\ the relevant terms here are
given by \refs{\ADHS,\od}
\eqnn\fm
$$\eqalignno{
&f_{\Delta_1 \Delta_2 \Delta_3 \Delta_4} (v) =
{\Gamma(\Delta_1)\Gamma(\Delta_2)\Gamma(\Delta_3)\Gamma(\Delta_4) \over 
\Gamma(\Delta_3+\Delta_4)}\, {}_2F_1(\Delta_2,\Delta_3;\Delta_3+\Delta_4;1-v) \, , \cr
&a_{\Delta_1\Delta_2\Delta_3\Delta_4}(u,v)\cr&={\ }-\log u\,{(-1)^s\over s!} {\Gamma (\Delta_1)\Gamma (\Delta_2)\Gamma (\Sigma-\Delta_3)\Gamma (\Sigma-\Delta_4)\over m!\,n!\,\Gamma (\Delta_1+\Delta_2)}\cr
&{\hskip 1 cm}\times \sum_{m,n=0}^{\infty}{(\Delta_1)_m(\Sigma-\Delta_3)_m(\Delta_2)_{m+n}(\Sigma-\Delta_4)_{m+n}\over m!\, n!\, (s+1)_m(\Delta_1+\Delta_2)_{2m+n}}
\, u^m(1-v)^n. & \fm}
$$ 

The part singular in $u$ given by $f_{\delta_1 \delta_2 \delta_3 \delta_4}(v)$ will be used to restrict the possible form of general $p$ large $N$ amplitudes in the final part of chapter 5.
Note that in \Df\ we only use $f_{\delta_1\delta_2\delta_3\delta_4}$ with $\delta_1+\delta_2=\delta_3+\delta_4$ so we could eliminate $\delta_1$.

\subsec{Known Results about Superconformal Ward Identities}

It was found that the two and three point functions of CPO's in $\N=2,4$ SYM are not
renormalized. This can be shown using Intriligator's reduction formula \Intr
$${\partial\over\partial g}\left<\varphi_1(x_1)\ldots\varphi_n(x_n)\right>=\int d^4y
\left<\L(y)\varphi_1(x_1)\ldots\varphi_n(x_n)\right>.$$
The $\N=4$ SYM on-shell Lagrangian $\L$ is part of the stress tensor
multiplet with a numerical coefficient of $g^2$. Therefore in the path integral formalism the derivative with respect to the
coupling constant brings the Lagrangian down which is equivalent to inserting the stress
tensor multiplet into the correlator and picking the Lagrangian term with a suitable
superspace integration. 
It was shown that only nilpotent invariants can contribute to
this and since there are no nilpotent invariants for 4 or less points \nilpo\ it follows that 2 and 3
point functions are not renormalized.

Also, for $\N=4$ surprising nonrenormalization theorems for extremal and next-to-extremal correlation functions were found \Ext.
The fields in the correlator are chiral primary $\half$-BPS operators
transforming in the $R$-symmetry representation with Dynkin labels $[0,p_i,0]$ and having
dimension $p_i$. The condition of extremality for an $N$-point correlation function is
$\sum_{i=1}^{N-1}p_i=p_N$ or $E=0$ in \SigE. Later this was generalized to a similar case, the
next-to-extremal case, where $\sum_{i=1}^{N-1}p_i=p_N+2$.
Initially it was observed that extremal three point functions in $\N=4$ SYM coincide in
the zero and the strong coupling limit \Sei. The
nonrenormalization of extremal $n$-point functions was conjectured from supergravity
considerations \Ext.  It was shown that these correlators
receive no $g^2$-corrections \Ext\ and no instanton corrections in the field theory \Ext. The
non-renormalization was also shown using $\N=2$ harmonic superspace techniques \Ext.

It is unknown if more nonrenormalization theorems exist for SYM theories.
A well known fact is that conformal symmetry fixes two and three point functions in
their functional form. Four point functions may contain an arbitrary function of the two
conformal invariants of four points. It is not fully known how much further superconformal
symmetry restricts correlation functions. The conformal invariants extend to
superconformal invariants and additionally one can construct nilpotent superconformal invariants
for $n$ points with $n\geq 5$ \nilpo. This does not lead to any further constraints. But the
fields are actually constrained fields in irreducible representations of the $R$-symmetry. 
This together with the symmetry can impose further
restrictions on the correlation functions. What is needed is a computation of the superconformal Ward identities.
Once the superconformal Ward identities are solved, results obtained 
will automatically be compatible with superconformal symmetry.

We will consider here four point functions of $\half$-BPS chiral primary operators (CPO's). 
They are short and therefore cannot receive anomalous dimensions. For an analysis of their 
three point functions refer to \Sei, the vanishing of perturbative corrections was shown 
in \refs{\HFS,\OST} and also in the harmonic superspace approach in \HSW.
The first four point function studied were the ones for $\half$-BPS CPO's 
with dimension $2$ which contains the energy momentum tensor. Perturbative results can be found in 
\pert\ and in the supergravity regime in \Arut. In \scft\ the same approach as in this thesis was taken. 
For dimensions $\Delta=3,4$ the results can be found 
in \refs{\hpert,\ADHS,\Degen}.  Using these results the analysis of the operator product expansion (OPE) 
was performed in \refs{\Hokone,\OPEN,\OPEW,\Edent,\Except,\BKRS,\HMR}.
Similar results to ours were previously obtained by Heslop and Howe \Howe\ based on 
expansions in terms of Schur polynomials for $SU(2,2|2)$ and 
$PSU(2,2|4)$\foot{In eq. (49) of \Howe\ 
$S_{020}(Z) = (X_1-Y_1)(X_1-Y_2)(X_2-Y_1)(X_2-Y_2)$ which appears
as an overall factor in the Schur polynomial for long representations.}.
Using the formalism of harmonic superspace \ASok\ also provides a method
for deriving superconformal identities which are equivalent to those obtained
here.

The analysis we perform in general in this thesis was first undertaken in \scft\ for the simplest case of $[0,2,0]$-CPO's and then carried 
on for the case of $p=3$ in \ADHS. 
Unfortunately it 
turned out to be not easy to generalize to CPO's of higher dimension. The tensors involved for CPO's with 
higher dimension become increasingly complicated. For each correlation function in the Ward identities 
appropriate invariant tensors have to be constructed and it is not obvious how to perform an analysis for 
CPO's of arbitrary dimension. We will solve this issue by the introduction of null vectors as explained earlier.

\subsec{Known large $N$ results}

In this thesis we will endeavour to work out the consequences of superconformal
symmetry for the four point correlation functions of BPS operators. As a result
our considerations are lacking in dynamical input since we do not consider
any details of $\N=2$ or $\N=4$ superconformal theories. The results of our
analysis will demonstrate that the details of the dynamics resides in the function $\K$ as in \Gsol\ or \Gsolt.
In particular cases results have been obtained using perturbation theory \pert\
or with the AdS/CFT correspondence \refs{\Arut,\ADHS,\Degen}. We here summarise
some of the results obtained in \refs{\Arut,\ADHS,\Degen} in the context
of this paper.

The large $N$ results obtained through the AdS/CFT correspondence are expressible 
in terms of functions $\oD_{n_1n_2n_3n_4}(u,v)$
which satisfy various identities listed in \scft,\ADHS\ and appendix G.
When $p=2$
\eqn\Htwo{
\H^{(2)}(u,v;\lam,\mun) = - {4\over N^2} \, u^2 \oD_{2422}(u,v) \, ,
}
and for $p=3$,
\eqn\Hthree{
\H^{(3)}(u,v;\lam,\mun) = - {9\over N^2} \, u^3 \big ( ( 1+\lam+\mun) \oD_{3533}
+ \oD_{3522} + \lam \, \oD_{2523} + \mun \, \oD_{2532} \big ) \, .
}
The expression for $p=4$ can be found in chapter 5 where we present a new simplified form obtained starting from the result in \Degen.

\subsec{Outline}

In detail the structure of this thesis is then as follows. 
In chapter 2
we derive the superconformal Ward identities for $\N=2$ superconformal
symmetry, discuss the solution and apply it by analysing the contributions
of different supermultiplets in the OPE. 
The
discussion is extended to the $\N=4$ case in chapter 3. For
the operator product expansion it is shown how there are potential
contributions from non-unitary semi-short supermultiplets. We explicitly show how they may be cancelled in such a way 
that only unitary multiplets remain given an appropriate operator content.
In chapter 4 we take into account the restrictions imposed by 
crossing symmetry making use of $\S_3$ representations. 
In chapter 5 we present a simplified expression for the $p=4$ amplitude and perform a conformal partial wave expansion 
to compute the averaged first order anomalous dimensions for the different representations.
Following up on an observation made in \od\ we present work in progress showing how the amplitudes for higher $p$ might be constrained by crossing symmetry and the structure 
of singularities in $u$. By generalizing a universal structure in the singularities as observed for $p=2,3,4$ \refs{\Arut,\ADHS,\Degen, \od} and using the crossing symmetry present we will be able to first reproduce
the results computed before and finally also work out the amplitude for general $p$.
Some final comments on the work done and future investigations are made in the conclusion.

Various technical issues are addressed in seven appendices. 
In
appendix A we discuss how derivatives involving the null vector $t_r$
are compatible with $t^2=0$. 
In appendix B we consider two variable harmonic
polynomials, depending on $\lam,\mun$ given by the two cross ratios of four points, which are used
in the expansion of general four point correlation functions. 
Appendix C
describes some differential operators which play an essential role
in our analysis whereas in appendix D we consider non unitary semi-short
representations for $PSU(2,2|4)$ which are important in our operator
product analysis.
Appendix E contains a simpler derivation of the Ward identities for the $\N=2$ case than presented in chapter 2. 
We pursue the more complicated route in chapter 2 since only this easily generalizes to the $\N=4$ case in chapter 3.
In appendix F we present a Mathematica program used for computation of conformal partial wave expansions in  the large $N$ limit
of the $p=4$ case.
Finally, in appendix G we list a number of standard identities for $\oD$ functions used in chapter 5.
\vfill
\eject

\newsec{$\N=2$}

\subsec{Superconformal Ward Identities}

The algebraic complications involved in the analysis of Ward identities
are much simpler for $\N=2$ superconformal symmetry. In this case the
$R$-symmetry group is just $U(2)$ and discussion of the representations
is much easier. In order to facilitate the comparison with the $\N=4$
case later we consider BPS chiral primary operators which belong
to representations of  $SU(2)_R$ symmetry for $R=n$, an integer. The
BPS condition requires that the scale dimension $\Delta =2n$. Such fields
form superconformal primary states for a short supermultiplet with necessarily
unrenormalised scale dimensions. The fields in this case are represented by
symmetric traceless tensors $\vphi_{r_1 \dots r_n}$ with $r_i=1,2,3$.
To derive the Ward identities we need to consider just the superconformal 
transformations at the lowest levels of the multiplet. First
\eqn\ptph{
\de \vphi_{r_1\dots r_n} = \hep \, \tau_{(r_n} \psi_{r_1\dots r_{n-1})} +
\bpsi{}_{(r_1\dots r_{n-1}} \tau_{r_n)} \, \hbep \, ,
}
where $\psi_{i r_1\dots r_{n-1}\alpha}, \, \bpsi{}_{i r_1\dots r_{n-1}\dal}$
are spinor fields, traceless and symmetric on the indices $r_1\dots r_{n-1}$,
 satisfying irreducibility constraints, with $i=1,2$ and $\tau_r$ the usual Pauli matrices,
\eqn\tpsi{
\tau_r \, \psi_{r r_1\dots r_{n-2}} = 0 \, , \qquad
\bpsi{}_{r_1\dots r_{n-2}r} \, \tau_r = 0 \, .
}
Thus both $\psi$ and $\bpsi$ belong to $SU(2)_R$ representations with 
$R=n-\half$. In \ptph\ we have
\eqn\scep{
\hep_i{}^{\! \alpha}(x)= \ep_i{}^{\! \alpha} - i \, \bta{}_{i\dal}
\tx^{\dal\alpha} \, , \qquad \hbep{}^{\, i \dal}(x) = \bep{}^{\, i \dal}
+ i \, \tx^{\dal\alpha} \eta^i{}_{\! \alpha} \, .
}
where $\ep_i{}^{\! \alpha}, \bta{}_{i\dal}, \bep{}^{\, i \dal},
\eta^i{}_{\! \alpha}$ are the $R=\half$ anticommuting parameters for an 
$\N=2$ superconformal transformation. In addition to \ptph\ we use
\eqn\ptps{\eqalign{
\de \psi_{r_1\dots r_{n-1}\alpha} ={}&
i \pr_{\alpha\dal}\, \vphi_{r_1\dots r_{n-1}s} \, \tau_s \hbep{}^{\, \dal}
+ 4n \, \vphi_{r_1\dots r_{n-1}s} \, \tau_s \eta_\alpha \cr
&{}+ J_{r_1 \dots r_{n-1}\alpha\dal} \, \hbep{}^{\, \dal} - {n-1\over 2n-1}\,
\tau_{(r_1} J_{r_2 \dots r_{n-1})s\alpha\dal }\, \tau_s \hbep{}^{\, \dal}\, ,
\cr}
}
where $J_{r_1 \dots r_{n-1}\alpha\dal}$, a symmetric traceless rank $n-1$
tensor, is a $R=n-1$ current. Using \ptps\ together with its conjugate 
we may verify closure of the superconformal algebra acting on 
$\vphi_{r_1\dots r_n}$,
\eqn\clos{
[\de_2 , \de_1] \vphi_{r_1\dots r_n} = - v{\cdot \pr} \vphi_{r_1\dots r_n}
- n(\si + \bsi)\, \vphi_{r_1\dots r_n} + n \, t_{(r_n|s}\,
\vphi_{r_1\dots r_{n-1})s} \, ,
}
where $v^a$, which is quadratic in $x$, and $\si , \bsi, t_{rs}=-t_{sr}$,
which are linear in $x$, are constructed from 
$\hep_1, \hbep_1, \hep_2,\hbep_2$.

Similarly to the definition of $\vphi$ in \tphi\ we define $$\psi^{(n-1)}_\alpha (x,t) = \psi_{r_1\dots r_{n-1}\alpha}(x)\, 
t_{r_1} \dots t_{r_{n-1}}$$ and $$\bpsi{}^{\,(n-1)}_\dal (x,t) =
\bpsi{}_{r_1\dots r_{n-1}\dal}(x)\, t_{r_1} \dots t_{r_{n-1}}$$ while
$$J^{(n-1)}_{\alpha\dal} (x,t) = J_{r_1 \dots r_{n-1}\alpha\dal}(x) \,
t_{r_1} \dots t_{r_{n-1}}.$$ With this notation \ptph\ may be rewritten as
\eqn\varp{
\de \vphi^{(n)}(t) = \hep \,\tau{\cdot t} \, \psi^{(n-1)}(t) + 
\bpsi{}^{\,(n-1)}(t)\, \tau{\cdot t}\, \hbep \, ,
}
and \ptps\ becomes
\eqn\vars{\eqalign{
\de \psi^{(n-1)}_\alpha (t) = {}& {1\over n} \,
\tau{\cdot {\pr \over \pr t}}\, i \pr_{\alpha\dal} \vphi^{(n)}(t) \, 
\hbep{}^{\, \dal}
+ 4 \, \tau{\cdot {\pr \over \pr t}} \vphi^{(n)}(t) \eta_\alpha \cr
&{}+ \bigg ( 1 - {1\over 2n-1}\,
\tau{\cdot t} \, \tau{\cdot {\pr \over \pr t}} \bigg ) 
J^{(n-1)}_{\alpha\dal}(t) \, \hbep{}^{\, \dal} \, . \cr}
}
A precise form for differentiation with respect to $t_r$ satisfying \nul\
is given in appendix A. The conditions \tpsi\ are now
\eqn\condit{
\tau{\cdot {\pr \over \pr t}} \, \psi^{(n-1)}_\alpha (t) = 0 \, , \qquad
\bpsi{}^{\,(n-1)}(t)\, \tau{\cdot {\overleftarrow{\pr \over \pr t}}} =0 \, .
}

$F(u,v;t)$ is also a $SU(2)_R$ scalar which was specified in \Fourp, clearly
there is a freedom to modify it by suitable powers of $u$ or $v$ at the expense
of changing the terms involving $r_{ij}$ in \Fourp. The choice made on \Fourp\
has some convenience in the later discussion.

The fundamental superconformal Ward identities arise from expanding
\eqn\ward{
\de \l \psi^{(n_1-1)}_\alpha (x_1,t_1) \, \vphi^{(n_2)}(x_2,t_2)\, 
\vphi^{(n_3)}(x_3,t_3)\, \vphi^{(n_4)}(x_4,t_4 )\r =0 \, ,
}
using \varp\ and \vars. This gives, suppressing the arguments $t_i$ for
the time being,
\eqn\exw{\eqalign{
& {1\over n_1}\, i \pr_{1\alpha\dal}\, \tau{\cdot {\pr \over \pr t_1}}
\l \vphi^{(n_1)} (x_1) \, \vphi^{(n_2)}(x_2)\, 
\vphi^{(n_3)}(x_3)\, \vphi^{(n_4)}(x_4)\r \, \hbep{}^{\, \dal}(x_1) \cr
& \qquad {} + 4 \, \tau{\cdot {\pr \over \pr t_1}}
\l \vphi^{(n_1)} (x_1) \, \vphi^{(n_2)}(x_2)\, 
\vphi^{(n_3)}(x_3)\, \vphi^{(n_4)}(x_4)\r \, \eta_\alpha \cr
& {}+ \bigg ( 1  - {1\over 2n_1-1}\,
\tau{\cdot t_1} \, \tau{\cdot {\pr \over \pr t_1}} \bigg ) 
\l J^{(n_1-1)}_{\alpha\dal}(x_1)  \, \vphi^{(n_2)}(x_2)\,
\vphi^{(n_3)}(x_3)\, \vphi^{(n_4)}(x_4)\r \, \hbep{}^{\, \dal}(x_1) \cr 
&{}+ \l \psi^{(n_1-1)}_\alpha (x_1) \, \bpsi{}^{\,(n_2-1)}_\dal (x_2) \, 
\vphi^{(n_3)}(x_3)\, \vphi^{(n_4)}(x_4)\r \, \tau{\cdot t_2} \,
\hbep{}^{\, \dal}(x_2) \cr
&{}+ \l \psi^{(n_1-1)}_\alpha (x_1) \, \vphi^{(n_2)}(x_2) \, 
\bpsi{}^{\,(n_3-1)}_\dal (x_3) \, \vphi^{(n_4)}(x_4)\r \,\tau{\cdot t_3} \,
\hbep{}^{\,\dal}(x_3)\cr
&{}+ \l \psi^{(n_1-1)}_\alpha (x_1) \, \vphi^{(n_2)}(x_2)\,
\vphi^{(n_3)}(x_3)\, \bpsi{}^{\, (n_4-1)}_\dal(x_4)\r \, \tau{\cdot t_4} \,
\hbep{}^{\, \dal}(x_4)  = 0 \, . \cr }
}
To apply this we make use of general expressions compatible with conformal
invariance for each four point function which appears. Thus
\eqn\fourpsi{\eqalign{
& \l \psi^{(n_1-1)}_\alpha (x_1) \, \bpsi{}^{\,(n_2-1)}_\dal (x_2) \, 
\vphi^{(n_3)}(x_3)\, \vphi^{(n_4)}(x_4)\r \cr
&\quad {}= 2i \, { r_{23}^{\, \, \Sigma - 2n_2 - 2n_3}  \,
r_{34}^{\, \, \Sigma - 2n_3 - 2n_4}\over
r_{13}^{\, \, 2n_1}\,  r_{24}^{\, \, \Sigma - 2n_3} }\, \bigg ( 
{1\over r_{12}} \x_{12\alpha\dal}\, R_2 + {1\over r_{13}r_{24}}
(\x_{13} \tx_{34} \x_{42})_{\alpha\dal}\, S_2 \bigg ) \, , \cr
& \l \psi^{(n_1-1)}_\alpha (x_1) \, \vphi^{(n_2)}(x_2) \,
\bpsi{}^{\,(n_3-1)}_\dal (x_3) \, \vphi^{(n_4)}(x_4)\r \cr
&\quad {}= 2i \, { r_{23}^{\, \, \Sigma - 2n_2 - 2n_3}  \,
r_{34}^{\, \, \Sigma - 2n_3 - 2n_4}\over
r_{13}^{\, \, 2n_1}\,  r_{24}^{\, \, \Sigma - 2n_3} }\, \bigg (
{1\over r_{13}} \x_{13\alpha\dal}\, R_3 + {1\over r_{14}r_{23}}
(\x_{14} \tx_{42} \x_{23})_{\alpha\dal}\, S_3 \bigg ) \, , \cr
& \l \psi^{(n_1-1)}_\alpha (x_1) \, \vphi^{(n_2)}(x_2)\,
\vphi^{(n_3)}(x_3)\, \bpsi{}^{\, (n_4-1)}_\dal(x_4)\r \cr
&\quad {}= 2i \, { r_{23}^{\, \, \Sigma - 2n_2 - 2n_3}  \,
r_{34}^{\, \, \Sigma - 2n_3 - 2n_4}\over
r_{13}^{\, \, 2n_1}\,  r_{24}^{\, \, \Sigma - 2n_3} }\, \bigg (
{1\over r_{14}} \x_{14\alpha\dal}\, R_4 + {1\over r_{13}r_{24}}
(\x_{13} \tx_{32} \x_{24})_{\alpha\dal}\, S_4 \bigg ) \, , \cr }
}
where $R_n,S_n$ are functions of $u,v$ and also scalars formed from $t_i$
(to verify completeness of the basis chosen in \fourpsi\ we use relations 
such as $\x_{13} \tx_{34} \x_{42} + \x_{14} \tx_{43} \x_{32} 
= r_{34} \, \x_{12}$). In addition we have
\eqn\fourJ{\eqalign{
& \l J^{(n_1-1)}_{\alpha\dal}(x_1)  \, \vphi^{(n_2)}(x_2)\,
\vphi^{(n_3)}(x_3)\, \vphi^{(n_4)}(x_4)\r \cr
&\quad {}= 2i \, { r_{23}^{\, \, \Sigma - 2n_2 - 2n_3}  \,
r_{34}^{\, \, \Sigma - 2n_3 - 2n_4}\over
r_{13}^{\, \, 2n_1}\,  r_{24}^{\, \, \Sigma - 2n_3} }\, \Big (
\X_{1[23]\alpha\dal} \, I + \X_{1[43]\alpha\dal} \, J \Big ) \, , \cr}
}
for 
\eqn\defX{
\X_{i[jk]} = {\x_{ij}\tx_{jk}\x_{ki}\over r_{ij}\, r_{ik}} =
{1\over r_{ij}} \x_{ij} - {1\over r_{ik}} \x_{ik} \, ,
}
which transforms under conformal transformations as a vector at $x_i$ and
is antisymmetric in $jk$. 

Using \fourpsi\ and \fourJ\ in \exw, noting that
\eqn\deep{
i \pr_{1\alpha\dal}\, {1\over r_{13}^{\, \, 2n_1}} \, \hbep{}^{\, \dal}(x_1)
+ 4n_1 \, {1\over r_{13}^{\, \, 2n_1}} \, \eta_\alpha = 
- 4n_1 i \, {1\over r_{13}^{\, \, 2n_1+1}} \, \hbep{}^{\, \dal}(x_3) \, ,
}
and $\pr_{1\alpha\dal} u = 2u \, \X_{1[23]\alpha\dal}, \,
\pr_{1\alpha\dal} v = 2v \, \X_{1[43]\alpha\dal}$, we may decompose \exw\
into independent contributions involving $\hbep(x_1)$ and
$\hbep(x_3)$ (note that $\x_{12} \hbep(x_2)/r_{12} = \x_{13}\hbep(x_2)/r_{13}
+ \X_{1[23]}  \hbep(x_1)$ and also for $x_2\to x_4$)
giving two linear relations,
\eqna\Wardrel
$$\eqalignno{
& { 1\over n_1} \, \tau{\cdot {\pr \over \pr t_1}} \big (
\X_{1[23]} \,  u\pr_u  + \X_{1[43]} \, v\pr_v \big ) F
+ \bigg ( 1  - {1\over 2n_1-1}\,
\tau{\cdot t_1} \, \tau{\cdot {\pr \over \pr t_1}} \bigg ) \big (
\X_{1[23]} \, I  + \X_{1[43]} \, J \big ) \cr
&{}+ \X_{1[23]} \, R_2 \, \tau{\cdot t_2} + \X_{1[43]} \, R_4 \,\tau{\cdot t_4}
+  \big ( u \X_{1[23]} -v \X_{1[43]} \big ) (  S_2 \, \tau{\cdot t_2}
-  S_4 \, \tau{\cdot t_4} ) = 0 \, , & \Wardrel{a} \cr
& \bigg ( - 2 \, \tau{\cdot {\pr \over \pr t_1}} F 
+ \big ( R_2 +(u-v) S_2 \big ) \tau{\cdot t_2} + R_3 \, \tau{\cdot t_3} + 
\big ( R_4 +(1-u+ v) S_4 \big ) \tau{\cdot t_4} \bigg ) 
{1\over r_{13}}\x_{13} \cr
&{}+  \big ( v S_2 \, \tau{\cdot t_2} + S_3 \, \tau{\cdot t_3}
- v S_4 \, \tau{\cdot t_4} \big ) {1\over r_{14}r_{23}} \, 
\x_{14} \tx_{42} \x_{23} = 0 \, . & \Wardrel{b} \cr}
$$
It is easy to decompose \Wardrel{a,b}\ into independent equations but crucial
simplifications are obtained essentially by diagonalising each $2\times 2$
spinorial equation in terms  of new variables $x,\zz$
which, as mentioned in the introduction, are the eigenvalues of 
$\x_{12}\,\x_{42}{}^{\!\!-1} \x_{43} \, \x_{13}{}^{\!\! -1}$. These are 
related to the conformal invariants $u,v$ defined in \defuv\ by \eiguv.
In \Wardrel{b} the spinorial matrix
$\tx_{41}{}^{\! -1} \tx_{42} \,\tx_{32}{}^{\! -1} \tx_{31}$ may be replaced
by $1/(1-x)$ and in \Wardrel{a} we may effectively replace $ \X_{1[23]} \to
1/x$ and $\X_{1[43]} \to - 1/(1-x)$ and in each case also for $x\to \zz$.
Using 
\eqn\part{
{\pr \over \pr x} = \zz \, {\pr \over \pr u} - (1-\zz) \, {\pr \over \pr v} \, ,
}
and the definitions
\eqn\hatRJ{
{T_2} = R_2 + x S_2 \, , \quad {T_3} = R_3 + {1\over 1-x} S_3 \, ,
\quad {T_4} = R_4 + (1-x) S_4 \, , \quad
K = {1\over x} I - {1\over 1-x} J \, ,
}
we may then obtain from \Wardrel{a,b}
\eqna\redeq
$$\eqalignno{
{ 1\over n_1} \, \tau{\cdot {\pr \over \pr t_1}} \, {\pr \over \pr x} F
= {}& - \bigg ( 1  - {1\over 2n_1-1}\,
\tau{\cdot t_1} \, \tau{\cdot {\pr \over \pr t_1}} \bigg ) K
- {1\over x} \, {T_2}\, \tau{\cdot t_2} 
+ {1\over 1-x } \, {T_4}\, \tau{\cdot t_4} \, , & \redeq{a} \cr
2 \, \tau{\cdot {\pr \over \pr t_1}} F = {}& {T_2}\, \tau{\cdot t_2}
+ {T_3}\, \tau{\cdot t_3} + {T_4}\, \tau{\cdot t_4}  \, . 
& \redeq{b} \cr}
$$
together with the corresponding equations obtained by $x\to \zz$ in 
\redeq{a,b}\ with also $T_i\to {\bar T}_i,\, K\to {\bar K}$, which are defined
just as in \hatRJ\ for $x\to \zz$.

The equations in \redeq{a,b}\ are equations for $\pr F/\pr t_1$. The 
integrability conditions, which are required by virtue of
$(\tau{\cdot \pr_{t_1}})^2=0$, are satisfied since we have, for $i=2,3,4$,
\eqn\cons{
\tau{\cdot {\pr \over \pr t_1}} \, {T_i} = 0 \, , \qquad
{T_i} \, \tau{\cdot {\overleftarrow{\pr \over \pr t_i}}} =0 \, ,
}
as a consequence of \condit. To reduce \redeq{a,b} into equations which
ultimately allow $T_i$ and $K$ to be eliminated
we first write, since $T_i$ is a $2\times 2$ matrix,
\eqn\UVS{
T_i \, \tau{\cdot t_i} = \tau {\cdot V_i} + W_i \, 1 \, , 
}
where $W_i$ and $V_{i,r}$ are respectively a scalar and a vector. From 
the results of appendix A we further decompose $V_i$ uniquely in the form
\eqn\VU{
V_i = {1\over n_1}\, {\pr \over \pr t_1} U_i + \hV_i \, , \qquad
t_1 {\cdot V_i} = U_i \, , \quad t_1 {\cdot \hV_i} = 0 \, .
}
The first equation in \cons\ then separates into $SU(2)$ scalar and vector
equations,
\eqn\VS{
{\pr \over \pr t_1} {\cdot V_i} = 0 \, , \qquad i 
{\pr \over \pr t_1} \times V_i + {\pr \over \pr t_1} W_i = 0 \, ,
}
where we may let $V_i \to \hV_i$ without change since both $\pr_1^{\,2} U_i$ 
and $\pr_1 \times \pr_1 U_i$ are zero. From \VS\ we may then find
\eqn\VSt{
L_1 W_i = i n_1 \hV_i \, ,
}
where we define the $SU(2)_R$ generators by
\eqn\genL{
L_i = t_i \times {\pr \over \pr t_i} \, .
}

Substituting \UVS\ into \redeq{a} gives
\eqn\Fx{
{\pr \over \pr x} F = - {1\over x} \, U_2  + {1\over 1-x} \, U_4 \, ,
}
and
\eqn\eqJ{
{n_1 \over 2n_1-1}\, K = - {1\over x} \, W_2 + {1\over 1-x} \, W_4 \, ,
}
which is just an equation giving $K$, and also
\eqn\eqJV{
- {1 \over 2n_1-1}\, iL_1 K = - {1\over x} \, \hV_2 + 
{1\over 1-x} \, \hV_4 \, .
}
It is easy to see that this follows from \eqJ\ as a consequence of \VSt.
Similarly substituting \UVS\ into \redeq{b} gives three equations
\eqn\FYn{
2n_1 \, F = \sum_{i=2}^4 \, U_{i} \, , 
}
and
\eqn\Vz{
\sum_{i=2}^4 \, \hV_{i} =0 \, , 
}
as well as
\eqn\Sz{
\qquad \sum_{i=2}^4 \, W_{i} =0 \, .
}
Clearly \Vz\ follows from \Sz. 

An essential constraint may also be obtained
from the second equation in \cons\ which gives
\eqn\contw{
(n_i + 1 ) T_i \, \tau {\cdot t_i} = - ( T_i \, \tau {\cdot t_i} )
\, i \tau {\cdot \overleftarrow L}_i \, .
}
With the decomposition \UVS\ this leads to
\eqna\relSU
$$\eqalignno{
(n_i +1) W_i = {}& -i L_i {\cdot V_i} \, ,  & \relSU{a} \cr
(n_i +1) V_i ={}& - L_i \times V_i - i L_i W_i \, . & \relSU{b} \cr}
$$
Contracting  \relSU{b} with $L_i$, and using $L_i \times L_i = - L_i,
\ L_i {\!}^2 W_i = n_i(n_i+1) W_i$, gives \relSU{a}. In addition we have from
$T_i(\tau{\cdot t_i})^2 = 0 $
\eqn\ttT{
t_i{\cdot V_i}=0 \, , \qquad i \, t_i \times V_i = t_i \, W_i \, .
}
With the aid of the results in appendix A we may obtain $(2n_1+1)\, \pr_i {\cdot (}
t_i \, W_i - i \, t_i \times V_i ) = (2n_i+3) \big ( (n_i+1) W_i + i L_i {\cdot V_i}
\big )$ so that \ttT\ implies \relSU{a}. Similarly, since $\pr_i \times (t_i\times V_i)
= (\pr_i \times t_i) \times V_i) + \pr_i (t_i {\cdot V_i}) -
\pr_i {\cdot (}t_i \, V_i)$, we have from appendix A 
$(2n_i+1) \pr_i \times (t_i\times V_i) =
- (2n_i+3) ( L_i \times V_i + (n_i+4) V_i )$ and
$(2n_i+1) \pr_i \times (t_i W_i) = - (2n_i+3) L_i W_i $. Hence it is clear that \ttT\ 
also implies \relSU{b}\foot{The converse follows using $t_i {\cdot L_i} =0 , \
t_i \times L_i = t_i \, t_i{\cdot \pr_i}$ and $t_i \times (L_i \times V_i)
= - t_i \times V_i$.}.

Using \VU\ and \VSt\ for $\hV_i$ in \relSU{a} we obtain
\eqn\LLU{
\big ( L_1 \! \cdot L_i + n_1 ( n_i +1) \big ) W_i =
\half \big ( (L_1 + L_i)^2 + (n_1+n_i)(n_1+n_i+1) \big ) W_i = - i
{\pr \over \pr t_1} {\cdot L_i } \, U_i \, .
}
$U_i(u,v;t)$, which is defined by \UVS, is a homogeneous polynomial
in $t_1,t_i$ of ${\rm O}(t_1^{n_1}, t_i^{n_i})$ such that the
$SU(2)_R$ representation with $R_{(1i)} = n_1+n_i$ is absent. In 
consequence the operator $(L_1 + L_i)^2 + (n_1+n_i)(n_1+n_i+1)$, which
commutes with $\pr_1{\cdot L_i}$, in \LLU\ may be inverted to give $W_i$ in 
terms of $U_i$. Alternatively we may obtain from \ttT
\eqn\LUW{
i \, t_i \times \pr_1 U_i = - t_1 \times L_1 W_i + n_1 \, t_i W_i \, .
}

To analyse these equations further we now consider the decomposition of $F$ 
and also $U_i$ in terms of $SU(2)_R$ scalars.
We first assume the $n_i$ are ordered so that
\eqn\order{
n_1 \le n_2 \le n_3 \le n_4 \, , 
}
and further assume
\eqn\inn{
\qquad n_4 = n_1 +n_2 + n_3 - 2E \, ,
}
for integer $E=0,1,2\dots$, where $E$ is a measure of how close the
correlation function is to the extremal case. With \order\ and \inn\ 
$F$, which is ${\rm O}(t_1^{n_1},t_2^{n_2},
t_3^{n_3},t_4^{n_4})$, can in general be written in the form
\eqn\exF{
F(u,v;t) = \big ( t_1 {\cdot t}_4 \big )^{n_1-E} \big ( t_2 {\cdot t}_4 
\big )^{n_2-E} \big ( t_1 {\cdot t}_2 \big )^{E} 
\big ( t_3 {\cdot t}_4 \big )^{n_3} \, \F(u,v;\lam,\mun) \, ,
}
where $\F$ is a polynomial in $\lam,\mun$, defined in \deflu,
with all terms $\lam^p\mun^q$ satisfying $p+q\le E$.
If $E> n_1$ then all terms in $\F$ must contain a factor $\mu^{E-n_1}$ to
cancel negative powers of $t_1 {\cdot t}_4$ in \exF. Since $t_i$ 
are three dimensional vectors $t_{1[r} t_{2s} t_{3t} t_{4u]} =0$ so that
$\lam,\mun$ are not independent but obey the relation
\eqn\conlm{
\Lambda \equiv \lam^2 + \mun^2  + 1 - 2\lam \mun - 2\lam - 2 \mun = 0 \, .
}
This may be solved in terms of a single variable $\alpha$ by
\eqn\defa{
\lam = \alpha^2 \, , \qquad \mun = (1-\alpha)^2 \, ,
}
so that 
\eqn\FhF{
\F(u,v;\lam,\mun) = {\hF}(x,\zz;\alpha) \, . 
}
${\hat \F}(x,\zz;\alpha)$ is symmetric in $x,\zz$ and, for $E \le n_1$, 
is a polynomial in $\alpha$ of degree $2E$, so 
that there are ${2E+1}$ independent coefficients, while if $E > n_1$ then it
must be of the form $(1-\alpha)^{2(E-n_1)}p(\alpha)$  with $p$ a polynomial
of degree $n_1$, so that the number of coefficients is $2n_1+1$.
These results correspond exactly of course to the number of 
$SU(2)_R$ invariants which can be formed in the four point function,
subject to \inn, together with \order, that can be found using standard 
$SU(2)$ representation multiplication rules. 

A similar expansion to \exF\ can be given for each $U_i$ 
\eqn\exF{\eqalign{
U_i(x,\zz;t)  = {}& \big ( t_1 {\cdot t}_4 \big )^{n_1-E} 
\big ( t_2 {\cdot t}_4 \big )^{n_2-E}
\big ( t_1 {\cdot t}_2 \big )^{E} \big ( t_3 {\cdot t}_4 \big )^{n_3} \, 
\U_i(x,\zz;\lam,\mun) \, ,\cr
& \U_i(x,\zz;\lam,\mun) = {\hU}_i(x,\zz;\alpha) \, . \cr}
}
The analysis of \eqJ\ and \Sz\ depends on using \LLU, or \LUW, as shown in 
appendix C, to relate $W_i$ and $U_i$. Defining
\eqn\WW{
W_i = i \,
t_2 {\cdot (t_3 \times t_4)} \, \big ( t_1 {\cdot t}_4 \big )^{n_1-E} \!
\big ( t_2 {\cdot t}_4 \big )^{n_2-E} \! \big ( t_1 {\cdot t}_2 \big )^{E-1} 
\big ( t_3 {\cdot t}_4 \big )^{n_3-1} \W_i \, ,
}
then we obtain
\eqn\crit{
2(2n_1-1) \W_i = \hD_i \hU{}_{i}  \, ,
}
where $\hD_i$ are linear operators given by
\eqn\DDD{
\hD_2 = {\d \over \d \alpha} + {2(E - n_1)\over 1-\alpha} \, , \quad
\hD_3 = {\d \over \d \alpha} - {2n_1\over \alpha} +
{2(E-n_1)\over 1-\alpha} \, , \quad
\hD_4 = {\d \over \d \alpha} + {2 E\over 1-\alpha} \, .
}

The superconformal identities \Fx, \FYn\ and \Sz\ then become
\eqna\superid
$$\eqalignno{
{\pr \over \pr x} \hF = {}& - {1\over x} \, \hU_2 + 
{1\over 1-x}\, \hU_{4} \, , & \superid{a} \cr
2n_1 \, \hF = {}& \hU_{2} + \hU_{3} + \hU_{4} \, , 
& \superid{b} \cr
\hD_2 \, \hU_{2} + {}& \hD_3 \,\hU_{3} + \hD_4 \, \hU_{4} = 0 \, .
& \superid{c} \cr}
$$
By acting on \superid{b} with $\hD_3$ and using \superid{c} we may obtain
\eqn\FDD{
\hD_3 \hF = - {1\over \alpha}\, \hU_{2} - 
{1\over \alpha(1-\alpha)} \, \hU_{4} \, ,
}
and substituting in \superid{a} gives
\eqn\diF{
\Big ( x {\pr \over \pr x} - \alpha \hD_3 \Big ) \hF =
\Big ( {x\over 1-x} + {1\over 1-\alpha} \Big ) \hU_{4} \, .
}      
The right hand side of \diF\ vanishes when $\alpha=1/x$ leaving an
equation for $\hF$ alone. With the explicit
form for $\hD_3$ in \DDD\ we have
\eqn\resF{
{\pr \over \pr x} \bigg ( x^{2n_1} \Big ( 1 - {1\over x} \Big )^{2(n_1-E)}
\hF \Big ( x,\zz; {1\over x} \Big ) \bigg ) = 0 \, .
}
Together with its partner or conjugate equation involving $\pr/\pr \zz$ 
\resF\ provides the final result for the constraints due to superconformal 
identities for the four point function when $\N=2$.

For the $\N=2$ case we may also require instead of \inn
\eqn\inc{
n_4 = n_1 +n_2 + n_3 - 2E - 1 \, ,
}
since $F$ can then be written as
\eqn\exFc{
F(u,v;t) = \big ( t_1 {\cdot t}_4 \big )^{n_1-E}
\big ( t_2 {\cdot t}_4 \big )^{n_2-E-1}\big ( t_1 {\cdot t}_2 \big )^{E} 
\big ( t_3 {\cdot t}_4 \big )^{n_3-1} \, t_2 \, {\cdot \, t_3 \times t_4}\
{\hat \F}(x,\zz;\alpha) \, .
}
There is an essentially unique expression in \exFc, with a single function 
${\hat \F}$ as a consequence of identities for the various possible 
vector cross products for null vectors  which take the form
\eqn\idthree{\eqalign{
t_1 \, {\cdot \, t_2 \times t_3} \ t_2{\cdot t_4} = {}& \half
(\lam - \mun + 1) \, t_2 \,{\cdot \, t_3 \times t_4} \ 
t_1{\cdot t_2} \, ,\cr
t_1 \, {\cdot \, t_2 \times t_4} \ t_2{\cdot t_3} = {}& \half
(\lam - \mun - 1) \, t_2 \,{\cdot \, t_3 \times t_4} \ 
t_1{\cdot t_2}\,,\cr 
t_1 \, {\cdot \, t_3 \times t_4} \ t_2{\cdot t_4} \, \mun = {}& \half
(\lam + \mun - 1) \, t_2 \,{\cdot \, t_3 \times t_4} \
t_1{\cdot t_4}\, . \cr} 
}
Since, as shown in appendix B, effectively $t_1 {\cdot t}_4 \, 
t_2 \, {\cdot \, t_3 \times t_4} = {\rm O}(1-\alpha)$ we can take
in \exFc, if $n_1-E\ge1$, $(1-\alpha){\hat \F}(x,\zz;\alpha)$ to be a polynomial 
of degree $2E+1$. If $n_1-E<1$ then ${\hat \F}(x,\zz;\alpha)$ must contain
a factor $(1-\alpha)^{2(E-n_1)-1}$. It is easy to see that the number
of independent coefficients matches with the number of independent
terms in the four point function obtained by counting possible
representations in each case.

There is a similar expansion as \exFc\ for $U_i$.
Instead of \WW\ and \crit\ we now have
\eqn\crittwo{
W_i = {i \over 2n_1-1} \, 
\big ( t_1 {\cdot t}_4 \big )^{n_1-E-1} \!
\big ( t_2 {\cdot t}_4 \big )^{n_2-E} \!
\big ( t_1 {\cdot t}_2 \big )^{E} \big ( t_3 {\cdot t}_4 \big )^{n_3} \mun
\, \hD_i \hU_{i} \, ,
}
with $\hD_i$ exactly as in \DDD. In consequence the superconformal
identities reduce to \superid{a,b,c} and we may derive the final result
\resF, albeit with $E$ given by \inc.

\subsec{Solution of Identities}

Although in the $\N=2$ case the identities can be solved rather trivially
we show here how they may be put in a form which makes the connection
with the operator product expansion, and the possible supermultiplets
which may contribute to it, rather obvious. For the purposes of analysing
the operator product expansion for $x_1\sim x_2$ we use the expression of the four point function in terms of the function
$G(u,v;t)$ in \Fourp, so we write
\eqn\fourp{\eqalign{
\l \vphi^{(n_1)} (x_1,t_1)& \, \vphi^{(n_2)}(x_2,t_2)\,
\vphi^{(n_3)}(x_3,t_3)\, \vphi^{(n_4)}(x_4,t_4 )\r \cr
&{} = {1 \over r_{12}^{\, \, n_1+n_2}\,  r_{34}^{\, \, n_3+n_4} }
\bigg ( {r_{24}\over r_{14}} \bigg )^{\! n_1-n_2} 
\bigg ( {r_{14}\over r_{13}} \bigg )^{\! n_3-n_4} G(u,v;t) \, ,\cr}
}
where
\eqn\GF{
G(u,v;t) = u^{n_1+n_2} \, v^{n_1+n_4-n_2-n_3} F(u,v;t) \, .
}

For application of the superconformal Ward identities here it is
convenient here to replace the variable $\alpha$ by $y$ where
\eqn\defy{
y = 2\alpha -1 \, ,
}
and $x,\zz$ by $z,\bz$ given by
\eqn\defz{
z= {2\over x} -1 \, , \qquad \bz= {2\over \zz} -1 \, .
}
Assuming now
\eqn\exG{
G(u,v;t) = \big ( t_1 {\cdot t}_4 \big )^{n_1-E} \big ( t_2 {\cdot t}_4 
\big )^{n_2-E} \big ( t_1 {\cdot t}_2 \big )^{E} 
\big ( t_3 {\cdot t}_4 \big )^{n_3} \, \G(u,v;y) \, ,
}
the solution of \resF\ and its conjugate equation, maintaining the
symmetry under $z \leftrightarrow \bz$, becomes
\eqn\Gf{
\G ( u,v; z) = u^{n_1+ n_2 -2E} f(\bz) \, , \qquad
\G ( u,v; \bz ) = u^{n_1+ n_2 -2E} f(z) \, ,
}
where $f$ is an unknown single variable function. Since $\G(u,v;y)$ 
is just a polynomial
in $y$ \Gf\ requires
\eqn\Gsol{
\G(u,v;y) = u^{n_1+n_2-2E}\, {(y-\bz)f(\bz) - (y-z) f(z) \over z-\bz}
+ (y-z)(y-\bz)\, \K(u,v;y) \, ,
}
where $\K(u,v;y)$ is undetermined and contains the dynamical part of the correlation function.
If $\G(u,v;y)$ is a polynomial of
degree $2E$ in $y$ then clearly $\K$ is a polynomial of degree $2E-2$.

\subsec{OPE Analysis}
The operator product expansion applied to this correlation function is realised
by expanding it in terms of conformal partial waves 
$\G_\Delta^{(\ell)}(u,v;\Delta_{21}, \Delta_{43})$, which represent the contribution to a four point function for
four scalar fields, with scale dimensions $\Delta_i$, from an operator of
scale dimension $\Delta$ and spin $\ell$, and all its conformal descendants.
Explicit expressions, in four dimensions, were found in \Dos\
which are simple in terms of the variables
$x,\zz$ defined in \eiguv. For the definition refer to \ope.

For this case the expansion is also over the contributions for differing $SU(2)_R$
$R$-representations and has the form, if $n_1 \ge E$,
\eqn\OPEG{
\G(u,v;y) = \sum_{R=n_4-n_3}^{n_1+n_2} \sum_{\Delta,\ell} a_{R,\Delta,\ell} \, 
P_{R+n_3-n_4}^{(2n_1-2E, 2n_2-2E)}(y)
\ \G_\Delta^{(\ell)}(u,v;2(n_2-n_1), 2(n_4-n_3) ) \, ,
}
with $P_n^{(a,b)}$ a Jacobi polynomial. For $a$ a negative integer
$P_n^{(a,b)}(y) \propto (1-y)^{-a}$ and $n+a \ge 0$. Hence when $n_1<E$ we
require a similar expansion to \OPEG\ but with $R=n_2-n_1, \dots , n_1+n_2$
and then $\G(u,v;y) \propto \mun^{E- n_1}$ as required in \exG\ to avoid negative
powers of $t_1 {\cdot t}_4$. The different terms appearing in the sum in \OPEG\ 
then determine the necessary spectrum of operators required by this correlation 
function. The symmetry properties of this operator product expansion follow
from \OPEG\ and $P_n^{(a,b)}(y)=(-1)^n P^{(b,a)}_n(-y)$.

We first consider the case when $n_1=n_2=n_3=n_4=n$, so that $E=n$.
To apply \Gf\ we first consider the expansion in terms of Legendre polynomials
(to which the Jacobi polynomial reduce in this case),
\eqn\Gexp{
\G(u,v;y) = \sum_{R=0}^{2n} a_R(u,v)  P_R(y) \, , \qquad 
\K(u,v;y) = \sum_{R=0}^{2n-2} A_R(u,v)  P_R(y) \, .
}
The $P_R(y)$ in \Gexp\ correspond to the $2n+1$ possible $SU(2)_R$ 
invariants for the four point function \fourp\ and, as a consequence of 
results in appendix B, the coefficients  $a_R$ represent the contribution
to the correlation function from operators belonging just to the $SU(2)_R$ 
$R$-representation in the operator product  expansion for 
$\vphi^{(n)} (x_1,t_1) \, \vphi^{(n)}(x_2,t_2)$.

From \Gsol\ it is easy to
see that the single variable function $f$ involves terms linear in $y$ and
so contributes only  for $R=0,1$  giving
\eqn\af{
a^f_0 = { z f(z) - \bz f(\bz ) \over z - \bz} \, ,
\qquad a^f_1 = - { f(z) - f(\bz ) \over z - \bz} \, .
}

Using the expansion in \Gexp\ for $\K$ in \Gsol\ and standard recurrence
relations for Legendre polynomials gives corresponding expressions for 
$a_R$. For the terms involving $A_R$ we have
\eqn\asol{\eqalign{
a^{A_R}_{R+2} = {}& {(R+1)(R+2) \over (2R+1)(2R+3)}\, A_{R} \, , \qquad
a^{A_R}_{R-2} =  {(R-1)R \over (2R-1)(2R+1)}\, A_{R} \, , \cr
a^{A_R}_{R+1} = {}& -{2(R+1) \over 2R+1}\, {1-v \over u} \, A_{R} \, , \qquad
a^{A_R}_{R-1} = -{2R \over 2R+1}\, {1-v \over u} \, A_{R} \, , \cr
a^{A_R}_{R} = {}& \bigg ( 2{1+v\over u} - {1\over 2} 
+ {1 \over 2(2R-1)(2R+3)}\bigg )  A_{R} \, . \cr}
}
For $R\ge 2$ $a_R$ is therefore given in terms $A_{R\pm 2},A_{R\pm 1},A_R$
while for $R=0,1$, with \af, we have
\eqn\asoln{\eqalign{
a_0 ={}&  a_0^f + \Big ( 2\, {1+v\over u} - 
{2\over 3}\Big ) A_0 - {{2\over 3}}\, {1-v\over u}\, A_1 
+ {{2\over 15}}\,  A_2 \, , \cr 
a_1 = {}&  a_1^f  - 2{1-v\over u}\, A_0 +
\Big (2\, {1+v\over u} - {2\over 5}\Big ) A_1 - {{4\over 5}}\, {1-v \over u}\, 
A_2 + {{6\over 35}} \, A_3 \, . \cr}
}
In \asol\ and \asoln\ any contributions involving $A_R$ for $R>2n-2$
should be dropped.

The significance of the results given by \asol\ and \asoln\ is that they
correspond exactly to the $\N=2$ supermultiplet structure of operators
appearing in the operator product expansion. Each $a_R(u,v)$ may then be expanded
in terms of $\G_\Delta^{(\ell)}(u,v)\equiv \G_\Delta^{(\ell)}(u,v;0,0)$
\eqn\pwave{
a_R(u,v) = \sum_{\Delta,\ell} b_{R,\Delta,\ell} \, 
\G_\Delta^{(\ell)}(u,v) \, .
}
The conformal partial waves $\G_\Delta^{(\ell)}(u,v)$ satisfy crucial recurrence 
relations \scft,
\eqnn\recurG
$$\eqalignno{
- 2\,{1-v \over u} \, \G_{\Delta}^{(\ell)}(u,v) = {}& 
4 \, \G_{\Delta-1}^{(\ell+1)}(u,v) + \G_{\Delta-1}^{(\ell-1)}(u,v)
+ a_s \, \G_{\Delta+1}^{(\ell+1)}(u,v) 
+ \quar a_{t-1} \, \G_{\Delta-1}^{(\ell+1)}(u,v) \, , \cr
\Big ( 2\, {1+v \over u} - 1 \Big ) \, \G_{\Delta}^{(\ell)}(u,v) 
= {}&  4 \, \G_{\Delta-2}^{(\ell)}(u,v) + 4a_s \, \G_{\Delta}^{(\ell+2)}(u,v)
+ \quar a_{t-1} \, \G_{\Delta}^{(\ell-2)}(u,v)  \cr
\noalign{\vskip -4pt}
&\qquad {} + \quar a_s a_{t-1} \, \G_{\Delta+2}^{(\ell)}(u,v) \, , &\recurG \cr}
$$
where
\eqn\defst{
s = \half ( \Delta + \ell ) \, , \quad t = \half ( \Delta - \ell ) \, ,
\qquad a_s = {s^2 \over (2s-1)(2s+1)} \, .
}
In \recurG\ $a_{t-1}>0$ if $\Delta > \ell +3$.

\subsec{Long Operators}

If $A_R$ is restricted to a single partial wave so that
\eqn\azG{
A_R \to \G_{\Delta+2}^{(\ell)} \, .
}
then, using \recurG\ with \asol, 
\eqnn\longR
$$\eqalignno{
& \hskip 3cma^{A_R}_{R'} \to a_{R'}^{\vphantom g}
\big (\A^\Delta_{R,\ell} \big)  = \sum_{(\Delta',\ell')} b_{(\Delta';\ell')}
\, \G^{(\ell')}_{\Delta'}\, ,\cr & |R'-R|=2 , \  (\Delta';\ell') =
(\Delta+2;\ell) \, , \ \ |R'-R|=1 , \  (\Delta';\ell') =
(\Delta+3,\Delta+1;\ell\pm 1 ) \, , \cr & R'-R=0 , \  (\Delta';\ell') =
(\Delta+4,\Delta;\ell), \,  (\Delta';\ell') = (\Delta+2;\ell\pm2, \ell) \,
. &\longR \cr} 
$$ 
This gives exactly the expected
contributions corresponding to those operators present in 
a long $\N=2$ supermultiplet, which we may denote $\A^\Delta_{R,\ell}$, 
whose lowest dimension operator has dimension $\Delta$, spin $\ell$ belonging 
to the $SU(2)_R$ $R$-representation. From \asol\ and the positivity 
constraints for \recurG\ we may then easily see  that in \pwave\
$b_{(\Delta',\ell')} >0$ for $\Delta>\ell+1$.
For a unitary representation, so that all states in
$\A^\Delta_{R,\ell}$ have positive norm, (we consider here multiplets whose
$U(1)_R$ charge is zero) the requirement is
\eqn\ineq{
\Delta \ge  2 R + \ell + 2\, .
}
Since $\G_{\Delta}^{(\ell)}(u,v) = u^{{1\over 2}(\Delta-\ell)}F(u,v)$
with $F(u,v)$ expressible as a power series in $u,1-v$ we must have
from \azG\ for $u\sim 0$,
\eqn\unit{
A_R(u,v) \sim u^{R+2+\ep} \, , \qquad \ep \ge 0 \, .
}

\subsec{Semishort Operators}

The contribution of the single variable function $f$ \af\ represents 
operators just with twist $\Delta- \ell =2$. From the results in \Dos\
we have
\eqn\ttwo{
\G_{\ell+2}^{(\ell)}(u,v) = u \, { g_{\ell+1}(x) - g_{\ell+1}(\zz) \over x-\zz} 
= - 2 \, { g_{\ell+1}(x) - g_{\ell+1}(\zz) \over z-\bz}\, ,
}
for
\eqn\gell{
g_\ell(x)= (-\half x)^{\ell-1} x F(\ell,\ell;2\ell;x) = -{2 \over z^{\ell}}\, 
F \Big ( \half \ell, \half \ell+\half ; \ell+\half ;{1\over z^2} \Big ) \, , 
}
where $F$ is just an ordinary hypergeometric function\foot{$g_\ell(x)
\propto Q_{\ell-1}(z)$ with $Q_\nu$ an associated Legendre function.}. As shown 
in \scft\ $g_\ell$ satisfies
\eqn\recurg{
z \, g_\ell(x) = - g_{\ell-1} (x) - a_\ell \, g_{\ell+1}(x) \, .
}
In general we therefore expand the single variable function $f$ in \af\ in the
form
\eqn\expaf{
f(z) = \sum_{\ell=0}^\infty b_\ell \, g_\ell(x) \, .
}
For this to be possible $f(z)$ must be analytic in $1/z$, or equivalently in $x$.
If we consider just $f\to 2 g_{\ell+2}$ and use \recurg\ in \af\
then $a_R^f \to a_R(\C_{0,\ell})$ where 
\eqn\afl{
a_1(\C_{0,\ell}) = \G_{\ell+3}^{(\ell+1)} \, , \qquad
a_0(\C_{0,\ell}) = \G_{\ell+2}^{(\ell)} + 
a_{\ell+2} \, \G_{\ell+4}^{(\ell+2)} \, .
}
These results for $a_0,a_1$ then correspond to the contributions of operators
belonging to a semi-short $\N=2$ supermultiplet $\C_{0,\ell}$ whose lowest
dimension operator is a $SU(2)_R$ singlet with spin $\ell$ and
$\Delta = \ell+2$, i.e. at the unitarity threshold \ineq.

In general we denote by $\C_{R,\ell}$ the semi-short multiplet
whose lowest dimension operator has spin $\ell$, belongs to the 
representation $R$, and has $\Delta=2R+\ell$, so that the bound \ineq\
is saturated. At the unitarity threshold given by \ineq\ a long 
multiplet $\A^\Delta_{R,\ell}$ may be decomposed into
two semi-short supermultiplets $\C_{R,\ell}$ and $\C_{R+1,\ell-1}$, 
\short. This is reflected in the contributions to the four point function
since, with $a_{R'}^{\vphantom g}(\A^\Delta_{R,\ell})$ defined by \azG\ 
and \longR,
\eqn\ACC{
a_{R'}(\A^{2 R + \ell + 2}_{R,\ell}) = 
4 \, a_{R'}(\C_{R,\ell}) + {R+1 \over 2R+1} \,a_{R'}(\C_{R+1,\ell-1}) \, .
}
where we take
\eqn\sem{\eqalign{
a_R({\C_{R,\ell}}) = {}& \G_{2R+\ell+2}^{(\ell)}
+ \quar a_R\, \G_{2R+\ell+4}^{(\ell)} 
+ a_{R+\ell+2} \,\G_{2R+\ell+4}^{(\ell+2)} \, , \cr
a_{R-1}({\C_{R,\ell}}) = {}& {R\over 2R+1} \Big \{ \G_{2R+\ell+3}^{(\ell+1)} 
+ \quar \, \G_{2R+\ell+3}^{(\ell- 1)} +
\quar a_{R+\ell+2} \, \G_{2R+\ell+5}^{(\ell+1)} \Big \} \, ,\cr
a_{R-2}({\C_{R,\ell}}) = {}& 
{(R-1)R \over 4(2R-1)(2R+1)}\, \G_{2R+\ell+4}^{(\ell)} \, , \cr
a_{R+1}({\C_{R,\ell}}) = {}& 
{ R+1\over 2R+1}\, \G_{2R+\ell+3}^{(\ell+1)} \, .  \cr}
}
For $R=0$ \sem\  coincides with \afl. Thus the contribution of any
semi-short supermultiplet $\C_{R,\ell}$, $R=0,1,\dots ,2n-1$, to the 
four point function may be obtained by combining the results for 
long supermultiplets at unitarity threshold with \afl. There is no
reason why any particular  $\C_{R,\ell}$, except $\C_{0,0}$ which 
contains the energy momentum tensor and the conserved $SU(2)_R$ current,
should be present but if $f(z)$ is non zero it is necessary for there 
to be at least one semi-short contribution involving operators with 
protected dimensions.

\subsec{Short Operators}

A special case arises if we set $\ell=-1$ in the result for semishort operators. Formally, as shown in \short,
$\C_{R,-1} \simeq \B_{R+1}$ where $\B_R$ denotes the short supermultiplet
whose lowest dimension operator belongs to the $R$-representation with
$\Delta=2R, \, \ell=0$, obeying the full $\N=2$ shortening conditions.
The conformal partial waves as shown in \scft\ satisfy
\eqn\Gell{
(\quar )^{\ell-1} \G^{(-\ell)}_\Delta = - \G^{(\ell-2)}_\Delta \, , \qquad
\G^{(-1)}_\Delta = 0 \, ,
}
and hence from \sem\ we have
\eqn\CB{
a_{R'}(\C_{R,-1}) = {R+1 \over 2R+1} \, a_{R'} ( \B_{R+1} ) \, ,
}
where
\eqn\aBR{\eqalign{
a_R(\B_R) = {}& \G_{2R}^{(0)} \, , \qquad
a_{R-1}(\B_R) = {R\over 2R+1} \, \G_{2R+1}^{(1)} \, , \cr
a_{R-2}(\B_R) = {}& {(R-1)R\over 4(2R-1)(2R+1)} \, \G_{2R+2}^{(0)} \, . \cr}
}
The operators whose contributions appear in \aBR\ are just those
expected for the short supermultiplet $\B_R$ and there are possible 
contributions to the four point function for $R=1,2,\dots ,2n$. Since
\eqn\Gzero{
\G^{(0)}_0 (u,v) = 1 \, ,
}
then it is easy to see from \aBR\ that
\eqn\azero{
a_R(\B_0) = a_R(\I) = \de_{R0} \, ,
}
where $ a_R(\I)$ denotes the contribution of the identity in the operator
product expansion. Besides \CB\ we may also note that
\eqn\CBtwo{
a_{R'}(\C_{R,-2}) = -4 \, a_{R'} ( \B_R ) \, .
}
For $R=0$ this is in accord with \azero\ since $\G^{(-2)}_0 = -4$.

\subsec{(Next-to-) Extremal Correlators}

Apart from the case of the correlation function for four identical operators
as considered  above there are other solutions of the superconformal Ward 
identities which are of interest corresponding to extremal and next-to-extremal
correlation functions \Ext. The extremal case corresponds to taking $E=0$ in \inn.
There is then a unique $SU(2)_R$  invariant coupling which also follows from the
requirement that $\F$ in \exF, or $\G$ where in \GF\
$G(u,v;t) = (t_1 {\cdot t_4})^{n_1}(t_2 {\cdot t_4})^{n_2} 
(t_3 {\cdot t_4})^{n_3} \G(u,v)$, must be independent of $\lam,\mun$ 
and hence equivalently also of $\alpha$. In this case the result 
\resF\ for $\pr_x$ and its conjugate for $\pr_\zz$ simply imply
\eqn\Gext{
\G(u,v) = C \, u^{n_1+n_2} \, ,
}
where $C$ is independent of both $x,\zz$ and thus a constant. 
To interpret this in terms of the operator
product expansion for $\vphi^{(n_1)} (x_1,t_1) \vphi^{(n_2)}(x_2,t_2)$
we may use the result from \Dos,
\eqn\Gextr{
\G_{\Delta_1+\Delta_2}^{(0)}(u,v; \Delta_{21}, \Delta_1 + \Delta_2 ) 
= u^{{1\over 2}(\Delta_1+\Delta_2)} \, .
}
The result \Gext\ then shows that the only operators contributing
to the operator product expansion in the extremal case
have $\Delta=2(n_1+n_2), \ \ell=0$ and necessarily $R=n_1+n_2$. Such
operators can only be found as the lowest dimension operator
in the short supermultiplet $\B_{n_1+n_2}$.

For the next-to-extremal case we set $E=0$ in \inc. The solution 
of \resF\ can be conveniently expressed as
\eqn\next{
( 1 - z  )\, \G  ( u,v;z  ) = u^{n_1+n_2-1} \, f(\bz) \, ,
}
where in \GF\ we have
$G(u,v;t) = (t_1 {\cdot t_4})^{n_1}(t_2 {\cdot t_4})^{n_2-1}
(t_3 {\cdot t_4})^{n_3-1}\, t_2 \, {\cdot \, t_3 \times t_4} \,
\G(u,v;y)$. For $E=0$, $(1-y)\G(u,v;y)$ is linear in $y$ and from
\next\ and its conjugate we may find
\eqn\nextG{
(1-y)\G(u,v;y) = u^{n_1+n_2-1} \, {(y-\bz)f(\bz) - (y-z)f(z)
\over z - \bz} \, ,
}
so that $\G$ is determined just by the single variable function $g$
in this case. Since $\K$ does not appear this implies that no dynamical 
information is contained in the correlation function confirming the nonrenormalization property.

For the next-to-extremal correlation function there are just two 
independent $SU(2)_R$ invariant couplings.
In a similar fashion to \Gexp, we have an expansion, from appendix B,
in terms of two Jacobi polynomials
\eqn\Gextexp{
(1-y)\, \G(u,v;y) = a_{n_1+n_2-1}(u,v)\, P_0^{(2n_1-1,2n_2-1)}(y)
+  a_{n_1+n_2}(u,v)\, P_1^{(2n_1-1,2n_2-1)}(y) \, ,
}
where $a_R$, $R=n_1+n_2-1, \, n_1+n_2$ represent the contribution of 
the two possible $R$-representations of $SU(2)_R$ in this case. From
\nextG\ we obtain
\eqn\Solg{\eqalign{
a_{n_1+n_2-1} = {}& - {1\over n_1+n_2} \, u^{n_1+n_2} \, 
{ f(z) - f(\bz) \over z - \bz} \, , \cr
a_{n_1+n_2} = {}& u^{n_1+n_2} \bigg ( { z f(z)
- \bz f(\bz ) \over z - \bz } 
+ {n_1-n_2 \over n_1+n_2} \, { f(z) - f(\bz) \over z - \bz} \bigg )
\, . \cr}
}
To interpret this in terms of the operator product expansion we may
use, extending \ttwo,
\eqn\text{\eqalign{
& \G_{\Delta_1+\Delta_2+\ell}^{(\ell)}(u,v;\Delta_{21}, \Delta_1+\Delta_2+2 )
= u^{{1\over 2}(\Delta_1+\Delta_2)}\, {g_{\ell+1}(x;\Delta_1,\Delta_2) - 
g_{\ell+1}(\zz;\Delta_1,\Delta_2) \over x-\zz} \, ,\cr
& g_\ell(x;\Delta_1,\Delta_2 )= (-\half x)^{\ell-1} x F(\ell+\Delta_2 -1 ,\ell;
2\ell+ \Delta_1+\Delta_2 - 2 ;x) \, . \cr}
}
In consequence only operators with twist $\Delta_1+\Delta_2$ can contribute for 
the solution for $a_R$ given by \Solg. If in \Solg\ let we $f(z) \to 2 
g_{\ell+2}(x;2n_1,2n_2)$ then 
$a_R \to a_R(\C_{n_1+n_2-1,\ell})$ where
\eqn\semin{\eqalign{
a_{n_1+n_2}& (\C_{n_1+n_2-1,\ell}) = {1\over n_1+n_2} \,
\G^{(\ell+1)}_{2n_1+2n_2+\ell+1} \, , \cr
a_{n_1+n_2-1}& (\C_{n_1+n_2-1,\ell}) \cr
= {}& \G^{(\ell)}_{2n_1+2n_2+\ell}
+ (n_2-n_1){(\ell+1)(2n_1+2n_2+\ell)\over (n_1+n_2+\ell)
(n_1+n_2+\ell+1)(n_1+n_2)}\, \G^{(\ell+1)}_{2n_1+2n_2+\ell+1} \cr
&{}+ {(\ell+2)(2n_1+\ell+1)(2n_2+\ell+1)(2n_1+2n_2+\ell)\over 
(n_1+n_2+\ell+1)^2(2n_1+2n_2+2\ell+1)(2n_1+2n_2+2\ell+3)}\, 
\G^{(\ell+2)}_{2n_1+2n_2+\ell+2} \, , \cr}
}
where $\G^{(\ell')}_{2n_1+2n_2+\ell'}$ are as in \text\ with $\Delta_i \to 2n_i$.
The contributions appearing in \semin\ correspond to those expected  from the
semi-short supermultiplet $\C_{n_1+n_2-1,\ell}$. Using $\C_{R,-1} \simeq \B_{R+1}$
again we may obtain the contribution for the short multiplet $\B_{n_1+n_2}$,
\eqn\ss{
a_{R}(\C_{n_1+n_2-1,{-1}}) = {1\over n_1+n_2} \, a_{R}(\B_{n_1+n_2})\,,
}
giving
\eqn\ash{\eqalign{
a_{n_1+n_2}(\B_{n_1+n_2}) = {}& \G^{(0)}_{2n_1+2n_2} \, , \cr
a_{n_1+n_2-1}(\B_{n_1+n_2}) = {}& {4n_1n_2 \over (n_1+n_2)(2n_1+2n_2+1)}\,
\G^{(1)}_{2n_1+2n_2+1} \, . \cr}
}
For the next-to-extremal correlation function therefore only the protected
short and semi-short supermultiplets $\B_R$ and $\C_{R-1,\ell}$ can
contribute to the operator product expansion.

\subsec{Summary}

By analysis \refs{\Non,\bpsN} of three point functions the 
possible $\N=2$ supermultiplets which may appear in the operator product
expansion of two $\N=2$ short supermultiplets is determined by the 
decomposition, for $n_2\ge n_1$,
\eqn\decomp{
\B_{n_1} \otimes \B_{n_2} \simeq \bigoplus_{n=n_2-n_1}^{n_2+n_1} \B_n
\oplus \bigoplus_{\ell\ge 0} \bigg (\bigoplus_{n=n_2-n_1}^{n_2+n_1-1} 
\C_{n,\ell} 
\oplus \bigoplus_{n=n_2-n_1}^{n_2+n_1-2} \A^\Delta_{n,\ell}\bigg ) \, ,
}
where for $\A^\Delta_{n,\ell}$ all $\Delta > 2n+\ell+2$ is allowed.
By considering also the corresponding result for $\B_{n_3}\otimes\B_{n_4}$
in all cases discussed above the general solution of the 
$\N=2$ superconformal identities accommodates all possible $\N=2$ 
supermultiplets which may contribute to the four point function in the
operator product expansion according to \decomp. In the
extremal case it is clear that only $\B_{n_1+n_2}$ contributes while
for the next-to-extremal case long multiplets which undergo
renormalisation are also excluded.
\vfill
\eject

\newsec{$\N=4$}

\subsec{Superconformal Ward Identities}

We here describe an analysis of the superconformal Ward identities
for the four point function of $\N=4$ chiral primary operators
belonging to the $SU(4)_R$ $[0,p,0]$ representation with $\Delta=p$
represented by symmetric traceless fields
$\vphi_{r_1, \dots , r_p}(x)$, $r_i=1,\dots, 6$. As in \tphi\ we 
define $\vphi^{(p)}(x,t)$,  homogeneous of degree $p$ in $t$, in terms
of a six dimensional null vector $t_r$. The superconformal transformation
of $\vphi^{(p)}(x,t)$ is then expressible in the form
\eqn\susp{
\de \vphi^{(p)}(x,t) = - \hep(x) \, \gamma {\cdot t} \, 
\psi^{(p-1)}(x,t) + \bpsi {}^{(p-1)}(x,t)\, \bga {\cdot t} \,\hbep(x) \, ,
}
where the conformal Killing spinors $\hep^\alpha_i(x), \,
\hbep{}^{i\dal}(x)$ are as in \scep, with $i=1,2,3,4$ and 
$\ga_r{}^{\! ij} = - \ga_r{}^{\! ji}, \, \bga_{rij} = \half \vep_{ijkl} 
\ga_r{}^{\! kl}$ are $SU(4)$ gamma matrices, 
$\ga_r \bga_s + \ga_s \bga_r = - 2 \de_{rs} 1$, $\gamma_r{\!}^\dagger =
- \bga_r$. In \susp\ $\psi^{(p-1)}{}_{i\alpha}(x,t)$, 
$\bpsi{}^{(p-1)}{}^i{}_\dal(x,t)$ are homogeneous spinor fields of degree 
$p-1$ in $t$ and satisfy constraints similar to \condit
\eqn\conp{
\gamma {\cdot {\pr \over \pr t}} \, \psi^{(p-1)}_\alpha (x,t) = 0 \, , 
\qquad \bpsi{}^{\,(p-1)}(x,t)\, 
\bga {\cdot {\overleftarrow{\pr \over \pr t}}} =0 \, ,
}
which are necessary for them to belong to $SU(4)_R$ representations
$[0,p-1,1], \, [1,p-1,0]$.
At the next level the superconformal transformations involve a current
belonging to the $[1,p-1,1]$ representation which corresponds to
a homogeneous field of degree $p-1$ with one $SU(4)_R$ vector index
$J^{(p-1)}{}_{r\alpha\dal}(x,t)$ satisfying
\eqn\conJ{
t_r J^{(p-1)}{}_{r\alpha\dal}(x,t) = 0 \, , \qquad
{\pr \over \pr t_r } J^{(p-1)}{}_{r\alpha\dal}(x,t) = 0 \, .
}
The superconformal transformation of $\psi^{(p-1)}(x,t)$, neglecting
$\hep$ terms, is then
\eqn\varp{\eqalign{
\de \psi^{(p-1)}_\alpha (x,t) = {}& {1\over p} \,
\bga {\cdot {\pr \over \pr t}}\, i \pr_{\alpha\dal} \vphi^{(p)}(x,t) \,
\hbep{}^{\, \dal} (x)
+ 2 \, \bga {\cdot {\pr \over \pr t}} \vphi^{(p)}(x,t)\, \eta_\alpha \cr
&{}+ \bigg ( 1 + {1\over 2p+2}\,
\bga{\cdot t} \, \gamma{\cdot {\pr \over \pr t}} \bigg )
J^{(p-1)}{}_{r \alpha\dal}(x,t) \, \bga_r \hbep{}^{\, \dal}(x) \, . \cr}
}
Superconformal transformations which generate the full BPS multiplet
listed in \Class\ can be obtained similarly to \scft\ but the superconformal
Ward identities depend only on \susp\ and \varp. 

The general four point function of chiral primary operators can be
written in an identical form to \Fourp,
\eqn\Fourpf{\eqalign{
\l \vphi^{(p_1)}& (x_1,t_1) \, \vphi^{(p_2)}(x_2,t_2)\,
\vphi^{(p_3)}(x_3,t_3)\, \vphi^{(p_4)}(x_4,t_4 )\r \cr
&{} = { r_{23}^{\, \, \Sigma - p_2 - p_3}  \,
r_{34}^{\, \, \Sigma - p_3 - p_4}\over
r_{13}^{\, \, p_1}\,  r_{24}^{\, \, \Sigma - p_3} }\, F(u,v;t) \, ,
\quad \Sigma = \half ( p_1+p_2+p_3+p_4 )  \, . \cr}
}
The derivation of superconformal Ward identities initially follows an
almost identical path as that in chapter 2 leading to \redeq{a,b}.
With similar definitions to \fourpsi, \fourJ, taking $2n_i\to p_i$, and 
\hatRJ\ we find
\eqna\redeqf
$$\eqalignno{
{ 1\over p_1} \, \bga{\cdot {\pr \over \pr t_1}} \, {\pr \over \pr x} F
= {}& - \bigg ( 1  + {1\over 2p_1+2 }\,
\bga{\cdot t_1} \, \gamma {\cdot {\pr \over \pr t_1}} \bigg ) \bga{\cdot K}
- {1\over x} \, {T_2}\, \bga{\cdot t_2}
+ {1\over 1-x } \, {T_4}\, \bga{\cdot t_4} \, , & \redeqf{a} \cr
\bga{\cdot {\pr \over \pr t_1}} F = {}& {T_2}\, \bga{\cdot t_2}
+ {T_3}\, \bga{\cdot t_3} + {T_4}\, \bga{\cdot t_4}  \, .
& \redeqf{b} \cr}
$$
Instead of \cons\ we have the constraints, which follow from \conp\ and \conJ,
\eqn\consa{
\gamma{\cdot {\pr \over \pr t_1}} \, {T_i} = 0 \, , \quad
{T_i} \, \bga{\cdot {\overleftarrow{\pr \over \pr t_i}}} =0 \, , \qquad
t_1 {\cdot K} = {\pr \over \pr t_1}  {\cdot K} = 0 \, . 
}

As with \UVS\ we exhibit the dependence on $SU(4)$ gamma matrices by
writing
\eqn\TV{
T_i \, \bga{\cdot t_i} = \bga {\cdot V_i} + {\ts {1\over 6}} \,
\bga_{[r} \gamma_s \bga_{u]} W_{i,rsu} \, .
}
Since we take\foot{Note that $(\gamma_1 \bga_2 \gamma_3 \bga_4
\gamma_5 \bga_6)^\dagger = - \gamma_1 \bga_2 \gamma_3 \bga_4
\gamma_5 \bga_6$.} $\gamma_{[r}\bga_s\gamma_u\bga_v\gamma_w \bga_{z]} =
i \vep_{rsuvwz}$,
\eqn\seld{
 \bga_{[r} \gamma_s \bga_{u]} = - {\ts {1\over 6}} \, 
i \vep_{rsuvwz} \bga_v \gamma_w \bga_z \, ,
}
so that we must require the self-duality condition
\eqn\sdual{
W_{i,rsu} = {\ts {1\over 6}} \, i \vep_{rsuvwz} W_{i,vwz} \, .
}
Imposing the first equation in \consa\ we have
\eqn\consis{
{\pr \over \pr t_1} {\cdot V_i} = 0 \, , \qquad
\pr_{1[r} V_{i,s]} = \pr_{1u} W_{i,rsu} \, .
}
Just as in \VU\ we write,
\eqn\VUp{
V_i = {1\over p_1}\, {\pr \over \pr t_1} U_i + \hV_i \, , \qquad
t_1 {\cdot V_i} = U_i \, , \quad t_1 {\cdot \hV_i} = 0 \, .
}
so that in \consis\ we may let $V_i \to \hV_i$.

Using \TV\ with \VUp, and $\bga{\cdot t_1} \gamma {\cdot \pr_1} 
\bga{\cdot K} = - p_1 \, \bga{\cdot K} + \half 
\bga_{[r} \gamma_s \bga_{u]} L_{1,rs} K_u$,
\redeqf{a,b} may be decomposed into three pairs of equations,
\eqn\FU{
{\pr \over \pr x} F = -{1\over x}\, U_2 + {1\over 1-x}\, U_4 \, , \qquad
\qquad \  p_1 F = \sum_{i=2}^4 U_i  \, ,
}
and
\eqn\KV{
{p_1+2 \over 2p_1 +2} \, K_r = -{1\over x}\, \hV_{2,r} + {1\over 1-x}\, 
\hV_{4,r}\, , \qquad \sum_{i=2}^4 \hV_{i,r} = 0 \, ,
}
and
\eqn\KS{
{3 \over 2p_1 +2} \, \big ( L_{1,[rs} K_{u]} \big )_{\rm sd} 
= -{1\over x}\, W_{2,rsu} + {1\over 1-x}\, W_{4,rsu} \, , \qquad
\sum_{i=2}^4 W_{i,rsu} = 0 \, ,
}
where we define for $i=1,2,3,4$
\eqn\defL{
L_{i,rs} = t_{ir} \pr_{is} - t_{is} \pr_{ir} \, ,
}
and for any $X_{rsu} = X_{[rsu]}$ the self dual part, satisfying \sdual, is given by
\eqn\Xsd{
\big ( X_{rsu} \big )_{\rm sd} = \half \, X_{rsu} + {\ts {1\over 12}} \,
i \vep_{rsuvwz} X_{vwz} \, .
}

Since $2t_{1s} \pr_{1[r} V_{i,s]} = - p_1 \hV_{i,r}$ we may obtain from 
\consis\ 
\eqn\VT{
p_1 \hV_{i,r} = -L_{1,su}W_{i,rsu} \, ,
}
which gives $\hV_{i,r}$ in terms of $W_{i,rsu}$.
Furthermore from \consa\ $L_{1,rs} K_s = K_r$ and using also, as a 
consequence of the commutation relations for $L_1$, $[L_{1,rs},L_{1,ru}] 
= - 4 L_{1,su}$ we have $L_{1,rs}L_{1,ru} K_s = 3 K_u$. With, in addition,  
$\half L_{1,rs}L_{1,rs}\, K_u = - (p_1-1)(p_1+3) K_u$ we may then obtain
\eqn\LLK{
3L_{1,rs}L_{1,[rs} K_{u]} = - 2p_1(p_1+2) K_u \, .
}
Since also $\vep_{rsuvwz} L_{1,su}L_{1,vw} K_z = 0$ it is clear from \VT\
and \LLK\ that eqs. \KS\ imply \KV. However, if we define 
\eqn\Wbar{
{\bar W}_{i,rsu} = 3\big ( L_{1,[rs} \hV_{i,u]} \big )_{\rm sd} - (p_1+2)
\, W_{i,rsu} \, ,
}
with $\hV_{i,u}$ determined by \VT, then as a consequence of
\KV\ and \KS\ we must also require
\eqn\KW{
{1\over x}\, {\bar W}_{2,rsu} = {1\over 1- x} \, {\bar W}_{4,rsu} \, .
}

{}From the second equation in \consa\ we may obtain
$\half L_{i,rs} (T_i \bga{\cdot t_i}) \gamma_r \bga_s 
= (p_i+4) T_i \bga{\cdot t_i}$ which leads to the relations
\eqna\LVW
$$\eqalignno{
& L_{i,rs} V_{i,s}- L_{i,su} W_{i,rsu} = (p_i+4) V_{i,r} \, , &\LVW{a} \cr
& 3 \big( L_{i,[rs} V_{i,u]} \big )_{\rm sd} 
+ 3L_{i,[r|v}W_{i,su]v} = (p_i+4) W_{i,rsu} \, , &\LVW{b} \cr}
$$
where $L_{i,[u|v} W_{i,rs]v}$ is self dual as a consequence
of \sdual. We also have from $T_i \bga{\cdot t_i} \, \gamma{\cdot t_i} = 0$
\eqn\Ttt{
t_i {\cdot V_i} = 0 \, , \qquad t_{i[r} V_{i,s]} + W_{i,rsu} t_{iu} = 0 \, .
}
For consistency we note that $\pr_{is} ( t_{i[r} V_{i,s]} + W_{i,rsu} t_{iu})
= 0$ is identical with \LVW{a}. Furthermore using \sdual\ we have
$(\pr_{i[r} \, W_{i,su]v} t_{iv})_{\rm sd} = \half \big (
\pr_{i[r} \, W_{i,su]v} t_{iv} - \pr_{i v} \, t_{i[r} W_{i,su]v} \big ) + \thir
\pr_{iv} ( W_{i,rsu} t_{iv} )$ and, from appendix A, $(p_i+2)\pr_{iv} 
( W_{i,rsu} t_{iv} ) = {(p_i+3)(p_i+4)}\, W_{i,rsu}$ while acting on $W_{i,suv}$
similarly $(p_i+2)\pr_{i[r} t_{iv]} = - (p_i+3)\, \half L_{i,rs}$. Hence we have
demonstrated that 
$\big (\pr_{i[r} ( t_{i s} V_{i,u]} + W_{i,su]v} t_{iv} \big )_{\rm sd}=0$
is identical to \LVW{b} so that this equation is also implied by \Ttt.

Combining \Ttt\ with \VUp\ gives the essential equation
\eqn\WUp{
t_{i[r} \pr_{1s]} U_i + p_1 \, t_{i[r} \hV_{i,s]} + p_1 \, W_{i,rsu} t_{iu} = 0 \, ,
}
where $p_1 \hV_{i,s}$ is determined by \VT.

As in \exF\ we may expand the correlation function $F$, as defined in
\Fourpf, in terms of $SU(4)$ invariants
\eqn\exFp{
F(u,v;t) = \big ( t_1 {\cdot t}_4 \big )^{p_1-E} \big ( t_2 {\cdot t}_4
\big )^{p_2-E} \big ( t_1 {\cdot t}_2 \big )^{E}
\big ( t_3 {\cdot t}_4 \big )^{p_3} \, \F(u,v;\lam,\mun) \, ,
}
where we assume
\eqn\ppp{
p_1 \le p_2 \le p_3 \le p_4 \, , \qquad 2E = p_1 +p_2 + p_3 - p_4 \, .
}
In \exFp\ $\F(u,v;\lam,\mun)$ is a polynomial in $\lam,\mun$ consistent
with $F(u,v;t) = {\rm O}(t_1^{p_1},t_2^{p_2},t_3^{p_3},t_4^{p_4})$ and 
hence $E\ge 0$ is a integer. For $p_1 \ge E$ then 
$\F$ is expressible as a polynomial of degree $E$ in $\lam,\mun$, i.e.
a linear expansion in the $\half (E+1)(E+2)$ independent monomials 
$\lam^p\mun^q$ with $p+q \le E$. 
For $p_1 < E$ it is necessary also that $ q \ge E - p_1$ giving
only $\half (p_1 +1)(p_1 + 2)$ independent terms. It is easy to see that 
this matches the number of invariants that may be constructed
by finding common representations in $[0,p_1,0] \otimes [0,p_2,0]$ and
$[0,p_3,0] \otimes [0,p_4,0]$ using the tensor product result
\eqn\Oprod{
[0,p_1,0] \otimes [0,p_2,0] \simeq \bigoplus_{r=0}^{p_1} 
\bigoplus_{s=0}^{p_1-r} [r,p_2-p_1+2s,r] \, .
}
Hence representations $[r,p_2-p_1+2s,r]$ may contribute for
$s=0,\dots , n-r , \ r =0, \dots , n$ with $n=E$ if $p_1 \ge E$, otherwise 
$n=p_1$.

In an exactly similar fashion to \exFp\ we may express $U_i(x,\zz;t)$
in terms of $\U_i(x,\zz;\lam,\mun)$ so that \FU\ becomes
\eqn\FUc{
{\pr \over \pr x} \F = -{1\over x}\, \U_2 + {1\over 1-x}\, \U_4 \, , \qquad
\qquad \  p_1 F = \U_2 + \U_3 + \U_4   \, .
}

Furthermore we may also decompose $W_{i,rsu}(x,\zz;t)$ for $i=2,3,4$
in terms of four independent self dual tensors,
\eqn\exWp{\eqalign{
W_{i,rsu}= - & \big ( t_1 {\cdot t}_4 \big )^{p_1-E} 
\big ( t_2 {\cdot t}_4 \big )^{p_2-E} \big ( t_1 {\cdot t}_2 \big )^{E-2}
\big ( t_3 {\cdot t}_4 \big )^{p_3-1} \, \cr
&{} \times \Big ( \big ( t_{1[r} t_{2s} t_{3u]} \big )_{\rm sd} \,
t_2 {\cdot t_4} \, \A_i
+ \big ( t_{1[r} t_{4s} t_{2u]} \big )_{\rm sd} \,
t_2 {\cdot t_3} \, \B_i \cr
&\quad {} + \big ( t_{1[r} t_{3s} t_{4u]} \big )_{\rm sd} \,
t_2 {\cdot t_3} \, t_2 {\cdot t_4} \, {1\over t_3{\cdot t_4}} \, \C_i + 
\big ( t_{2[r} t_{3s} t_{4u]} \big )_{\rm sd} \,
t_1 {\cdot t_2} \, \W_i\Big ) \, , \cr}
}
with $\A_i,\B_i,\C_i$ and $\W_i$ polynomials in $\lam,\mun$ of degree $E-2$
and $E-1$, if $p_1 \ge E$. From its definition in \TV\ we must have
\eqn\ABC{
\C_2 = \B_3 = \A_4 = 0 \, .
}
The result \KS\ then requires
\eqn\sumW{
\A_2 + \A_3 = 0 \, , \quad \B_2 + \B_4 = 0 \, , \quad \C_3 + \C_4 = 0 \, ,
\quad \W_2 + \W_3 + \W_4 = 0 \, .
}

We may similarly decompose $\hV_{i,r}$ in the form
\eqn\exVp{\eqalign{
\hV_{i,r}= {} & \big ( t_1 {\cdot t}_4 \big )^{p_1-E-1}
\big ( t_2 {\cdot t}_4 \big )^{p_2-E} \big ( t_1 {\cdot t}_2 \big )^{E-1}
\big ( t_3 {\cdot t}_4 \big )^{p_3-1} \, \cr
&{} \times \big ( ( t_{2r} \, t_1 {\cdot t_4} -  t_{4r} \, t_1 {\cdot t_2})
t_3 {\cdot t_4} \, \I_i
+ ( t_{3r} \, t_1 {\cdot t_4} -  t_{4r} \, t_1 {\cdot t_3})
t_2 {\cdot t_4} \, \J_i
+ t_{1r}\, t_2 {\cdot t_4} t_3 {\cdot t_4} \, \V_i \big ) \, , \cr}
}
where we impose $t_1{\cdot \hV_i}=0$. The coefficient of $t_{1r}$ is determined 
by the requirement $\pr_1{\cdot \hV_i}=0$,
\eqn\OV{
(p_i+2)\V_i =  - \O_\lam \I_i + \I_i  - 
(\lam \O_\lam - \mun \O_\mun)  \J_i + \J_i \, ,
}
with differential operators
\eqn\defO{\eqalign{
\O_\lam = {}& ( \lam + \mun -1) \, {\pr \over \pr \lam} + 2
\mun {\pr \over \pr \mun} + p_1 - 2E +1 \, , \cr
\O_\mun = {}& 2 \lam {\pr \over \pr \lam} + ( \lam + \mun - 1) \bigg (
{\pr \over \pr \mun} + { p_1 - E \over \mun} \bigg )  - p_1 +1 \, . \cr}
}
Using \VT\ we get
\eqn\solIJ{\eqalign{
6p_1 \, \I_i = {}& (p_1+2) ( \lam\, \A_i - \mun\, \B_i ) - 
(\lam \O_\lam - \mun \O_\mun) \W_i \, , \cr
6p_1 \, \J_i = {}& - (p_1+2) ( \A_i - \mun\, \C_i ) + \O_\lam \W_i  \, . \cr}
}
{}From \OV\ we then obtain
\eqn\VABC{
6p_1 \, \V_i = \tau \big ( (\O_\lam + 1 ) \B_i - (\O_\mun + 1 ) \A_i
- ( \lam (\O_\lam + 1 ) - \mun (\O_\mun + 1 ) ) \C_i \big ) \, .
}

As a consequence of \WUp\ the coefficients in \exFp\ are not independent but
we have  relations which determine $\A_i, \B_i , \C_i$ for each $i$,
\eqn\relABC{\eqalign{
(p_1+1)\, \A_2 = {}& 3\, {\pr \over \pr \lam} \U_2 + \half ( \O_\lam - p_1) \W_2 \, , \cr
(p_1+1)\, \B_2 = {}& - 3 \bigg ( {\pr \over \pr \mun} +{p_1 -E \over \mun} \bigg )\U_2 
+ \half ( \O_\mun - p_1) \W_2 \, , \cr
(p_1+1)\,\lam  \A_3 = {}&  3 \bigg ( \lam {\pr \over \pr \lam} + 
\mun {\pr \over \pr \mun}
- E  \bigg )\U_3 + \half ( \lam \O_\lam - \mun \O_\mun - p_1) \W_3 \, , \cr
(p_1+1)\, \lam \C_3 = {}& 3 \bigg ( {\pr \over \pr \mun} +{p_1 -E \over \mun} \bigg )\U_3 
-  \half ( \O_\mun +  p_1) \W_3 \, , \cr
(p_1+1)\,\mun  \B_4 = {}&- 3 \bigg ( \lam {\pr \over \pr \lam} + 
\mun {\pr \over \pr \mun}
- E  \bigg )\U_4 -  \half ( \lam \O_\lam - \mun \O_\mun + p_1) \W_4 \, , \cr
(p_1+1)\, \tau \C_4 = {}& - 3 \, {\pr \over \pr \lam} \U_4 - \half 
( \O_\lam + p_1) \W_4 \, . \cr}
}
Combining this with \sumW\ and also the result in \FUc\ for $p_1 \F$ leads to
\eqn\FUW{\eqalign{
\bigg ( \mun {\pr \over \pr \mun} + p_1 -E  \bigg ) \F = {}& \U_4 + {\ts {1\over 6}}
(\lam - \mun -1 ) \W_4 - {\ts {1\over 3}} \, \mun \W_2 \, , \cr
\bigg ( \lam {\pr \over \pr \lam} + \mun {\pr \over \pr \mun} - E  \bigg ) \F = {}&
- \U_2 + {\ts {1\over 6}} (\lam - \mun -1 ) \W_2 - {\ts {1\over 3}} \, \W_4 \, . \cr}
}
If ${\bar W}_{i,rsu}$ given by \Wbar\ is defined in terms of 
${\bar \A}_i, {\bar \B}_i, {\bar \C_i}$ and ${\bar \W}_i$ as in \exWp\ then
it is easy to see that ${\bar \W}_i = -(p_1+2) \W_i$ and, as a consequence of
\solIJ\ and \relABC,
\eqnn\ABCW
$$\eqalignno{
2(p_1+1){\bar \A}_i ={}&  \bigg ( {\pr \over \pr \lam} 
\big ( \lam \O_\lam - \mun \O_\mun + p_1 + 2 \big ) 
+ \Big ( \lam {\pr \over \pr \lam} + \mun {\pr \over \pr \mun} - E +1 \Big )
\big ( \O_\lam - p_1 - 2 \big ) \bigg ) \W_i \, , \cr
&{} \mun {\bar \C}_i - {\bar \A}_i = \O_\lam \W_i \, , \qquad 
\mun {\bar \B}_i - \lam {\bar \A}_i = (\lam \O_\lam - \mun \O_\mun) \W_i \, . 
& \ABCW \cr}
$$
Hence \KW\ reduces to just
\eqn\WW{
{1\over x} \, \W_2 = {1\over 1-x} \, \W_4 \, .
}

Just as  the invariants 
$u,v$ are expressed in terms of $x,\zz$ it is convenient to write
$\lam,\mun$ in a similar  form involving new variables $\alpha,\bet$,
\eqn\defab{
\lam = \alpha \bet \, , \qquad \mun = (1-\alpha)(1-\bet) \, .
}
In terms of these variables $\alpha,\bet$, \FUW\ becomes
\eqn\relF{
\alpha(1-\alpha) {\pr \over \pr \alpha} \F + E \, \alpha \F - p_1 \, \F =
- (1-\alpha) \U_2 - \U_4 + {\ts {1\over 6}} (\alpha - \bet ) \big ( (1-\alpha) \W_2
+ \W_4 \big ) \, , 
}
together with the conjugate equation obtained for $\alpha \leftrightarrow \bet$.
If this is used together with \FUc\ for $\pr_x \F$ we may eliminate $\U_2$ to
obtain
\eqn\WarF{\eqalign{
\bigg ( x {\pr \over \pr x} - \alpha {\pr \over \pr \alpha}& - E \, 
{\alpha \over 1-\alpha} + p_1 \, {1\over 1-\alpha} \bigg ) \F \cr 
= {}& \bigg ( {x \over 1-x} + {1\over 1-\alpha} \bigg ) \U_4 - {\ts {1\over 6}} 
(\alpha - \bet ) \bigg ( \W_2 + {1\over 1-\alpha} \W_4 \bigg ) \cr
= {}& {1-\alpha x \over (1-\alpha)(1-x) } \Big ( \U_4 - {\ts {1\over 6}} 
(\alpha - \bet ) \W_4 \Big ) \, , \cr}
}
where we have used \WW. Writing $\F(u,v;\lam,\mun)={\hat \F}(x,\zz;\alpha,\bet)$
evidently
\eqn\Wwp{
\bigg ( x {\pr \over \pr x} - \alpha {\pr \over \pr \alpha} + E \, + 
(p_1 - E) \, {1\over 1-\alpha} \bigg ) {\hat \F}(x,\zz;\alpha,\bet) \Big |_{
\alpha= {1\over x}}  = 0 \, ,
}
which is solved by writing
\eqn\solW{
u^E ( 1-v)^{p_1 - E} {\hat \F}\big (x,\zz;{1\over x},\bet \big ) =
f(\zz,\bet ) \, .
}
Together with the conjugate equation in which $\alpha \to \bet$ \solW\ is the
basic solution of the superconformal Ward identities in this context.

\subsec{Solution of Identities}

As previously
it is more convenient for consideration of the operator product expansion
to change from $F(u,v;t)$ to $G(u,v;t)$, defined in \Fourp. 
Writing $G(u,v;t)$ in a similar fashion to \exFp\ then 
the corresponding function $\G$ is given in terms of $\F(u,v;\lam,\mun)$ by
\eqn\GFp{
\G(u,v;\lam,\mun) = u^{{1\over 2}(p_1+p_2)}\, v^{p_1-E}\F(u,v;\lam,\mun) \, .
}
For the applications in this section it is convenient to write
\eqn\Gyy{
\G(u,v;\lam,\mun) = \hG(u,v;y,\yz)=\hG(u,v;\yz,y) \, ,
}
where $\hG$ depends on the variables
\eqn\yy{
y = 2\alpha-1 \, , \qquad \yz=2\bet-1 \, .
}
The solution \solW\ then gives, with $z,\bz$ defined in \defz,
\eqn\solid{
\hG(u,v;z,\yz) = u^{{1\over 2}(p_1+p_2)-E}f(\bz,\yz) \, , \qquad 
\hG(u,v;\bz,\yz) = u^{{1\over 2}(p_1+p_2)-E} f(z,\yz) \, . 
}
For consistency, since $f(z,z)=f(\bz,\bz)$,  we must have
\eqn\real{
f(z,z) = k \, .
}
$\G(u,v;y,\yz)$ is a symmetric polynomial in $y,\yz$ with degree $p$. Since it
must also be symmetric in $z,\bz$ \solid\ implies
\eqnn\Gsolt
$$\eqalignno{
\hG(u,v;y,\yz) = {}& -k + {(y-z)(\yz-\bz)\big ( f(z,\yz) + f(\bz,y) \big )
- (y-\bz)(\yz- z) \big ( f(z,y)+ f(\bz,\yz) \big )\over (z-\bz)(y-\yz)}\cr
&{} + (y-z)(y-\bz)(\yz-z)(\yz-\bz)\, \K(u,v;\lam,\mun) \, , & \Gsolt \cr}
$$
with $\K(u,v;\lam,\mun)={\hat \K}(u,v;y,\yz)$ defining an undetermined symmetric 
polynomial in $y,\yz$ of degree $p-2$. Since this is the part of the correlation function 
not constrained by the Ward identities, it must contain the full dynamical information.

\subsec{OPE Analysis}

In general the conformal partial wave expansion and the decomposition into
contributions for differing $SU(4)_R$ representations further into conformal
partial waves is realised by writing
for $p_1 \ge E$.
\eqn\OPEf{ \eqalign{
\hG(u,v;y,\yz) = {}& \sum_{0\le m \le n \le E} a_{nm}(u,v) \,
P_{nm}^{(p_1-E, p_2-E )}(y,\yz) \cr
= {}& \sum_{0\le m \le n \le E}\sum_{\Delta,\ell} a_{nm,\Delta,\ell} \,
P_{nm}^{(p_1-E, p_2-E )}(y,\yz) \ \G_\Delta^{(\ell)}(u,v;p_2-p_1,p_4-p_3) \, ,\cr}
}
where $\G_\Delta^{(\ell)}$ are described in the introduction in \GF\ and $P_{nm}^{(a,b)}(y,\yz)$
are symmetric polynomials of degree $n$ (i.e. for an expansion in
terms of the form $(y\yz)^s(y^t+\yz^t)$, $s+t\le n$) which are discussed in 
appendix B and which are given in terms of Jacobi polynomials
\eqn\poly{
P^{(a,b)}_{nm}(y,\yz) =
{ P^{(a,b)}_{n+1}(y) P^{(a,b)}_{m\vphantom{1}}(\yz) - 
P^{(a,b)}_{m\vphantom{1}} (y) P^{(a,b)}_{n+1} (\yz) 
\over y - \yz} = - P^{(a,b)}_{m-1\, n+1}(y,\yz)  \, ,
}
In \OPEf\  $a_{nm,\Delta,\ell}$ then corresponds to the presence of an operator 
in the operator product expansion ifor $\vphi^{(p_1)}$ and  $\vphi^{(p_2)}$ 
belonging to the $SU(4)_R$ representation with Dynkin labels 
$[n-m,p_1+p_2-2E+2m,n-m]$ and  with scale dimension $\Delta$, spin $\ell$. 
The expansion \OPEf\ also extends to $p_1 < E$ save that then $m \ge E - p_1$ 
and $P^{(a,b)}_{nm}(y,\yz) \propto \mun^{E-p_1}$.

To take account of the constraint \real\ we write
\eqn\ff{
f(z,y) = k + (y-z) \hf(z,y) \, ,
}
with $\hf(z,y)$ a free function, polynomial in $y$ of degree $p-1$. In terms of this new function $\hf(z,y)$ we can write $\hG(u,v;y,\yz)$ as

\eqnn\Gsoltt
$$\eqalignno{
\hG(u,v;y,\yz) = {}& k+{(y-z)(\yz-\bz)(\yz-z)f(z,\yz) + (y-z)(\yz-\bz)(y-\bz)f(\bz,y)\over (z-\bz)(y-\yz)}\cr
&{} -{(y-\bz)(\yz- z)(y-z)f(z,y)+(y-\bz)(\yz- z)(\yz-\bz)f(\bz,\yz)\over (z-\bz)(y-\yz)}\cr
&{} + (y-z)(y-\bz)(\yz-z)(\yz-\bz)\, \K(u,v;\lam,\mun) \, . & \Gsoltt \cr}
$$

The decomposition of $\hG(u,v;y,\yz)$ into the contributions for
different possible $SU(4)_R$ representations is given by \OPEf\
where  $a_{nm}$ are for this case the coefficients corresponding to
the representation with Dynkin labels $[n-m,2m,n-m]$. For this case
in \poly\ $P^{(0,0)}_n(y) = P_n(y)$, conventional Legendre polynomials. 

We consider the contribution of the two variable function
$\K$ in \Gsolt\ which is expanded, for 
$P_{nm}(y,\yz) \equiv P^{(0,0)}_{nm}(y,\yz)$, as
\eqn\Kexp{
{\hat \K}(u,v;y,\yz) = \sum_{0\le m \le n \le p-2} \!\! A_{nm}(u,v) \, 
P_{nm}(y,\yz) \, , 
}
with $\half(p-1)p$ terms. In this case the Legendre recurrence relations
give
\eqnn\longr
$$\eqalignno{
a_{n{-2}\, m{-2}}^{A_{n m}}={}&{(m-1) m\, n (n+1)
\over(2m-1)(2m+1)(2n+1)(2n+3)}\, A_{n m} \, ,\cr
a_{n{-2}\, m{+2}}^{A_{n m}}={}& {(m+1)(m+2)n(n+1)
\over(2m+1)(2m+3)(2n+1)(2n+3)}\, A_{n m}  \, , \cr
a_{n{+2}\, m{-2}}^{A_{n m}}={}&{(m-1)m(n+2)(n+3)
\over(2m-1)(2m+1)(2n+3)(2n+5)}\, A_{n m} \, ,\cr
a_{n{+2}\, m{+2}}^{A_{n m}}={}&{(m+1)(m+2)(n+2)(n+3)
\over(2m+1)(2m+3)(2n+3)(2n+5)}\, A_{n m} \, , \cr
a_{n{-2}\, m{-1}}^{A_{n m}}={}& - {2 m n(n+1)\over(2m+1)(2n+1)(2n+3)}\,
{1-v\over u}\, A_{n m}  \, , \cr
a_{n{-2}\, m{+1}}^{A_{n m}}={}&-{2 (m+1) n(n+1)\over (2m+1)(2n+1)(2n+3)}\,
{1-v\over u}\, A_{n m}  \, , \cr
a_{n{-1}\, m{-2}}^{A_{n m}}={}&-{2 (m-1)m(n+1)\over(2m-1)(2m+1)(2n+3)} \,
{1-v\over u}\, A_{n m}  \, , \cr
a_{n{-1}\, m{+2}}^{A_{n m}}={}&-{2(m+1)(m+2)(n+1)\over(2m+1)(2m+3)(2n+3)}\,
{1-v\over u}\, A_{n m}  \, , \cr
a_{n{+2}\, m{-1}}^{A_{n m}}={}&-{2m(n+2)(n+3)\over(2m+1)(2n+3)(2n+5)}\,
{1-v\over u}\, A_{n m}  \, , \cr
a_{n{+2}\, m{+1}}^{A_{n m}}={}&-{2(m+1)(n+2)(n+3)\over(2m+1)(2n+3)(2n+5)}\,
{1-v\over u}\, A_{n m}  \, , \cr
a_{n{+1}\, m{-2}}^{A_{n m}}={}&-{2(m-1)m(n+2)\over(2m-1)(2m+1)(2n+3)}\,
{1-v\over u}\, A_{n m}  \, , \cr
a_{n{+1}\, m{+2}}^{A_{n m}}={}&-{2(m+1)(m+2)(n+2)\over(2m+1)(2m+3)(2n+3)}\,
{1-v\over u}\, A_{n m}  \, , \cr
a_{n{-1}\, m{-1}}^{A_{n m}}={}&{4 m(n+1)\over(2m+1)(2n+3)}\,
{(1-v)^2\over u^2}\, A_{n m}  \, , \cr
a_{n{-1}\, m{+1}}^{A_{n m}}={}& {4(m+1)(n+1)\over(2m+1)(2n+3)}\,
{(1-v)^2\over u^2}\, A_{n m}  \, , \cr
a_{n{+1}\, m{-1}}^{A_{n m}}={}& {4m(n+2)\over(2m+1)(2n+3)}\,
{(1-v)^2\over u^2}\, A_{n m}  \, , \cr
a_{n{+1}\, m{+1}}^{A_{n m}}={}& {4(m+1)(n+2)\over(2m+1)(2n+3)}\,
{(1-v)^2\over u^2}\, A_{n m}  \, , \cr
a_{n{-2}\, m}^{A_{n m}}={}& {2 n(n+1)\over (2n+1)(2n+3)} \, B_m\, A_{n m}  \, ,
\qquad\ \,  a_{n{+2}\, m}^{A_{n m}}={2(n+2)(n+3)\over (2n+3)(2n+5)}
\, B_m\, A_{n m}  \, , \cr
a_{n\, m{-2}}^{A_{n m}}={}& {2 (m-1)m \over(2m-1)(2m+1)} \, 
B_{n+1}\, A_{n m} \, ,
\quad\   a_{n\, m{+2}}^{A_{n m}}={2(m+1)(m+2)\over(2m+1)(2m+3)} \, 
B_{n+1}\, A_{n m}  \, ,\cr
a_{n{-1}\, m}^{A_{n m}}={}&-{4(n+1)\over 2n+3} \, B_m {1-v\over u}\, 
A_{n m}  \, , \qquad \ \, a_{n{+1}\, m}^{A_{n m}}=-{4(n+2)\over 2n+3}\, 
B_m {1-v\over u}\, A_{n m}  \, , \cr
a_{n\, m{-1}}^{A_{n m}}={}&-{4 m\over 2m+1} \, B_{n+1} {1-v\over u}\, 
A_{n m}  \, , \qquad a_{n\, m{+1}}^{A_{n m}}= -{4(m+1)\over 2m+1} 
\, B_{n+1} {1-v\over u}\, A_{n m}  \, , \cr
a_{n\, m}^{A_{n m}}={}& 4 B_m B_{n+1} \, A_{n m}  \, , & \longr \cr}
$$
where
\eqn\defBm{
B_m = {1+v \over u} - {m^2+m-1\over (2m-1)(2m+3)} \, .
}
For $m=n,n-1,n-2,n-3$, and also if $n=0,1,2$, \poly\ may be used 
to combine terms to ensure that we only have $a_{n'm'}^{A_{n m}}$ for
$0\le m' \le n'$. For $m=n=0$ this prescription gives
\eqn\Azero{\eqalign{
a_{22} = {}& {4\over 15}\, A_{00} \, , \qquad a_{21} = - {4\over 5} \, 
{1-v\over u} \,  A_{00} \, , \qquad a_{20} = {4\over 15} \Big (
3\, {1+v\over u}- 1 \Big ) \,  A_{00} \, , \cr
a_{11} = {}& {4\over 15} \Big ( 10\, {(1-v)^2\over u^2}
-5\, {1+v\over u} + 1 \Big ) \,  A_{00} \, , \quad
a_{10} = - {4\over 3} \Big ( 2\, {1+v\over u}- 1 \Big ){1-v\over u}\,
A_{00} \, , \cr
a_{00} = {}& {4\over 15} \Big ( 15\, {(1+v)^2\over u^2} - 
5\, {(1-v)^2\over u^2} - 8\, {1+v\over u} + 1 \Big ) \,  A_{00} \, , \cr}
}
which is equivalent to the results in  \scft. Similarly for $n=1,m=0,1$
the resulting $a_{n'm'}^{A_{n m}}$ correspond to those in  \ADHS.

To analyse the contributions arising from the function $\hf(z,y)$
this may be expanded as
\eqn\expf{
\hf(z,y) = \sum_{n=0}^{p-1} f_n(z)\, P_n(y) \, .
}
Using this in \ff\ and \Gsolt\ then $f_n$ gives rise to the following
contributions to $a_{nm}$ just for $m=0,1$,
\eqn\afour{\eqalign{
a^{f_n}_{n+1\,m} = {}&{(n+1)(n+2)\over (2n+1)(2n+3)}\, 
F_{nm}(z,\bz) \, , \qquad
a^{f_n}_{n-3\,m} =  {(n-1)n \over (2n-1)(2n+1)}\, F_{nm}(z,\bz) \, , \cr
a^{f_n}_{n\,m} = {}& -{n+1 \over 2n+1}\,  (z+\bz) F_{nm}(z,\bz)
\, , \qquad a^{f_n}_{n-2\,m} = -{n \over 2n+1}\, (z+\bz) F_{nm}(z,\bz) \, , \cr
a^{f_n}_{n-1\,m} = {}& \bigg ( z\bz + {1\over 2} 
+ {1 \over 2(2n-1)(2n+3)}\bigg )  F_{nm}(z,\bz) \, , \cr}
}
where
\eqn\FF{
F_{n1}(z,\bz) = - {f_n(z)-f_n(\bz) \over z-\bz} \, , \qquad
F_{n0}(z,\zz) = {zf_n(z)-\bz f_n(\bz) \over z-\bz} \, .
}
For low $n$ the results need to be modified but these can be obtained
from \afour\ by taking into account the symmetry relation in \poly. For 
$n=0$, $a^{f_0}_{11},a^{f_0}_{10}$ are as in \afour\ but for $a^{f_0}_{00}$
we need to take
\eqn\afz{
a^{f_0}_{00} - a^{f_0}_{-11} \to
a^{f_0}_{00} = - {(z^2-{1\over 3}) f_0(z) - (\bz^2-{1\over 3}) f_0(\bz)
\over z - \bz } \, ,
}
while for $n=1$, $a^{f_1}_{21},a^{f_1}_{20},a^{f_1}_{11},a^{f_1}_{10}$
are given by \afour\ but
\eqn\afone{
a^{f_1}_{00} = {\bz(z^2-{1\over 3}) f_1(z) - 
z (\bz^2-{1\over 3}) f_1(\bz) \over z - \bz } +{4\over 15}\, F_{11} \, . 
}

\subsec{Long Operators}

The solution of the superconformal identities given by \longr\ may now be naturally interpreted in terms of the operator
product expansion. If in \longr\ we consider a single conformal partial
wave for $A_{nm}$ by letting
\eqn\AZG{
A_{nm} \to \G_{\Delta+4}^{(\ell)} \, ,
}
then, if $\A^\Delta_{[q,p,q],\ell}$ denotes a long superconformal
multiplet  whose lowest state has spin $\ell$, scale dimension
$\Delta$ and which belongs to a $SU(4)_R$ representation with Dynkin
labels $[q,p,q]$,  we obtain
\eqn\delL{
a_{n'm'}^{A_{n m}} \to a_{n'm'}^{\vphantom g}
\big (\A^\Delta_{nm,\ell} \big ) \, , \qquad \A^\Delta_{nm,\ell}
\equiv \A^\Delta_{[n-m,2m,n-m],\ell} \, .
}
The non zero results obtained from \longr\ with
\AZG\ may be conveniently expressed in the form
\eqn\Ared{
a_{n+i\,m+j}^{\vphantom g}\big (\A^\Delta_{nm,\ell} \big ) 
= N_{n+1,i} N_{m,j} \, {A}{}^{nm}_{|i|\, |j|} \, , \qquad 
i,j = \pm 2, \pm 1, 0 \, ,
}
for
\eqn\AA{\eqalign{
N_{m,2} = {}& {(m+1)(m+2)\over (2m+1)(2m+3)} \, , \quad
N_{m,1} = {m+1\over 2m+1}\, , \quad  N_{m,0} = 1 \, , \cr
N_{m,-1} = {}& {m\over 2m+1}\, , \qquad\quad
N_{m,-2} = {(m-1)m\over (2m-1)(2m+1)} \, . \cr}
}
and using \recurG\ we have
\eqn\expA{\eqalign{
& \hskip 3cm a_{n+i\,m+j}^{\vphantom g}\big (\A^\Delta_{nm,\ell} \big ) = 
\sum_{(\Delta';\ell')} b_{(\Delta';\ell')} \, \G^{(\ell')}_{\Delta'} \, ,\cr
& |i|=|j|=2 \, , \  (\Delta';\ell') = (\Delta+4; \ell) \, , \cr
& |i|=2, \, |j|=1 , \ |i|=1, \, |j|=2 \, , \  
(\Delta';\ell') = (\Delta+5,\Delta+3; \ell \pm 1 )\, ,\cr 
& |i|=|j|=1 \, , \   (\Delta';\ell') =  
(\Delta+6,\Delta+4,\Delta+2; \ell \pm 2,\ell)\, , \cr
& |i|=1, \, j=0 , \ i=0, \, |j|=1 \, , \  
(\Delta';\ell') = (\Delta+7,\Delta+1;\ell \pm 1 ), \, 
(\Delta+5,\Delta+3; \ell\pm 3, \ell \pm 1 ) \, , \cr
& i=j=0 \, , \   (\Delta';\ell') =  (\Delta+8,\Delta;\ell), \,
(\Delta+6,\Delta+2; \ell \pm 2,\ell), \,
(\Delta+4; \ell\pm 4,\ell \pm 2,\ell) \, .  \cr}
}
In consequence $a_{n'm'}^{\vphantom g}(\A^\Delta_{nm,\ell})$ 
corresponds to the contribution in the operator product expansion applied 
to the correlation function for all expected
operators belonging to $\A^\Delta_{nm,\ell}$. In \expA\
$b_{(\Delta';\ell')}  > 0$ if $\Delta > \ell+1$. 
If $m\le n \le m+3$ the results are modified since we then obtain 
from \longr\ contributions with $m'>n'$. In this case, for $n'\ge m'$ and 
$a_{m'-1\,n'+1}^{\vphantom g} \big (\A^\Delta_{nm,\ell} \big )$ non 
zero, we should take
\eqn\anm{
a_{n'\,m'}^{\vphantom g} \big (\A^\Delta_{nm,\ell} \big ) - 
a_{m'-1\, n'+1}^{\vphantom g} \big (\A^\Delta_{nm,\ell} \big ) \to 
a_{n'\,m'}^{\vphantom g} \big (\A^\Delta_{nm,\ell} \big ) \, .
}
Furthermore any contribution with $m'=n'+1$ should be dropped. 
Using this result and \anm\ we may then easily show that
\eqn\Azer{
a_{n'\,m'}^{\vphantom g} \big (\A^\Delta_{n\, n+1,\ell} \big ) = 0 \, ,
}
and for later reference we also note the symmetry relation
\eqn\Asym{
a_{n'\,m'}^{\vphantom g} \big (\A^\Delta_{n\, m,\ell} \big ) =
a_{m'-1\,n'+1}^{\vphantom g} \big (\A^\Delta_{m-1\, n+1,\ell} \big ) \, .
}

The unitarity condition for a long multiplet $\A^\Delta_{nm,\ell}$ requires 
\eqn\unitl{
\Delta \ge 2n + \ell + 2 \, , 
}
and so, as in \unit, using \AZG\ we must have for $u\sim 0$,
\eqn\unit{
A_{nm}(u,v) \sim u^{n+3+\ep} \, , \qquad \ep \ge 0 \, .
}

\subsec{Semi-Short and Non-Unitary Operators}

To analyse the contribution of the single variable functions $f_n$ in \expf\
we use the result \ttwo\ for the conformal partial wave for twist
two operators as well as
\eqn\tzero{
\G_{\ell}^{(\ell)}(u,v)\big |_{\Delta_1=\Delta_2=\Delta_3=\Delta_4}
= \half \, { \bz\, g_{\ell}(x) - z \, g_{\ell} (\zz) \over z-\bz}\, , 
}
with $g_\ell$ as in \gell, for twist zero. Taking 
\eqn\fg{
f_{n+1}(z) \to \half g_{\ell+2}(x) \, , \qquad n=0,1,2,\dots \, ,
}
in \FF\ and \expf\ then leads to results corresponding to only 
twist zero and twist two operators. These operators can be interpreted as 
belonging to a multiplet $\D_{n\,0,\ell}$, where in general we denote by 
$\D_{n\,m ,\ell}\equiv \D_{[n-m,2m,n-m],\ell}$ the semi-short supermultiplet
in which the lowest dimension operator has $\Delta = 2m + \ell$, or
twist $2m$, and belongs to the $[n-m,2m,n-m]$ $SU(4)_R$ representation.
These non unitary super multiplets are discussed in appendix D.
For $\D_{n\,m ,\ell}$ the conformal partial waves may be expressed in general
in the form
\eqn\decD{
a_{n+i \, m+j} \big (\D_{n\,m,\ell}\big ) = N_{n+1,i}N_{m,j} \,
D^{nm}_{|i|\, j } \, , \qquad D^{nm}_{|i|\, 2} = 0 \, .
}
Corresponding to \fg\ we then have 
\eqn\single
{\eqalign{
D^{n0}_{21} = {}& \quar \, \G_{\ell+3}^{(\ell+1)}\, , \qquad \qquad
D^{n0}_{20} = \quar \big (\G_{\ell+2}^{(\ell)} +
a_{\ell+2} \, \G_{\ell+4}^{(\ell+2)} \big ) \, , \cr
D^{n0}_{11} = {}&
\G_{\ell+2}^{(\ell+2)} + \quar\big ( \G_{\ell+2}^{(\ell)} +
a_{\ell+2} \, \G_{\ell+4}^{(\ell+2)} \big )\, , \cr
D^{n0}_{10} = {}&
\G_{\ell+1}^{(\ell+1)} + a_{\ell+2}\, \G_{\ell+3}^{(\ell+3)} +
\quar \big ( \G_{\ell+1}^{(\ell-1)} +  b_\ell \,\G_{\ell+3}^{(\ell+1)} +
a_{\ell+2} a_{\ell+3}\, \G_{\ell+5}^{(\ell+3)} \big )  \, , \cr
D^{n0}_{01} = {}&
\G_{\ell+1}^{(\ell+1)} + a_{\ell+2} \, \G_{\ell+3}^{(\ell+3)}
+ \quar b_n \, \G_{\ell+3}^{(\ell+1)} \, ,  \cr
D^{n0}_{00} = {}& \G_{\ell}^{(\ell)} + b_\ell \, \G_{\ell+2}^{(\ell+2)}
+ a_{\ell+2} a_{\ell+3}\, \G_{\ell+4}^{(\ell+4)}
+ \quar b_n \big ( \G_{\ell+2}^{(\ell)}
+ a_{\ell+2} \, \G_{\ell+4}^{(\ell+2)} \big ) \, , \cr}
}
whereas $a_\ell$ is as in \defst\ and
\eqn\bell{
b_\ell = a_{\ell+2}+a_{\ell+1}=
{2\ell^2 + 6 \ell +3 \over (2\ell+1)(2\ell+5)} \, .
}
A list of relevant representations for differing dimensions contained in
$\D_{n\,0 ,\ell}\equiv \D_{[n,0,n],\ell}$ is listed in appendix D, the
twist zero and twist two representations correspond with those necessary for
\single.  For $f_0$ these results are modified. From \afz\ only twist two
contributions are required since, taking now $f_0(z) \to 2 g_{\ell+3}(x)$,
\eqn\ttwob{\eqalign{
a_{00}\big (\C_{00,\ell}\big ) = {}&
\G_{\ell+2}^{(\ell)} + {2(\ell+2)(\ell+3)\over 3(2\ell+3)(2\ell+7)} \,
\G_{\ell+4}^{(\ell+2)} + a_{\ell+3}a_{\ell+4} \, 
\G_{\ell+6}^{(\ell+4)} \, , \cr
a_{10}\big (\C_{00,\ell}\big ) = {}& {2\over 3} \Big (
\G_{\ell+3}^{(\ell+1)} + a_{\ell+3}  \,
\G_{\ell+5}^{(\ell+3)} \Big ) \, ,
\quad a_{11}\big (\C_{00,\ell}\big ) =
{2\over 3}  \, \G_{\ell+4}^{(\ell+2)}  \, .\cr}
}
Here we denote by $\C_{nm,\ell}\equiv \C_{[n-m,2m,n-m],\ell}$ the semi-short
supermultiplet in which the lowest dimension operator has $\Delta = 2n+\ell+2$
and belongs to the $[n-m,2m,n-m]$ $SU(4)_R$ representation.

The multiplets $\D_{[q,p,q],\ell}$  fail to satisfy the unitarity condition
\unitl\ on $\Delta$ and so their contributions as in \single\ must be
cancelled in a unitary theory.
This may be achieved by a corresponding long multiplet contribution.
When  $\Delta =  2m + \ell$ or $\Delta =  2n + \ell +2$ the long multiplet
$\A^\Delta_{nm,\ell}$ can be decomposed into semi-short multiplets resulting in
\eqn\DD{
a_{n'm'}^{\vphantom g}\big ( \A^{2m+\ell}_{n\, m,\ell} \big )
= 16 \, a_{n'm'}^{\vphantom g} \big ( \D_{n\, m,\ell} \big ){}
+ {4(m+1)\over 2m+1} \, a_{n'm'}^{\vphantom g}
\big ( \D_{n\,m{+1},\ell-1} \big ) \, ,
}
and, at the unitarity threshold \unitl,
\eqn\CC{
a_{n'm'}^{\vphantom g}\big ( \A^{2n+\ell+2}_{n\, m,\ell} \big )
= 16 \, a_{n'm'}^{\vphantom g} \big ( \C_{n\, m,\ell} \big ){}
+ {4(n+2)\over 2n+3} \, a_{n'm'}^{\vphantom g}
\big ( \C_{n{+1}\,m,\ell-1} \big ) \, .
}
When $n=m$ we have the special case
\eqn\CD{
a_{n'm'}^{\vphantom g}\big ( \A^{2n+\ell}_{n\,n,\ell} \big )
= 16 \, a_{n'm'}^{\vphantom g} \big ( \D_{n\,n,\ell} \big ){}
+ {(n+1)(n+2)\over (2n+1)(2n+3)} \,
a_{n'm'}^{\vphantom g} \big ( \C_{n+1\,n+1,\ell-2} \big ) \, .
}
The results \DD, \CC\ and \CD\ reflect a decomposition of long multiplets
at particular values of $\Delta$ as described in appendix D.
{}From \DD\ we may obtain $a_{n'm'}^{\vphantom g}\big ( \D_{n\, m,\ell} \big)$
iteratively starting from \single. With the  notation in \decD\ the results
are 
\eqnn\resD
$$\eqalignno{
D^{nm}_{2\,-2} = {}& {\ts{1\over 16}}\, \G^{(\ell)}_{2m+\ell+4} \, ,
\qquad\qquad D^{nm}_{2\,1} = \quar\, \G^{(\ell+1)}_{2m+\ell+3} \, , \cr
D^{nm}_{2\,-1} = {}& {\ts{1\over 16}}\big ( \G^{(\ell-1)}_{2m+\ell+3} +
4\, \G^{(\ell+1)}_{2m+\ell+3} + 
a_{m+\ell+2}\, \G^{(\ell+1)}_{2m+\ell+5} \big )\,, \cr
D^{nm}_{2\,0} = {}& {\ts{1\over 16}}\big ( 4\, \G^{(\ell)}_{2m+\ell+2} +
a_{m}\, \G^{(\ell)}_{2m+\ell+4} + 4a_{m+\ell+2}\, 
\G^{(\ell+2)}_{2m+\ell+4} \big )\,, \cr
D^{nm}_{1\,-2} = {}& {\ts{1\over 16}}\big ( \G^{(\ell-1)}_{2m+\ell+3} + 
4 \, \G^{(\ell+1)}_{2m+\ell+3} + \quar a_{m+1}\, \G^{(\ell-1)}_{2m+\ell+5} 
+ a_{m+\ell+2} \,\G^{(\ell+1)}_{2m+\ell+5}\big )\,,\cr
D^{nm}_{1\, 1} = {}& {\ts{1\over 4}}\big ( 
\G^{(\ell)}_{2m+\ell+2} + 4 \, \G^{(\ell+2)}_{2m+\ell+2} + \quar a_{m}\,
\G^{(\ell)}_{2m+\ell+4} + a_{m+\ell+2} \,\G^{(\ell+2)}_{2m+\ell+4}\big )\,,\cr
D^{nm}_{1\,-1} = {}& {\ts{1\over 16}}\big (
\G^{(\ell-2)}_{2m+\ell+2} + \quar a_{m+1}\, \G^{(\ell-2)}_{2m+\ell+4} \cr
&\quad {} + 8 \, \G^{(\ell)}_{2m+\ell+2} + ( b_{m+\ell} + a_m ) 
\G^{(\ell)}_{2m+\ell+4} 
+ \quar a_{m+1}a_{m+\ell+2}\, \G^{(\ell)}_{2m+\ell+6} \cr 
&\quad {} + 16 \, \G^{(\ell+2)}_{2m+\ell+2} + 8 a_{m+\ell+2}\,
\G^{(\ell+2)}_{2m+\ell+4} + a_{m+\ell+2}a_{m+\ell+3} \,
\G^{(\ell+2)}_{2m+\ell+6}\big ) \,, \cr
D^{nm}_{1\, 0} = {}& {\ts{1\over 16}}\big (
4\,\G^{(\ell-1)}_{2m+\ell+1} + 2 a_{m}\, \G^{(\ell-1)}_{2m+\ell+3} +
\quar a_{m}a_{m+1}\G^{(\ell-1)}_{2m+\ell+5}\cr
&\quad {} + 16\, \G^{(\ell+1)}_{2m+\ell+1} 
+ 4 ( b_{m+\ell} + a_m ) \G^{(\ell+1)}_{2m+\ell+3} 
+ 2 a_{m}a_{m+\ell+2}\, \G^{(\ell+1)}_{2m+\ell+5} \cr
&\quad {} + 16a_{m+\ell+2} \, \G^{(\ell+3)}_{2m+\ell+3} + 
4a_{m+\ell+2}a_{m+\ell+3} \, \G^{(\ell+3)}_{2m+\ell+5}\big ) \,, \cr
D^{nm}_{0\,-2} = {}& {\ts{1\over 16}}\big ( 4\, \G^{(\ell)}_{2m+\ell+2} +
\quar a_{m+1} \, \G^{(\ell-2)}_{2m+\ell+4} + b_n\, \G^{(\ell)}_{2m+\ell+4} \cr
&\quad {} + 4 a_{m+\ell+2}\, \G^{(\ell+2)}_{2m+\ell+4}
+ \quar a_{m+1} a_{m+\ell+2} \,\G^{(\ell)}_{2m+\ell+6}\big )\,,\cr
D^{nm}_{0\,1} = {}& {\ts{1\over 4}}\big (4\, \G^{(\ell+1)}_{2m+\ell+1} +
\quar a_m \, \G^{(\ell-1)}_{2m+\ell+3} + b_n \, \G^{(\ell+1)}_{2m+\ell+3} \cr
&\quad {} + 4 a_{m+\ell+2}\, \G^{(\ell+3)}_{2m+\ell+3} 
+ \quar a_m  a_{m+\ell+2} \,\G^{(\ell+1)}_{2m+\ell+5}\big )\, ,\cr
D^{nm}_{0\, -1} = {}& {\ts{1\over 16}}\big ( \quar a_{m+1}\, 
\G^{(\ell-3)}_{2m+\ell+3} + 4\,\G^{(\ell-1)}_{2m+\ell+1} + 
(a_{m} + b_n ) \, \G^{(\ell-1)}_{2m+\ell+3} 
+ \quar a_{m+1}b_{m+\ell}\, \G^{(\ell-1)}_{2m+\ell+5} \cr
&\quad {} + 16\, \G^{(\ell+1)}_{2m+\ell+1}
+ 4 ( b_{m+\ell} + b_n ) \G^{(\ell+1)}_{2m+\ell+3}
+ a_{m+\ell+2}(a_{m}+b_n) \, \G^{(\ell+1)}_{2m+\ell+5} \cr
&\quad {} + \quar a_{m+1} a_{m+\ell+2} a_{m+\ell+3}\, \G^{(\ell+1)}_{2m+\ell+7}\cr
&\quad {} + 16 a_{m+\ell+2}\, \G^{(\ell+3)}_{2m+\ell+3} +
4 a_{m+\ell+2}a_{m+\ell+3}\, \G^{(\ell+3)}_{2m+\ell+5} \big )  \, ,  \cr
D^{nm}_{0\, 0} = {}& {\ts{1\over 16}}\big (  a_{m}\,
\G^{(\ell-2)}_{2m+\ell+2} + \quar a_m a_{m+1} \,\G^{(\ell-2)}_{2m+\ell+4} +
16\, \G^{(\ell)}_{2m+\ell} + 4(a_{m} + b_n ) \, \G^{(\ell)}_{2m+\ell+2} \cr
&\quad {} +  a_{m}(b_{m+\ell}+b_n)\, \G^{(\ell)}_{2m+\ell+4}
+ \quar  a_{m} a_{m+1} a_{m+\ell+2}\, \G^{(\ell)}_{2m+\ell+6} \cr
&\quad {} + 16 b_{m+\ell} \, \G^{(\ell+2)}_{2m+\ell+2}
+ 4 a_{m+\ell+2}(a_{m}+b_n) \, \G^{(\ell+2)}_{2m+\ell+4} 
+  a_{m} a_{m+\ell+2} a_{m+\ell+3}\, \G^{(\ell+2)}_{2m+\ell+6}\cr
&\quad {} +
16 a_{m+\ell+2}a_{m+\ell+3}\, \G^{(\ell+4)}_{2m+\ell+4} \big )  \, . & \resD \cr
}
$$

The corresponding results for the semi-short multiplet $\C_{n\, m,\ell}$
may be obtained from those for $\D_{n\, m,\ell}$ given above by taking
\eqn\relDC{
a_{n'\, m'}^{\vphantom g} \big ( \C_{n\, m,\ell} \big ) =
a_{m'-1\, n'+1}^{\vphantom g} \big ( \D_{m-1\, n+1,\ell} \big ) \, .
}
Using \Asym\ then \CC\ easily follows from \DD. We may also verify
that \CD\ is satisfied.  Combining \DD\ for $m=n+1$ with \Azer\ we 
may then obtain
\eqn\CDe{\eqalign{
a_{n'\,m'}^{\vphantom g}\big (& \A^{2n+\ell}_{n\,n,\ell} \big ) 
- a_{m'-1\,n'+1}^{\vphantom g}\big ( \A^{2n+\ell}_{n\,n,\ell} \big )
=  16 \big ( a_{n'm'}^{\vphantom g} \big ( \D_{n\,n,\ell} \big ) 
- a_{m'-1\, n'+1 }^{\vphantom g} \big ( \D_{n\,n,\ell} \big ) \big ) \cr
&{} + {(n+1)(n+2)\over (2n+1)(2n+3)}\Big ( -
a_{m'-1\, n'+1}^{\vphantom g} \big ( \C_{n+1\,n+1,\ell-2} \big ) +
a_{n'm'}^{\vphantom g} \big ( \C_{n+1\,n+1,\ell-2} \big )\Big )  \, , \cr}
}
which for $n'\ge m'$, and noting the requirement \anm, gives exactly \CD.

In general the results from \relDC\ can be expressed as
\eqn\decC{
a_{n+i \, m+j} \big (\C_{n\,m,\ell}\big ) = N_{n+1,i}N_{m,j} \,
C^{nm}_{i\, j } \, , \qquad C^{nm}_{2\, j} = 0 \, .
}
For general $n,m$ the necessary operators are just those given in table 4
of \short. For $m=n$ the relation \relDC\ combined with \resD\ in this
case and applying the corresponding results to \anm\ gives
\eqnn\resC
$$\eqalignno{
C^{nn}_{1\,1} = {}& \G^{(\ell+2)}_{2n+\ell+4} \, , \cr
C^{nn}_{1\,0} = {}& \G^{(\ell+1)}_{2n+\ell+3} + \quar a_n \, 
\G^{(\ell+1)}_{2n+\ell+5} + a_{n+\ell+3}\, \G^{(\ell+3)}_{2n+\ell+5} \, , \cr
C^{nn}_{0\,0} = {}& \G^{(\ell)}_{2n+\ell+2} + \quar a_n \, 
\G^{(\ell)}_{2n+\ell+4} + (b_{n+\ell+1}-a_{n+1}) \, 
\G^{(\ell+2)}_{2n+\ell+4} \, , \cr
&{} +{\ts {1\over 16}} a_n a_{n+1}\, \G^{(\ell)}_{2n+\ell+6} 
+ \quar a_n a_{n+\ell+3}\, \G^{(\ell+2)}_{2n+\ell+6} 
+ a_{n+\ell+3}a_{n+\ell+4} \, \G^{(\ell+4)}_{2n+\ell+6} \, , \cr
C^{nn}_{1\,-1} = {}&\quar \, \G^{(\ell)}_{2n+\ell+4} + \G^{(\ell+2)}_{2n+\ell+4} 
+ \quar a_{n+\ell+3}\, \G^{(\ell+2)}_{2n+\ell+6} \, , \cr
C^{nn}_{0\,-1} = {}& \quar \, \G^{(\ell-1)}_{2n+\ell+3} +  
\G^{(\ell+1)}_{2n+\ell+3} + {\ts{1\over 16}} a_{n+1}\, 
\G^{(\ell-1)}_{2n+\ell+5} +  \quar b_{n+\ell+1}\, \G^{(\ell+1)}_{2n+\ell+5} \, , \cr
&{} +a_{n+\ell+3}\, \G^{(\ell+3)}_{2n+\ell+5} 
+ {\ts {1\over 16}} a_{n+1}a_{n+\ell+3} \, \G^{(\ell+1)}_{2n+\ell+7} 
+ \quar a_{n+\ell+3}a_{n+\ell+4} \, \G^{(\ell+3)}_{2n+\ell+7} \, , \cr
C^{nn}_{-1\,-1} = {}& {\ts {1\over 16}} \,\G^{(\ell-2)}_{2n+\ell+4} + \quar  \, 
\G^{(\ell)}_{2n+\ell+4} + \G^{(\ell+2)}_{2n+\ell+4} 
+ {\ts {1\over 16}} (b_{n+\ell+1}-a_{n+2}) \, \G^{(\ell)}_{2n+\ell+6} \cr
&{} + \quar a_{n+\ell+3} \, \G^{(\ell+2)}_{2n+\ell+6} 
+ {\ts {1\over 16}}a_{n+\ell+3}a_{n+\ell+4} \, \G^{(\ell+2)}_{2n+\ell+8} \, , \cr
C^{nn}_{1\,-2} = {}& \quar \, \G^{(\ell+1)}_{2n+\ell+5} \, , \cr
C^{nn}_{0\,-2} = {}& \quar \, \G^{(\ell)}_{2n+\ell+4} + 
{\ts{1\over 16}} a_{n+1}\, \G^{(\ell)}_{2n+\ell+6} 
+ \quar a_{n+\ell+3}\, \G^{(\ell+2)}_{2n+\ell+6} \, , \cr
C^{nn}_{-1\,-2} = {}& {\ts{1\over 16}}\,\G^{(\ell-1)}_{2n+\ell+5} + \quar \, 
\G^{(\ell+1)}_{2n+\ell+5} + {\ts {1\over 16}} a_{n+\ell+3}\, 
\G^{(\ell+1)}_{2n+\ell+7} \, , \cr
C^{nn}_{-2\,-2} = {}& {\ts {1\over 16}} \, \G^{(\ell)}_{2n+\ell+6} \, . &\resC \cr}
$$
The necessary operators correspond exactly to those listed in \short\ (see table
3) as present in the semi-short supermultiplet for this case.
For $n=0$ \resC\ reproduces \ttwo. We may also note that, since for $m\ge 1$,
$\quar < a_m \le {1\over 3}$ and $b_n > \half$, all coefficients
in \resC\ are positive as required by unitarity.

\subsec{Short Operators}

As in the $\N=2$ case the semi-short results also include the contributions
for short BPS multiplets when extended to negative $\ell$. Formally as shown
in \short\ $\C_{[q,p,q],-1} \simeq \B_{[q+1,p,q+1]}$ where $\B_{[q,p,q]}$
denotes the BPS supermultiplet whose lowest state has spin zero, $\Delta
= 2q+p$, and belongs to the $SU(4)_R$ $[q,p,q]$ representation. For $q>0$
the lowest state is annihilated by $\quar$ of the $Q$ and also $\bar Q$
supercharges whereas when $q=0$ we have a $\half$-BPS multiplet with $\half$
the $Q$ and $\bar Q$ supercharges annihilating the lowest state. As
earlier we identify, for $n\ge m$, $\B_{n\,m} \equiv \B_{[n-m,2m,n-m]}$ and 
we then have
\eqn\CBfour{
a_{n'm'}^{\vphantom g}\big ( \C_{n\, m,-1} \big ) = {n+1\over 2n+1} \,
a_{n'm'}^{\vphantom g}\big ( \B_{n+1\, m} \big ) \, ,
}
where
\eqn\decB{
a_{n+i \, m+j} \big (\B_{n\,m}\big ) = N_{n+1,i}N_{m,j} \,
B^{nm}_{i\, j } \, , \qquad B^{nm}_{2\, j} = B^{nm}_{1\, j} = 0 \, .
}
For general $n,m$ we have
\eqn\BB{
B^{nm}_{i\, j } = B^{nm}_{i\, |j| } \, ,
}
and
\eqn\resB{\eqalign{
B^{nm}_{0\,2} = {}& \quar \, \G^{(0)}_{2n+2} \, , \cr
B^{nm}_{-2\,2} = {}& {\ts {1\over 16}} \, \G^{(0)}_{2n+4}  \, , \cr
B^{nm}_{-1\,2} = {}& \quar \, \G^{(1)}_{2n+3} \, , \cr
B^{nm}_{0\,1} = {}&  \G^{(1)}_{2n+1} + \quar a_{n+1}\, \G^{(1)}_{2n+3} \, , \cr
B^{nn}_{-2\,1} = {}& \quar \, \G^{(1)}_{2n+3} + {\ts{1\over 16}} a_{n+2}\,
\G^{(1)}_{2n+5} \, , \cr
B^{nm}_{-1\,1} = {}& \quar  \,\G^{(0)}_{2n+2} + \G^{(2)}_{2n+2}+{\ts{1\over 16}}
a_{n} \,\G^{(0)}_{2n+4} + \quar a_{n+2} \, \G^{(2)}_{2n+4}\, , \cr
B^{nn}_{0\, 0} = {}& \G^{(0)}_{2n} + \quar(b_{m-1}-a_n ) \, \G^{(0)}_{2n+2}
+  a_{n+1}\, \G^{(2)}_{2n+2} + {\ts{1\over 16}} a_n  a_{n+1}\, 
\G^{(0)}_{2n+4} \, , \cr
B^{nm}_{-2\,0} = {}&\quar  \,\G^{(0)}_{2n+2} +  {\ts{1\over 16}}
(b_{m-1}-a_{n+1} )  \,\G^{(0)}_{2n+4} + \quar a_{n+2}\, \G^{(2)}_{2n+4} +
{\ts {1\over 64}} a_{n+1} a_{n+2}\, \G^{(0)}_{2n+6}  \, , \cr
B^{nm}_{-1\,0} = {}& \G^{(1)}_{2n+1} + \quar b_{m-1}  \,\G^{(1)}_{2n+3}
+ a_{n+2} \,  \G^{(3)}_{2n+3} + {\ts {1\over 16}} a_{n} a_{n+2} \,
\G^{(1)}_{2n+5} \, . \cr}
}
Again all coefficients are positive and the necessary operators are
exactly as expected for this supermultiplet (see table 2 in \short). For
$n=m+1$ the multiplet is truncated with, in \decB, the following non zero,
\eqn\resBm{\eqalign{
B^{m+1\,m}_{0\,1} = {}&  \G^{(1)}_{2m+3} \, , \cr
B^{m+1\,m}_{0\,0} = {}& \G^{(0)}_{2m+2} + \quar a_m \, \G^{(0)}_{2m+4}
+  a_{m+2}\, \G^{(2)}_{2m+4}  \, , \cr
B^{m+1\,m}_{-1\,0} = {}& \G^{(1)}_{2m+3} + \quar a_m \, \G^{(1)}_{2m+5}
+ a_{m+3} \,  \G^{(3)}_{2m+5} \, , \cr
B^{m+1\,m}_{0\,-1} = {}&  \G^{(1)}_{2m+3} + 
\quar a_{m+2}\, \G^{(1)}_{2m+5} \, , \cr
B^{m+1\,m}_{-1\,-1} = {}& \quar  \,\G^{(0)}_{2m+4} + 
\G^{(2)}_{2m+4}+{\ts{1\over 16}} a_{m+1} \,\G^{(0)}_{2m+6} + 
\quar a_{m+3} \, \G^{(2)}_{2m+4}\, , \cr
B^{m+1\,m}_{-2\,-1} = {}& \quar \, \G^{(1)}_{2m+5} + {\ts{1\over 16}} a_{m+3}\,
\G^{(1)}_{2m+7} \, , \cr
B^{m+1\, m}_{0\, -2} = {}& \quar  \, \G^{(0)}_{2m+4} \, , \qquad
B^{m+1\, m}_{-1\, -2} =  \quar  \, \G^{(1)}_{2m+5} \, , \qquad
B^{m+1\, m}_{-2\, -2} =  {\ts{1\over 16}}  \, \G^{(0)}_{2m+6} \, . \cr}
}
The necessary operators correlate again with those expected for this
$\quar$-BPS multiplet (see table 5 in \short).

If we consider the semi-short multiplet for $\ell=-2$ we get
\eqn\CBfs{
a_{n'm'}^{\vphantom g}\big ( \C_{n\, m,-2} \big ) = -4 \,
a_{n'm'}^{\vphantom g}\big ( \B_{n \, m} \big ) \, ,
}
which allows results for $a_{n'm'}^{\vphantom g}\big ( \B_{n \, m} \big )$
to be derived for $m=n$ in addition to $m<n$ as given by \CBfour.
However in this case there is a further decomposition into 
contributions corresponding to $\half$-BPS multiplets.
Such $\half$-BPS contributions are obtained in \decB\  by letting 
$\B_{nm} \to {\hat \B}_{nn}$ and $B^{nm}_{ij} \to {\hat B}{}^{nn}_{ij}$ where
\eqn\Bhalf{\eqalign{
{\hat B}{}^{nn}_{0\,0} = {}& \G^{(0)}_{2n} \, , \qquad\quad\quad \,
{\hat B}{}^{nn}_{0\,-1} = \G^{(1)}_{2n+1}\, , \qquad \quad 
{\hat B}{}^{nn}_{-1\,-1} = \G^{(2)}_{2n+2} \, , \cr
{\hat B}{}^{nn}_{0\,-2} = {}& \quar \, \G^{(0)}_{2n+2} \, , \qquad 
{\hat B}{}^{nn}_{-1\,-2} =  \quar \, \G^{(1)}_{2n+3} \, , \qquad
{\hat B}{}^{nn}_{-2\,-2} =  {\ts{1\over 16}} \, \G^{(0)}_{2n+4} \, , \cr}
}
(the relevant operators here correspond to table 1 in \short). With the
result given in \Bhalf\ we can then write in \CBfs
\eqn\CBh{
a_{n'm'}^{\vphantom g}\big ( \B_{n\, n} \big ) = 
a_{n'm'}^{\vphantom g}\big ( {\hat \B}_{n\, n} \big )
- {(n+1)(n+2)\over 4(2n+1)(2n+3)} \,
a_{n'm'}^{\vphantom g}\big ( {\hat \B}_{n+1\, n+1} \big ) \, .
}
{}From \Gzero\ and \Gident\ it is also easy to see that
\eqn\Id{
a_{nm}^{\vphantom g}\big ( {\hat \B}_{0\, 0} \big ) =
a_{nm}^{\vphantom g}\big ( \I \big ) \, .
}
Any $\half$-BPS contribution $a_{n'm'}
\big ({\hat \B}_{n\, n} \big )$ may then be isolated by considering 
appropriate linear combinations of 
$a_{n'm'}\big ( \C_{n\, n,-2} \big )$ together with
$a_{n'm'}\big ( \I \big )$.

\subsec{Identity Operator}

We consider the contribution resulting from the constant $k$
in \Gsolt\ and \ff. It is easy to see that this gives only
\eqn\azero{
a^k_{00} = k \, .
}
The constant $k$ clearly corresponds to the identity operator,
\eqn\Gident{
a_{nm}(\I) = \de_{n0} \de_{m0} \, .
}

\subsec{(Next-to-) Extremal Case}

We also consider the extremal and next-to-extremal cases. When $E=0$
$\G$ is independent of $y,\yz$ and so must also be the function $f$ in
\solid. From \real\ and \Gextr\ we then get the solution
\eqn\extp{
\G(u,v) = u^{{1\over 2}p_+} \, k  \, ,
}
where we define
\eqn\ppp{
p_\pm = p_2 \pm p_1 \, .
}
Noting that
\eqn\extPG{
P^{(p_1,p_2)}_{00}(y,\yz) = \half (p_+ +2) \, , \qquad
\G^{(0)}_{p_+}(u,v;p_-,p_+) = u^{{1\over 2}p_+} \, ,
}
it is clear that the only operator which is necessary in the operator product 
expansion  has $\Delta=p_+$ and is spinless belonging to the $[0,p_+,0]$
representation. This is of course may be identified with the contribution of
just the $\half$-BPS operator belonging to the short $\B_{[0,p_+,0]}$
supermultiplet so that for the extremal case, up to a constant factor,
\eqn\exts{
a_{nm}\big({\B}_{[0,p_+,0]} \big ) = \de_{n0} \de_{m0}\
\G^{(0)}_{p_+}\, . }
The correlation function in this case has the very simple form
\eqn\Fourext{\eqalign{
\l \vphi^{(p_1)}(x_1,t_1)&  \, \vphi^{(p_2)}(x_2,t_2)\,
\vphi^{(p_3)}(x_3,t_3)\, \vphi^{(p_4)}(x_4,t_4 )\r \big |_{p_4=p_1+p_2+p_3}\cr
&{} = {\big ( t_1 {\cdot t}_4 \big )^{p_1} \big ( t_2 {\cdot t}_4 \big )^{p_2} 
\big ( t_1 {\cdot t}_3 \big )^{p_3}  \over
r_{14}^{\, \, p_1}\ r_{24}^{\, \, p_2}\  r_{34}^{\, \, p_3 } }\, k \, . \cr}
}

For the next-to-extremal case, $E=1$, we have a similar solution to that
given by \Gsolt\ and \ff, but with no arbitrary $\K$ term and $\hf$ a single
variable function of $z$,
\eqn\next{\eqalign{
\hG(u,v; y, \yz) = {}& u^{{1\over 2}p_+ -1} \Big ( k - {1\over z - \bz}
\big ( (y-z)(\yz - z) \hf(z) - (y- \bz)(\yz - \bz) \hf(\bz) \big ) \Big ) \cr
= {}& \sum_{0\le m \le n \le 1} \! \! a_{nm}(u,v) \, P^{(p_1-1,p_2-1)}_{nm}
(y,\yz) \, , \cr}
}
where we have expanded in terms of the different possible $SU(4)_R$ 
representations. From this we obtain
\eqn\deco{\eqalign{
{\ts {1\over 16}}p_+ (p_+ +1)& (p_+ +2)\, a_{11} = \ha_{11} = F_0 \, , \cr
{\ts {1\over 8}}(p_+ +1)& (p_+ +2)\, a_{10} = \ha_{10} = 
F_1 + {p_- \over p_+} \, F_0 \, ,\cr
\half p_+ \, a_{00} = \ha_{00} = {}& k \, u^{{1\over 2}p_+-1} + 
F_2 + {2p_-\over p_+ +2} \, 
F_1 + { p_-{\!\!}^2 - (p_+ +2)  \over (p_+ +1)(p_+ +2)} \, F_0 \, , \cr}
}
for
\eqn\Fn{
F_n(z,\bz) = - (-1)^n \, u^{{1\over 2}p_+ -1}\, { z^n \hf(z) - \bz^n \hf(\bz)
\over z - \bz } \, .
}

Keeping only the term in \deco\ involving $k$ we may easily from \extPG\ see 
that this represents the contribution of just the $\half$-BPS chiral
primary operator belonging to the $\B_{[0,p_+-2,0]}$ supermultiplet
so that in the next-to-extremal case we have
\eqn\azz{
\ha_{nm}\big ({\B}_{[0,p_+-2,0]} \big ) = 
\de_{n0} \de_{m0} \, \G^{(0)}_{p_+-2}\, .
}
If in \deco\ and \Fn\ we let $\hf(z) \to 2 g_{\ell+3}(x;p_1,p_2)$ and use the
definitions in \text\ we obtain the contributions for the semi-short
supermultiplet $\C_{[0,p_+-2,0],\ell}$,
\eqnn\semsh
$$\eqalignno{ 
\ha_{11}\big ( \C_{[0,p_+-2,0],\ell}\big ) = {}&
\G^{(\ell+2)}_{p_++\ell+2} \, , \cr
\ha_{10}\big (\C_{[0,p_+-2,0],\ell}\big ) = {}&
\G^{(\ell+1)}_{p_+ +\ell+1} + b_{\ell+2} \,\G^{(\ell+3)}_{p_++\ell+3} 
+ {4(\ell+2) p_- (p_+ + \ell+1) \over p_+ ( p_+ +2 \ell+ 2) ( p_+ +2 \ell+ 4)}\,
\G^{(\ell+2)}_{p_+ +\ell+2}  \, ,\cr
\ha_{00}\big (\C_{[0,p_+-2,0],\ell}\big ) 
= {}& 
\G^{(\ell)}_{p_+ +\ell}  + b_{\ell+2} b_{\ell+3} \,\G^{(\ell+4)}_{p_+ +\ell+4}
+ c_{\ell+2} \, \G^{(\ell+2)}_{p_++\ell+2} \cr
&{}+ {8(\ell+1)p_-(p_++\ell+1)\over (p_++2)(p_++2\ell)(p_++2\ell+4)}\,
\G^{(\ell+1)}_{p_+ +\ell+1} \cr
&{}+ {8(\ell+2)p_-(p_++\ell+2)\,b_{\ell+2}\over 
(p_++2)(p_++2\ell+2)(p_++2\ell+6)}\, \G^{(\ell+3)}_{p_+ +\ell+3} \, , & \semsh \cr}
$$
for
\eqn\bell{\eqalign{
b_\ell = {}& {4\,(\ell+1)(p_1+\ell)(p_2+\ell)(p_++\ell-1) \over
(p_+ + 2\ell -1)(p_+ + 2\ell)^2 (p_+ + 2\ell+1)} \, , \cr
c_\ell ={}& {2\, \ell(p_+ \! +\ell-1)\over(p_+ \! +1)(p_+ \! +2\ell-3)
(p_+ \! +2\ell+1)} \bigg (
p_+ \! -1 +{p_-{\!\!}^2 \big ( 8(\ell-1)(p_+ \! +\ell) - p_+(p_+ \! -1)\big ) \over
(p_+ \! +2)(p_+ \! +2\ell-2)(p_+ \! +2\ell) } \bigg ) \, . \cr}
}
The necessary operators required for \semsh\ correspond exactly with those in this 
semi-short supermultiplet (see table 3 in \short).

Just as previously we may extend these  \semsh\ to $\ell=-1,-2$ to obtain
results for short multiplets. Thus
\eqn\CBB{\eqalign{
\ha_{nm}\big ( \C_{[0,p_+-2,0],-1}\big ) = {}&
\ha_{nm}\big ( \B_{[1,p_+-2,1]}\big ) \, , \cr
\ha_{nm}\big ( \C_{[0,p_+-2,0],-2}\big ) = {}&
\ha_{nm}\big ( \B_{[0,p_+,0]}\big ) - 4 \, 
\ha_{nm}\big ( \B_{[0,p_+-2,0]}\big ) \, , \cr}
}
where, together with \azz,
\eqn\Bext{\eqalign{
\ha_{11}\big ( \B_{[1,p_+-2,1]}\big ) ={}& \G^{(1)}_{p_++1} \, , \cr
\ha_{10}\big ( \B_{[1,p_+-2,1]}\big ) ={}& \G^{(0)}_{p_+} 
+{4p_- \over p_+(p_++2)} \, \G^{(1)}_{p_++1} + b_1 \, \G^{(2)}_{p_++2} \, , \cr
\ha_{00}\big ( \B_{[1,p_+-2,1]}\big ) ={}& b_1 \bigg ( \G^{(1)}_{p_++1} 
+{8p_- (p_+ + 2) \over p_+(p_++2)(p_++4) } \, \G^{(2)}_{p_++2} + b_2 \,
\G^{(3)}_{p_++3} \bigg ) \, , \cr}
}
and
\eqn\Bex{
\ha_{11}\big ( \B_{[0,p_+,0]}\big ) = \G^{(0)}_{p_+} \, , \quad
\ha_{10}\big ( \B_{[0,p_+,0]}\big ) = b_0 \, \G^{(1)}_{p_++1} \, , \quad
\ha_{00}\big ( \B_{[0,p_+,0]}\big ) = b_0b_1 \, \G^{(2)}_{p_++2} \, .
}
The necessary operators here correspond to table 5 and table 1 in \short.

\subsec{Summary}

The results obtained above show that the operator product expansion for
$\half$-BPS operators can be decomposed into short, semi-short and long
supermultiplets. For $p_-=p_2-p_1 \ge 0$,
\eqn\decompB{\eqalign{
\B_{[0,p_1,0]} \otimes \B_{[0,p_2,0]} \simeq {}&
\bigoplus_{0 \le m \le n \le p_1} \!\! \B_{[n-m,p_-+ 2m,n-m]} \cr
&{} \oplus \bigoplus_{\ell\ge 0} 
\bigoplus_{0\le m \le n \le p_1-1} \!\! \C_{[n-m,p_-+ 2m,n-m],\ell} \cr
&{} \oplus  \bigoplus_{\ell\ge 0}  \bigoplus_{0 \le m \le n \le p_1-2} \!\!
\A^\Delta_{[n-m,p_-+ 2m,n-m],\ell}  \, , \cr}
}
in accordance with the results of Eden and Sokatchev \bpsN. In \decompB\
we identify $\B_{[0,0,0]} \simeq \I$, corresponding to the unit operator
in the operator product expansion. It immediately follows from \decompB\
that long supermultiplets, with non zero anomalous dimensions, cannot
contribute to extremal and next-to-extremal correlation functions.
\vfill
\eject

\newsec{Crossing Symmetry}

The operator product expansion provides the strongest constraints when 
combined with crossing symmetry. For a correlation function for four identical
chiral primary operators the correlation function is invariant under
permutations of all $x_i, t_i$ for all $i=1,2,3,4$. Permutations 
of the form $(ij)(kl)$ act trivially so we may restrict to permutations 
leaving $x_4, t_4$ invariant so that crossing symmetry transformations  
correspond to the permutation group $\S_3$, which is of order 6. 
The action of each permutation on the essential conformal 
invariants $u,v$ or $x,\zz$ or $y, \bz$ and also on the $R$-symmetry
invariants $\lam,\mun$ or $\alpha,\bet$ or $y,\yz$ is given in table 1,
where the transformations of $\zz$ are identical to those of $x$, and
similarly for $\bz,\bet,\yz$.

\vskip 6pt
\vbox{\tabskip=0pt \offinterlineskip
\hrule
\halign{&\vrule# &\strut \ \hfil#\  \cr
height2pt&\omit&&\omit&&\omit&&\omit&&\omit&&\omit&&\omit&
&\omit&&\omit&&\omit&&\omit&&\omit&&\omit&\cr
& $~e$ && $~(12)$ && $~(13)$ && $~(23)$ && $(123)$ && $(132)$ &&&
& $~e$ && $~(12)$ && $~(13)$ && $~(23)$ && $(123)$ && $(132)$ &\cr
height2pt&\omit&&\omit&&\omit&&\omit&&\omit&&\omit&&\omit& 
&\omit&&\omit&&\omit&&\omit&&\omit&&\omit&\cr
\noalign{\hrule}
height4pt&\omit&&\omit&&\omit&&\omit&&\omit&&\omit&&\omit&
&\omit&&\omit&&\omit&&\omit&&\omit&&\omit&\cr
&\hfil$u$\hfil&
&$\hfil{u\over v}\hfil$ &&\hfil$v$\hfil &&\hfil${1\over u}\hfil$ &
&\hfil${v\over u}$\hfil  &&\hfil${1\over v}$\hfil&&&
&\hfil $\lam$\hfil &&\hfil$\mun$\hfil &&\hfil${\lam \over \mun}$\hfil &
&\hfil${1\over \lam}$\hfil &&\hfil${1\over \mun}$\hfil  &&
\hfil${\mun\over \lam}$\hfil  &\cr
height4pt&\omit&&\omit&&\omit&&\omit&&\omit&&\omit&&\omit&&\omit&&\omit&
&\omit&&\omit&&\omit&&\omit&\cr
\noalign{\hrule}
height4pt&\omit&&\omit&&\omit&&\omit&&\omit&&\omit&&\omit&&\omit&&\omit&
&\omit&&\omit&&\omit&&\omit&\cr
&\hfil $v$\hfil &&\hfil${1\over v}$\hfil &&\hfil$u$\hfil &&\hfil${v\over u}$\hfil&
&\hfil${1\over u}$\hfil &&\hfil${u\over v}$\hfil &&&
&\hfil $\mun$\hfil &&\hfil$\lam$\hfil &&\hfil${1\over \mun}$\hfil &
&\hfil${\mun\over \lam}$\hfil &&
\hfil${\lam\over \mun}$\hfil  &&\hfil${1\over \lam}$\hfil  &\cr
height4pt&\omit&&\omit&&\omit&&\omit&&\omit&&\omit&&\omit&
&\omit&&\omit&&\omit&&\omit&&\omit&&\omit&\cr
\noalign{\hrule}
height4pt&\omit&&\omit&&\omit&&\omit&&\omit&&\omit&&\omit&&\omit&&\omit&
&\omit&&\omit&&\omit&&\omit&\cr
&\hfil $x$\hfil &&\hfil${x\over x-1}$\hfil&&\hfil$1-x$\hfil&&\hfil${1\over x}$\hfil&
&\hfil${x-1\over x}$\hfil &&\hfil${1\over 1-x}$\hfil &&&
&\hfil $\alpha$\hfil&&\hfil$1-\alpha$\hfil &&\hfil${\alpha\over \alpha-1}$\hfil &
&\hfil${1\over \alpha}$\hfil &&
\hfil${1\over 1-\alpha}$\hfil  &&\hfil${\alpha-1\over \alpha}$\hfil  &\cr
height4pt&\omit&&\omit&&\omit&&\omit&&\omit&&\omit&&\omit&&\omit&&\omit&&\omit&
&\omit&&\omit&&\omit&\cr
\noalign{\hrule}
height4pt&\omit&&\omit&&\omit&&\omit&&\omit&&\omit&&\omit&&\omit&&\omit&
&\omit&&\omit&&\omit&&\omit&\cr
&\hfil $z$\hfil &&\hfil$-z$\hfil &&\hfil${z+3\over z-1}$\hfil&
&\hfil${3-z\over 1+z}$\hfil &
&\hfil${3+z\over 1-z}$\hfil &&\hfil${z-3\over z+1}$\hfil &&&
&\hfil $y$\hfil &&\hfil$-y$\hfil &&\hfil${y+3\over y-1}$\hfil &
&\hfil${3-y\over 1+y}$\hfil &&
\hfil${y+3\over y-1}$\hfil  &&\hfil${y-3\over y+1}$\hfil  &\cr
height4pt&\omit&&\omit&&\omit&&\omit&&\omit&&\omit&&\omit&&\omit&&\omit&
&\omit&&\omit&&\omit&&\omit&\cr}
\hrule}
Table 1. Symmetry transformations of variables under crossing.

\subsec{$\N=4$}

For the $\N=4$ case with $p_i=p$ the crossing symmetry conditions on the
correlation function $\G(u,v;\lam,\mun)$ are generated by considering just
$(12)$ and $(13)$ which give
\eqn\cross{
\G(u,v;\lam,\mun) =  \G(u/v,1/v;\mun,\lam ) = 
\Big ( {u\over v}\, \mun \Big )^p \G(v,u;\lam/\mun,1/\mun) \, .
}
The general construction of such invariant correlation functions follows
by determining polynomials in $\lam,\mun$ which transform according
to the irreducible representations of $\S_3$. We first consider symmetric
polynomials satisfying
\eqn\polylm{
S_p(\lam,\mun) = S_p(\mun,\lam) = \mun^p S_p(\lam/\mun,1/\mun) \, .
}
As described by Heslop and Howe \Howe, for any given $p$, $\S_3$ acts on
the $\half(p+1)(p+2)$ monomials $\lam^r\mun^s, \, r+s\le p$, giving chains of 
length $6$ or $3$ or $1$ which may be added to give minimal polynomial 
solutions of \polylm. If the chain contains a monomial $(\lam\mun)^r$, for 
$0\le r \le [\half p]$, where $[x]$ denotes the integer part of $x$, then this 
term is invariant under the action of the permutation $(12)$ and the chain is 
of length 3, except if $p$ is divisible by 3 then $(\lam\mun)^{p/3}$ satisfies
\polylm\ by itself and so forms a chain of length 1. All other chains are of 
length 6. With this counting the number of independent such minimal symmetric 
polynomials is, 
\eqn\countN{
N_p = \cases{(n+1)3n+1 \, , \quad &\hbox{$p=6n\, ;$}  \cr
(n+1)(3n+q) \, , \quad &\hbox{$p=6n+q, \ q=1,2,3,4,5\, .$} \cr}
}

We list the first few non trivial cases in table 2, of course 
$S_0(\lam,\mun)=1$.

\vskip 6pt
\hbox{
\vbox{\tabskip=0pt \offinterlineskip
\hrule
\halign{&\vrule# &\strut \ \hfil#\  \cr
height2pt&\omit&&\omit&&\omit&\cr
&$p$&& polynomial\hfil &&$(i,j)$\hfil&\cr
height2pt&\omit&&\omit&&\omit&\cr
\noalign{\hrule}
height4pt&\omit&&\omit&&\omit&\cr
&$1$&&$\lam+\mun+1$\hfil&&$(0,0)$\hfil&\cr
height4pt&\omit&&\omit&&\omit&\cr
\noalign{\hrule}
height4pt&\omit&&\omit&&\omit&\cr
&$2$&&$\matrix{\scs\lam^2+\mun^2+1\hfill\cr\scs\lam\mun+\lam+\mun\hfill}$\hfil&
&${\scs(0,0),(1,0)}$\hfil&\cr
height4pt&\omit&&\omit&&\omit&\cr
\noalign{\hrule}
height4pt&\omit&&\omit&&\omit&\cr
&$3$&&$\matrix{\scs\lam^3+\mun^3+1\hfill\cr\scs\lam^2\mun+\lam\mun^2+\lam^2+
\mun^2+\lam+\mun\hfill\cr\scs\lam\mun\hfill}\hfil$&
&$\matrix{\scs(0,0),(1,0)\hfill\cr\scs(0,1)\hfill}$\hfil&\cr
height4pt&\omit&&\omit&&\omit&\cr
\noalign{\hrule}
height4pt&\omit&&\omit&&\omit&\cr
&$4$&&$\matrix{\scs\lam^4+\mun^4+1\hfill\cr\scs\lam^3\mun+\lam\mun^3+\lam^3+
\mun^3+\lam+\mun\hfill\cr
\scs\lam^2\mun^2+\lam^2+\mun^2\hfill\cr\scs\lam^2\mun+\lam\mun^2
+\lam\mun\hfill}$\hfil&&
$\matrix{\scs(0,0),(1,0),(2,0)\hfill\cr\scs(0,1)\hfill}$\hfil&\cr
height4pt&\omit&&\omit&&\omit&\cr
\noalign{\hrule}
height4pt&\omit&&\omit&&\omit&\cr
&$5$&&$\matrix{\scs\lam^5+\mun^5+1\hfill\cr\scs\lam^4\mun+\lam\mun^4+\lam^4+
\mun^4+\lam+\mun\hfill\cr\scs
\lam^3\mun^2+\lam^2\mun^3+\lam^3+\mun^3+\lam^2+\mun^2\hfill\cr
\scs\lam^3\mun+\lam\mun^3+\lam\mun\hfill\cr
\scs\lam^2\mun^2+\lam^2\mun+\lam\mun^2\hfill}$\hfil&&
$\matrix{\scs(0,0),(1,0),(2,0)\hfill\cr\scs(0,1),(1,1)\hfill}$\hfil&\cr
height4pt&\omit&&\omit&&\omit&\cr}
\hrule}
\vbox{\tabskip=0pt \offinterlineskip
\hrule
\halign{&\vrule# &\strut \ \hfil#\  \cr
height2pt&\omit&&\omit&&\omit&\cr
&$p$&& polynomial\hfil &&$(i,j)$\hfil&\cr
height2pt&\omit&&\omit&&\omit&\cr
\noalign{\hrule}
height4pt&\omit&&\omit&&\omit&\cr
&$6$&&$\matrix{\scs\lam^6+\mun^6+1\hfill\cr\scs\lam^5\mun+\lam\mun^5+\lam^5+
\mun^5+\lam+
\mun\hfill\cr\scs\lam^4\mun^2+\lam^2\mun^4+\lam^4+\mun^4+\lam^2+\mun^2\hfill\cr
\scs\lam^3\mun^3+\lam^3+\mun^3\hfill\cr
\scs\lam^4\mun+\lam\mun^4+\lam\mun\hfill\cr
\scs\lam^3\mun^2+\lam^2\mun^3+\lam^3\mun+\lam\mun^3+\lam^2\mun+\lam\mun^2\hfill\cr
\scs\lam^2\mun^2\hfill}$\hfil&&
$\matrix{\scs(0,0),(1,0),(2,0),(3,0)\hfill\cr\scs(0,1),(1,1)\hfill\cr
\scs(0,2)\hfill}$\hfil&\cr
height4pt&\omit&&\omit&&\omit&\cr
\noalign{\hrule}
height4pt&\omit&&\omit&&\omit&\cr
&$7$&&$\matrix{\scs\lam^7+\mun^7+1\hfill\cr\scs\lam^6\mun+\lam\mun^6+\lam^6+
\mun^6+\lam+\mun\hfill\cr\scs
\lam^5\mun^2+\lam^2\mun^5+\lam^5+\mun^5+\lam^2+\mun^2\hfill\cr
\scs\lam^4\mun^3+\lam^3\mun^4+\lam^4+\mun^4+\lam^3+\mun^3\hfill\cr
\scs\lam^5\mun+\lam\mun^5+\lam\mun\hfill\cr
\scs\lam^4\mun^2+\lam^2\mun^4+\lam^4\mun+\lam\mun^4+\lam^2\mun+\lam\mun^2\hfill\cr
\scs\lam^3\mun^3+\lam^3\mun+\lam\mun^3\hfill\cr
\scs\lam^3\mun^2+\lam^2\mun^3+\lam^2\mun^2\hfill}$\hfil&&
$\matrix{\scs(0,0),(1,0),(2,0),(3,0)\hfill\cr\scs(0,1),(1,1),(2,1)\hfill\cr
\scs(0,2)\hfill}$\hfil&\cr
height4pt&\omit&&\omit&&\omit&\cr}
\hrule\vskip21pt}
}

Table 2. Symmetric polynomials.

An alternative basis for $S_p$, valid for general $p$, may be obtained 
by constructing from $\lam,\mun$ two invariants $I_1,I_2$ under $\S_3$ 
and then introducing for any $p$ a factor
to ensure that \polylm\ holds. With suitable restrictions
the result becomes a polynomial expressible in the form
\eqn\Syy{\eqalign{
{S}_{p,(i,j)}(\lam,\mun) = {}& (\lam+\mun+1)^{p}\, I_1( \lam,\mun )^i
I_2( \lam,\mun )^j \, , \cr
I_1( \lam,\mun ) ={}& { \lam\mun+\lam+\mun \over (\lam+\mun+1)^2 } \, , 
\qquad I_2( \lam,\mun ) ={\lam\, \mun\over (\lam+\mun+1)^3} \, , \cr
i={}& 0,1,\dots ,[\half p] \, , \qquad
j = 0,1,\dots ,[{\ts{1\over 3}}(p-2i)] \, . \cr}
}
Lists of possible $(i,j)$ for $p$ up to 7 are given in table 2. 
This result may also be easily expressed as symmetric polynomial in $y,\yz$
by using
\eqn\lmyy{\eqalign{
\lam+\mun+1 = {}& \half ( y\yz+3 ) \, , \qquad \lam\mun = {\ts {1\over 16}}
(1-y^2)(1-\yz^2) \, , \cr
\Lambda = {}& (\lam+\mun+1)^2 - 4(\lam\mun+\lam+\mun)= 
{\ts {1\over 4}} ( y-\yz)^2 \, , \cr}
}
where $\Lambda$ is defined in \conlm.
Completeness of the basis provided by \Syy\ is straightforwardly demonstrated
by showing that it gives the same number of independent polynomials $N_p$
as given in \countN.

For the antisymmetric representation of $\S_3$ we require,
\eqn\repA{
a  \toinf{(12)} -a \, , \qquad a \toinf{(123)} a \, .
}
while the two-dimensional mixed symmetry representation of $\S_3$ is defined 
on a basis $(b,c)$ where
\eqn\mix{
\pmatrix{b\cr c} \toinf{(12)} \pmatrix{-1&0\cr0&1}\pmatrix{b\cr c} \, ,
\qquad
\pmatrix{b \cr c} \toinf{(123)} \pmatrix{-{1\over 2}&-{\sqrt 3 \over 2}\cr
{\sqrt 3 \over 2}&-{1\over 2}}  \pmatrix{b \cr c} \, .
}
It is easy to see that the tensor products formed by $a\,a'$ and 
$b \, b' + c\, c'$ are 
symmetric while $(b\, c' + c\, b', b\, b' + c\, c')$ is a basis for a
mixed symmetry representation and $b\, c' - c \, b'$ is antisymmetric.

For functions of $\lam,\mun$ \repA\  is satisfied by
\eqn\anti{
a(\lam,\mun) = {(\lam-\mun)(\lam-1)(\mun-1) \over (\lam+\mun+1)^3} \, .
} 
For $p\ge 3$, $a(\lam,\mun)S_{p,(i,j)}(\lam,\mun)$ is a polynomial if we allow
$i=0,1,\dots [\half(p-3)]$ and $j = 0,1,\dots ,[{\ts{1\over 3}}(p-2i-3)]$
giving $N_{p-3}$ antisymmetric polynomials.
For the mixed symmetry transformations in \mix\ there essentially two 
independent possibilities
\eqn\mixl{
b_1(\lam,\mun)= {\lam - \mun\over  \lam+\mun+1} \, ,
\qquad c_1(\lam,\mun) = {\lam+\mun - 2 \over \sqrt 3(\lam+\mun+1)} \, .
}
and
\eqn\mixlm{
b_2(\lam,\mun)= {\lam - \mun \over (\lam+\mun+1)^2} \, , 
\qquad c_2(\lam,\mun) = - {\lam+\mun - 2\lam\mun \over \sqrt 3(\lam+\mun+1)^2} \, .
}
By considering
$(b_r(\lam,\mun),c_r(\lam,\mun))S_{p,(i,j)}(\lam,\mun)$ for $p\ge r$, $r=1,2$,
for appropriate $i,j$ we obtain $N_{p-r}$ polynomial mixed symmetry
representations of $\S_3$. Together with the symmetric polynomials $S_{p,(i,j)}$
and $a S_{p,(i,j)}$ these provide a complete basis for two variable polynomials
in $\lam,\mun$ of order $p$ since 
$N_p + 2(N_{p-1}+N_{p-2})+N_{p-3}= \half(p+1)(p+2)$. We may also note that these 
polynomials form a closed set under multiplication since
\eqnn\mult
$$\eqalignno{
b_1{}^{\!2} + c_1{}^{\!2} = {}& {\ts{4\over 3}} - 4I_1 \, , \ \ \, \qquad
\sqrt 3 ( 2 b_1c_1 ,  b_1{}^{\!2} - c_1{}^{\!2}) 
= 2( b_1 - 3 b_2,  c_1 - 3 c_2 ) \, , \cr
b_1 b_2 + c_1 c_2 = {}& {\ts{2\over 3}}I_1 - 6I_2 \, , \qquad
\sqrt 3 ( b_1c_2 + c_1 b_2 ,  b_1 b_2  - c_1 c_2 ) 
= 2( I_1 b_1 - b_2,  I_1 c_1 - c_2 )  \, , \cr
b_2{}^{\!2} + c_2{}^{\!2} = {}&  {\ts{4\over 3}} I_1{\!}^2 - 4 I_2 \, , \qquad
\sqrt 3 ( 2 b_2c_2 ,  b_2{}^{\!2} - c_2{}^{\!2}) 
=  2( 3 I_2 b_1 - I_1 b_2,  3I_2 c_1 - I_1 c_2 ) \, , \cr
\sqrt 3(b_1c_2 - c_1 b_2) = {}& 2 a \, , \qquad\qquad\qquad
a^2 = I_1 {}^{\!2} - 4 I_1{}^{\!3} +18 I_1 I_2 - 4 I_2 - 
27 I_2{}^{\!2} \, . &\mult \cr}
$$
where $I_1,I_2$ are the invariants defined in \Syy.

The superconformal Ward identities require
\eqn\scc{
\G(u,v;\lam,\mun)\big |_{\bet = {1\over \zz}} = f(x,\alpha) \, ,
}
so that \cross\ gives
\eqn\crossf{
f(x,\alpha) = f \Big ( {x\over x-1}, 1 - \alpha \Big ) =
\bigg ( {x(\alpha-1) \over 1-x} \bigg )^p f \Big ( 1-x , 
{\alpha \over \alpha-1} \Big ) =  ( x\alpha  )^p 
f\Big ( {1\over x}, {1\over \alpha } \Big ) \, .
}
To obtain an extension to a fully crossing symmetric correlation function
we may consider for any $S_p$ satisfying \polylm
\eqn\crossG{
\G(u,v;\lam,\mun) = S_p\Big ( u \, \lam , {u\over v} \,\mun \Big ) \, ,
}
which obeys \cross\ as a consequence of \polylm. From \scc\ we obtain
\eqn\fsym{
f(x,\alpha) = S_p \Big ( x \alpha, {x(1-\alpha) \over x-1} \Big ) \, ,
}
which automatically satisfies \crossf.

The function $f(x,\alpha)$
is required to be a general solution of the crossing symmetry conditions 
given by 
\eqn\crossff{
f(x,\alpha) = f \Big ( {x\over x-1}, 1 - \alpha \Big ) =
\bigg ( {x(\alpha-1) \over 1-x} \bigg )^p f \Big ( 1-x , 
{\alpha \over \alpha-1} \Big ) =  ( x\alpha  )^p 
f\Big ( {1\over x}, {1\over \alpha } \Big ) \, .
}
which is also a polynomial of degree $p$ in $\alpha$.
It is also analytic in $x$ in the neighbourhood of $x=0$ with singularities 
only at $x=1,\infty$. If we write
\eqn\Fnew{
f(x,\alpha) = P(x,\alpha)^p g(x,\alpha) \, , \qquad
P(x,\alpha) = {x^2\alpha - 2x\alpha + 2x  - 1 \over x-1 } \, ,
}
then $g$ is an invariant under the action of $\S_3$, as displayed in Table 1.
Determining a general form for $g$ is then reducible to finding a basis for
all possible independent invariants which may be formed from $x$ and 
$\alpha$. Crossing symmetry in one variable $\alpha$ will be studied in the case of $\N=2$. 
We will need the results for the different representations of $\S_3$ here. Basically they are obtained from the two variable case 
by setting $\bet=\alpha$.
The symmetric or invariant representation is
\eqn\Sinvv{
s(\alpha) = {\alpha^2(1-\alpha)^2\over (\alpha^2-\alpha+1)^3}.
}
The antisymmetric representation is given by
\eqn\antiyy{
a(\alpha) = (2\alpha-1)\,{(\alpha-2)(\alpha-1)\alpha(\alpha+1)\over
(\alpha^2-\alpha+1)^3}.
}
The 2 mixed symmetry representation solutions are
\eqn\mixzz{\eqalign{
b_1(\alpha) =  {}& {2\alpha-1\over\alpha^2-\alpha+1},\cr
c_1(\alpha) = {}& {1\over \sqrt 3}\,{2\alpha^2-2\alpha -1 \over \alpha^2-\alpha+1},\cr}
}
and
\eqn\mixyy{\eqalign{
b_2(\alpha) =  {}&  (2\alpha-1){\alpha(\alpha-1)\over
(\alpha^2-\alpha+1)^2},\cr
c_2(\alpha) = {}& \sqrt 3 \, {\alpha(\alpha-1)\over (\alpha^2-\alpha+1)^2}.\cr}
}

Since the action of $\S_3$ on any polynomial in $\alpha$ may be 
decomposed, up to functions of the invariant $s(\alpha)$, into contributions 
linear in $1$, $a(\alpha)$ and $(b_r(\alpha),c_r(\alpha)), \,  r=1,2$, as 
given in \Sinvv, \antiyy, \mixzz, \mixyy, then a basis for such invariants 
is obtained, in addition to the separate invariants $s(x), \, s(\alpha)$,
by combining these non trivial irreducible representations with corresponding
representations involving $x$ to give
\eqn\SSinv{
A ( x , \alpha) = a( x^{-1}) \, a ( \alpha ) \, ,
}
where $a( x^{-1}) = - a(x)$, and also
\eqn\Smix{\eqalign{
S_1 ( x,\alpha ) ={}&  b_1( x^{-1})\, b_1( \alpha ) + c_1( x^{-1})\, c_1( \alpha )\cr
= {}& {4\over 3} - {2(x \alpha  -1)^2 \over (\alpha^2-\alpha+1)
(x^2-x+1)} \, , \cr
S_2 ( x,\alpha ) ={}&  b_2( x^{-1})\, b_2( \alpha ) + c_2( x^{-1})\, c_2( \alpha )\cr
= {}&  {2 \alpha(1-\alpha) \, x(1-x) \over 
(\alpha^2-\alpha+1)^2(x^2-x+1)^2} \, ( x\alpha - 2\alpha - 2x + 1) \, ,\cr
S_3 ( x,\alpha ) ={}&  b_2( x^{-1})\, b_1( \alpha ) + c_2( x^{-1})\, c_1( \alpha )\cr
= {}& {2 x(1-x) \over (\alpha^2-\alpha+1)(x^2-x+1)^2} 
\, ( x\alpha^2 - 2x\alpha + 2\alpha  - 1) \, , \cr
S_4 ( x,\alpha ) ={}&  b_1( x^{-1})\, b_2( \alpha ) + c_1( x^{-1})\, c_2( \alpha )\cr
= {}& {2 \alpha(1-\alpha) \over (\alpha^2-\alpha+1)^2(x^2-x+1)}
\, ( x^2\alpha - 2x\alpha + 2x  - 1) \, .
\cr}
}
These are not independent since
\eqn\relAS{\eqalign{
& A ( x , \alpha) = {\ts {3\over 4}} \big ( S_1 ( x,\alpha )S_2 ( x,\alpha )
- S_3 ( x,\alpha ) S_4 ( x,\alpha ) \big ) \, , \cr
& S_2( x,\alpha) - \half \, S_1( x,\alpha)S_3(x,\alpha) - {\ts {1\over 3}}\,
S_3(x,\alpha) = - 2s(x) \, , \cr 
& S_2( x,\alpha) - \half \, S_1( x,\alpha)S_4(x,\alpha) - {\ts {1\over 3}}\,
S_4(x,\alpha) = - 2s(\alpha)  \, , \cr
& 2 \big ( S_3(x,\alpha) + S_4(x,\alpha) \big ) - 6 S_2( x,\alpha) +
S_1 ( x,\alpha )^2 - {\ts {2\over 3}} S_1 ( x,\alpha ) = {\ts {8\over 9}}\, .\cr}
}
A crucial further constraint arises from \real\
which here requires that $f(x,x^{-1})$ is a constant. Since $P(x,x^{-1})=3$
we also require that $g$ depends on invariants $s_r(x,\alpha)$ such that
$s_r(x,x^{-1})$ are constants. Taking account of the relations in \relAS\ 
there are then two independent solutions which we take as
\eqnn\solss
$$\eqalignno{
s_1(x,\alpha) = {}&  2 \, {S_3(x,\alpha)s(\alpha)\over 
S_4(x,\alpha)^2} = { R(x,\alpha) \over P(x,\alpha)^2} \, , \qquad
s_2(x,\alpha) = 8 \, {s(x)s(\alpha)^2\over S_4(x,\alpha)^3}
= { Q(x,\alpha) \over P(x,\alpha)^3} \, , \cr
R(x,\alpha) = {}&  {x(x\alpha^2 - 2x\alpha + 2\alpha  - 1) \over 1-x }  \, , 
\qquad\quad Q(x,\alpha) = {x^2 \alpha(1-\alpha) \over x-1} \, , & \solss \cr}
$$
where $R(x,x^{-1})=3, \, Q(x,x^{-1})=1$. It is then evident that $g$ in
\Fnew\ must be of the form
\eqn\gexp{
g = \sum_{{i,j\ge0\atop 2i+3j \le p}} c_{ij} s_1{}^{\! i}\,  s_2{}^{\! j} \, .
}
Noting that
\eqn\relb{ \eqalign{
P(x,\alpha) = {}& \Big ( u \,\lam + {u\over v}\, \mun + 1 \Big )\Big |_{\bet  
= {1\over \zz}} \, , \qquad
Q(x,\alpha)= {u^2\over v} \, \lam\mun \Big  |_{\bet = {1\over \zz}}  \, , \cr
R (x,\alpha)= {}& \Big ( {u^2\over v}\,  \lam\mun + u \,\lam + {u\over v}\, \mun 
\Big ) \Big  |_{\bet = {1\over \zz}}  \, ,  \cr}
}
it is easy to see, as a consequence of \Syy, that $I_r(u\lam, u \mun/v)
|_{\bet = 1/\zz} = s_r(x,\alpha)$ and hence the expression given by 
\Fnew\ and \gexp\ for the 
function $f$ may always be extended to a fully crossing symmetric result 
for the full correlation function $\G$ of the form \crossG\ with 
$S_p(\lam,\mun)= (\lam+\mun+1)^p \sum_{i,j} c_{ij} \, I_1(\lam,\mun)^i\, 
I_2(\lam,\mun)^j$ and where $f$ satisfies \fsym. With appropriate 
coefficients for the independent terms in $S_p$ \crossG\ corresponds to the 
results of free field theory. 
In general, using the formalism of harmonic superspace, the Intriligator
insertion technique \Intr\ demonstrates that only $\K$ as in \Gsol\ or \Gsolt, 
can depend on the coupling $g$, and so are
dynamical. The functions $f(x)$ or $f(x,\alpha)$ are then identical with
the free theory, or $g=0$, results.

The remaining part of the correlation function may also be expressed
in terms of $\S_3$ representations. It is convenient  to define from \mixl\ 
and \mixlm\ $(b_r{\!}'(u,v),c_r{\!}'(u,v)) = (b_r(1/u,v/u),c_r(1/u,v/u))$.
We may then write
\eqnn\factf
$$\eqalignno{
( \alpha x & -1  ) ( \alpha \zz - 1 ) ( \bet x -1  ) ( \bet \zz - 1 ) 
= {\ts {1\over 16}} u^2 (y-z)(y-\bz)(\yz-z)(\yz-\bz) \cr
={}& v + \lam^2 uv + \mun^2 u + \lam \, v(v-1-u) + \mun (u+v-1) + 
\lam \mun \, u(u-1-v)\cr
= {}& (\lam+\mun+1)^2(u+v+1)^2 \Big ({\ts {1\over 3}} I_1 (u,v) 
+ {\ts {1\over 3}} I_1 (\lam,\mun) - 2 I_1 (u,v) I_1 (\lam,\mun)  \cr
&\hskip 4.1cm {}- \half 
\big ( b_1{\!}'(u,v) \, b_2(\lam,\mun) +  c_1{\!}'(u,v) \, c_2(\lam,\mun) \cr 
& \hskip 4.5cm {}+ b_2{\!}'(u,v) \, b_1(\lam,\mun) 
+ c_2{\!}'(u,v) \, c_1(\lam,\mun) \cr
& \hskip 4.5cm {}  - 3 \, b_2{\!}'(u,v) \, b_2(\lam,\mun)
- 3\,  c_2{\!}'(u,v) \, c_2(\lam,\mun) \big ) \Big ) \, . & \factf \cr}  
$$
The function $\K$ in \Gsolt\ must then satisfy the crossing symmetry relations
\eqn\crossK{
\K(u,v;\lam,\mun) =  \K(u/v,1/v;\mun,\lam ) =
\Big ( {u\over v}\Big )^{p+2} \mun^{p-2}\, \K(v,u;\lam/\mun,1/\mun) \, .
}

\subsec{$\N=2$}

For $\N=2$ there are further restrictions as a consequence of \conlm. Taking
$p\to 2n$ we construct, instead of \polylm\ since $\lam,\mun$ are expressible
in terms of just $\alpha$ by \defa, the single variable polynomials $f_n$ of
degree $2n$, satisfying under the action of $\S_3$
\eqn\Sy{\eqalign{
f_n(\alpha) = {}& f_n(1-\alpha) = (\alpha-1)^{2n} f_n \Big ( 
{\alpha\over \alpha-1} \Big ) = \alpha^{2n} f_n \Big ( {1\over \alpha } \Big ) \cr
= {}&  (\alpha-1)^{2n} f_n \Big ( {1 \over 1- \alpha} \Big ) =
(\alpha-1)^{2n} f_n \Big ( {\alpha -1 \over \alpha} \Big ) \, . \cr}
}
As shown by Heslop and Howe, \Howe, the sum of terms produced by the
action of $\S_3$ as given by \Sy\ starting from $\alpha^{r}$
generates a linearly independent set of polynomials
for $r=0,1,\dots [{\ts{1\over 3}}n]$, giving $[{\ts{1\over 3}}n]+1$ solutions
for $f_n$. Alternatively an equivalent basis is provided by
\eqn\Soly{
{S}_{n,j}(\alpha) = (\alpha^2-\alpha+1)^n s(\alpha)^j \, , 
\qquad j = 0,1, \dots , [{\ts{1\over 3}}n] \, ,
}
where $s(\alpha)$ is the $\S_3$ invariant
\eqn\Sinv{
s(\alpha) = {\alpha^2(1-\alpha)^2\over (\alpha^2-\alpha+1)^3} =
4\, {(1-y^2)^2 \over (y^2+3)^3}  \, .
}
The solutions given by \Soly\ correspond to \Syy\ for $i=0$ since $\Lambda=0$ 
in this case. A general polynomial solution of \Sy\ is then given by
\eqn\fgen{
f_n(\alpha) = (\alpha^2-\alpha+1)^n P(s(\alpha)) \, ,
}
with $P(s)$ a polynomial of degree $[{\ts{1\over 3}}n]$.

We may also consider other representations of $\S_3$.
For the antisymmetric representation, as in \anti, we may define
\eqn\antiy{
a(\alpha) = (2\alpha-1)\,{(\alpha-2)(\alpha-1)\alpha(\alpha+1)\over
(\alpha^2-\alpha+1)^3} = 4 \, {y(y^2-1)(y^2-9)\over (y^2+3)^3 } \, ,
}
so that $a(\alpha){S}_{n,j}(\alpha)$ is then a polynomial for $n\ge 3$ and
$j = 0,1, \dots , [{\ts{1\over 3}}n]-1$. For the mixed symmetry representation 
there are two essential solutions which can be written in the form
\eqn\mixz{\eqalign{
b_1(\alpha) =  {}& {2\alpha-1\over\alpha^2-\alpha+1} = {4} \,  {y\over y^2+3} \, , \cr
c_1(\alpha) = {}& {1\over \sqrt 3}\,{2\alpha^2-2\alpha -1 \over \alpha^2-\alpha+1} =
{2\over \sqrt 3} \, {y^2 -3 \over y^2+3} \, , \cr}
}
and
\eqn\mixy{\eqalign{
b_2(\alpha) =  {}&  (2\alpha-1){\alpha(\alpha-1)\over
(\alpha^2-\alpha+1)^2} = {4} \, {y(y^2-1) \over (y^2+3)^2} \, , \cr
c_2(\alpha) = {}& \sqrt 3 \, {\alpha(\alpha-1)\over (\alpha^2-\alpha+1)^2} = 
4\sqrt 3 \, {y^2-1 \over (y^2+3)^2} \, . \cr}
}
It is easy to see that $(b_r(\alpha),c_r(\alpha)) {S}_{n,j}(\alpha)$
are polynomials for $j = 0,1, \dots , [{\ts{1\over 3}}(n-r)]$ if $n\ge r$
for $r=1,2$. The basis provided by ${S}_{n,j}(\alpha)$, 
$(b_r(\alpha),c_r(\alpha)) {S}_{n,j}(\alpha)$, $r=1,2$ and
$a(\alpha){S}_{n,j}(\alpha)$ is then complete in that it gives $2n+1$
linearly independent polynomials, allowing for the expansion of any
arbitrary polynomial of degree $2n$, $2\big([{\ts{1\over 3}}n]+
[{\ts{1\over 3}}(n-1)]+[{\ts{1\over 3}}(n-2)]\big)+5=2n+1$.

For the $\N=2$ case instead of \cross\ we have
\eqn\crosstwo{
\G(u,v;\lam,\mun) =  \G(u/v,1/v;\mun,\lam ) =
\Big ( {u^2\over v^2}\, \mun \Big )^n \G(v,u;\lam/\mun,1/\mun) \, ,
}
where, with $\lam,\mun$ constrained as in \defa, the superconformal Ward
identities are
\eqn\scct{
\G(u,v;\lam,\mun)\big |_{\alpha = {1\over \zz}} = f(x) = 
f \Big ( {x\over x-1} \Big ) = \bigg ( {x\over x-1} \bigg )^{\! 2n}\! f(1-x) 
= x^{2n} f \Big ( {1\over x} \Big ) \, ,
}
where we also exhibit the crossing symmetry relations for the single variable
function $f$. The corresponding solution to  \crossG\ is given by
\eqn\crossGt{
\G(u,v;\lam,\mun) = S_n\Big ( u^2 \, \lam , {u^2\over v^2} \,\mun \Big ) \, ,
}
which implies
\eqn\fSn{
f(x) = S_n \Big ( x^2 , {x^2\over (1-x)^2} \Big ) \, .
}
In this case if we consider the contribution of individual factors in the basis 
given by \Syy\ to $f(x)$ as expected from \crossGt\ and \fSn\ we have
\eqn\rela{ \eqalign{
P= {}& \Big ( u^2 \lam + {u^2\over v^2}\, \mun + 1 \Big )\Big |_{\alpha  
= {1\over \zz}} \!  =  p^2 \, , \quad
Q = {u^4\over v^2} \, \lam\mun \Big  |_{\alpha = {1\over \zz}} \! = q^2 \, , \cr
R = {}& \Big ( {u^4\over v^2}\,  \lam\mun + u^2 \lam + {u^2\over v^2}\, \mun 
\Big ) \Big  |_{\alpha = {1\over \zz}} \! = 2pq  \, ,  \cr}
}
where
\eqn\defpq{
p(x) = {x^2 -x +1 \over 1-x} \, , \qquad q(x) = {x^2\over 1-x} \, .
}
so that we have the relation $R^2=4PQ$. In consequence we may restrict in 
\Syy\ to those polynomials with $i=0,1$.

Conversely we may argue that for the $\N=2$ case all single variable 
functions $f(x)$ may be expressible in terms of $S_n$ as in \fSn\ 
and therefore may be extended to a fully crossing symmetric form for
$\G(u,v;\lam,\mun)$ as exhibited in \crossGt. To demonstrate this
we suppose all solutions of the crossing symmetry relations in \scct\ for
$f$ are solvable by writing
\eqn\fgr{
f(x) = p(x)^{2n} g\big ( s(x) \big ) \, , \qquad s(x) = {q(x) \over p(x)^3} \, ,
}
for some function $g$ of the crossing invariant $s$ given by $\Sinv$. 
Note that for $x\to 0, \, s \sim x^2$, $x \to 1 , \, s \sim (1-x)^2$ and for 
$x \to \infty, \, s \sim 1/x^2$. From the superconformal representation 
theory for the corresponding
contributions to the operator product expansion $f(x)$ should be analytic
in the neighbourhood of $x=0$ with singularities only at $x=1,\infty$. In
consequence $g(s)$ must be a polynomial which is then restricted to have
maximal degree $[{2\over 3}n]$ to avoid singularities when $x^2-x+1=0$. 
It is then easy to see that $f$
can be written as a polynomial in $P,Q$ with terms also linear in $R$, as 
defined in \rela, which is consistent with \fSn\ where $S_n$ has an  expansion
in terms of $S_{n,(i,j)}$ with $i=0,1$ and $j$ restricted as in \Syy.

The remaining part of the correlation function may also be expressed
in terms of $\S_3$ representations. It is convenient as for $\N=4$ to define from \mixl\ 
and \mixlm\ $(b_r{\!}'(u,v),c_r{\!}'(u,v)) = (b_r(1/u,v/u),c_r(1/u,v/u))$.
We may then write for the factor which appears in the
solution of the superconformal identities in \Gsol
\eqn\factwo{
(\alpha x - 1 ) (\alpha \zz - 1 ) = (\alpha^2-\alpha+1)(u+v+1) \big (
{\ts {1\over 3}} - \half ( b_1{\!}'(u,v) \, b_1(\alpha) + 
c_1{\!}'(u,v) \, c_1 (\alpha) ) \big ) \, .
} 

\subsec{(Next-to-)Extremal Case}

It is also of interest to extend the considerations of crossing symmetry to
the next-to-extremal case when $p_1=p_2=p_3=p$, $p_4=3p-2$.  In this case
$\G$, defined by \GFp, must satisfy for the permutations $(12)$ and $(23)$
\eqn\Gcross{
\G(u,v;\lam,\mun) = v^{p-1} \G\Big ({u\over v},{1\over v};\mun, \lam\Big ) \, , 
\qquad \G(u,v;\lam,\mun) = u^{2p-1}\lam\, 
\G\Big ({1\over u},{v\over u};{1\over \lam}, {\mun \over \lam} \Big )  \, .
}
The solution \next\ can be rewritten as
\eqn\Gext{
\G(u,v;\lam,\mun) = u^{p -1} \bigg ( k + {x\zz\over x - \zz}
\Big ( \big (\alpha -1/x\big )\big (\bet  - 1/x\big ) f(x) - 
\big (\alpha  - 1/\zz\big )\big (\bet - 1/\zz \big ) f(\zz) \Big ) \bigg ) \, ,
}
and then \Gcross\ requires
\eqn\fcross{
f(x) = - f \Big ( {x \over x-1} \Big ) \, , \qquad f(x) = - x^2 f \Big ( {1\over x}
\Big ) + k \, x \, .
}
A particular solution of \fcross\ is given by
\eqn\fsol{
f(x) = {k\over 3}\, \Big ( x - {x \over x-1} \Big ) \, .
}
To obtain a general solution of \fcross\ it is then sufficient to seek the 
general solution $f_0(x)$ of \fcross\ with $k=0$. Using results obtained
above this is
\eqn\fgen{
f_0(x) = {(x-2)x(x+1)(2x-1)\over (x-1)(x^2-x+1)} \, h\big ( s(x) \big ) \, ,
}
where $s$ is the invariant defined by \fgr\ and \defpq. This introduces 
unphysical singularities for $x^2-x+1=0$ unless cancelled by $h$. However, for
compatibility with semi-short representations, $h(s)$ must be analytic in $s$
for $s\sim 0$ (if $h(s)=1/s$, which cancels the singularity at $x^2-x+1=0$, then
$f_0(x) \sim 1/x$ for $x\to 0$). Hence we conclude that there is no possible
solution of the form \fgen\ and hence we only have \fsol. In this case
\eqn\Gwex{
\G(u,v;\lam,\mun) =  {\ts {1\over 3}}k  \, u^{p-1} \Big ( 1 + \lam \, u + 
\mun \, {u\over v} \Big ) \, .
}
\vfill
\eject

\newsec{Large $N$ Results}

\subsec{Simplification of the Amplitude for $p=4$}

For $p=4$ the results obtained in \Degen\ are expressible in terms of two functions $\F(u,v), \tF(u,v)$ in the following way
\eqn\HFtFP{\eqalign{
\H^{(4)}(u,v;\sigma,\tau)={}&{u\over v}\F(u,v)+\sigma^2{u^3\over v^2}\F(1/v,u/v)+\tau^2{u\over v^2}\F(v,u)\cr
&{}+\sigma\tau{u^3\over v^2}\tF(u,v)+\sigma{u^2\over v}\tF(v,u)+\tau{u^2\over v^3}\tF(1/v,{u/v}).
}}
The functions $\F(u,v), \tF(u,v)$ both satisfy the same crossing symmetry constraint
\eqn\cs{\F(u,v)={1\over v}\F(u/v,1/v),\quad \tF(u,v)={1\over v}\tF(u/v,1/v).}
This ensures that $\H^{(4)}(u,v;\sigma,\tau)$ satisfies the crossing symmetry relations
\eqn\csh{\eqalign{
\H^{(4)}(u,v;\sigma\tau)={}&{1\over v^2}\H^{(4)}(u/v,1/v;\tau,\sigma)\cr
={}&\left({u\over v}\right)^2 \tau^{2}\H^{(4)}(v,u;\sigma/\tau,1/\tau)
}}
The results from \Degen\ give $\F(u,v), \tF(u,v)$ in terms of $\oD$ functions \Df\ (note $\F(u,v)=\alpha_1(u,v)$, $\tF(u,v)=\beta_3(v,u)$) which were discussed in the introduction 
and identities for which can be found in appendix G. 
Using identities \didso, \didst{b} we directly obtain
\eqn\oldab{\eqalign{
\F(u,v)&=-{4\over N^2}(u^2\oD_{4446}+2u\oD_{3346}+2\oD_{2246}),\cr
\tF(u,v)&=-{16\over N^2}v(\oD_{4446}-\oD_{3355}+5\oD_{3344}+\oD_{4244}+\oD_{2444}+\oD_{3243}+\oD_{2343}+\oD_{2242}).
}}
Using \didst{a} and \didst{b} we can rewrite
\eqn\rewf{
\F(u,v)=-{4\over N^2}u^3v(\oD_{4644}+2\oD_{4633}+2\oD_{4622}).
}
To simplify $\tF(u,v)$ we first use \didst{b} to pull a factor of $u$ out of $\tF(u,v)$
\eqn\tfsone{
\tF(u,v)={}-{16\over N^2}uv(\oD_{6444}-u\oD_{5533}+5\oD_{4433}+\oD_{4424}+\oD_{4442}+\oD_{3423}+\oD_{3432}+\oD_{2422}).
}
Now we use \didtutd{c} for $\oD_{5533}$ to obtain
\eqn\tfstwo{
\tF(u,v)={}-{16\over N^2}uv(\oD_{6444}-\oD_{4444}+4\oD_{4433}+\oD_{4424}+\oD_{4442}+\oD_{3423}+\oD_{3432}+\oD_{2422}).
}
Using \didst{a} for $\oD_{6444}$ and \didtutd{b} for $\oD_{3432}, \oD_{4442}, \oD_{4424}, \oD_{2422}$
we can simplify $\tF(u,v)$ to 
\eqn\tfsthree{
\tF(u,v)=-{16\over N^2}uv(v(\oD_{4644}+\oD_{4552}+\oD_{3533}+\oD_{3542}+\oD_{2532})+3\oD_{4433}-\oD_{4444}-\oD_{5443}).
}
Now we use \dids\ for $3\oD_{4433}$ and \didref\ and \didst{a}\ for $\oD_{5443}$ to obtain
\eqn\tfsfour{
\tF(u,v)=-{16\over N^2}uv^2(\oD_{4644}+\oD_{4552}+\oD_{4543}+\oD_{3533}+\oD_{3542}+\oD_{2532}).
}
We use \didtutd{a}\ for $\oD_{2532}$ and obtain
\eqn\tfsfive{
\tF(u,v)=-{16\over N^2}uv^2(\oD_{4644}+\oD_{4552}+\oD_{4543}+\oD_{3533}+\oD_{2633}).
}
From \didref\ for $\oD_{4552}$ and \didtutd{a}\ for $\oD_{4543}$ we get
\eqn\tfssix{
\tF(u,v)=-{16\over N^2}uv^2(\oD_{4644}+\oD_{3643}+\oD_{3634}+\oD_{2633}).
}
Now we use \didtutd{a}\ for $\oD_{3643}$ and obtain
\eqn\tfsseven{
\tF(u,v)=-{16\over N^2}uv^2(\oD_{4644}-2\oD_{2633}+\oD_{2734}+\oD_{3634}+\oD_{2633}).
}
Finally, we use \dids\ for $\oD_{2734}+\oD_{3634}$. This gives us the final result
\eqn\tFf{
\tF(u,v)=-{16\over N^2}uv^2(\oD_{4644}-\oD_{2644}+2\oD_{2633}).
}
Using this result we can write down the full amplitude in a simplified and manifestly crossing symmetric form
\eqnn\Hfour
$$\eqalignno{
\H^{(4)}(u,v;\lam,\mun) = - {4\over N^2} \, u^4 & \Big ( (1+\lam^2 + \mun^2
+ 4 \lam + 4 \mun + 4 \, \lam\mun ) \oD_{4644} \cr 
&{}+ 2 ( \oD_{4633} + \oD_{4622} ) + 2\lam^2 ( \oD_{3634} + \oD_{2624} )
+2 \mun^2 ( \oD_{3643} + \oD_{2642} ) \cr
&{} - 4\, \lam( \oD_{4624} -2 \oD_{3623} ) - 4 \, \mun( \oD_{4642} -2 \oD_{3632} ) \cr
&{} - 4\, \lam\mun( \oD_{2644} -2 \oD_{2633} ) \Big ) \, . & \Hfour \cr}
$$
Since $\K(u,v;\lam,\mun) = {1\over 16}u^2 \H(u,v;\lam,\mun)$ it is easy to
verify both the crossing symmetry conditions \crossK\ using $\oD$ identities.
Furthermore the results given by \Htwo, \Hthree\ and \Hfour, in which overall
factors of $u^p$ are present, are manifestly
compatible with the unitarity conditions flowing from \Kexp\ and \unit\ 
since the leading log. term $\oD_{n_1n_2n_3n_4}(u,v)$ is $\log u$ itself.
When expressed in terms of conformal partial waves $\G^{(\ell)}_{\Delta+4}$
it is easy to see in each case that only contributions with minimum twist 
$\Delta - \ell = 2p$ are required. Hence \Htwo, \Hthree\ and \Hfour\ require the
presence of operators belonging to long multiplets which have anomalous dimensions 
with twist, at zeroth order in $1/N$, $\Delta - \ell = 2(p+t), \ t=0,1,2,\dots $
for the lowest scale dimension operators in each multiplet. The condition
$\Delta - \ell = 2p$ is stronger than that required by unitarity \unit, with
$n\le p-2$, which shows that for any representation some low twist multiplets
decouple (thus for the singlet case twist 2 is absent as it disappears in the large $N$
limit but twist 4 multiplets, which are necessary in the $p=2$ correlation function, 
decouple from the correlation functions for $p=3,4$).  

\subsec{Computation of First Order Anomalous Dimensions}

For $p_i=p$ in the large $N$ limit the leading result for the free contribution of single trace 
operators may be, with a suitable normalisation, simply obtained from 
disconnected graphs in free field theory
\eqn\GlN{
\G^{(p)}_0(u,v;\lam,\mun) = 1 + (\lam u)^p + \Big ( \mun \, {u\over v} \Big )^p \, .
}
The definitions \solid\ and \real\ then give
\eqn\flN{
f_0^{(p)}(z,y) = 1 + \Big ( {1+y \over 1+z} \Big )^p + \Big ( {1-y \over 1-z} \Big )^p \, ,
\qquad k=3 \, .
}
Using \Gsolt\ we can then determine, assuming
$\K^{(p)}(u,v;\lam,\mun) = {1\over 16}u^2 \H^{(p)}(u,v;\lam,\mun)$, the free field expression
\eqn\Hztwo{
\H_0^{(2)}(u,v;\lam,\mun)= 1 + {1\over v^2} \, ,
}
\eqnn\Hzthree
$$\eqalignno{ 
\H_0^{(3)}(u,v;\lam,\mun)= {1\over v^3} \Big (& \half(\lam+\mun) u (1+v^3)
+ \half(\lam- \mun) \big ( -3u(1-v^3) + 2(1-v)(1+v^3) \big ) \cr 
&{}+ u (1+v^3) - 1 + 2v+2v^3 - v^4 \Big ) \, , & \Hzthree \cr}
$$
\eqnn\Hzfour
$$\eqalignno{
\H_0^{(4)}(u,v;\lam,\mun)= {1\over v^4}&  \Big ( \lam\mun \, u^2(1+v^4) \cr
&{}  + \half (\lam-\mun)^2
\big ( 2(1-v)^2(1+v^4) - 5u (1+v^5) + 3uv (1+v^3) + 4u^2(1+v^4) \big ) \cr
&{}+ \half (\lam^2 - \mun^2) \big ( u(1-v)(1+v^4) - 2u^2 (1-v^4) \big ) \cr
&{} + \half (\lam+\mun) \big ( - (1-v)^2 + u (1+v) \big ) (1+v^4)  \cr
&{} + \half (\lam-\mun) \big ( (1-v) (- 3(1+v)+7u) (1+v^4) + 8v(1-v)(1+v^3)
- 4u^2 (1-v^4) \big ) \cr
&{}+ 1+v^6 - 3v(1-v)(1-v^3) - 2u (1-v)(1-v^4) + u^2(1+v^4) 
\Big ) \, . & \Hzfour \cr}
$$
In each case the crossing symmetry relation $\H_0^{(p)}(u,v;\lam,\mun) =
\H_0^{(p)}(u/v,1/v;\mun,\lam)/v^2$ is satisfied but the corresponding one for
$u\leftrightarrow v$ is not since it is necessary to take account of the
function $f_0^{(p)}(z,y)$ then as well.

To obtain the anomalous scale dimensions in detail it is necessary to decompose 
both \Htwo, \Hthree, \Hfour\ and \Hztwo, \Hzthree, \Hzfour\ in terms of different 
representations, as in \Kexp, and then to expand each term in conformal partial waves.
The expressions \Hztwo, \Hzthree\ and \Hzfour\ require 
contributions with twist zero and above but the corresponding low twist operators
in long supermultiplets, for which there are no anomalous dimensions, 
are cancelled by semi-short multiplets which are required by the expansion
of $f_0(z,y)$. For $p=2,3$ and $4$ a detailed discussion is contained in \scft,\ADHS\ and \od
(although some details are different the analysis is equivalent to the
the results that would be obtained by expanding $\H$ as given by \Hztwo\ and \Hzthree).
For $p=4$ we first decompose $\H$ into contributions for each $SU(4)$ representation \od
\eqnn\Hreps
$$\eqalignno{
A_{22}={}&-{2\over 5N^2}u^4\left(\oD_{4644}+\oD_{2633}-\oD_{2644}-{1\over 3}(\oD_{1643}+\oD_{1634})\right),\cr
A_{21}={}&-{2\over 5N^2}u^4\left(\oD_{3634}-\oD_{3643}-\oD_{1634}+\oD_{1643}\right),\cr
A_{20}={}&-{4\over 5N^2}u^4\left(-\oD_{2633}-{1\over 3}(\oD_{1643}+\oD_{1634})\right),\cr
A_{11}={}&-{4\over 5N^2}u^4\left(8\oD_{4644}+{20\over 3}\oD_{4633}-{1\over 3}\oD_{2633}-{14\over 3}\oD_{2644}-(\oD_{1643}+\oD_{1634})\right),\cr
A_{10}={}&-{2\over 3N^2}u^4\left(4(\oD_{3623}-\oD_{3632})+5(\oD_{3634}-\oD_{3643})-\oD_{1634}+\oD_{1643}\right),\cr
A_{00}={}&-{4\over N^2}u^4\biggl({5\over 2}\oD_{4644}+{10\over 3}\oD_{4633}+2\oD_{4622}+{1\over 30}\oD_{2633}-{5\over 6}\oD_{2644}\cr
&{}{\hskip 1.5 cm}-{1\over 10}(\oD_{1643}+\oD_{1634})\biggr).& \Hreps \cr
}$$
This simplified form is achieved by using the following identities for $\oD$-functions \od\ which can be derived from the basic identities in appendix G
\eqnn\did
$$\eqalignno{
\oD_{3634}(u,v)+\oD_{3643}(u,v)={}&-\oD_{2644}(u,v)+\oD_{2633}(u,v),\cr
\oD_{2624}(u,v)+\oD_{2642}(u,v)={}&-2\oD_{2633}(u,v)-\oD_{1634}(u,v)-\oD_{1643}(u,v)+2{1+v\over u v^3},\cr
\oD_{3623}(u,v)+\oD_{3632}(u,v)={}&-\oD_{2633}(u,v)+{1\over u^2v^2},\cr
\oD_{4624}(u,v)+\oD_{4642}(u,v)={}&-2\oD_{4633}(u,v)+\oD_{2644}(u,v)-2\oD_{2633}(u,v)+{1\over u^2 v^2},\cr
\oD_{2624}(u,v)-\oD_{2642}(u,v)={}&-\oD_{1634}(u,v)+\oD_{1643}(u,v)-2{1-v\over u v^3},\cr
\oD_{4624}(u,v)-\oD_{4642}(u,v)={}&\oD_{3623}(u,v)-\oD_{3632}(u,v)-\oD_{3634}(u,v)+\oD_{3643}(u,v).\cr
& & \did \cr
}$$
Although $\oD_{1634}+\oD_{1643}=\oD_{4361}+\oD_{3461}$ and we might use the standard identity \dids\ this would lead to $\Delta_4=0$ and the appearance of $\oD_{3371}$. This will introduce two singularities.
Still a similar relation taking account of the singularities in this case can be found in \od.
We have omitted terms in the amplitudes which do not contain any $\log u$ factors and which thus do not contribute to anomalous dimensions.
The $L^2$-eigenfunctions we used are consistent with the $p_{nm}$'s in appendix B, which differ from the $Y_{nm}$'s by a normalization factor of $3,10$ for $n=1,2$ respectively.
We split the amplitude into two parts, one regular in $u$ and one containing a factor of $\log u$
\eqn\spl{A_{nm}(u,v)=\half \log u\ {\hat A_{nm}(u,v)}+\ldots}
Now we perform a conformal partial wave expansion of ${\hat A_{nm}}$ using the computer program in appendix F
\eqn\cpwe{\hat A_{nm}(u,v)=\sum_{\tau\geq4,\ell} A_{nm,\tau\ell}u^{\tau}G_{\ell+2\tau+4}^{(\ell)}(u,v).}
Here we set $\tau=\half (\Delta-\ell)$ since only even twists occur.
We factor off a universal part
\eqn\cpweu{A_{nm,\tau\ell}=(\tau-3)_8\ 2^{\ell-5}{(\tau!)^2(\ell+\tau+1)!^2\over 45(2\ell+2\tau+2)!(2\tau)!}a_{nm,\tau\ell}.}
There is an arbitrary choice of normalization here which agrees with the one used in \od.
The results we obtain for the expansion coefficients are
\eqnn\cpwer
$$\eqalignno{
a_{22,\tau\ell}={}&-{4\over 75N^2}\big((\ell+1)(\ell+2\tau+2)(25(\ell+\tau+1)(\ell+\tau+2)+57\tau(\tau+1)-90)\cr
\noalign{\vskip-3pt${\hskip 1.4 cm}+60(\tau-1)\tau(\tau+1)(\tau+2)\big)$,}
a_{21,\tau\ell}={}&-{4\over 25N^2}(\ell+1)(\ell+2\tau+2)\big(25(\ell+\tau+1)(\ell+\tau+2)+9\tau(\tau+1)-30\big),\cr
a_{20,\tau\ell}={}&-{8\over 15N^2}(\ell+1)(\ell+2\tau+2)\big(5(\ell+\tau+1)(\ell+\tau+2)-3\tau(\tau+1)\big),\cr
a_{11,\tau\ell}={}&-{8\over 25N^2}\big((\ell+1)(\ell+2\tau+2)(25(\ell+\tau+1)(\ell+\tau+2)+97\tau(\tau+1)-1140)\cr
&{}{\hskip 1.4 cm}+160(\tau-3)\tau(\tau+1)(\tau+4)\big),\cr
a_{10,\tau\ell}={}&-{4\over 3N^2}(\ell+1)(\ell+2\tau+2)\big(5(\ell+\tau+1)(\ell+\tau+2)+21\tau(\tau+1)-270\big),\cr
a_{00,\tau\ell}={}&-{4\over 5N^2}\big((\ell+1)(\ell+2\tau+2)(5(\ell+\tau+1)(\ell+\tau+2)+37\tau(\tau+1)-450)\cr
&{}{\hskip 1.3 cm}+100(\tau-3)(\tau-2)(\tau+3)(\tau+4)\big).\cr &{\ }&\cpwer\cr
}$$
The final quantity we compute is $\left<\eta_{nm,\tau\ell}\right>={a_{nm,\tau\ell}\over a^{(0)}_{nm,\tau\ell}}$. $\left<\eta\right>$ is the expectation value of the first order anomalous dimension 
of all the operators in one particular representation of the amplitude. 
In particular for only one operator present this actually directly gives the first order anomalous dimension.
For $A^{(0)}$ we take the free amplitudes computed in \od (setting $a=b=c=0$ since these are proportional to $1/N^2$)
\eqn\azzero{\eqalign{
A^{(0)}_{nm,\tau\ell}=2^{\ell-5}{(\tau !)^2(\ell+\tau+1)!^2\over 45(2\ell+2\tau+2)!(2\tau)!}a^{(0)}_{nm,\tau\ell},
}}
\eqnn\azero
$$\eqalignno{
a^{(0)}_{22,\tau\ell}={}&{1\over 3}(\tau-1)\tau(\tau+1)(\tau+2)(\ell+1)(\ell+2\tau+2)\cr&{}\times (\ell+\tau)(\ell+\tau+1)(\ell+\tau+2)(\ell+\tau+3),\cr
a^{(0)}_{21,\tau\ell}={}&(\tau-2)\tau(\tau+1)(\tau+3)(\ell+1)(\ell+2\tau+2)\cr&{}\times (\ell+\tau-1)(\ell+\tau+1)(\ell+\tau+2)(\ell+\tau+4),\cr
a^{(0)}_{20,\tau\ell}={}&{2\over 3}(\tau-2)(\tau-1)(\tau+2)(\tau+3)(\ell+1)(\ell+2\tau+2)\cr&{}\times (\ell+\tau-1)(\ell+\tau)(\ell+\tau+3)(\ell+\tau+4),\cr
a^{(0)}_{11,\tau\ell}={}&2(\tau-3)\tau(\tau+1)(\tau+4)(\ell+1)(\ell+2\tau+2)\cr&{}\times (\ell+\tau-2)(\ell+\tau+1)(\ell+\tau+2)(\ell+\tau+5),\cr
a^{(0)}_{10,\tau\ell}={}&{5\over 3}(\tau-3)(\tau-1)(\tau+2)(\tau+4)(\ell+1)\cr
&{}\times(\ell+2\tau+2)(\ell+\tau-2)(\ell+\tau)(\ell+\tau+3)(\ell+\tau+5),\cr
a^{(0)}_{00,\tau\ell}={}&(\tau-3)(\tau-2)(\tau+3)(\tau+4)(\ell+1)(\ell+2\tau+2)\cr&{}\times (\ell+\tau-2)(\ell+\tau-1)(\ell+\tau+4)(\ell+\tau+5). &\azero\cr
}$$
This together with the results of our expansion gives the following expressions for $\left<\eta\right>$
\eqnn\et
$$\eqalignno{
\left<\eta_{22,\tau\ell}\right>={}&-{4\over 25N^2}{(\tau-3)(\tau-2)(\tau+3)(\tau+4)\over (\ell+\tau)(\ell+\tau+1)(\ell+\tau+2)(\ell+\tau+3)
(\ell+1)(\ell+2\tau+2)}\cr 
&{}\times\big((\ell+1)(\ell+2\tau+2)(25(\ell+\tau+1)(\ell+\tau+2)+57\tau(\tau+1)-90)\cr
&{}{\hskip .7 cm}+60(\tau-1)\tau(\tau+1)(\tau+2)\big),\cr
\left<\eta_{21,\tau\ell}\right>={}&-{4\over 25N^2}{(\tau-3)(\tau-1)(\tau+2)(\tau+4)\over (\ell+\tau-1)(\ell+\tau+1)(\ell+\tau+2)(\ell+\tau+4)}\cr
&{}\times\big(25(\ell+\tau+1)(\ell+\tau+2)+9\tau(\tau+1)-30\big),\cr
\left<\eta_{20,\tau\ell}\right>={}&-{4\over 5N^2}{(\tau-3)\tau(\tau+1)(\tau+4)\over (\ell+\tau-1)(\ell+\tau) (\ell+\tau+3)(\ell+\tau+4)}\cr
&{}\times\big(5(\ell+\tau+1)(\ell+\tau+2)-3\tau(\tau+1)\big),\cr
\left<\eta_{11,\tau\ell}\right>={}&-{4\over 25N^2}{(\tau-2)(\tau-1)(\tau+2)(\tau+3)\over (\ell+\tau-2)(\ell+\tau+1)(\ell+\tau+2)(\ell+\tau+5)(\ell+1)(\ell+2\tau+2)}\cr
&{}\times\big((\ell+1)(\ell+2\tau+2)(25(\ell+\tau+1)(\ell+\tau+2)+97\tau(\tau+1)-1140)\cr
&{}{\hskip .7 cm}+160(\tau-3)\tau(\tau+1)(\tau+4)\big),\cr
\left<\eta_{10,\tau\ell}\right>={}&-{4\over 5N^2}{(\tau-2)\tau(\tau+1)(\tau+3)\over (\ell+\tau-2)(\ell+\tau)(\ell+\tau+3)(\ell+\tau+5)}\cr
&{}\times\big(5(\ell+\tau+1)(\ell+\tau+2)+21\tau(\tau+1)-270\big),\cr
\left<\eta_{00,\tau\ell}\right>={}&-{4\over 5N^2}{(\tau-1)\tau(\tau+1)(\tau+2)\over (\ell+\tau-2)(\ell+\tau-1)(\ell+\tau+4)(\ell+\tau+5)(\ell+1)(\ell+2\tau+2)}\cr
&{}\times\big((\ell+1)(\ell+2\tau+2)(5(\ell+\tau+1)(\ell+\tau+2)+37\tau(\tau+1)-450)\cr
&{}{\hskip .7 cm}+100(\tau-3)(\tau-2)(\tau+3)(\tau+4)\big).\cr
&{\ }&\et\cr
}$$
Notice that the $\tau$-dependence in the numerator of the first factor can easily be seen to be
\eqn\td{u_{nm}=(\tau-n-1)(\tau-m)(\tau+m+1)(\tau+n+2).}
Now we define
\eqn\v{v_r={(\tau-r)(\tau+r+1)\over (\ell+\tau-r+1)(\ell+\tau+r+2)}.}
Using this we can write the results in the following relatively simple way for $\tau=4,5,\ldots$
\eqnn\etf
$$\eqalignno{
\left<\eta_{22,\tau\ell}\right>={}&-{4\over N^2}{u_{22}\over(\ell+1)(\ell+2\tau+2)}\bigg(1 + {12\over 25} v_0 + {4\over 5} v_1 + {3\over 25} v_0 v_1\bigg),\cr
\left<\eta_{21,\tau\ell}\right>={}&-{4\over N^2}{u_{21}\over(\ell+1)(\ell+2\tau+2)}\bigg(1 + {4\over 25} v_0 - {4\over 5} v_2- {9\over 25} v_0 v_2\bigg),\cr
\left<\eta_{20,\tau\ell}\right>={}&-{4\over N^2}{u_{20}\over(\ell+1)(\ell+2\tau+2)}\bigg(1 - {2\over 5} v_1- {6\over 5} v_2 + {3\over 5} v_1 v_2\bigg),\cr
\left<\eta_{11,\tau\ell}\right>={}&-{4\over N^2}{u_{11}\over(\ell+1)(\ell+2\tau+2)}\bigg(1 + {2\over 25} v_0 + {14\over 5} v_3 + {63\over 25} v_0 v_3\bigg),\cr
\left<\eta_{10,\tau\ell}\right>={}&-{4\over N^2}{u_{10}\over(\ell+1)(\ell+2\tau+2)}\bigg(1 - {4\over 25} v_1+ {84\over 25} v_3 - {21\over 5} v_1 v_3\bigg),\cr
\left<\eta_{00,\tau\ell}\right>={}&-{4\over N^2}{u_{00}\over(\ell+1)(\ell+2\tau+2)}\bigg(1 + {4\over 5} v_2+{28\over 5} v_3 + {63\over 5} v_2 v_3\bigg).\cr
&{\ }&\etf\cr}$$
For comparison we can also rewrite the results for $p=2$
$$\eqalignno{
\left<\eta_{00,\tau\ell}\right>={}&-{4\over N^2}{u_{00}\over(\ell+1)(\ell+2\tau+2)},
}$$
and $p=3$
$$\eqalignno{
\left<\eta_{11,\tau\ell}\right>={}&-{4\over N^2}{u_{11}\over(\ell+1)(\ell+2\tau+2)}\bigg(1 + {1\over 2} v_0\bigg),\cr
\left<\eta_{10,\tau\ell}\right>={}&-{4\over N^2}{u_{10}\over(\ell+1)(\ell+2\tau+2)}\bigg(1 - v_1\bigg),\cr
\left<\eta_{00,\tau\ell}\right>={}&-{4\over N^2}{u_{00}\over(\ell+1)(\ell+2\tau+2)}\bigg(1 + 5 v_2\bigg),
}$$
in this form.
This suggests a universal form for these anomalous dimensions for general $p$.
In particular it seems to emerge a universal behaviour for large $\ell$
$$\left<\eta_{nm,\tau\ell}\right>=-{4\over N^2}{u_{nm}\over(\ell+1)(\ell+2\tau+2)}(1+\O(\ell^{-2})).$$

\subsec{Conjectures for General Chiral Four Point Functions}

It was observed in \od\ that the structure of the singularities in $u$ is universal in the sense, that up to order $p-1$ in $u$ the terms in the amplitude are given by a universal function $\F$ 
and thus for increasing $p$ new singular terms appear, but the existing ones are not modified. 
Also, from this it follows that we expect the free field part of the amplitude to have the same form with opposite sign. The reason is that these singular contributions correspond to long multiplets 
which have no corresponding $\log u$ terms and thus do not receive anomalous dimensions. Assuming that all long multiplets do receive anomalous dimensions we expect these protected contributions to be cancelled by the free field part of the amplitude.
We will use this observation as an assumption together with crossing symmetry and will work out the constraints it imposes on the general $p$ amplitude. 
This is still work in progress since some of our assumptions about the singularity structure
are not well justified yet. 

We will attempt to write down expressions for the general large $N$ amplitude 
for a four point function of four identical single trace $\half$-BPS operators 
belonging to the $SU(4)$ $[0,p,0]$ representation.
The following will be based on what has been observed  \od\ in the examples for 
$p=2,3,4$ which have been explicitly calculated using the AdS/CFT 
correspondence \refs{\Arut,\ADHS,\Degen}. We believe that after making use of the freedom of decomposing long multiplets into semi-short ones
the free part $\H_0$ of the amplitude $G$ defines a function identical to $\F$ but with opposite sign
\eqn\Gfuniv{
\H^{(p)}_0(u,v;\sigma,\tau)={p^2\over N^2} \, \F(u,v;\sigma,\tau),
}
such that the singular contributions of the free and the dynamical part cancel and all remaining contributions from long operators possess $\log u$ terms generating anomalous dimensions as well.
Given this, it would provide the opportunity to compute $\F(u,v;\sigma,\tau)$ 
from the free field amplitude. One technical complication is the ambiguity of the distinction between $\H,\hf$ in \Gsolt, 
which is not really understood yet.

While developing these ideas, Mathematica has proven very useful as a tool to test 
our ideas and check what amplitudes would be the result for $p=2,\ldots,9$ long before we had explicit expressions for the coefficients.

Crossing symmetry plays an essential role in what follows.  
We first exhibit a basis for crossing symmetric polynomials in the variables 
$\sigma,\tau$ of degree $n$, i.e they may be expanded in monomials 
$\sigma^g \tau^h$ with $g+h\le n$, which are defined by
\eqn\csr{
S^{(n)}(\sigma,\tau)=S^{(n)}(\tau,\sigma)=\tau^n S^{(n)}(\sigma/\tau,1/\tau)\, .
}
A simple basis for these polynomials is given by 
\eqn\csb{
S^{(n)}_{ab}(\sigma,\tau)=\cases{\sigma^a \tau^a + \sigma^a \tau^{n-2a}
+ \sigma^{n-2a} \tau^a , &$a=b\,$,\cr 
\sigma^a \tau^b + \sigma^b \tau^a + \sigma^a \tau^a , &$2a+b=n$,\cr
\sigma^{{1\over 3}n} \tau^{{1\over 3}n} , &${\ts{1\over 3}}n\in {\Bbb N}$,\cr
\sigma^a \tau^b \! + \sigma^b \tau^a \! 
+ \sigma^a \tau^{n-a-b} \! + \sigma^{n-a-b} \tau^a \! + \sigma^b \tau^{n-a-b}
\! + \sigma^{n-a-b} \tau^b , &otherwise,\cr}
}
where $a,b$ are integers satisfying
\eqn\ab{
0 \le b \le a \, , \qquad 2a+b \le n \, .
}
Notice that this basis is related to the one used in Table 2 in chapter 4 by $a=i+j,b=j$ and $n=p$ whereas here it will be $n=p-2$.
The first two cases in \csb\ are distinguished according to whether $n>3a$
or $3a>n$ respectively.
For any $n$  the set of possible $(a,b)$ are the points of an integer lattice 
inside or on a triangle with vertices $(0,0)$, $({1\over 3}n, {1\over 3}n)$
and $({1\over 2}n,0)$.
The number of independent crossing symmetric polynomials as in \countN\ is
$$(n-3[{\textstyle{1\over 6}}n])([{\textstyle{1\over 6}}n]+1) + \delta_{n,6[{\textstyle{1\over 6}}n]}.$$ From \csb\
it is trivial to verify that the number of terms $n^{(n)}_{ab}$ in the above symmetric
polynomials is in each case
\eqn\sab{
n^{(n)}_{aa\vphantom b}  = n^{(n)}_{a\, n-2a} = 3 \, , \quad
n^{(n)}_{{1\over 3}n \, {1\over 3}n} = 1 \, , \quad
n^{(n)}_{ab} = 6 \ \ \hbox{otherwise} \, ,
}
and then
\eqn\check{
\sum_{{0\le b \le a \atop \le n-a-b}} n^{(n)}_{ab} = \half (n+1)(n+2) \, ,
}
which is the number of independent polynomials in $\sigma,\tau$ of degree $n$.

The four point function for general $p$ can be reduced to an amplitude
$\H^{(p)}$, polynomial of degree $p-2$ in $\sigma,\tau$, satisfying
\eqn\csh{
\H^{(p)}(u,v;\sigma,\tau)={1\over v^2}\, \H^{(p)}(u/v,1/v;\tau,\sigma)=
\left({u\over v}\right)^{p-2}\! \tau^{p-2} \, \H^{(p)}(v,u;\sigma/\tau,1/\tau).
}
The results for $p=2,3,4$ \refs{\Arut,\ADHS,\Degen}\ can be reduced to the following form
\eqn\lna{
\H^{(p)}(u,v;\sigma,\tau)=-{p^2\over N^2} \ u^p  \!\!
\sum_{{0\le b \le a \atop 2a+b \le p-2}}
\sum_{i,j,k} \, c^{(p)}_{ijk,ab}\, T^{(p)}_{ijk,ab}(u,v;\sigma,\tau) \, , 
}
where $T^{(p)}_{ijk,ab}$ are completely crossing symmetric combinations of $\oD$ 
functions which are related to the crossing symmetric polynomials in \csb. For
$p=2,3,\dots$ and restricting $0 \le b \le a , \ 2a+b \le p-2$, corresponding to
\ab, we define
\eqn\T{\eqalign{ \!\!\!
T^{(p)}_{ijk,ab}(u,v;\sigma,\tau) & \cr
= {\ts{1\over 6}}n^{(p-2)}_{ab}&  \big (
\sigma^a\tau^b\, \oD_{i\,p+2\,jk}(u,v)+ \sigma^b\tau^a\, \oD_{i\, p+2\,kj}(u,v)\cr
&{} + \sigma^a\tau^{p-2-a-b}\, \oD_{j\, p+2\, ik}(u,v)+
\sigma^{p-2-a-b}\tau^a\, \oD_{j\, p+2\, ki}(u,v) \cr
&{}+ \sigma^b\tau^{p-2-a-b}\, \oD_{k\, p+2\, ij}(u,v)+
\sigma^{p-2-a-b}\tau^b\, \oD_{k\, p+2\, ji}(u,v) \big ) \, . \cr}
}
The crossing identities for $\oD$ functions \didso\ ensure that 
$u^p T^{(p)}_{ijk,ab}(u,v;\sigma,\tau)$ satisfies \csh\ without any additional
factors of $u,v$ being necessary. The factors ${\ts{1\over 6}}n^{(p-2)}_{ab}$
are introduced for later convenience, essentially since, for the boundary 
values of $a,b$, we have
\eqn\symT{\eqalign{
& T^{(p)}_{ijk,aa} = T^{(p)}_{ikj,aa} \, , \qquad
T^{(p)}_{ijk,a\,p-2-2a} = T^{(p)}_{kji,a\, p-2-2a} = T^{(p)}_{jki,aa} \, , \cr
& T^{(p)}_{ijk,{1\over3}(p-2)\,{1\over3}(p-2)} = 
T^{(p)}_{(ijk),{1\over3}(p-2)\,{1\over3}(p-2)} \, . \cr}
}
When $a,b$ satisfy the conditions in \symT\ then \T\ can be simplified in
appropriate cases,
\eqn\TT{\eqalign{
&{} T^{(p)}_{ijj,aa}(u,v;\sigma,\tau)= T^{(p)}_{jij,a\, p-2-2a}(u,v;\sigma,\tau)\cr
&{} = \sigma^a\tau^a\, \oD_{i\,p+2\,jj}(u,v)+ \sigma^a\tau^{p-2-2a}\, 
\oD_{j\, p+2\, ij}(u,v) 
+ \sigma^{p-2-2a} \tau^a\, \oD_{j\, p+2\, ji}(u,v)  \, , \cr
&{} T^{(p)}_{iii,{1\over3}(p-2)\,{1\over3}(p-2)} (u,v;\sigma,\tau)
= (\sigma \tau)^{{1\over3}(p-2)} \oD_{i\,p+2\,ii}(u,v) \, . \cr }
}

To determine contributions in the operator product expansion from long multiplets
corresponding to different $SU(4)$ representations for the lowest dimension operators
we may expand in terms of $SU(4)$ harmonics
\eqn\exH{
\H^{(p)}(u,v;\sigma,\tau) = \sum_{0\le m \le n\le p-2} A_{nm}(u,v)
\, Y_{nm}(\sigma,\tau) \, ,
}
where $nm$ corresponds to a $SU(4)$ representation with Dynkin labels
$[n-m,2m,n-m]$. $Y_{nm}(\sigma,\tau)$ are polynomials of degree $n$ in 
$\sigma,\,\tau$.
The condition that only long multiplets may contribute to $A_{nm}(u,v)$
satisfying the unitarity condition requires that for $u \sim 0$ we must
have
\eqn\unit{
A_{nm}(u,v)= {\rm O}(u^{n+1})  \, .
}

The $\oD$ functions that can appear in \T\ are constrained by the unitarity 
conditions \unit. To obtain these we first list a few more essential properties used 
here apart from the ones introduced in the introduction \Df. From standard relations for hypergeometric functions we have
\eqn\hyper{
n_1 f_{n_1n_2n_3n_4}(v)= f_{n_1+1\, n_2\, n_3+1\, n_4}(v)
+ f_{n_1+1\, n_2n_3\,n_4+1}(v) \, ,
}
and also
\eqn\crossf{
f_{n_1 n_2 n_3 n_4} (v) = v^{-n_2}f_{n_1 n_2 n_4 n_3} (1/v) \, .
}
The $\log u$ terms in \Df\ result in anomalous dimensions for operators
belonging to long multiplets which have twist $\Delta - \ell \ge 2p$ at zeroth
order in the $1/N$ expansion, where $\Delta$ is the scale dimension and $\ell$
the spin.  The terms involving negative powers $u^{-s+m}$ have no corresponding 
$\log u$ terms and would correspond in the operator product expansion to 
contributions from long multiplets which are unrenormalised. We assume
that these are all cancelled although there remain contributions from various
semi-short multiplets which cannot be combined to form a long multiplet, thus
all multiplets not satisfying a shortening condition gain anomalous dimensions
in the $1/N$ expansion as expected.
This is the essential assumption that leads to strong constraints on the 
expansion \lna.

Using the properties of the $SU(4)$ harmonics $Y_{nm}(\sigma,\tau)$ it is
clear that the unitarity condition \unit\ is satisfied if
for any contribution in \lna\ involving
\eqn\ssD{
\sigma^g \tau^h \, u^p \oD_{n_1 \, p+2 \, n_3 n_4}(u,v) \, ,
}
we require
\eqn\gh{
0 \le s \le p - g - h -1 \, .
}
With the assumed expansion given by \lna\ and \T\
the condition $s\ge 0$ ensures from \Df\ that only long multiplets
with twist $\Delta - \ell \ge 2p$ with non zero  anomalous dimensions in the 
large $N$ limit can contribute to the operator product expansion of the
chiral four point function.

Applying the condition \gh\ to \T\ gives the following inequalities
\eqn\ies{\eqalign{
p-2a\leq {}& i+j-k\leq p+2\, ,\cr
p-2b\leq {}& i+k-j\leq p+2\, ,\cr
2(a+b+2)-p \leq {}& j+k-i\leq p+2 \, .\cr}
}
We also impose
\eqn\ca{
i,j,k\leq p \, .
}
It is clear that there is a finite number of possibilities for $i,j,k$, note
that $p+4 \le i+j+k \le 3p$.
We should also note that the expansion \T\ is not unique since
\eqn\DDD{
\half( i+j+k - p -2) \oD_{i\, p+2\, j\,k+1} =
\oD_{i\, p+2\, j+1\,k+1} + \oD_{i+1\, p+2\, j\,k+1} + \oD_{i+1\, p+2\, j+1\,k}\,.
}
This is the only relation for $\oD$ functions of the form appearing in \T.
Correspondingly we may take
\eqn\TTT{
\half( i+j+k - p -2) T^{(p)}_{ijk,ab} = T^{(p)}_{i\,j+1\,k+1,ab} +
T^{(p)}_{i+1\,j\,k+1,ab}  + T^{(p)}_{i+1\,j+1\, k,ab} \, ,
}
which allows the expansion \lna\ to be simplified if each term in \TTT\ 
satisfies the constraints \ies\ and \ca. For this to be the case
we must have
\eqn\ijk{
i+j-k, i+k-j, j+k-i \le p \, , \qquad i,j,k \le p-1 \, .
}
Whenever \ijk\ is satisfied one of the terms appearing in \TTT\ may be omitted
in the general expansion.

If we use \ijk\ to remove terms with the lowest value of $i+j+k$  whenever
appropriate then for general $p$ the list of possible terms obtained from 
\ies, \ca\ modulo \ijk\ is
\eqn\pp{\eqalign{
&T^{(p)}_{pjj,ab}\, ,\ \quad j=a+b+2,\dots , p \, , \cr
&T^{(p)}_{iip,ab}\, ,\ \quad i=p-a,\dots , p-1 \, , \  a \ge 1 \, , \cr
&T^{(p)}_{ipi,ab}\, ,\ \quad i=p-b,\dots , p-1 \, , \  b \ge 1 \, , \cr
&T^{(p)}_{i \, i+k -p-2 \, k,ab}\, ,\ \ i=p-a+1,\dots , p \, , \ 
k=a+b+3,\dots , p \, , \ a \ge 1 \, , \cr
&T^{(p)}_{i j\, i+j -p-2 ,ab}\, ,\ \ \, i=p-b+1,\dots , p \, , \
j=a+b+3,\dots , p \, , \ b \ge 1 \, , \cr
&T^{(p)}_{j+k -p-2 \, j k,ab}\, ,\ \ j=p-a+1,\dots , p \, , \
k=p-b+1 ,\dots , p \, , \ a,b \ge 1 \, . \cr}
}
When $a=b$ or $b=p-2-2a$ it is necessary also to take into account the
symmetry conditions in \symT\ to obtain an independent basis.
If $N^{(p)}_{ab}$ are the number of possibilities for each $a,b$ then we have
\eqn\num{\eqalign{
N^{(p)}_{ab} = {}& (a+b+1)(p-a-b-1) + ab \, , \quad
0 \le b< a < \half ( p-2-b) \, , \cr
N^{(p)}_{aa} = {}&  (a+1)(p-{\ts {3\over 2}}a - 1) \, , \qquad
0 \le a < {\ts {1\over 3}} (p-2)  \, , \cr
N^{(p)}_{a\, p-2-2a} = {}& (a+1)(p-{\ts {3\over 2}}a -1) \, , \qquad
{\ts {1\over 3}} (p-2) < a \le \half ( p-2) \, , \cr
N^{(p)}_{{1\over 3}(p-2) \, {1\over 3}(p-2)} = {}& 
{\ts {1\over 18}}(p+1)(p+4) \, , \ \quad \qquad p=2 \ \hbox{mod} \ 3 \, . \cr}
}

The crucial assumption, in addition to \T, is that the leading terms in the
expansion of $\H^{(p)}$ in powers of $u$, which do not involve any $\log u$ terms, 
are universal, i.e. we have for any $p=2,3, \dots$
\eqn\Huniv{
\H^{(p)}(u,v;\sigma,\tau)=-{p^2\over N^2} \, \F(u,v;\sigma,\tau) 
+ {\rm O}(u^{p}) \, ,
}
where $\F$ is independent of $p$. It satisfies the same crossing properties
under $u\to u/v, \, v \to 1/v$ and $\sigma \leftrightarrow \tau$ as $\H^{(p)}$
in \csh. In the expansion \lna\ for $\H^{(p)}$, with $T^{(p)}_{ijk,ab}$  given
by \T, then only the leading singular terms in the expansion of the $\oD$
functions shown in \Df\ contribute to $\F$ in \Huniv. This ensures that it is
expressible as an expansion with the form
\eqn\expF{\eqalign{
\F(u,v;\sigma,\tau) = {}& \sum_{n\ge 1} u^n \F_n(v;\sigma,\tau) \cr
= {}& \sum_{g,h\ge 0} \sum_{l,m} d_{lm,gh} \,
\sigma^g \tau^h u^{g+h+1} f_{l+m-g-h-3\, g+h+3 \,  lm }(v) \, , \cr}
}
where \crossf\ has been used to restrict to terms of the form
$u^{n} f_{l+m-n-2\, n+2 \,lm}(v)$.

In general for each contribution in $\F$ proportional to $\sigma^g \tau^h$ 
we may expect a series involving $n=g+h+1,g+h+2, \dots$, but as verified 
later it is possible to restrict to just the minimal form shown in \expF\
when only $n=g+h+1$ for each $g,h$. This is an essential assumption 
for determining the coefficients $c^{(p)}_{ppp,ab}$ for which we currently have no
justification other than we observe that it is possible to impose this restriction and obtain results consistent with
known numerical results.
Subject to using \hyper\ we may also require 
\eqn\kl{
l,m \le g+h+2 \, , \quad  l+m \ge g+h+4 \, , \qquad 0\le m-l \le g-h \ \hbox{or} \
0 \le l-m \le h-g \, .
}
Apart from the relation \hyper\ the functions $u^{n} f_{l+m-n-2\, n+2 \,lm}(v)$
appearing in \expF\ are assumed to be linearly independent. With the restricted
form in \expF\ then, as shown later, we are both able to determine $\F$ from
\Huniv\ and using the explicit form of $\F$, up to terms of ${\rm O}(u^{p-1})$, 
to find all the coefficients $c^{(p)}_{ijk,ab}$  which appear in the expansion \lna.

A simple consequence of \Huniv\ and \lna, which does not require any restrictions
of the form for $\F$, is that for $p=3,4,\dots$,
\eqn\univ{\eqalign{
u^p \!\!\! & \sum_{{0\le b \le a \atop 2a+b \le p-2}}
\sum_{i,j,k} \, c^{(p)}_{ijk,ab}\, T^{(p)}_{ijk,ab}(u,v;\sigma,\tau)\cr
&{} = u^{p-1} \!\!\! \sum_{{0\le b \le a \atop 2a+b \le p-3}}
\sum_{i,j,k} \, c^{(p-1)}_{ijk,ab}\, T^{(p-1)}_{ijk,ab}(u,v;\sigma,\tau)
+ {\rm O}(u^{p-1}) \, , \cr}
}
where only the leading singular terms displayed explicitly in \Df\ need be 
considered.  These equations are invariant under $\sigma \leftrightarrow \tau$
and $u\to u/v, \, v \to 1/v$.
In \univ\ all $\oD$ functions with $s=1,2, \dots$ are relevant on 
the right hand side but only those with $s=2,3,\dots$ on the left hand side.
It is important to note that \univ\ does not constrain,
$c^{(p)}_{ppp,ab}$, the coefficient for $\oD_{p\,p+2\,pp}$ which is present
for any $a,b$.

\subsec{Applications for Low $p$}
We now show how the above suggestions work out in practice for low $p$, initially
using only \univ.

For $p=2$ $\H^{(2)}$ is independent of $\sigma,\tau$ and there is just one
possible $\oD$ function which is in accord with
the simplification of results obtained from AdS/CFT,
\eqn\Htwo{
\H^{(2)}(u,v)= -{4\over N^2}\, u^2\oD_{2422}(u,v) \, .
}
For $p=3$ we must have again $a=b=0$ and there are just two crossing symmetric 
forms
\eqn\Hthree{\eqalign{
\H^{(3)}& (u,v;\sigma,\tau) =-{9\over N^2}u^3 \Big (
c^{(3)}_{322,00}T^{(3)}_{322,00}+c^{(3)}_{333,00}T^{(3)}_{333,00} \Big ) \, ,\cr
={}&-{9\over N^2}\, u^3 \Big (c^{(3)}_{333,00} (1+\sigma+\tau)\oD_{3533}+
c^{(3)}_{322,00}\big (\oD_{3522}+\sigma\oD_{2523}+\tau\oD_{2532}\big )\Big )\, .
\cr}
}
The equations \univ\ give just one condition arising from
$u^3\oD_{3522}(u,v) \sim u f_{1322}(v)$ and in \Htwo\
$u^2\oD_{2422}(u,v) \sim u f_{1322}(v)$ so that we require
\eqn\cth{
c^{(3)}_{322,00} = 1 \, .
}
The known results also give
\eqn\cp{
c^{(3)}_{333,00} = 1 \, .
}

For $p=4$, for comparison with previous results, we write
\eqn\Hfour{\eqalign{
\H^{(4)}(u,v;\sigma,\tau)={}&
-{16\over N^2}\, u^4\Big (c^{(4)}_{422,00}T^{(4)}_{422,00}+
c^{(4)}_{433,00}T^{(4)}_{433,00}+c^{(4)}_{444,00}T^{(4)}_{444,00}\cr&
\qquad\qquad+c^{(4)}_{323,10}T^{(4)}_{323,10}+c^{(4)}_{424,10}T^{(4)}_{424,10}
+c^{(4)}_{444,10}T^{(4)}_{444,10} \Big )\, ,\cr}
}
where from \symT\ and \TTT\ $T^{(4)}_{334,10} = T^{(4)}_{433,10} = \half
(T^{(4)}_{323,10} - T^{(4)}_{424,10})$ so that such contributions are discarded.
Expanding \Hfour\ then gives
\eqn\Hexp{\eqalign{
\H^{(4)}(u,v;\sigma,\tau)
= -{16\over N^2}\, u^4 \Big (&  
\big (c^{(4)}_{444,00}(1+\sigma^2+\tau^2)+c^{(4)}_{444,10}
(\sigma+\tau+ \sigma\tau)\big )\oD_{4644}\cr
&{} + c^{(4)}_{323,10}
\big (\sigma\oD_{3623}+\tau\oD_{3632}+\sigma\tau\oD_{2633}\big )\cr
&{} + c^{(4)}_{424,10}
\big (\sigma\oD_{4624}+\tau\oD_{4642}+\sigma\tau\oD_{2644}\big )\cr
&{} +c^{(4)}_{422,00}\big (\oD_{4622}+\sigma^2\oD_{2624}+\tau^2\oD_{2642}\big )\cr
&{} +c^{(4)}_{433,00}\big (\oD_{4633}+\sigma^2\oD_{3634}+\tau^2\oD_{3643}\big )
\Big ) \, . \cr}
}
Applying \univ\ then gives for the 1 terms
\eqn\one{
c^{(4)}_{422,00} = \half \, , \qquad c^{(4)}_{422,00} - c^{(4)}_{433,00} = 0 \, ,
}
and from the $\sigma$ terms
\eqn\sig{
c^{(4)}_{323,10}\, f_{1423}(v) +  c^{(4)}_{424,10}\, f_{2424}(v) =
f_{2433}(v) +  f_{1423}(v) \, .
}
Using \hyper\ this is easily solved giving
\eqn\cfour{
c^{(4)}_{323,10} =  2 \, , \qquad c^{(4)}_{424,10} = - 1 \, .
}
The remaining coefficients which are undetermined by \univ\ are
\eqn\cfourp{
c^{(4)}_{444,10} =  1 \, , \qquad c^{(4)}_{444,00} =  {\ts {1\over 4}}  \, .
}
For the basis corresponding to \pp\ then, instead of \Hexp, we should take
\eqn\Hexpt{\eqalign{
\H^{(4)}(u,v;\sigma,\tau)
= -{16\over N^2}\, u^4 \Big ( & 
{\ts {\sum_{j=2}^4}} c^{(4)}_{4jj,00} T^{(4)}_{4jj,00} \cr
&{} + 2 c^{(4)}_{433,10} T^{(4)}_{433,10} + 
c^{(4)}_{424,10} T^{(4)}_{424,10} + c^{(4)}_{444,10} T^{(4)}_{444,10}  \Big ) \, .
\cr}
}
Here we introduce a factor 2 for the $c^{(4)}_{433,10}$ terms to count equal 
contributions from $T^{(4)}_{433,10}$ and $T^{(4)}_{343,10}$. This ensures
uniformity with later general results. 
In this case \one\ and \cfour\ are unchanged but instead of \cfour\ we should
take
\eqn\cfoura{
c^{(4)}_{433,10} =  2 \, , \qquad c^{(4)}_{424,10} = 1 \, .
}

On the basis of the results for $p=2,3,4$\refs{\Arut,\ADHS,\Degen} we determine the first few terms in 
the function $\F$ introduced in \Huniv,
\eqn\Funiv{\eqalign{
\F(u,v;\sigma,\tau) = {}& u f_{1322}(v) \cr
&{}+ \sigma \, u^2 \big ( f_{1423}(v) + f_{2433}(v) \big ) 
+ \tau \,  u^2 \big ( f_{1432} (v) + f_{2433}(v) \big ) \cr
&{} + \sigma^2 u^3 \half \big ( f_{1524}(v) + f_{2534}(v) + \half f_{3544}(v)\big )\cr
&{} + \tau^2 u^3 \half \big ( f_{1542}(v) + f_{2543}(v) + \half f_{3544}(v) \big )\cr
&{}+ \sigma \tau \, u^3  \big ( 2f_{1533}(v) + f_{3544}(v) \big ) + {\rm O}(u^4)\, .  \cr}
}
This result is in accord with the assumed form in \expF.

For $p=5$ we have the general form
\eqn\Hfive{\eqalign{
\H^{(5)}(u,v;\sigma,\tau)
=  -{25\over N^2}\, u^5 \Big ( & {\ts \sum_{j=2}^5} 
c^{(5)}_{5jj,00}T^{(5)}_{5jj,00}\cr&
{} + {\ts \sum_{j=3}^5} c^{(5)}_{5jj,10}T^{(5)}_{5jj,10} 
+ c^{(5)}_{445,10}T^{(5)}_{445,10} \cr&
{} +c^{(5)}_{524,10}T^{(5)}_{524,10}
+ c^{(5)}_{535,10}T^{(5)}_{535,10}\cr&
{}+ 3 c^{(5)}_{544,11}T^{(5)}_{544,11} +
3 c^{(5)}_{553,11}T^{(5)}_{553,11}+ c^{(5)}_{555,11}T^{(5)}_{555,11}\Big )\, .\cr}
}
Here we note from \symT\ that $T^{(5)}_{ijk,11}$ is completely symmetric
in $i,j,k$ and ${1\over 3}T^{(5)}_{333,11} = T^{(5)}_{443,11} = 
T^{(5)}_{544,11} + \half T^{(5)}_{553,11}$, so neither of these terms are present
in the expansion in \Hfive. By using \TTT, we also
eliminate terms involving $T^{(5)}_{434,10}, \, T^{(5)}_{423,10}$. As in 
\Hexpt\ we introduce factors to take account of the sum over identical terms
related by \symT.
With the basis in \Hfive\ and the results for $\H^{(4)}$ we may readily
solve the equations \univ\ giving
\eqn\resfive{\eqalign{
& c^{(5)}_{522,00} = c^{(5)}_{533,00} = {\ts {1\over 6}} \, , \qquad
c^{(5)}_{544,00} = {\ts {1\over 12}} \, , \cr
& c^{(5)}_{533,10} = c^{(5)}_{544,10} = 1 \, , \qquad
c^{(5)}_{524,10} = c^{(5)}_{535,10} = \half \, , \qquad
c^{(5)}_{445,10} = {\ts {3\over 4}} \, , \cr
& c^{(5)}_{544,11} = 3 \, , \qquad c^{(5)}_{553,11} = 1 \, . \cr}
}
Only $c^{(5)}_{555,00},c^{(5)}_{555,10},c^{(5)}_{555,11}$ are undetermined at this
stage. For the expansion \expF\ we may now obtain
\eqn\Ffour{\eqalign{
\F_4(v;\sigma,\tau) = {}& \big ( c^{(5)}_{555,00} - {\ts {1\over 36}} \big)
f_{4655}(v) \cr &
{}+ (\sigma+\tau+\sigma^2+\tau^2) \big ( c^{(5)}_{555,10}- \quar \big )f_{4655}(v) \cr
&{}+ \sigma \tau \big ( c^{(5)}_{555,11}-  1 \big )f_{4655}(v)  \cr
&{}+ \sigma^3 \big ( {\ts {1\over 6}} f_{1625}(v) + {\ts {1\over 6}} f_{2635}(v)
+ {\ts {1\over 12}} f_{3645}(v) + 2 c^{(5)}_{555,00}  f_{4655}(v) \big ) \cr
&{}+ \tau^3 \big ( {\ts {1\over 6}} f_{1652}(v) + {\ts {1\over 6}} f_{2653}(v)
+ {\ts {1\over 12}} f_{3654}(v) + 2 c^{(5)}_{555,00}  f_{4655}(v) \big ) \cr
&{}+ \sigma^2\tau \big ( f_{1634}(v) + \half f_{2644}(v)
+ \quar f_{3645}(v) + c^{(5)}_{555,10}  f_{4655}(v) \big ) \cr
&{}+ \sigma\tau^2 \big ( f_{1643}(v) + \half f_{2644}(v)
+ \quar f_{3654}(v) + c^{(5)}_{555,10}  f_{4655}(v) \big ) \, . \cr}
}
The restrictions, imposed by the additional constraint that just the leading 
term for $n=g+h+1$ appears in the expansion of $\F$ for each $g,h$ as assumed in
\expF, is easily achieved in \Ffour\ by setting
\eqn\solfour{
c_{555,00}^{(5)} = {\ts {1\over 36}} \, , \qquad
c_{555,10}^{(5)} = \quar \, , \qquad c_{555,11}^{(5)} = 1 \, .
}

For $p=6$ we have the following expansion in terms of independent crossing 
symmetric functions
\eqn\Hsix{\eqalign{
\H^{(6)}(u,v;\sigma,\tau)
=  -{36\over N^2}\, u^6 \Big ( & {\ts \sum_{j=2}^6}
c^{(6)}_{6jj,00}T^{(6)}_{6jj,00}\cr&
{} + {\ts \sum_{j=3}^6} c^{(6)}_{6jj,10}T^{(5)}_{6jj,10}
+ c^{(6)}_{556,10}T^{(6)}_{556,10} \cr&
{} + c^{(6)}_{624,10}T^{(6)}_{624,10}
+ c^{(6)}_{635,10}T^{(6)}_{635,10} +  c^{(6)}_{636,10}T^{(6)}_{636,10} \cr&
{}+ {\ts \sum_{j=4}^6} c^{(6)}_{6jj,11}T^{(6)}_{6jj,11} +
2c^{(6)}_{556,11}T^{(6)}_{556,11} \cr
& {} + 2c^{(6)}_{635,11}T^{(6)}_{635,11}
+ 2c^{(6)}_{646,11}T^{(6)}_{646,11} +  c^{(6)}_{466,11}T^{(6)}_{466,11} \cr
&{} + 2c^{(6)}_{635,20}T^{(6)}_{635,20}
+ c^{(6)}_{646,20}T^{(6)}_{646,20} +  c^{(6)}_{525,20}T^{(6)}_{525,20} \Big )\, .\cr}
}
Here for $a=b=1$ and $a=2,b=0$ we have used \symT\ to reduce the number
of necessary terms. In this case the equations give
\eqn\csix{\eqalign{
& c^{(6)}_{622,00}= c^{(6)}_{633,00} = {\ts {1\over 24}} \, , \quad
c^{(6)}_{644,00} = {\ts {1\over 48}}  \, , \quad 
c^{(6)}_{655,00} = - {\ts {1\over 48}} + c^{(5)}_{555,00}  \, , \cr
& c^{(6)}_{633,10}= c^{(6)}_{644,10} = {\ts {1\over 3}} \, , \quad
c^{(6)}_{655,10} = - {\ts {1\over 12}} + c^{(5)}_{555,10} \, , \cr
& c^{(6)}_{624,10}= c^{(6)}_{635,10} = {\ts {1\over 6}} \, , \quad
c^{(6)}_{646,10}= {\ts {1\over 12}} \, ,
\quad  c^{(6)}_{556,10}= {\ts {1\over 12}} + c^{(5)}_{555,00} \, , \cr
& c^{(6)}_{644,11}= {\ts {3\over 2}} \, , \quad c^{(6)}_{655,11} =
{\ts {1\over 2}} + c^{(5)}_{555,11} \, , \quad 
c^{(6)}_{466,11} = {\ts {1\over 4}} \, , \cr
& c^{(6)}_{635,11}= c^{(6)}_{646,11} = {\ts {1\over 2}} \, , \quad
c^{(6)}_{556,11}= {\ts {3\over 4}} + c^{(5)}_{555,10} \, , \cr
& c^{(6)}_{644,20}=  {\ts {3\over 8}} \, , \quad 
c^{(6)}_{655,20}= {\ts {1\over 8}} + c^{(5)}_{555,10} \,
\quad c^{(6)}_{525,20}= c^{(6)}_{646,20} = {\ts {1\over 4}} \, , \quad
c^{(6)}_{635,20} = {\ts {1\over 4}} \, . \cr }
}

We may now extend the results given by \Funiv\ and \Ffour, assuming \solfour, to
obtain
\eqnn\Ffive
$$\eqalignno{
\F_5(v;\sigma,\tau) = {}& \big ( c^{(6)}_{666,00} - {\ts {1\over 576}} \big)
f_{5766}(v) \cr &
{}+ (\sigma+\tau+\sigma^3+\tau^3) \big ( c^{(6)}_{666,10}- 
{\ts {1\over 36}}  \big )f_{5766}(v) \cr
&{}+ \sigma \tau (1+\sigma +\tau) \big ( c^{(6)}_{666,11}-  \quar \big )
f_{5766}(v)  \cr
&{}+ (\sigma^2 +\tau^2) \big ( c^{(6)}_{666,20}-  {\ts{1\over 16}} \big )
f_{5766}(v)  \cr  
&{}+ \sigma^4 \big ( {\ts {1\over 24}} f_{1726}(v) + {\ts {1\over 24}}
f_{2736}(v) + {\ts {1\over 48}} f_{3746}(v) + {\ts {1\over 144}} f_{4756}(v)
+ c^{(6)}_{666,00} f_{5766}(v) \big ) \cr
&{}+ \tau^4 \big ( {\ts {1\over 24}} f_{1762}(v) + {\ts {1\over 24}}
f_{2763}(v) + {\ts {1\over 48}} f_{3764}(v) + {\ts {1\over 144}} f_{4765}(v)
+ c^{(6)}_{666,00} f_{5766}(v) \big )
\cr &{}+ \sigma^3\tau \big ( {\ts {1\over 3}} f_{1735}(v) + {\ts {1\over 3}}
f_{2745}(v) + {\ts {1\over 18}} f_{4756}(v) + c^{(6)}_{666,10}f_{5766}(v) \big ) \cr
&{}+ \sigma\tau^3 \big ( {\ts {1\over 3}} f_{1753}(v) + {\ts {1\over 3}}
f_{2754}(v) + {\ts {1\over 18}} f_{4765}(v) + c^{(6)}_{666,10}f_{5766}(v) \big ) \cr
&{}+ \sigma^2\tau^2 \big ( {\ts {3\over 4}} f_{1744}(v) + {\ts {3\over 8}}
f_{3755}(v) + c^{(6)}_{666,20}  f_{5766}(v) \big ) \, . & \Ffive \cr}
$$
Again the same fashion as \solfour\ and in accord with \expF\ we also obtain
\eqn\solfive{
c_{666,00}^{(6)} = {\ts {1\over 576}} \, , \qquad
c_{666,10}^{(6)} = {\ts {1\over 36}} \, , \qquad c_{666,11}^{(6)} = 
{\ts {1\over 4}} \, , \qquad c_{666,20}^{(6)} = {\ts {1\over 16}} \, . 
}

\subsec{General Solutions}

We here discuss how the equations which follow from \Huniv, assuming
\lna\ with $T^{(p)}_{ijk,ab}$ restricted as in  \pp, can be solved if
we also suppose that the only contributions in the expansion for $\F$ 
are restricted to the form shown in \expF. The general expansion has the
form
\eqnn\Hexpall
$$\eqalignno{
\H^{(p)}=-{p^2\over N^2} \ u^p \!\!\!\!\sum_{{0\le b \le a \atop 2a+b \le p-2}} 
\!\!\! \bigg ( & 
\sum_{j=a+b+2}^p \! c^{(p)}_{pjj,ab} \, T^{(p)}_{pjj,ab} +
\sum_{i=p-a}^{p-1}\, c^{(p)}_{iip,ab} \, T^{(p)}_{iip,ab} +
\sum_{i=p-b}^{p-1}\, c^{(p)}_{ipi,ab} \, T^{(p)}_{ipi,ab} \cr
&{} + \sum_{i=p-a+1}^p \sum_{k=a+b+3}^p c^{(p)}_{\,i\, i+k-p-2 \, k,ab} \,
T^{(p)}_{i\, i+k-p-2 \, k,ab} \cr
&{} + \sum_{i=p-b+1}^p \sum_{j=a+b+3}^p c^{(p)}_{\,i j\, i+j-p-2,ab} \, 
T ^{(p)}_{i\, j\, i+j-p-2,ab} \cr
& {}+ \sum_{j=p-a+1}^p \sum_{k=p-b+1}^p \, c^{(p)}_{j+k-p-2\, jk,ab}\, 
T^{(p)}_{j+k-p-2\, jk,ab} \bigg ) \, . & \Hexpall \cr}
$$

We first consider terms independent of $\sigma,\tau$ which arise only from
$\sum_{j=2}^p c^{(p)}_{pjj,00} T^{(p)}_{pjj,00}$ where $T^{(p)}_{pjj,00} \to
\oD_{p\, p+2 \, jj}$. Requiring 
$\F(u,v;\sigma,\tau) = u f_{1322}(v) + {\rm O}(\sigma,\tau)$ as in \Funiv\ we 
obtain
\eqn\abz{
\sum_{j=2}^p \sum_{m=0}^{p-j} {(-1)^m\over m!}(p-j-1)! \, 
c^{(p)}_{pjj,00} u^{j+m-1} f_{j-1+m\, j+1+m\, j+m \, j+m}(v) = 
u f_{1322}(v) \, ,
}
which requires
\eqn\solz{
(p-2)! \, c^{(p)}_{p22,00} = 1 \, , \qquad 
\sum_{m=0}^{k-2} {(-1)^m\over m!} \, c^{(p)}_{p\, k-m\,k-m,00} = 0 \, , \  \
k =3, \dots , p \, .
}
This is easily solved giving
\eqn\solcz{
c^{(p)}_{pjj,00} = {1\over (p-2)!\, (j-2)!} \, , \qquad j=2,\dots, p \, .
}

We next consider the calculation of the coefficients $c^{(p)}_{ijk,ab}$ for 
$a\ge 1, \, b=0$. These are determined in \Huniv\ by the terms in the 
expansion \expF\ with  $g=a, \, h=0$. Contributions proportional to 
$\sigma^g$ first arise in an expansion in powers of $u$ of $\H^{(p)}$ 
from $\sum_{j=2}^p \, c^{(p)}_{pjj,00}\, T^{(p)}_{pjj,00}$,
with $T^{(p)}_{pjj,00}(u,v;\sigma,\tau) \to \sigma^p \oD_{j\,p+2\,jp}(u,v)$,
for $p=g+2$.
Using \Huniv\ with \solcz\ this gives the relevant contribution to $\F$
for this case,
\eqn\Fsa{
\F(u,v;\sigma,\tau) \big |_{\sigma^g} = \sigma^g u^{g+1} {1\over g!} 
\sum_{j=2}^{g+2} {1\over (j-2)!} \, f_{j-1\, g+3\, j \, g+2}(v) \, .
}
Assuming this form for $\F$ in general then for $p\ge a+3$ \Huniv\ requires, 
for the contributions which involve powers $u^n$ with $n<p$ arising only from the
$T^{(p)}_{pjj,a0}, \, T^{(p)}_{i\, i+k-p-2\, k,a0}$ and $T^{(p)}_{iip,a0}$ terms
in \Hexpall, keeping just the first term in \T, 
\eqn\recura{\eqalign{
& \sum_{j=a+2}^p c^{(p)}_{pjj,a0} \sum_{m=0}^{p-j} {(-1)^m\over m!}(p-j-m)! \, 
u^{j+m-1} f_{j-1+m\, j+1+m\, j+m \, j+m}(v)\cr
&{}+ \sum_{i=p-a+1}^p \sum_{k=a+3}^p c^{(p)}_{i\, i+k-p-2 \, k,a0} 
\sum_{m=0}^{p-k+1} {(-1)^m\over m!}(p-k+1-m)! \cr
\noalign{\vskip - 6pt}
& \hskip 4.8cm {} \times  u^{k+m-2} 
f_{i+k-p-2+m\, k+m \,i+k-p-2+m\, k+m}(v)  \cr
&{}+ \sum_{i=p-a}^{p-1}\,
c^{(p)}_{iip,a0} \,  u^{p-1} f_{i-1\, p+1 \, i \, p }(v) \cr
&{} =  u^{a+1} {1\over a!}
\sum_{j=2}^{a+2} {1\over (j-2)!} \, f_{j-1\, a+3\, j \, a+2}(v) \, . \cr}
}
To analyse \recura\ we consider first all terms proportional to $u^{a+1}$ 
when we obtain
\eqn\abase{\eqalign{
& (p-a-2)! \, c^{(p)}_{p\,a+2\,a+2,a0} \, f_{a+1\, a+3\, a+2 \, a+2}(v) \cr
&{}+ (p-a-2)! \! \sum_{i=p-a+1}^p  c^{(p)}_{i\,i+a-p+1\,a+3,a0} \, 
f_{i+a-p+1\, a+3\, i+a-p+1 \, a+3}(v) \cr
&{}= {1\over a!}
\sum_{j=2}^{a+2} {1\over (j-2)!} \, f_{j-1\, a+3\, j \, a+2}(v) \, . \cr}
}
Applying \hyper\ repeatedly for $f_{j-1\, a+3\, j \, a+2}(v)$ we then get
\eqn\solca{\eqalign{
c^{(p)}_{i\,i+a-p+1\,a+3,a0} = {}& {1\over (p-a-2)! \, a! \, (i+a-p-1)!} \, ,
\quad  i = p-a+1, \dots , p \, , \cr
c^{(p)}_{p\,a+2\,a+2,a0} = {}& {a+1\over (p-a-2)! \, a!^2} \, . \cr}
}
{}From contributions in \recura\ proportional to $u^{k-1} f_{k-1\, k+1\, kk}(v)$,
for $k=a+3,\dots, p-1$, and $u^{k-2} f_{i+k-p-2\, k\, i+k-p-2 \, k}(v)$, 
for $k=a+4, \dots , p, \, i = p-a+1, \dots , p$, we get
\eqn\relcc{
\sum_{m=0}^{k-a-2} {(-1)^m\over m!} \, c^{(p)}_{p\, k-m\,k-m,a0} = 0 \, , \quad
\sum_{m=0}^{k-a-3} {(-1)^m\over m!} \, c^{(p)}_{i\, i+k-p-2-m\,k-m,a0} = 0 \, ,
}
which in conjunction with \solca\ may be solved giving
\eqn\solcaf{\eqalign{
c^{(p)}_{pjj,a0} = {}& {a+1\over (p-a-2)! \, a!^2\, (j-a-2)!} \, , \quad
j = a+2, \dots , p-1 \, , \cr
c^{(p)}_{i\,i+k-p-2\,k,a0} = {}& 
{1\over (p-a-2)! \, a! \, (i+a-p-1)! \,(k-a-3)!}\,,\cr
\noalign{\vskip - 3pt}
&\qquad k=a+3, \dots , p, \, i = p-a+1, \dots , p \, . \cr}
}
Using \solcaf\ the remaining part of \recura\ becomes
\eqn\recurf{\eqalign{
& \bigg ( c^{(p)}_{ppp,a0}  - {a+1 \over (p-a-2)!^2 \, a!^2} \bigg )
\,  f_{p-1\, p+1\, pp }(v)\cr
&{}- {1\over (p-a-2)!^2 \, a!}  
\sum_{j=p-a+1}^p  {1\over (j+a-p-1 )!} \, f_{j-1\, p+1 \, j-1 \, p+1 }(v)  \cr
&{}+ \sum_{i=p-a}^{p-1}\,
c^{(p)}_{iip,a0} \, f_{i-1\, p+1 \, i \, p }(v)  = 0 \, . \cr}
}
With the aid of \hyper\ again we finally obtain for this case
\eqn\solcat{\eqalign{
c^{(p)}_{ppp,a0} = {}& {1\over (p-a-2)!^2 \, a!^2} \, , \cr
c^{(p)}_{iip,a0} = {}& {p-a-1\over (p-a-2)!^2 \, a! \, (i+a-p)!}\,, \quad
i = p-a, \dots , p-1 \, . \cr}
}
For $p=2a+2$ the symmetry conditions \symT\ ensure that in the expansion
\Hexpall\ we may require $c^{(p)}_{pjj,a0}= c^{(p)}_{jjp,a0}$
and $c^{(p)}_{i\, i+k-p-2\, k,a0}$ is symmetric in $i,k$. 
With these constraints instead of \recura\ we have now
\eqn\recurs{\eqalign{
& \sum_{j=a+2}^p c^{(p)}_{pjj,a0} \bigg ( \sum_{m=0}^{p-j} 
{(-1)^m\over m!}(p-j-m)! \,  u^{j+m-1} f_{j-1+m\, j+1+m\, j+m \, j+m}(v) \cr
\noalign{\vskip - 10pt}
& \hskip 8cm {} + u^{p-1} f_{j-1\, p+1 \, j \, p }(v) \bigg ) \cr
&{}+ \!\! \sum_{i,k=a+3}^p \!\! c^{(p)}_{i\, i+k-p-2 \, k ,a0} \sum_{m=0}^{p-k+1} 
{(-1)^m\over m!}(p-k+1-m)! \cr
\noalign{\vskip - 6pt}
& \hskip 4.5cm {} \times  u^{k+m-2}
f_{i+k-p-2+m\, k+m \, i+k-p-2+m\, k+m}(v)  \cr
&{} =  u^{a+1} {1\over a!}
\sum_{j=2}^{a+2} {1\over (j-2)!} \, f_{j-1\, a+3\, j \, a+2}(v) \, . \cr}
}
The solution of \recurs\ is essentially as before giving in this case
\eqn\Sols{\eqalign{
c^{(p)}_{ppp,a0} = {}& {1\over a!^4} \, , \qquad
c^{(p)}_{pjj,a0} = {a+1\over a!^3 \, (j-a-2)!}\,, \quad
j = a+2, \dots , p-1 \, . \cr
c^{(p)}_{i\,i+k-p-2\,k,a0} = {}&
{1 \over a!^2 \, (i-a-3)! \,(k-a-3)!}  \, , \quad i,k = a+3, \dots , p \, . \cr}
}
These results are just as expected from \solcaf\ and \solcat\ after 
substituting $p=2a+2$. The results manifestly satisfy the necessary symmetry
conditions.

For completeness it is also necessary to analyse the other contributions
which are present in crossing symmetric expressions exhibited in \T\ for 
$T^{(p)}_{pjj,a0}, \, T^{(p)}_{i\, i+k-p-2\, k,a0}$ and $T^{(p)}_{iip,a0}$.
The six terms in \T\ form three pairs related by $\sigma\leftrightarrow \tau$
under which our equations are invariant.
For those terms proportional to $\sigma^{p-2-a}$ we obtain
\eqnn\recurp
$$\eqalignno{
\sigma^{p-2-a} &  \bigg ( \sum_{j=a+2}^p c^{(p)}_{pjj,a0} \, u^{p-1} 
f_{j-1\, p+1\, j \, p}(v)\cr 
&{}+ \! \sum_{i=p-a+1}^p \sum_{k=a+3}^p 
c^{(p)}_{i\,i+k-p-2\, k,a0} \sum_{m=0}^{p-i+1} {(-1)^m\over m!}(p-i+1-m)! \cr
\noalign{\vskip - 4pt}
& \hskip 4.7cm {} \times  u^{i+m-2} 
f_{i+k-p-2+m\, i+m \, i+k-p-2+m\, i+m}(v)  \cr
&{}+ \sum_{i=p-a}^{p-1} c^{(p)}_{iip,a0}\, \sum_{m=0}^{p-i} 
{(-1)^m\over m!}\,(p+1-i-m)!
\,  u^{i+m-1} f_{i-1+m\, i+1+m\, i+m \, i+m}(v) \bigg ) \cr
&{} = \sigma^{p-2-a} u^{p-1-a} {1\over (p-2-a)!}
\sum_{j=2}^{p-a} {1\over (j-2)!} \, f_{j-1\, p-a+1\, j \, p-a}(v) \cr
&{} =  \F(u,v;\sigma,\tau) \big |_{\sigma^{p-2-a}} \, ,& \recurp\cr}
$$
using \Fsa. The identity shown in \recurp\ is obtained by following the identical 
procedure as in calculations described  above after using
\eqn\symc{
c^{(p)}_{i\, i+k-p-2\, k,a0} \leftrightarrow c^{(p)}_{k\, i+k-p-2\, i,a0} \, , \quad
c^{(p)}_{pjj,a0} \leftrightarrow c^{(p)}_{jjp,a0} \, , 
\ \ \hbox{for} \ \ a\to p-a-2 \, ,
}
which are easily seen to be a property of the solutions \solcaf\ and \solcat. 
The result \recurp\ is then in accord with expectation from \Huniv.

For the remaining terms we consider those proportional to $\sigma^{p-2-a}
\tau^a$ for which, up to terms of ${\rm O}(u^p)$, we have just
\eqn\sta{
\sigma^{p-2-a}\tau^a u^{p-1} \bigg ( \sum_{j=a+2}^p c^{(p)}_{pjj,a0} \,
f_{j-1\, p+1\, jp}(v) + \sum_{i=p-a}^{p-1} c^{(p)}_{iip,a0} \,
f_{i-1\, p+1\, p\, i }(v) \bigg ) \, .
}
These are identified as required by \Huniv\ with the following term
in $\F$, after using the expressions \solcaf\ and \solcat,
\eqn\Fnew{\eqalign{
\F(u,v;\sigma,\tau) \big |_{\sigma^g\tau^h} = \sigma^g\tau^h u^{g+h+1}
& \bigg ( \sum_{j=h+2}^{g+h+1} {h+1 \over g!\, h!^2 \, (j-h-2)!} \,
f_{j-1\, g+h+3 \, j \, g+h+2}(v) \cr
&{}+ \sum_{j=g+2}^{g+h+1} {g+1 \over g!^2 h!\, (j-g-2)!} \,
f_{j-1\, g+h+3 \, g+h+2\, j}(v) \cr
& + {1\over g!^2\, h!^2}\, 
f_{g+h+1\, g+h+3 \, g+h+2\, g+h+2}(v)  \bigg ) \, . \cr}
}
This result is obtained from \sta\ for $g>h$, the corresponding result
for $g<h$ is obtained by using the symmetry under $\sigma \leftrightarrow
\tau$, for $g=h$ it is necessary to use \symc. For $h=0$ \Fnew\ coincides
with \Fsa. Although \Fnew\ is not immediately of the form expected from
\expF\ and \kl, it can be reduced to it by application of \hyper.

With the determination of $\F$ in general in \Fnew\ we can now determine
the remaining coefficients in \Hexpall. For terms proportional to 
$\sigma^a\tau^b$ in \Huniv\ we have, corresponding to \recura, 
\eqn\recurab{\eqalign{
& \sum_{j=a+b+2}^p c^{(p)}_{pjj,ab} \sum_{m=0}^{p-j} {(-1)^m\over m!}(p-j-m)! \, 
u^{j+m-1} f_{j-1+m\, j+1+m\, j+m \, j+m}(v)\cr
&{}+ \sum_{i=p-a+1}^p \sum_{k=a+b+3}^p c^{(p)}_{i\, i+k-p-2 \, k,ab} 
\sum_{m=0}^{p-k+1} {(-1)^m\over m!}(p-k+1-m)! \cr
\noalign{\vskip - 6pt}
& \hskip 4.8cm {} \times  u^{k+m-2} 
f_{i+k-p-2+m\, k+m \,i+k-p-2+m\, k+m}(v)  \cr
&{}+ \sum_{i=p-b+1}^p \sum_{j=a+b+3}^p c^{(p)}_{i\,\, j\, i+j-p-2,ab} 
\sum_{m=0}^{p-j+1} {(-1)^m\over m!}(p-j+1-m)! \cr
\noalign{\vskip - 6pt}
& \hskip 4.8cm {} \times  u^{j+m-2} 
f_{i+j-p-2+m\, j+m \, j+m\, i+j-p-2+m}(v)  \cr
&{}+ \sum_{i=p-a}^{p-1}\,
c^{(p)}_{iip,ab} \,  u^{p-1} f_{i-1\, p+1 \, i \, p }(v) + \sum_{i=p-b}^{p-1}\,
c^{(p)}_{ipi,ab} \,  u^{p-1} f_{i-1\, p+1 \, p \, i }(v) \cr
&{} =  u^{a+b+1}
\bigg ( \sum_{j=b+2}^{a+b+1} {b+1 \over a!\,b!^2 (j-b-2)!} \,
f_{j-1\, a+b+3 \, j \, a+b+2}(v) \cr
& \hskip 1.8cm {} + \sum_{j=a+2}^{a+b+1} {a+1 \over a!^2 b! (j-a-2)!} \,
f_{j-1\, a+b+3 \, a+b+2\,\, j}(v) \cr
&\hskip 1.8cm {} + {1\over a!^2 b!^2}\, 
f_{a+b+1\, a+b+3 \, a+b+2\, a+b+2}(v)  \bigg ) \, . \cr}
}
This may be analysed in a similar fashion to previously. For terms
proportional to $u^{a+b+1}$,
\eqn\abbase{\eqalign{
& (p-a-b-2)! \,\, c^{(p)}_{p\,a+b+2\,a+b+2,ab} \, 
f_{a+b+1\, a+b+3\, a+b+2 \, a+b+2}(v)
\cr &{}+ (p-a-b-2)! \!\! \sum_{i=p-a+1}^p \! c^{(p)}_{i\,i+a+b-p+1\,a+b+3,ab} \, 
f_{i+a+b-p+1\, a+b+3\, i+a+b-p+1 \, a+b+3}(v) 
\cr &{}+ (p-a-b-2)! \!\! \sum_{i=p-b+1}^p \! c^{(p)}_{i\,a+b+3\,i+a+b-p+1,ab}
\,  f_{i+a+b-p+1\, a+b+3\, a+b+3\,i+a+b-p+1}(v) \cr
&{}= \sum_{j=b+2}^{a+b+1} {b+1 \over a!\,b!^2 (j-b-2)!} \,
f_{j-1\, a+b+3 \, j \, a+b+2}(v) \cr
&\quad {} + \sum_{j=a+2}^{a+b+1} {a+1 \over a!^2 b! (j-a-2)!} \,
f_{j-1\, a+b+3 \, a+b+2\,\, j}(v) + {1\over a!^2 b!^2}\, 
f_{a+b+1\, a+b+3 \, a+b+2\, a+b+2}(v) \, . \cr}
}
Using \hyper\ this may be decomposed to give
\eqnn\solcab
$$\eqalignno{
c^{(p)}_{i\,a+b+i-p+1\,a+b+3,ab} = {}& {1\over (p-a-b-2)! \, a! \,b!^2 (i+a-p-1)!}
\, , \quad  i = p-a+1, \dots , p \, ,\cr
c^{(p)}_{i\,a+b+3\,a+b+i-p+1,ab} = {}& {1\over (p-a-b-2)! \, a!^2\,
b!\,(i+b-p-1)!}
\, , \quad  i = p-b+1, \dots , p \, , \cr
c^{(p)}_{p\,a+b+2\,a+b+2,ab} = {}& {a+b+1\over (p-a-b-2)! \, a!^2\,b!^2} \, . &
\solcab \cr}
$$
To obtain \solcab\ we have used the identity $\sum_{m=0}^k (n-1+m)!/m! =
(n+k)!/n\, k!$. For terms proportional to $u^n$ in \recurab\ for
$n=a+b+2, \dots ,p-2$ we get
\eqn\relcab{\eqalign{
\sum_{m=0}^{k-a-b-2} {(-1)^m\over m!} \, c^{(p)}_{p\, k-m\,k-m,ab} = {}& 0 \, ,
\qquad k = a+b+3 , \dots , p-1 \, , \cr
\sum_{m=0}^{k-a-b-3} {(-1)^m\over m!} \, 
c^{(p)}_{i\, i+k-p-2-m\,k-m,ab} ={}&  0 \, , \qquad{}
\cases{i=p-a+1, \dots, p\cr k=a+b+4,\dots ,p} \, , \cr
\sum_{m=0}^{j-a-b-3} {(-1)^m\over m!} \, 
c^{(p)}_{i\, j-m \, i+j-p-2-m,ab} ={}&  0 \, , \qquad
\cases{i=p-b+1, \dots, p\cr j=a+b+4,\dots ,p} \, . 
\cr}
}
Combining with \solcab\ then gives
\eqn\solcabf{\eqalign{
& c^{(p)}_{pjj,ab} = {a+b+1\over (p-a-b-2)! \, a!^2\,b!^2 (j-a-b-2)!} \, , \quad
j = a+b+2, \dots , p-1 \, , \cr
& c^{(p)}_{i\,i+k-p-2\,k,ab} = 
{1\over (p-a-b-2)! \, a! \,b!^2\, (i+a-p-1)! \,(k-a-b-3)!}\,,\cr
\noalign{\vskip - 3pt}
&\hskip 3cm k=a+b+3, \dots , p, \, i = p-a+1, \dots , p \, ,\cr
& c^{(p)}_{i\, j\, i+j-p-2,ab} =  
{1\over (p-a-b-2)! \, a!^2 \,b!\, (i+b-p-1)! \,(j-a-b-3)!}\,,\cr
\noalign{\vskip - 3pt}
&\hskip 3cm j=a+b+3, \dots , p, \, i = p-b+1, \dots , p \, . \cr}
}
For the remaining terms in \recurab\ proportional to $u^{p-1}$ after using the
result \solcabf\ we have
\eqn\recurt{\eqalign{
& \bigg ( c^{(p)}_{ppp,ab}  - {a+b+1 \over (p-a-b-2)!^2 \, a!^2 \,b!^2} \bigg )
\,  f_{p-1\, p+1\, pp }(v)\cr
&{}- {1\over (p-a-b-2)!^2 \, a!\, b!^2}
\sum_{j=p-a+1}^p  {1\over (j+a-p-1 )!} \, f_{j-1\, p+1 \, j-1 \, p+1 }(v)  \cr
&{}- {1\over (p-a-b-2)!^2 \, a!^2\, b!}
\sum_{j=p-b+1}^p  {1\over (j+b-p-1 )!} \, f_{j-1\, p+1 \, p+1 \, j-1}(v)  \cr
&{}+ \sum_{i=p-a}^{p-1}\, c^{(p)}_{iip,ab} \, f_{i-1\, p+1 \, i \, p }(v)
+ \sum_{i=p-b}^{p-1}\, c^{(p)}_{ipi,ab} \, f_{i-1\, p+1 \, p \, i}(v)   
= 0 \, . \cr}
}
This may be solved giving
\eqn\solcabt{\eqalign{
c^{(p)}_{ppp,ab} = {}& {1\over (p-a-b-2)!^2 \, a!^2\, b!^2} \, , \cr
c^{(p)}_{iip,ab} = {}& {p-a-1\over (p-a-b-2)!^2 \, a! \, b!^2\, (i+a-p)!}\,, \quad
i = p-a, \dots , p-1 \, , \cr
c^{(p)}_{ipi,ab} = {}& {p-b-1\over (p-a-b-2)!^2 \, a!^2 \, b!\, (i+b-p)!}\,, \quad
i = p-b, \dots , p-1 \, . \cr}
}
The coefficients satisfy the crucial relations
\eqn\symca{\eqalign{
& c^{(p)}_{\, i\, i+j-p-2\, j,ab} \leftrightarrow c^{(p)}_{j\, i+j-p-2\, i,ab} \, , 
\qquad
c^{(p)}_{\,i\, j\, i+j-p-2,ab} \leftrightarrow c^{(p)}_{\,i+j-p-2\,j\, i,ab} \, , \cr
& c^{(p)}_{iip,ab} \leftrightarrow c^{(p)}_{pii,ab} \, , \qquad 
c^{(p)}_{ipi,ab} \leftrightarrow c^{(p)}_{ipi,ab} \, , \ \ 
\hbox{for} \ \ a\to p-2-a-b \, , \cr}
}
and
\eqn\symcb{\eqalign{
& c^{(p)}_{\,i\, i+j-p-2\, j,ab} \leftrightarrow c^{(p)}_{i+j-p-2\, i\, j,ab} \, ,
\qquad
c^{(p)}_{\,i\, j\, i+j-p-2,ab} \leftrightarrow c^{(p)}_{j\, i \, i+j-p-2,ab} \, , \cr
& c^{(p)}_{iip,ab} \leftrightarrow c^{(p)}_{iip,ab} \, , \qquad
c^{(p)}_{ipi,ab} \leftrightarrow c^{(p)}_{pii,ab} \, , \ \
\hbox{for} \ \ b\to p-2-a-b \, . \cr}
}
There is also a similar relation for $a\leftrightarrow b$ which can be obtained
by combining \symca\ and \symcb.
These relations require for the undetermined coefficient so far
\eqn\solcc{\eqalign{
c^{(p)}_{j+k-p-2\, j \, k ,ab} = {}&
{1\over (p-a-b-2)!^2 \, a! \,b!\, (j+a-p-1)! \,(k+b-p-1)!}\,,\cr
\noalign{\vskip - 3pt}
&\qquad j=p-a+1, \dots , p, \, k = p-b+1, \dots , p \, . \cr}
}
The results \solcabf, \solcabt\ and \solcc\ hence determine the expansion
of $\H^{(p)}$ for general $p$£.
The symmetry conditions \symca\ and \symcb\ are necessary to ensure that
the other terms in the expression for $T^{(p)}_{ijk,ab}$, defined in \T,
contribute as required to the terms in $\F$ proportional to $\sigma^g\tau^h$
with $g=p-2-a-b, \, h=b$ and $g=a, \, h=p-2-a-b$. The results given by 
\solcabf, \solcabt\ and \solcc\ also satisfy
\eqnn\symcab
$$\eqalignno{
c^{(p)}_{\,i\, j\, i+j-p-2,aa} = {}& c^{(p)}_{\,i \,i+j-p-2\, j,aa} \, , \qquad
c^{(p)}_{j+k-p-2 \, jk ,aa} = c^{(p)}_{j+k-p-2\,kj ,aa} \, , \cr
c^{(p)}_{\,iip,aa}  = {}& c^{(p)}_{\, ipi,aa} \, , \cr
c^{(p)}_{\,i j\, i+j-p-2,a\, p-2-2a} = {}& c^{(p)}_{i+j-p-2\, j \, i,a\,p-2-2a} \, , 
\ \ c^{(p)}_{\, i \, i+k-p-2 \, k ,a\, p-2-2a} = 
c^{(p)}_{k\, i+k-p-2\,i,a\,p-2-2a} \, , \cr
c^{(p)}_{\,iip,a\, p-2-2a}  = {}& c^{(p)}_{pii,a\, p-2-2a} \, , & \symcab \cr}
$$
which shows that they remain valid in these cases when the symmetry requirements
in \symT\ hold.

\vfill\eject

\newsec{Conclusion and Future Investigations}

In this thesis we have derived $\N=2,4$ superconformal Ward identities 
for four point functions of $\half$-BPS operators. They were obtained by considering invariance of a correlation function
obtained from the pure CPO correlation function by action of one supercharge. In that sense the Ward identities are a manifestation of the
first order action of the supercharges. 

There could possibly exist further identities which require consideration of higher order actions of supercharges, although we have no reason to believe so.
A related question is whether all four point functions of all members of the multiplet generated from the CPOs we considered can be uniquely obtained
by action of a differential operator acting on the pure CPO four point function. For the case of three point functions for the case $p=2$ this was done by hand in \scft.
Ideally we would like to show that there is a set of differential operators which together with the superconformal constraints derived here
uniquely generate all four point functions for the short multiplets we considered.

Another area of possible future investigation is whether further constraints from the operator product expansion can be developed to extend
the requirements superconformal symmetry imposes. For example crossing symmetry allowed us to fix the form of the one variable function completely to free field form.
This argument is restricted to the case of three identical fields in the four point function.

Our results for the four point functions of $[0,p,0]$ $\half$-BPS operators in the large $N$ limit suggest a form for the averaged first order anomalous dimension for general $p$ at order $1/N$. To fix the concrete expression we would have to find a way to compute the constants present
in the results for $p=2,3,4$ for general $p$. It would be very interesting to compute the large $N$ amplitude for $p=5$ which could confirm this suggestion
and could possibly shed some light on the structure of the coefficients determining the averaged first order anomalous dimensions.

Finally, we followed up on an observation made in \od\ about the universal singularity structure in $u$ of the large $N$ amplitudes and showed how this might be exploited together with crossing symmetry to restrict the possible expressions for amplitudes with higher $p$.
By making one additional technical assumption about the details of the singularity structure we managed to derive explicit expressions for all coefficients of $\oD$ functions appearing in a general $p$ large $N$ four point amplitude of four identical $\half$-BPS single trace operators.

Our assumptions about the singularity structure are attempts to generalize from the cases explicitly known for $p=2,3,4$ \refs{\Arut,\ADHS,\Degen, \od}. They are based
on the expectation that all long operators should receive anomalous dimensions. Therefore we assume the cancellation of all terms with lower order than $p$ in $u$.
An important step further in this direction would be the computation of the free field amplitude
which would provide an independent derivation of the function $\F$ and a check on the first terms we computed for $\F$ from the dynamical part of the amplitude. 
To do this, also the ambiguity in the split of the amplitude into $\H,\hf$ has to be understood better.
Also, once these results were established one could attempt to use them to derive the structure of the general averaged first order anomalous dimension as mentioned above.
\vfill
\eject
\appendix{A}{Results for Null Vectors}

We discuss here some results for null vectors $t_r$ which are useful
in the text. For generality we allow $t$ to be $d$-dimensional. As
a consequence of \nul\ differentiation requires some care but for any
null vector $a$ we may define as usual
\eqn\diffo{
{\pr \over \pr t} (a{\cdot t})^n = n\, (a{\cdot t})^{n-1} a \, .
}
More generally for a set of null vectors $a_1,a_2,\dots, a_p $ we have
\eqn\diffm{\eqalign{
\! {\pr \over \pr t} \prod_{i=1}^p (a_i{\cdot t})^{n_i} = {}&
\sum_{i=1} ^p n_i\, (a_1{\cdot t})^{n_1} \dots (a_i{\cdot t})^{n_i-1}\dots
(a_p{\cdot t})^{n_p} a_i \cr
&{}- R \!\! \sum_{1\le i < j \le p} n_i n_j \, a_i{\cdot a_j} \, 
(a_1{\cdot t})^{n_1} \dots (a_i{\cdot t})^{n_i-1}\dots 
(a_j{\cdot t})^{n_j-1} \dots (a_p{\cdot t})^{n_p} t \, , \cr}
}
where
\eqn\defR{
R = {2\over 2N+d-4} \, , \qquad N = \sum_{i=1}^p n_i \, .
}
The right hand side of \diffm, with \defR, may be represented in the form
\eqn\tdiff{
\bigg ( {\pr\over \pr t } - t \, {1\over 2t{\cdot \pr} +d } \, \pr^2 \bigg )
\prod_{i=1}^p (a_i{\cdot t})^{n_i} \, ,
}
where the action of the derivatives is as usual, without regard to the 
constraint $t^2=0$. The resulting operator  is equivalent to a definition given in 
\Barg\ for an interior differential operator on the complex null cone.

{}From \diffm\ we may readily find
\eqn\Lap{
\bigg [ {\pr \over \pr t_r}, {\pr \over \pr t_s} \bigg ] 
\prod_{i=1}^p (a_i{\cdot t})^{n_i} = 0 \, , \quad
{\pr \over \pr t}{\cdot {\pr \over \pr t}} \prod_{i=1}^p (a_i{\cdot t})^{n_i}
= 0 \, , \quad t {\cdot {\pr \over \pr t}} \prod_{i=1}^p (a_i{\cdot t})^{n_i}
= N \, \prod_{i=1}^p (a_i{\cdot t})^{n_i} \, .
}
We also have
\eqn\dr{
\bigg [ {\pr \over \pr t_r}, \,  t_s \bigg ]
\prod_{i=1}^p (a_i{\cdot t})^{n_i} = \bigg ( \de_{rs} - {2\over 2N+d-2}
\, t_r {\pr \over \pr t_s} \bigg ) \prod_{i=1}^p (a_i{\cdot t})^{n_i} \, ,
}
which implies
\eqn\div{
{\pr \over \pr t_r} \bigg (  t_r \prod_{i=1}^p (a_i{\cdot t})^{n_i} \bigg )
= {(2N+d)(N+d-2)\over 2N+d-2} \,  \prod_{i=1}^p (a_i{\cdot t})^{n_i} \, .
}

Defining the generators of $SO(d)$ by 
\eqn\gend{
L_{rs} = t_r \pr_s - t_s \pr_r \, ,
}
then the above results give
\eqn\LLe{
L_{ru}L_{su} \prod_{i=1}^p (a_i{\cdot t})^{n_i} = - \big ( (N+d-3)\,
t_r \pr_s + (N-1)\, t_s \pr_r + N \, \de_{rs} \big ) 
\prod_{i=1}^p (a_i{\cdot t})^{n_i} \, ,
}
and 
\eqn\eiSO{
\half L_{rs} L_{rs} \prod_{i=1}^p (a_i{\cdot t})^{n_i} = - 
N (N+d-2) \prod_{i=1}^p (a_i{\cdot t})^{n_i}  \, ,
}
which reproduces the appropriate eigenvalue of the Casimir operator for the
representation formed by traceless rank $N$ tensors.

If $V_r(t)$ is homogeneous of degree $N$ then in general
\eqn\Vd{
V_r = {\hat V}_r + {1\over N+1}{\pr \over \pr t_r } \big ( t_s V_s \big ) \, ,
\qquad t_r {\hat V}_r = 0 \, ,
}
as used in \VU\ and \VUp. If $V_r$ also satisfies
\eqn\VD{
{\pr \over \pr t_r } V_s - {\pr \over \pr t_s } V_r = 0 \, , \qquad
{\pr \over \pr t_r } V_r = 0 \, ,
}
then, by contracting with $t_s$ and using \Lap, \dr, we easily see that
${\hat V}_r=0$.
As a further corollary if $V_r = \pr_s U_{rs}$, $U_{rs}=-U_{sr}$, 
$\pr_{[r} U_{su]} =0$ then $(N+1) V_r = \pr_r ( \half L_{su} U_{su})$ with
$L_{su}$ as in \gend. In general we have the decomposition
\eqn\Vdiv{\eqalign{
V_r = {}& {2N+d-4\over (2N+d-2)(N+d-3)}\, t_r \, \pr{\cdot V} \cr
&{} -{1\over (2N+d)(N+d-3)} \big ( (2N+d-2) \pr_s (t_r V_s - t_s V_r)
+ 2\, \pr_r (t{\cdot V}) \big ) \, . \cr}
}
\vfill
\eject
\appendix{B}{Two Variable Harmonic Polynomials}

For the expansion of four point functions in terms of $R$-symmetry
representations we consider here the eigenfunctions
of the $SO(d)$ Casimir operator
\eqn\Ls{
L^2 = \half L_{rs}L_{rs} \, ,
}
where the generators  are
\eqn\gen{
L_{rs} = t_{1r} \pr_{1s} - t_{1s} \pr_{1r} +  
t_{2r} \pr_{2s} - t_{2s} \pr_{2r} \, ,
}
formed by homogeneous functions of the null vectors $t_1,t_2,t_3,t_4$.
Obviously $L_{rs} t_1{\cdot t_2} = 0$ and hence $L^2 (t_1 {\cdot t_2})^k
(t_3 {\cdot t_4})^l f(\lam,\mun) = (t_1 {\cdot t_2})^k (t_3 {\cdot t_4})^l
L^2 f(\lam,\mun)$, where $\lam,\mun$ are given by \deflu. 
We therefore first consider eigenfunctions which are polynomials in $\lam,\mun$
\eqn\polyY{
Y(\lam,\mun) = \sum_{t\ge 0} \sum_{q=0}^t  c_{t,q} \, \lam^{t-q} \mun^q \, ,
}
satisfying
\eqn\eigL{
L^2 Y(\lam,\mun) = - 2C Y(\lam,\mun) \, .
}

With the aid of the given in  appendix A we may easily calculate the 
action of $L^2$ on a monomial formed from $\lam,\mun$,
\eqn\Llu{
L^2( \lam^p \mun^q) = - 2\big ( (d-2)(p+q) + 4 pq\big ) \lam^p \mun^q
+ 2(1-\lam-\mun)
\big ( p^2 \lam^{p-1} \mun^q + q^2 \lam^p \mun^{q-1} \big ) \, , 
}
or 
\eqn\LDd{
\half L^2 \to \D_d = (1-\lam-\mun)\Big ( {\pr \over \pr \lam} \lam
{\pr \over \pr \lam} + {\pr \over \pr \mun} \mun {\pr \over \pr \mun} \Big )
- 4 \lam\mun{\pr^2 \over \pr \lam\pr\mun} - (d-2) 
\Big ( \lam {\pr \over \pr \lam} +  \mun {\pr \over \pr \mun} \Big ) \, .  
}
Alternatively $\D_d$ may be written in the form
\eqn\Dad{
\D_d = {1\over w} \, {\underline \pr}^T w {\underline G}\, {\underline \pr}\, ,
\qquad {\underline G} = \pmatrix { \lam(1-\lam - \mun) & - 2\lam\mun\cr
- 2\lam\mun & \mun(1-\lam - \mun)} \, , \quad {\underline \pr} =
\pmatrix { \pr_\lam\cr \pr_\mun} \, ,
}
where, with $\Lambda=(\sqrt \lam + \sqrt \mun + 1) (\sqrt \lam + \sqrt \mun - 1)
(\sqrt \lam - \sqrt \mun + 1)(\sqrt \lam - \sqrt \mun - 1)$ as in \conlm,
\eqn\defw{
w = \Lambda^{{1\over 2}(d-5)} \, .
}
In general, for a polynomial as in \polyY\ with $t_{\rm max} = n$, we must have that 
$c_{n,q}$ forms an eigenvector for an $(n{+1}) \times (n{+1})$ 
matrix $M_n$,
\eqn\eigc{
M_{n,pq} c_{n,q} = C c_{n,p} \, , \quad M_{n,pq} = \de_{p\, q}
\big ( n(n+d-2)+ 2 p (n-p) \big ) + \de_{p \, q{-1}} \, q^2 +
\de_{p \, q{+1}} (n-q)^2 \, .
}
The coefficients $c_{t,q}$ with $t<n$ may then be obtained by solving
recurrence relations. For given $n$ there are $n+1$ eigenvectors 
solving \eigc\ and the corresponding eigenfunctions are
\eqn\eig{
Y_{nm}(\lam,\mun) \, , \quad C_{nm} =  n(n+d-3) + m(m+1) \, , \quad
n=0,1,2, \dots , \  m=0,\dots n \, .
}
As a consequence of \Dad\ and \defw\ the polynomials are orthogonal for
$d>5$ with respest to integration over $\lam,\mun \ge 0, \, 
\sqrt \lam + \sqrt \mun \le 1$ with weight $w$ (for a general discussion
of such two variable orthogonal polynomials see \refs{\Koo,\ortho}).

The polynomials $Y_{nm}$ are also eigenfunctions for higher order Casimir
invariants. Letting
\eqn\defQ{ 
\quar \, L_{rs}L_{st}L_{tu}L_{ur} - \half \, L^2 L^2 + \quar (d-2)(d-3)\, L^2  \to \Q
}
then acting on any $Y(\lam,\mun)$ we may express $\Q$ in a form similar to \Dad,
\eqn\Qact{\eqalign{
\Q ={}& - {1\over \Lambda^{{1\over 2}(d-5)}} 
\pmatrix { \pr_\lam{\!}^2 &  \pr_\mun{\!}^2} \Lambda^{{1\over 2}(d-3)}
\pmatrix { \lam^2 & - \lam\mun\cr - \lam\mun & \mun^2}
\pmatrix { \pr_\lam{\!}^2 \cr  \pr_\mun{\!}^2} \cr
&{} + (d-3) \,  {1\over \Lambda^{{1\over 2}(d-5)}} 
\pmatrix { \pr_\lam &  \pr_\mun} \Lambda^{{1\over 2}(d-5)}
\pmatrix { 2\lam & \lam+\mun-1\cr \lam+\mun-1 & 2\mun} 
\pmatrix { \pr_\lam\cr \pr_\mun} \cr 
&{} + (d-2) \,  {1\over \Lambda^{{1\over 2}(d-5)}} \big ( \pr_\lam
\, \Lambda^{{1\over 2}(d-3)} \, \pr_\mun + 
\pr_\mun \,\Lambda^{{1\over 2}(d-3)}\, \pr_\lam \big ) \, . \cr}
}
The harmonic polynomials then satisfy
\eqn\Qeig{
\Q \, Y_{nm} = - (n-m)(n+m+1)(n+m+d-3)(n-m+d-4) \, Y_{nm} \, .
}

Using \Llu\ it is  straightforward to construct the first few eigenfunctions 
satisfying \eigL\ by hand. With an arbitrary normalisation, we find for $n=0,1,2,3$,
\eqn\Ynm{\eqalign{
Y_{00}(\lam,\mun) = {}& 1  \, , \cr
Y_{10}(\lam,\mun) = {}& \lam - \mun \, , \cr
Y_{11}(\lam,\mun) = {}& \lam + \mun - {\ts {2\over d}}\, , \cr
Y_{20}(\lam,\mun) = {}& \lam^2 + \mun^2 - 2\lam\mun
- {\ts {2\over d-2}}(\lam + \mun) + {\ts {2\over (d-2)(d-1)}} \, , \cr
Y_{21}(\lam,\mun) = {}& \lam^2 - \mun^2 - {\ts {4\over d+2}}
(\lam - \mun) \, , \cr
Y_{22}(\lam,\mun) = {}& \lam^2 + \mun^2 + 4\lam\mun 
- {\ts {8\over d+4}}(\lam + \mun) + {\ts {8\over (d+2)(d+4)}} \, ,\cr
Y_{30}(\lam,\mun) = {}& \lam^3 - 3\lam^2\mun + 3\lam\mun^2 - \mun^3
- {\ts {6\over d}}(\lam^2 - \mun^2) + {\ts {12\over d(d+1)}}(\lam - \mun)
 \, , \cr
Y_{31}(\lam,\mun) = {}& \lam^3 - \lam^2\mun - \lam\mun^2 + \mun^3
- {\ts {8(d-1)\over (d+4)(d-2)}}(\lam^2 + \mun^2) +
{\ts {8(d-6)\over (d+4)(d-2)}}\, \lam\mun \cr
&{}+{\ts {4(3d+2)\over (d+1)(d+4)(d-2)}}(\lam + \mun) - 
{\ts {8\over (d+1)(d+4)(d-2)}} \, , \cr
Y_{32}(\lam,\mun) = {}& \lam^3 + 3\lam^2\mun - 3\lam\mun^2 - \mun^3
- {\ts {12\over d+6}}(\lam^2 - \mun^2) + 
{\ts {24\over (d+4)(d+6)}}(\lam - \mun)
 \, , \cr
Y_{33}(\lam,\mun) = {}& \lam^3 +9\lam^2\mun  + 9 \lam\mun^2 + \mun^3
- {\ts {18\over d+8}}(\lam^2 + \mun^2)- {\ts {72\over d+8}} \,\lam\mun \cr
&{}+{\ts {72\over (d+6)(d+8)}}(\lam + \mun) 
- {\ts {48\over (d+4)(d+6)(d+8)}} \, . \cr}
}
Up to an overall normalisation for $d=6$ each term may be identified
with terms in the projection operators constructed in \ADHS\ where $Y_{nm}$
corresponds to the $SU(4)\simeq SO(6)$ representation with Dynkin labels
$[n-m,2m,n-m]$. For $m=n$ in \eig\ we have $c_{n,q} = {n \choose q}^2$ 
and the recurrence relations may be easily solved giving
\eqn\Ynn{
Y_{nn} (\lam,\mun) = A_n F_4 (-n, n+\half d -1 ; 1,1; \lam,\mun) \, ,
}
where $F_4$ is one of Appell's generalised hypergeometric functions\foot{
$$F_4(a,b;c,c';x,y) = \sum_{m,n} {(a)_{m+n} (b)_{m+n} \over
(c)_m (c')_n m! \, n!} \, x^m y^n \, .
$$} and $A_n$ is some overall constant.

To obtain more general forms (see \Vretare) we used the variables $\alpha,\bet$
defined in \defab. Acting on $Y(\lam,\mun)= \P(\alpha,\bet)=
\P(\bet,\alpha)$
\eqn\LD{
\half L^2 \P(\alpha,\bet) = \cD_d \P(\alpha,\bet) \, ,
}
where, using \Llu\ or \LDd, we now have
\eqn\Dop{
\cD_d = {\pr \over \pr \alpha} \alpha(1-\alpha) 
{\pr \over \pr \alpha} + {\pr \over \pr \bet} \bet(1-\bet)
{\pr \over \pr \bet} + (d-4) {1\over \alpha- \bet} \Big (
\alpha(1-\alpha) {\pr \over \pr \alpha} - \bet(1-\bet)
{\pr \over \pr \bet} \Big ) \, .
}
Corresponding to \eigL\ and \eig\ we have
\eqn\eigD{
\cD_d \P_{nm}(\alpha,\bet) = - \big ( n(n+d-3) + m(m+1) \big ) 
\P_{nm}(\alpha,\bet) \, ,
}
where $\P_{nm}(\alpha,\bet)$ are generalised symmetric Jacobi polynomials.
For particular $d$ simplified formulae may be found in terms of well
know single variable Legendre polynomials $P_n$.  When $d=4$ it
is clear from \Dop\ that $\cD_4$ is just the sum of two independent 
Legendre differential operators  so that
\eqn\eigfour{
\P_{nm}(\alpha,\bet) = \half \big ( P_n(y) P_m(\yz) +
P_m(y) P_n(\yz) \big ) \, , \quad n\ge m \, , 
}
with $y,\yz$ defined in \yy.
For $d=6$ we may use the result
\eqn\DF{
\cD_6  {1\over \alpha -\bet } = {1\over \alpha -\bet } ( \cD_4 + 2 )  \, ,
}
to see that we can take the eigenfunctions to be of the form
\eqn\eigsix{
\P_{nm}(\alpha,\bet) =  p_{n{+1}m}(y,\yz) \, , \quad n\ge m \, , 
}
where
\eqn\defpnm{
p_{nm}(y,\yz) = - p_{mn}(y,\yz) = 
{ P_{n}(y) P_m(\yz) - P_m(y) P_{n} (\yz) \over y - \yz} \, .
}

It is also of interest to consider $d=8$ when we take
\eqn\PFe{
\P (\alpha,\bet) = {F(\alpha,\bet) \over (\alpha-\bet)^2}\, ,
}
and  the eigenvalue equation becomes
\eqnn\DFe
$$\eqalignno{
\cD_6 F(\alpha,\bet) &{} - {2\over (\alpha-\bet)^2} \Big ( \alpha(1-\alpha)
{\pr \over \pr \alpha} \big ( (\alpha-\bet)   F(\alpha,\bet) \big )
- \bet( 1- \bet ) {\pr \over \pr \bet} \big ( (\alpha-\bet) 
F(\alpha,\bet) \big ) \Big ) \cr
&{} = - (C+4) F(\alpha,\bet) \, . &  \DFe \cr}
$$
If we assume
\eqn\FPnm{
F(\alpha,\bet) = \sum_{n,m} a_{nm}\,  p_{nm}(y,\yz)\, ,
}
and use, from standard identities for Legendre polynomials,
\eqnn\recur
$$\eqalignno{ 
&{1\over y-\yz} \Big ( (1-y^2){\pr \over \pr y} - (1-\yz^2)
{\pr \over \pr \yz}\Big )\big ( (y-\yz) p_{nm}(y,\yz)\big ) \cr 
&\quad {} = {m(m+1)\over 2m+1} 
\big ( p_{nm{+1}}(y,\yz)- p_{nm{-1}}(y,\yz)\big ) -
{n(n+1)\over 2n+1} \big ( p_{n{+1}m}(y,\yz) - p_{n{-1}m}(y,\yz) \big )\cr 
&(y-\yz)  p_{nm}(y,\yz) = {1\over 2n+1} \big ( (n+1) p_{n{+1}m}(y,\yz) + n\,
p_{n{-1}m}(y,\yz) \big ) \cr 
& \qquad \qquad \qquad \qquad {}- {1\over 2m+1}\big ( (m+1) p_{nm{+1}}(y,\yz)
 + m \, p_{nm{-1}}(y,\yz) \big ) \, , & \recur \cr}
$$
then we may set up recurrence relations for $a_{nm}$ which for the
appropriate value of $C$ have just four terms.  For $C=n(n+1)+m(m+1)-6$
\FPnm\  gives a solution
\eqn\defq{\eqalign{
q_{nm}(y,\yz) = {1\over (y-\yz)^2} \bigg \{ & {n+1\over 2n+1} (n+m)(n-m-1)
p_{n{+1}m}(y,\yz) \cr &{} + {n\over 2n+1} (n+m+2)(n-m+1) p_{n{-1}m}(y,\yz) \cr
&{}- {m+1\over 2m+1} (n+m)(n-m+1) p_{nm{+1}}(y,\yz) \cr &
{}- {m\over 2m+1} (n+m+2)(n-m-1) p_{nm{-1}}(y,\yz) \bigg \} \, , \cr}
}
where $q_{nm}(y,\yz) = - q_{mn}(y,\yz), \ q_{nn}(y,\yz) = 
q_{n{+1}\, n}(y,\yz)=0$. Hence we can take
\eqn\eigeight{
\P_{nm}(\alpha,\bet) =  q_{n{+2}\, m}(y,\yz) \, . 
}

The above results for harmonic polynomials in $\lam,\mun$ are relevant
for discussing four point functions when each field belongs to the same
$SO(d)$ representation. For the more general case we also consider instead
of \eigL,
\eqn\eigLab{
L^2 \big ( (t_1{\cdot t_4})^a (t_2{\cdot t_4})^b 
Y^{(a,b)}(\lam,\mun)\big ) = -2C 
\big ( (t_1{\cdot t_4})^a (t_2{\cdot t_4})^b Y^{(a,b)}(\lam,\mun) \big )\, ,
}
where now the action of $L^2$ is determined by
\eqn\Lab{ \eqalign{
L^2 & \big ( (t_1{\cdot t_4})^a (t_2{\cdot t_4})^b  \lam^p \mun^q \big )\cr
& {}= (t_1{\cdot t_4})^a (t_2{\cdot t_4})^b \Big ( 
\big ( 2\D_d -(a+b)(a+b+d-2) - 4a p - 4bq \big ) ( \lam^p \mun^q ) \cr
&\qquad \qquad \qquad \qquad \ {}+ 2(1-\lam - \mun) \big ( bp \, 
\lam^{p-1} \mun^q  + a q \, \lam^p \mun^{q-1} \big ) \Big ) \, , \cr}
}
or
\eqn\defDab{
\half L^2 \big ( (t_1{\cdot t_4})^a (t_2{\cdot t_4})^b f(\lam,\mun) \big )
= (t_1{\cdot t_4})^a (t_2{\cdot t_4})^b \big ( \D^{(a,b)}_d
 - \half (a+b) (a+ b + d-2) \big ) f (\lam,\mun) \, ,
}
where
\eqn\LDa{
\D^{(a,b)}_d = \D_d + (1-\lam-\mun) \Big ( a \, {\pr \over \pr \mun} +
b\,{\pr \over \pr \lam} \Big ) -2a \, \lam {\pr \over \pr \lam} - 2b \,
\mun {\pr \over \pr \mun} \, .
}
This may also be written in the form \Dad\ with 
$w = \lam^b \mun^a \Lambda^{{1\over 2}(d-5)}$.
The possible eigenvalues for polynomial eigenfunctions with maximum 
power $p+q=n$ are then determined by the matrix
\eqnn\matM
$$\eqalignno{
M_{n,pq} = {}& \de_{p\, q}
\big ( n(n+d-2+a+b)+ \half (a+b)(a+b+d-2) + 2 p (n-p) + a(n-p) + b p\big ) \cr
&{} + \de_{p \, q{-1}} \, q(q+a)  + \de_{p \, q{+1}} (n-q)(n-q+b) \, . 
& \matM \cr}
$$
The eigenfunctions $Y^{(a,b)}_{nm}(\lam,\mun)$ for $m=0,1,\dots, n$
then have eigenvalues
\eqn\eigab{
C_{nm} =  \big ( n + \half (a+b)\big )\big (n+ \half (a+b) + d-3 \big ) + 
\big (m + \half (a+b)\big )\big (m + \half (a+b) + 1 \big ) \, .
}
For $d=6$ $Y^{(a,b)}_{nm}$ corresponds to the representation
$[n-m,a+b+2m,n-m]$. The simplest non trivial examples are
\eqn\Yab{\eqalign{
Y^{(a,b)}_{10}(\lam,\mun) = {}& \lam - \mun + {a-b \over a+b+d-2} \,, \cr
Y^{(a,b)}_{11}(\lam,\mun) = {}& {1\over b+1} \, \lam +
{1\over a+1} \, \mun - { 1\over a+b+\half d} \, . \cr}
}
Corresponding to \Ynn\ we have in general
\eqn\Ynnab{
Y^{(a,b)}_{nn} (\lam,\mun) = A_n 
F_4 (-n, n+ a+b+ \half d -1 ;b+1,a+1; \lam,\mun) \, .
}

Again more explicit results can be obtained by using the variables
$\alpha,\bet$. In \LDa\ the differential operator now becomes
\eqn\Dopab{
\cD^{(a,b)}_d = \cD_d^{\vphantom g} - \big (a\, \alpha - b(1-\alpha)\big ) 
{\pr\over \pr \alpha}- \big (a \, \bet - b(1-\bet )\big ) 
{\pr\over \pr \bet } \, ,
}
with $\cD_d$ given in \Dop.
Denoting the eigenfunctions of $\cD^{(a,b)}_d$ by $\P^{(a,b)}_{nm}
(\alpha,\bet)$ then previous results for $d=4,6$ for the eigenfunctions 
can be extended by using Jacobi polynomials $P_n^{(a,b)}$. For $d=4$
$\cD^{(a,b)}_4 = D^{(a,b)}_\alpha + D^{(a,b)}_\bet$ where $D^{(a,b)}$
is the ordinary differential operator defined by
\eqn\Dthreeab{
D^{(a,b)}_\alpha = {\d \over \d \alpha} \alpha(1-\alpha) {\d \over \d \alpha}
- a \, \alpha {\d \over \d \alpha} + b (1-\alpha) {\d \over \d \alpha} \, .
}
The eigenfunctions of $D^{(a,b)}_\alpha$ are just
$P^{(a,b)}_n(y)$, where $ y = 2\alpha-1$ and the eigenvalues are
$-n(n+a+b+1)$.
For $d=6$ the generalisation of \eigsix\ and \defpnm\ is then
\eqn\eigsixab{
\P^{(a,b)}_{nm}(\alpha,\bet) =  
{ P^{(a,b)}_{n+1}(y) P^{(a,b)}_m(\yz) - P^{(a,b)}_m(y) P^{(a,b)}_{n+1} (\yz) 
\over y - \yz} \, .
}

When $d=3$ the above results need to be considered separately since 
$\lam, \mun$ are not independent and satisfy the constraint \conlm.
The eigenfunctions $Y^{(a,b)}_{nm}(\lam,\mun)$ are also restricted
since $Y^{(a,b)}_{nm}(\lam,\mun) = 0$ for $m<n-1$ as a consequence
of \conlm. To obtain eigenfunctions of $L^2$ in general we make use
of the solution \defa\ which amounts to setting $\alpha= \bet$ in the
above, so that we are restricted just to single variable functions. 
Instead of \defDab\ we have
\eqn\Lab{ 
L^2 \big ( (t_1{\cdot t_4})^a (t_2{\cdot t_4})^b f(\alpha) \big )
= (t_1{\cdot t_4})^a (t_2{\cdot t_4})^b 
\big ( D^{(2a,2b)}_\alpha - (a+b)(a+b+1) \big )  f(\alpha) \, ,
}
using the definition \Dthreeab. In consequence
\eqn\Leig{
L^2 \big ( (t_1{\cdot t_4})^a (t_2{\cdot t_4})^b P^{(2a,2b)}_n(y) \big ) 
= - (n+a+b)(n+a+b+1) \, (t_1{\cdot t_4})^a (t_2{\cdot t_4})^b 
P^{(2a,2b)}_n(y) \, ,
}
corresponding to the $(n{+a}{+b})$-representation for $SU(2)\simeq SO(3)$. 
Hence for $d=3$ we may then take
\eqn\eigthreeab{
\P^{(a,b)}_{nn}(\alpha,\alpha) = P^{(2a,2b)}_{2n}( y ) \, , \qquad
\P^{(a,b)}_{n\, n{-1}}(\alpha,\alpha) = P^{(2a,2b)}_{2n-1}( y ) \, ,
}
with $\P^{(a,b)}_{nm}(\alpha,\alpha)=0$ for $m<n-1$.

For $d=3$ there are also eigenfunctions involving cross products. To
consider these we first define
\eqn\defT{
T_1 = t_1 \, {\cdot \, t_3 \times t_4} \, (t_1{\cdot t_4})^{a-1} 
(t_2{\cdot t_4})^b \, , \quad
T_2 = t_2 \, {\cdot \, t_3 \times t_4} \, (t_1{\cdot t_4})^{a}
(t_2{\cdot t_4})^{b-1} \, ,
}
and consider eigenfunctions of the form $T_1 f_1(\alpha) + T_2
f_2(\alpha)$. The action of $L^2$ on such functions is given by
\eqn\Lcross{\eqalign{
\big (L^2 &{} + (a+b-1)(a+b) \big )\, ( T_1 f_1 + T_2 f_2 ) \cr
&{}= \pmatrix { T_1 & T_2} \pmatrix { D^{(2a-1,2b+1)} - 2a & -2a \cr
- 2b &  D^{(2a+1,2b-1)} - 2b} \pmatrix {f_1 \cr f_2} \, . \cr}
}
However for $t_i$ three dimensional null vectors the basis given
by \defT\ is not independent since we have from \idthree
\eqn\TTid{
T_1 (1-\alpha ) + T_2 \alpha = 0 \, ,
}
so that $f_1, f_2$ are not unique. If we use this freedom to set $f_2=0$ 
the eigenvalue equation for $L^2$ reduces to
\eqn\eigf{\eqalign{
\Big ( &  D^{(2a-1,2b+1)} -2a + 2b \, {1-\alpha \over \alpha} - 
(a+b-1)(a+b) \Big ) f_1 \cr
&{}= {1\over \alpha} \Big (  D^{(2a-1,2b-1)} - (a+b-1)(a+b) \Big )
(\alpha f_1) = - C f_1 \, ,\cr}
}
which has solutions proportional to Jacobi polynomials,
\eqn\solcr{
f_1(\alpha) = {1\over \alpha}\, P_n^{(2a-1,2b-1)}(y) \, , \qquad
C = (n+a+b-1)(n+a+b) \, .
}
For $n\ge 1$ the apparent singularity for $\alpha \to 0$ may be removed
by using \TTid\ to give an appropriate non zero  $f_2$. 
The eigenfunctions for the solution in \solcr\  correspond to the $SU(2)$
$(n{+a}{+b}{-1})$-representation. Alternatively we may set $f_1=0$
and obtain the corresponding equation
\eqn\eigftwo{
{1\over 1-\alpha} \Big (  D^{(2a-1,2b-1)} - (a+b-1)(a+b) \Big )
\big ((1-\alpha) f_2 \big ) = - C f_2 \, .
}

\vfill
\eject
\appendix{C}{Calculation of Differential Operators}

A non trivial aspect in the derivation of the superconformal identities
is the determination of the differential operators \DDD\ which appear in
\crit. To sketch how these were obtained we first obtain, for any
dimension $d$ and arbitrary $f(\lam,\mun)$,
\eqnn\Ld
$$\eqalignno{
\half (k + & a + \half d - 2)\, L_{2[rs}\pr_{1u]} 
\big ( (t_1{\cdot t_2})^k 
(t_3{\cdot t_4})^l (t_1{\cdot t_4})^a (t_2{\cdot t_4})^b f \big ) \cr
&{}= - (t_1{\cdot t_2})^{k-2} (t_3{\cdot t_4})^{l-1} 
(t_1{\cdot t_4})^a (t_2{\cdot t_4})^b 
\big ( t_{1[r}t_{2s}^{\vphantom g}t_{3u]} \, t_2{\cdot t_4} \, \D_1 
+  t_{1[r}t_{4s}^{\vphantom g}t_{2u]} \, t_2{\cdot t_3}\, \D_2  \cr
&\qquad\qquad\qquad\qquad {}+ (k + a + \half d - 2) \, 
t_{2[r}t_{3s}^{\vphantom g}t_{4u]} \, 
t_1{\cdot t_2}\, ( \D_\lam - \D_\mun ) \big ) f \, , & \Ld \cr}
$$
where
\eqn\defDD{\eqalign{
\D_1 = {}& {\pr \over \pr \lam} \D_\mun +
\Big (\lam {\pr \over \pr \lam} + \mun {\pr \over \pr \mun} + 1 - k \Big )
\big (\D_\lam - \D_\mun \big ) \, , \cr
\D_2 = {}& - \Big ( {\pr \over \pr \mun}+ {a\over \mun} \Big ) \D_\lam
+ \Big (\lam {\pr \over \pr \lam} + \mun {\pr \over \pr \mun} +1 - k
\Big ) \big (\D_\lam - \D_\mun \big ) \, , \cr}
}
for
\eqn\DD{\eqalign{
\D_\lam = {}& \lam(1-\lam) {\pr^2 \over \pr \lam^2} - \mun^2
{\pr^2 \over \pr \mun^2} - 2\lam\mun\, {\pr^2 \over \pr \lam\pr\mun}\cr
&{}+ (b+1) {\pr \over \pr \lam} -  (a+b+\half d) 
\Big (\lam {\pr \over \pr \lam} + \mun {\pr \over \pr \mun} \Big )
+ k ( k+a+b+\half d -1 ) \, , \cr
\D_\mun = {}& \mun(1-\mun) {\pr^2 \over \pr \mun^2} - \lam^2
{\pr^2 \over \pr \lam^2} - 2\lam\mun \,{\pr^2 \over\pr\lam\pr\mun}\cr
&{}+ (a+1) {\pr \over \pr \mun} -  (a+b+\half d)
\Big (\lam{\pr \over \pr \lam} + \mun {\pr \over \pr \mun} \Big )
+ k ( k+a+b+\half d -1 ) \, . \cr}
}
In terms of \LDa\ and \LDd\ we have
\eqn\DDD{
\Delta^{(a,b)}_d \equiv \D_\lam + \D_\mun + ( \lam - \mun ) ( \D_\lam - \D_\mun ) 
= \D^{(a,b)}_d +  2k ( k+a+b+\half d -1 ) \, .
}
The operators in \defDD\ satisfy the identity
\eqn\simD{\eqalign{
& 2(\lam-\mun + 1 ) \D_1 + 2(\mun -\lam + 1 )\D_2 \cr
&\ = D_2 \, \Delta^{(a,b)}_d - \Lambda 
\bigg ( {\pr \over \pr \lam} + {\pr \over \pr \mun} + {a \over \mun} \bigg )
\big (\D_\lam - \D_\mun \big ) - (2k+2a-1)(\D_\lam - \D_\mun ) \, , \cr}
}
with $\Lambda$ as in \conlm\ and, as well as \DDD, defining
\eqn\Dtwo{
D_2 = ( \lam-\mun+1){\pr \over \pr \lam} + ( \lam-\mun - 1)\Big (
{\pr \over \pr \mun} + {a \over \mun} \Big ) \, .
}

When $d=3$, and $\Lambda =0$,  this result with \idthree\ leads to 
the simplified form for \Ld
\eqn\Dlth{\eqalign{
& 4(k+a-\half) \,{\pr \over \pr t_1}\cdot L_2  
\big ((t_1{\cdot t_2})^k (t_3{\cdot t_4})^l
(t_1{\cdot t_4})^a (t_2{\cdot t_4})^b {\hat f} \, \big ) \cr
&{}=  - t_2 \, {\cdot \, t_3 \times t_4} \,
(t_1{\cdot t_2})^{k-1} (t_3{\cdot t_4})^{l-1}(t_1{\cdot t_4})^a 
(t_2{\cdot t_4})^b  \,
\hD_2 \big ( D^{(2a,2b)} +2k(2k+2a+2b+1) \big ) {\hat f} \, , \cr}
}
letting $f(\lam,\mun) = {\hat f}(\alpha)$ and $D_2 \to \hD_2$ for
\eqn\Dtw{
\hD_2 = {\d \over \d \alpha} - {2a\over 1-\alpha} \, .
}
{}From \Lab\ the operator $D^{(2a,2b)} +2k(2k+2a+2b+1)$ acting on 
$ {\hat f}(\alpha)$ corresponds to  $L^2 + (2k+a+b)(2k+a+b+1)$. 
We may also note that
\eqn\Dinv{
\hD_2 D^{(2a,2b)} = {1\over 1-\alpha} \, D^{(2a-1,2b+1)} (1-\alpha) \hD_2 \, ,
}
and in \Dlth\ from \Lcross
\eqnn\LDD
$$\eqalignno{
& t_2 \, {\cdot \, t_3 \times t_4} \,(t_1{\cdot t_2})^{k-1} 
(t_3{\cdot t_4})^{l-1}(t_1{\cdot t_4})^a (t_2{\cdot t_4})^b \cr
\noalign{\vskip -4pt}
& \ {}\times {1\over 1- \alpha} \big (  D^{(2a-1,2b+1)} 
+2k(2k+2a+2b+1) \big ) \big ((1-\alpha) f \big ) & \LDD \cr
&{} = \big ( L^2 + (2k+a+b)(2k+a+b+1) \big )
t_2 \, {\cdot \, t_3 \times t_4} \,(t_1{\cdot t_2})^{k-1} 
(t_3{\cdot t_4})^{l-1}(t_1{\cdot t_4})^a (t_2{\cdot t_4})^b \, f \, . \cr}
$$

The equivalent results to \Dlth\ for $L_2 \to L_3$ and $L_2 \to L_4$
can be found by using the permutations $2\to 3 \to 4 \to 2$, along
with $a \to a'= k-l, \ b\to -a, \ k \to a+l$, ${l \to a+b+l}$ and
$\alpha \to \alpha' = -(1-\alpha)/\alpha$, and also $2\to 4 \to 3 \to 2$, 
along with in this case $a \to a''= -b, \, b\to l-k, \, k \to a+b+l, \, 
l \to b+k$ and
$\alpha \to \alpha'' = 1/(1-\alpha)$. From  \Dlth\ we then find
\eqn\Dltf{\eqalign{
\hD_3 = {}& \alpha^{2(a+l-1)}(1-\alpha)^{-2a}\hD'{}_{\!2} \,
\alpha^{-2(a+l)}(1-\alpha)^{2a} = {\d \over \d \alpha} + {2a\over 1-\alpha}
- {2(k+a) \over \alpha} \, , \cr
\hD_4 = {}& \alpha^{-2b}(1-\alpha)^{2(b+k-1)}\hD''{}_{\!\!2}  \,
\alpha^{2b}(1-\alpha)^{-2(b+k)} = {\d \over \d \alpha} + {2k\over 1-\alpha}\, . 
\cr}
}
Together with \Dtw, \Dltf\ is equivalent to \DDD.

For the analysis of the $\N=4$ superconformal identities a particular solution of
the constraints \consa\ is obtained by expressing $T_i$ in terms of scalar 
functions $Y_i(u,v;t)$ 
\eqn\TY{
T_i = - \, \bga{\cdot {\pr \over \pr t_1}} \, Y_i 
\ \gamma {\cdot {\overleftarrow{\pr \over \pr t_i}}}  \, . 
}
\TV\ and \VUp\ then give
\eqn\solY{\eqalign{
U_i ={}& \big ( L_{1,rs} L_{i,rs} + p_1 p_i \big ) Y_i \, , \qquad
W_{i,rsu} = 3 \big ( \pr_{1,[r}  L_{i,su]} Y_i \big )_{\rm sd} \, , \cr
\hV_{i,r} = {}& \pr_{1s} L_{i,rs} Y_i - {1\over p_1}\, \pr_{1r}\big (
\half L_{1,su} L_{i,su} Y_i \big ) \, . \cr}
}
Writing
\eqn\exYp{
Y_i(u,v;t) = \big ( t_1 {\cdot t}_4 \big )^{p_1-E} \big ( t_2 {\cdot t}_4
\big )^{p_2-E} \big ( t_1 {\cdot t}_2 \big )^{E}
\big ( t_3 {\cdot t}_4 \big )^{p_3} \, \Y_i(u,v;\lam,\mun) \, ,
}
then for $i=2$, using \Ld\ with $k=E, \, k+a=p_1, \, k+b=p_2$, we find
\eqn\Utwo{
\U_2 = \Delta^{(p_1-E,p_2-E)}_6 \Y_2 \, , \qquad 
\W_2 = 6 (\D_\lam - \D_\mun ) \Y_2 \, .
}
$\A_2$ and $\B_2$ are then given by \Ld\ and \defDD\ in terms of $\U_2, \W_2$ in 
accord with \relABC. The other results may be obtained by cyclic permutations. For
$2\to 3 \to 4 \to 2$, when $\lam \to \mun/\lam, \, \mun \to 1/\lam$ and $E\to
E - p_4-p_2$, then $\U_2 \to \mun^{p_1-E} \lam^{p_2-p_4-E} \U_3$, 
$\W_2 \to \mun^{p_1-E} \lam^{p_2-p_4-E+1} \W_3$,
$\A_2 \to \mun^{p_1-E} \lam^{p_2-p_4-E+2} \C_3$ and
$\B_2 \to \mun^{p_1-E} \lam^{p_2-p_4-E+2} \A_3$. For $2\to 4 \to 3 \to 2$, so that
$\lam \to 1/\mun, \, \mun \to \lam/\mun$ and $E\to - E +p_1+p_2$, in this case 
$\U_2 \to \mun^{-p_2}\, \lam^{p_2 -E} \U_4$, $\W_2 \to \mun^{1-p_2}\, \lam^{p_2 -E}\W_4$,
$\A_2 \to \mun^{2-p_2}\, \lam^{p_2 -E} \B_4$ and
$\B_2 \to \mun^{2-p_2}\, \lam^{p_2 -E} \C_4$.

However the representation \TY\ is not valid in general since it excludes
contributions involving the $\vep$-tensor. Nevertheless equivalent results
may be obtained by use of \WUp. With the expansion
\eqn\UVW{\eqalign{
& t_{2[r} \pr_{1s]} U_2 + p_1 \, t_{2[r} \hV_{2,s]} + p_1 \, W_{2,rsu} t_{2u} \cr
&{} = \big ( t_1 {\cdot t}_4 \big )^{p_1-E-1}
\big ( t_2 {\cdot t}_4 \big )^{p_2-E} \big ( t_1 {\cdot t}_2 \big )^{E-1}
\big ( t_3 {\cdot t}_4 \big )^{p_3-1} \cr
& \quad {} \times \bigg ( t_{2[r} t_{3s]} \, t_1 {\cdot t}_4 \, t_2 {\cdot t}_4 
\Big ( \pr_\lam \U_2 + p_1 \, \J_2 - {\ts {1\over 6}} p_1 ( \A_2 + \W_2 ) \Big ) \cr
& \qquad\quad {}+  t_{2[r} t_{4s]} \, t_1 {\cdot t}_4 \, t_2 {\cdot t}_3 
\Big ( \pr'{\!}_\mun \U_2 - p_1 {1\over \mun} ( \I_2 + \lam  \J_2) 
+ {\ts {1\over 6}} p_1 ( \B_2 + \W_2 ) \Big ) \cr
& \quad\qquad {}+ t_{1[r} t_{2s]} \, t_2 {\cdot t}_4 \, t_3 {\cdot t}_4
\Big ( {\tau \over p_1+1} \big ( ( \pr_\lam + \pr'{\!}_\mun ) ( E - \lam \pr_\lam
- \mun \pr_\mun ) + \pr_\lam \pr'{\!}_\mun \big ) \U_2 \cr
\noalign{\vskip-8pt}
&\hskip 7.4cm {}  - p_1 \, \V_2 
- {\ts {1\over 6}} p_1 \tau  ( \A_2 - \B_2 ) \Big )  \bigg ) \, ,  \cr }
}
where $\pr'{\!}_\mun = \pr_\mun + (p_1-E)/\mun$, then \WUp\ requires
\eqn\UUU{\eqalign{
6 \, \pr_\lam \U_2 = {}& - 6 p_1 \, \J_2 + p_1 ( \A_2 + \W_2 ) = 2 (p_1+1)\A_2
- (\O_\lam - p_1 ) \W_2 \, , \cr
6 \, \pr'{\!}_\mun \U_2 = {}& 6 p_1 {1\over \mun} ( \I_2 + \lam  \J_2) -
p_1 ( \B_2 + \W_2 ) = - 2 (p_1+1)\B_2 + (\O_\lam - p_1 ) \W_2 \, , \cr}
}
using \solIJ\ for $i=2$, which gives the first two equations in \relABC.
The remaining results in \relABC\ can be obtained by using permutations. 
In addition with \VABC\ we also obtain
\eqn\req{\eqalign{
(p_1+1) \big ( ( \O_\lam -p_1 +1 )& \B_2 - ( \O_\mun -p_1 +1 ) \A_2 \big ) \cr
= {}& {6} \big ( ( E -1 - \lam \pr_\lam - \mun \pr_\mun )
( \pr_\lam + \pr'{\!}_\mun ) + \pr_\lam \pr'{\!}_\mun \big ) \U_2 \cr
= {}& - {3} \big ( ( \O_\lam -p_1 +1  ) \pr'{\!}_\mun +
 ( \O_\mun -p_1 +1  ) \pr_\lam \big ) \U_2 \, . \cr} 
}
It is then straightforward to see that \req\ follows from \UUU\ using
$[\O_\lam, \O_\mun] = \O_\lam - \O_\mun$.

\vfill
\eject
\appendix{D}{Non Unitary Semi-short Representations}

In chapter 3 the analysis of the operator product expansion in general potentially
required contributions below the unitarity threshold on the scale dimension
$\Delta$. We show here how such truncations of the full representation
space arise for the superconformal algebra $PSU(2,2|4)$, following the
approach in \short\foot{This reproduces an analysis in \phd.}

The essential results are found by considering the chiral subalgebra $SU(2|4)$
(although no hermeticity conditions are imposed)
which has generators $Q^i{}_{\! \alpha}, \,S_i{}^{\!\alpha}$, $\alpha=1,2, \,
i = 1,\dots 4$, where
\eqn\algQS{
\big \{ Q^i{}_{\! \alpha} ,  S_j{}^{\!\beta} \big \} =   4 \big (
\de^i{}_{\! j} ( M_\alpha{}^{\! \beta} + \half  \,
\de_\alpha{}^{\! \beta} \hD ) - \de_\alpha{}^{\! \beta} R^i{}_{\! j} \big )   \, ,
}
as well as $\{ Q^i{}_{\! \alpha} , Q^j{}_{\! \beta} \} =
\{ S_i{}^{\!\alpha} , S_j{}^{\!\beta} \} = 0$. 
In \algQS\ $M_\alpha{}^{\! \beta}$ are generators of $SU(2)$ and
$R^i{}_{\! j}$ are generators of $SU(4)$,
$ \sum_i R^i{}_{\! i}=0$, with standard commutation relations.
$\hD$ is the dilation operator, with eigenvalues the scale dimension.
The commutators with $Q^i{}_{\! \alpha}$ and $S_i{}^{\!\alpha}$ are then
\eqn\MQS{\eqalign{
[M_\alpha{}^{\! \beta} , Q^i{}_{\! \gamma} ] = {}& \de_\gamma{}^{\! \beta}
Q^i{}_{\! \alpha} - \half \de_\alpha{}^{\! \beta} Q^i{}_{\! \gamma} \, , 
\qquad [M_\alpha{}^{\! \beta} ,  S_i{}^{\!\gamma} ] = - \de_\alpha{}^{\! \gamma}
S_i{}^{\!\beta} + \half  \de_\alpha{}^{\! \beta}  S_i{}^{\!\gamma} \, , \cr
[ R^i{}_{\! j} ,  Q^k{}_{\! \alpha} ] = {}&\de^k{}_{\! j}  Q^i{}_{\! \alpha}
- \quar  \de^i{}_{\! j}  Q^k{}_{\! \alpha} \, , \qquad \ \
[ R^i{}_{\! j} ,  S_k{}^{\!\alpha} ] =  - \de^i{}_{\! k} S_j{}^{\!\alpha}
+ \quar  \de^i{}_{\! j}   S_k{}^{\!\alpha} \, , \cr
[ \hD , Q^i{}_{\! \alpha}] ={}& \half  Q^i{}_{\! \alpha} \, , \hskip 3.05cm
[ \hD , S_i{}^{\!\alpha} ]  = - \half S_i{}^{\!\alpha} \, . \cr}
}
In terms of the usual $J_3, J_\pm$
\eqn\MJ{
\big [M_\alpha{}^{\! \beta}\big ] = \pmatrix{ J_3 & J_+ \cr J_- & -J_3 } \, ,
}
and it clear then that $( Q^i{}_{\! 1}, Q^i{}_{\! 2})$ and
$(S_i{}^{\!2},-S_i{}^{\! 1})$ form $j=\half$ doublets.
In terms of a standard Chevalley basis 
$E_r{}^{\!\pm}, \, H_r$, $r=1,2,3$, where $H_r{\!}^\dagger
=H_r, \, E_r{}^{\!+}{}^\dagger = E_r{}^{\!-}$ with commutators $[H_r,H_s]=0, \, 
[ E_r{}^{\! +}, E_s{}^{\! -} ] = \de_{rs} H_s, \, 
[H_r ,  E_s{}^{\! \pm}] = \pm K_{sr}  E_s{}^{\! \pm}$, for $[K_{rs}]$ the
$SU(4)$ Cartan matrix, 
\eqn\Car{
[K_{rs}] = \pmatrix{2& -1 & 0\cr -1& 2 &-1\cr 0&-1&2} \, ,
}
then we may take $R^1{}_{\! 2} = E_1{}^{\! +}, \,
R^2{}_{\! 3} = E_2{}^{\! +}, \, R^3{}_{\! 4} = E_3{}^{\! +}$ and 
$R^i{}_{\! i} = \quar (3H_1 + 2H_2 + H_3) - \sum_{r=1}^{i-1} H_r$.

For $SU(4)\otimes SU(2)$ highest weight states
$|p_1,p_2,p_3;j\rangle^{\rm hw} \equiv |p;j\rangle^{\rm hw}$ we have
\eqn\hw{\eqalign{
H_r |p;j\rangle^{\rm hw} = {}& p_r |p;j\rangle^{\rm hw} \, , \ \
J_3 |p;j\rangle^{\rm hw} = j |p;j\rangle^{\rm hw} \, , \ \
J_+ |p;j\rangle^{\rm hw} = E_r{}^{\!+} |p;j\rangle^{\rm hw} =0 \, , \cr}
}
from which states defining a representation with Dynkin labels $[p_1,p_2,p_3]j$
are constructed by the action of $E_r{}^{\! -},J_-$.
The representations of $SU(2|4)$ may then be formed from a highest weight state
which is also superconformal primary,
\eqn\DSw{
\hD |p;j\rangle^{\rm hw} = \Delta |p;j\rangle^{\rm hw} \, , \qquad
S_i{}^{\!\alpha} |p;j\rangle^{\rm hw} = 0 \, . 
}
The states of a generic supermultiplet, labelled
by $a^\Delta_{[p_1,p_2,p_3]j}$, are obtained by the action of the
supercharges, giving 
$\prod_{i,\alpha} \big ( Q^i{}_{\! \alpha} \big )^{n_{i\alpha}}
|p;j\rangle^{\rm hw}$ with $n_{i\alpha}=0,1$, together with the lowering
operators $E_r{}^{\!-}$. The possible $SU(4)\otimes SU(2)$ representations
$[p_1{\!}',p_2{\!}',p_3{\!}']j'$, with scale dimension $\Delta'$, forming  
the supermultiplet $a^\Delta_{[p_1,p_2,p_3]j}$ are obtained by adding the
$SU(4),SU(2)$ weights with $n_{i\alpha} =0,1$ so that
\eqn\weight{
p_r{\!}'=p_r+\sum_\alpha (n_{r\alpha}-n_{r{+1}\,\alpha}) \, , \quad
j{\,}'=j+\half \sum_i ( n_{i\,1} - n_{i\, 2} ) \, , \quad \Delta' = \Delta + \half
\sum_{i,\alpha} n_{i\alpha} \, ,
}
It is easy to see that ${\rm dim}\, a^\Delta_{[p_1,p_2,p_3]j} = 2^8
d(p_1,p_2,p_3)(2j+1)$, where $d(p_1,p_2,p_3)$ is the dimension of the $SU(4)$
representation with Dynkin labels $[p_1,p_2,p_3]$. If in \weight\ any
$p_r{\!}'$ or $j'$ are negative the Racah-Speiser algorithm, described in
\short, provides a precise prescription for removing such
$[p_1{\!}',p_2{\!}',p_3{\!}']j'$.

Shortening conditions arise for suitable $\Delta$ when descendant representations
satisfy the conditions \DSw\ to be superconformal primary. Since all 
$S_i{}^{\!\alpha}$ are obtained by commutators of $S_1{}^{\! 1}$ with $E_r{}^{\!+}$
and $J_+$ it is sufficient to impose only that $S_1{}^{\! 1}$ annihilates the
highest weight state of the representation. In such cases we may impose that
the appropriate combinations of $Q^i{}_{\! \alpha}$  
annihilate $|p;j\rangle^{\rm hw}$.
For application here it is convenient to define, acting on states $|\psi \rangle$ 
such that $J_3 |\psi \rangle = j |\psi \rangle$,
\eqn\Qti{
\tQ^i = Q^i{}_{\! 2} - {1\over 2j} \, Q^i{}_{\! 1} J_-   \, .
}
If $J_+ |\psi \rangle=0$ then $J_+ \tQ^i |\psi \rangle=0$ and
$J_3 \tQ^i |\psi \rangle = (j-\half) \tQ^i |\psi \rangle$. From \algQS\ we have
\eqn\sQ{
\half j \big \{ S_1{}^{\!1} , \tQ^ 1\big \} = 
\big ( 2j - J_3 - \half \hD + \quar (3H_1+2H_2+H_3) \big ) J_- \, , \quad
\half j \big \{ S_1{}^{\!1} , \tQ^2 \big \} = E_{1}{}^{\! -} J_- \, .
}
It is straightforward to show that $\tQ^1 |p;j\rangle^{\rm hw} \sim
|p_1+1,p_2,p_3;j-\half \rangle $. The shortening conditions considered in \short 
and previously are obtained by imposing
\eqn\shQ{
\tQ^i |p;j\rangle^{\rm hw} = 0 \ \cases  {\ i=1 \, , \cr \ i=1,2 &
if $\ p_1 =0 \, ,$\cr \ i=1,2,3 & if $\ p_1 = p_2 = 0 
\, , $\cr \ i=1,2,3,4 & if $\  p_1 = p_2 = p_3 =0 \, . $\cr}
}
In each case there is a consistency condition on $\Delta$ which can be found
by using \sQ\ and \DSw,
\eqn\sDel{
\Delta = 2 + 2j + \half (3p_1+2p_2+p_3) \, .
}
The corresponding supermultiplet is here denoted by $c_{[p_1,p_2,p_3]j}$. Detailed
results were given in \short, the $SU(4)\times SU(2)$ representations
present may be calculated as in \weight\ with the restriction $n_{i2} =0$
for those $i$ listed in \shQ\ for each case.

There are also additional shortening conditions of the form
\eqn\snew{
\Big ( \tQ^2 - {1\over p_1} \, \tQ^1 E_{1}{}^{\! -} \Big ) |p;j\rangle^{\rm hw} 
= 0 \, , \ p_1>0 \, , \quad
\tQ^1 \tQ^2 |0,p_2,p_3;j\rangle^{\rm hw} = 0 \, ,
}
where the left hand sides correspond to highest weight states 
$|p_1-1,p_2+1,p_3;j-\half \rangle$ and $|0,p_2+1,p_3;j- 1 \rangle$ respectively. 
Using \sQ\ these conditions then require
\eqn\sDelid{
\Delta = 2j + \half ( - p_1+2p_2+p_3) \, .
}
For $p_2=0$ the condition \snew\ extends also to
$\big ( \tQ^3 - {1\over p_1} \, \tQ^1 [E_{2}{}^{\! -}, E_{1}{}^{\! -}] \big ) 
|p;j\rangle^{\rm hw} = 0$.
The supermultiplet in each  case is denoted by $d_{[p_1,p_2,p_3]j}$.
The representations are obtained as in \weight\ with $n_{22}=0$, or if
$p_2=0$ then $n_{22}=n_{32}=0$.
For $p_1=0$ it is sufficient  to exclude $n_{12}=n_{22}=1$.

These semi-short representations lead to decompositions of the generic
multiplet,
\eqn\dem{\eqalign{
a^{2+2j+{1\over 2}(3p_1+2p_2+p_3)}_{[p_1,p_2,p_3]j}
\simeq {}& c_{[p_1,p_2,p_3]j} \oplus c_{[p_1+1,p_2,p_3]j{-{1\over 2}}} \, ,\cr
a^{2j+{1\over 2}(-p_1+2p_2+p_3)}_{[p_1,p_2,p_3]j}
\simeq {}& d_{[p_1,p_2,p_3]j} \oplus d_{[p_1-1,p_2+1,p_3]j{-{1\over 2}}} \, ,\cr
a^{2j+{1\over 2}(2p_2+p_3)}_{[0,p_2,p_3]j}
\simeq {}& d_{[0,p_2,p_3]j} \oplus c_{[0,p_2+1,p_3]j{-1}} \, .\cr}
}
Formally, as discussed in \short, we have
\eqn\bc{
c_{[p_1,p_2,p_3]-{1\over 2}} \simeq b_{[p_1+1,p_2,p_3]} \, , \qquad
c_{[p_1,p_2,p_3]-1} \simeq - b_{[p_1,p_2,p_3]} \, ,
}
where $b_{[p_1,p_2,p_3]}$ is the short supermultiplet formed by imposing
$Q^1{}_{\! \alpha}|p;0\rangle^{\rm hw}=0$ where we require $\Delta =
\half (3p_1+2p_2+p_3)$. Just as in \bc\ $d_{[p_1,p_2,p_3]-{1\over 2}}$
may be identified with a multiplet obtained from the highest weight
state $|p_1-1,p_2+1,p_3;0\rangle^{\rm hw}$, with $\Delta = \half (-p_1 + 2p_2 +p_3
+1)$, annihilated by 
$Q^2{}_{\! \alpha} - {1\over p_1}Q^1{}_{\! \alpha}E_1{}^{\! -}$.
Formally we have
\eqn\dcel{
d_{[p_1,p_2,p_3]j} \simeq c_{[-p_1-2,p_1+p_2+p_3+2,-p_3-2]j} \, ,
}
where, in accord with the Racah-Speiser algorithm described in \short, we may 
identify $SU(4)$ representations $[p_1,p_2,p_3] \simeq
[-p_1-2,p_1+p_2+p_3+2,-p_3-2]$ which are related by an even element of the Weyl 
group. This allows the detailed representation content and dimension for 
$d_{[p_1,p_2,p_3]j}$ to be determined from the results given in \short.

The generators of the superconformal group $PSU(2,2|4)$ are obtained by
extending those of $SU(2|4)$ to include the hermitian conjugates,
$\bQ_{i\dal} = Q^i{}_{\! \alpha}{}^\dagger, \, \bS{}^{i\dal} =
S_i{}^{\!\alpha}{}^\dagger , \, {\bar M}{}^\dbe{}_{\!\smash{\dal}} =
(M_\alpha{}^{\! \beta}){}^\dagger$, with an algebra obtained by conjugation
of that for $SU(2|4)$, assuming $\hD^\dagger = - \hD$ and
$(R^i{}_{\! j})^\dagger = R^j{}_{\! i}$. In addition
$\{ S_i{}^{\!\alpha} ,  \bQ_{j\dal} \} = \{ Q^i{}_{\! \alpha} , \bS{}^{j\dal}\}
= 0$ and $\{ Q^i{}_{\! \alpha} ,  \bQ_{j\dal} \} =  2\de^i{}_{\! j} 
(\si^a)_{\alpha\dal}P_a, \, 
\{ \bS{}^{i\dal} , S_j{}^{\!\alpha} \} = 2 \de^i{}_{\! j}
(\bsi^a)^{\dal\alpha} K_a$ where $P_a$ is the momentum operator and $K_a$ the
generator of special conformal transformations. The supermultiplets
are generated from highest weight superconformal primary states
$|p_1,p_2,p_3;j,\bj\rangle^{\rm hw}$, where $\bj$ is the $SU(2)$ 
quantum number for $\bJ_\pm, \bJ_3$ obtained from 
${\bar M}{}^\dbe{}_{\!\smash{\dal}}$, which is annihilated by $S_i{}^{\!\alpha},
\bS{}^{i\dal}$ and $K_a$. The representations are of course infinite 
dimensional, since they are generated by arbitrary powers of $P_a$, but
they are formed by a finite set of conformal primary representations,
annihilated by $K_a$, which are straightforwardly constructed from $SU(2|4)$
supermultiplet representations, as described above, combined with 
with their conjugates formed by the action of $\bQ_{i\dal}$. For these
$j\to \bj$ and $p_1 \leftrightarrow p_3$. The generic supermultiplet is
denoted $\A^\Delta_{[p_1,p_2,p_3](j,\bj)}$ where $[p_1,p_2,p_3](j,\bj)$
are the labels for the representation with lowest scale dimension $\Delta$.
The conformal primary states  form representations labelled by
$[p_1{\!}',p_2{\!}',p_3{\!}'](j{\,}',\bj{\,}')$, with scale dimension $\Delta'$,
which are given by
\eqn\weightL{\eqalign{
p_r{\!}'= {}& p_r +\sum_\alpha (n_{r\alpha}-n_{r{+1}\,\alpha}) +
\sum_{\dal} ( {\bar n}_{r{+1}\, \dal} - {\bar n}_{r \dal} ) \, , 
\cr
j{\,}'={}& j+\half \sum_i ( n_{i\, 1} - n_{i\, 2} ) \, , \qquad 
\bj{\,}'= \bj+\half \sum_i ( {\bar n}_{i\, 2} - {\bar n}_{i\, 1} ) \, , \cr 
\Delta' = {}& \Delta + \half
\sum_{i,\alpha} n_{i\alpha} + \half \sum_{i,\dal} {\bar n}_{i\dal}
\, , \qquad n_{i\alpha}, {\bar n}_{i\dal} = 0,1 \, . \cr}
}
The total dimension is $2^{16}d(p_1,p_2,p_3)(2j+1)(2\bj+1)$.

We here consider the case when shortening conditions are imposed for
both the $Q$ and $\bQ$ charges. Requiring \shQ\ with \sDel\ together with
its conjugate we have the semi-short multiplets,
\eqn\semiC{
\C_{[p_1,p_2,p_3](j,\bj)}\, , \qquad p_1-p_3 = 2(\bj-j)\, ,
\qquad \ \Delta = 2 + j+ \bj + p_1 + p_2 + p_3 \, ,
}
where we impose in \weightL\ $n_{12} = {\bar n}_{41}=0$, 
with further restrictions if $p_1$ or $p_3$ are zero. Requiring
\snew\ and \sDelid\ in both cases gives
\eqn\semiD{
\D_{[p_1,p_2,p_3](j,\bj)}\, , \qquad p_1-p_3 = 2(j-\bj)\, ,
\qquad \ \Delta = j+ \bj + p_2  \, ,
}
and we require in \weightL\ $n_{22}={\bar n}_{31}=0$. If $p_2=1$ then we
exclude $n_{32}={\bar n}_{21}=1$ while if $p_2=0$ then we require 
$n_{32}={\bar n}_{21}=0$ as well.
Corresponding to \dcel\ we have
\eqn\DCrel{
\D_{[p_1,p_2,p_3](j,\bj)} \simeq 
\C_{[-p_1-2,p_1+p_2+p_3+2,-p_3-2](j,\bj)} \, .
}
This result has essentially been used in \relDC.
We may also impose  \shQ\ with \sDel\ for the $Q$ charges and \snew\ and \sDelid\
for $\bQ$ giving
\eqn\semiE{
\E_{[p_1,p_2,p_3](j,\bj)}\, , \qquad p_1+p_3 = 2(\bj-j-1)\, ,
\qquad \ \Delta = 2+ j+ \bj + p_1 + p_2  \, ,
}
and we may also obtain a conjugate ${\bar \E}_{[p_1,p_2,p_3](j,\bj)}$.
Only for \semiC, where $\Delta$ is at the unitarity threshold, is there
a unitary representation. 

For relevance in chapter 3 we list the self-conjugate representations, when
$p_1=p_2, \, j=\bj$, arising in 
$\D_{[q,0,q](j,j)}$, obtained  by the action of equal powers of the $Q$
and $\bQ$ supercharges for each $\Delta$ 

\vskip 12pt
\hskip -1.7cm
\vbox{\tabskip=0pt \offinterlineskip
\hrule
\halign{&\vrule# &\strut \ \hfil#\  \cr
height2pt&\omit&&\omit&\cr
& $\ell$\hfil && $\ell+1$\hfil  &\cr
height2pt&\omit&&\omit&\cr
\noalign{\hrule}
height4pt&\omit&&\omit&\cr
& $~[q,0,q]_\ell$ &
&  $\matrix{\scs [q-1,0,q-1]_{\ell+1},[q-1,2,q-1]_{\ell+1}, 
3[q,0,q]_{\ell+1},[q+1,0,q+1]_{\ell+1}\cr
\scs [q-1,0,q-1]_{\ell-1},2[q,0,q]_{\ell-1},
[q+1,0,q+1]_{\ell-1}}$  &\cr
height4pt&\omit&&\omit&\cr}
\hrule}

\vskip -5pt
\hskip -1.7cm
\vbox{\tabskip=0pt \offinterlineskip
\hrule
\halign{&\vrule# &\strut \ \hfil#\  \cr
height2pt&\omit&\cr
& $\ell+2$\hfil &\cr
height2pt&\omit&\cr
\noalign{\hrule}
height4pt&\omit&\cr
&  $\matrix{\scs [q-2,2,q-2]_{\ell+2},
[q-1,0,q-1]_{\ell+2},2[q-1,2,q-1]_{\ell+2},4[q,0,q]_{\ell+2},[q,2,q]_{\ell+2},
[q+1,0,q+1]_{\ell+2}\cr
\scs [q-2,0,q-2]_\ell,[q-2,2,q-2]_\ell, 5[q-1,0,q-1]_\ell,2[q-1,2,q-1]_\ell,
8[q,0,q]_\ell,
[q,2,q]_\ell, 5[q+1,0,q+1]_\ell,[q+2,0,q+2]_\ell\cr
\scs [q,0,q]_{\ell-2}} $ &\cr
height4pt&\omit&\cr}
\hrule}

\vskip -5pt
\hskip -1.7cm
\vbox{\tabskip=0pt \offinterlineskip
\hrule
\halign{&\vrule# &\strut \ \hfil#\  \cr
height2pt&\omit&\cr
&$\ell+3$\hfil &\cr
height2pt&\omit&\cr
\noalign{\hrule}
height4pt&\omit&\cr
&$\matrix{\scs [q-1,0,q-1]_{\ell+3},[q-1,2,q-1]_{\ell+3},3[q,0,q]_{\ell+3},
[q+1,0,q+1]_{\ell+3}\cr
\scs [q-3,2,q-3]_{\ell+1},[q-2,0,q-2]_{\ell+1},4[q-2,2,q-2]_{\ell+1},
6[q-1,0,q-1]_{\ell+1},6[q-1,2,q-1]_{\ell+1}\cr
\scs 10[q,0,q]_{\ell+1},4[q,2,q]_{\ell+1},
6[q+1,0,q+1]_{\ell+1},[q+1,2,q+1]_{\ell+1},[q+2,0,q+2]_{\ell+1}\cr
\scs [q-1,0,q-1]_{\ell-1},[q-1,2,q-1]_{\ell-1},3[q,0,q]_{\ell-1}, 
[q+1,0,q+1]_{\ell-1}\cr}$&
\cr height4pt&\omit&\cr}
\hrule}

\vskip -5pt
\hfuzz = 10pt{
\hskip -1.7cm
\vbox{\tabskip=0pt \offinterlineskip
\hrule
\halign{&\vrule# &\strut \ \hfil#\  \cr
height2pt&\omit&\cr
& $\ell+4$ \hfil &\cr
height2pt&\omit&\cr
\noalign{\hrule}
height4pt&\omit&\cr
& $\matrix{\scs [q,0,q]_{\ell+4}\cr
\scs \!\! [q-2,0,q-2]_{\ell+2},[q-2,2,q-2]_{\ell+2},5[q-1,0,q-1]_{\ell+2},
2[q-1,2,q-1]_{\ell+2},8[q,0,q]_{\ell+2},[q,2,q]_{\ell+2},
5[q+1,0,q+1]_{\ell+2},[q+2,0,q+2]_{\ell+2}\!\!\! \cr
\scs [q-2,2,q-2]_{\ell},[q-1,0,q-1]_{\ell},2[q-1,2,q-1]_{\ell},
4[q,0,q]_{\ell},[q,2,q]_{\ell},[q+1,0,q+1]_{\ell}}$ & \cr
height4pt&\omit&\cr}
\hrule}
}

\vskip -18pt
\hskip -1.7cm
\vbox{\tabskip=0pt \offinterlineskip
\hrule
\halign{&\vrule# &\strut \ \hfil#\  \cr
height2pt&\omit&&\omit&\cr
& $\ell+5$\hfil &&$\ell+6$\hfil &\cr
height2pt&\omit&&\omit&\cr
\noalign{\hrule}
height4pt&\omit&&\omit&\cr
& $\matrix{\scs [q-1,0,q-1]_{\ell+3},2[q,0,q]_{\ell+3},[q+1,0,q+1]_{\ell+3}\cr
\scs[q-1,0,q-1]_{\ell+1},[q-1,2,q-1]_{\ell+1},3[q,0,q]_{\ell+1},
[q+1,0,q+1]_{\ell+1}}$&
& $[q,0,q]_{\ell+2}$ & \cr  
height4pt&\omit&&\omit&\cr}
\hrule}

\vskip -3pt
\noindent
Table 3. Diagonal representations for each $\Delta$ in
$\D_{[q,0,q]({1\over 2}\ell,{1\over 2}\ell)}$.

For application in the text we have the decompositions of self conjugate 
multiplets
\eqn\deCDE{\eqalign{
\A^{2+2j+p+2q}_{[q,p,q](j,j)}
\simeq {}& \C_{[q,p,q](j,j)} \oplus \C_{[q+1,p,q](j-{1\over 2},j)} \oplus
\C_{[q,p,q+1](j,j-{1\over 2})} \cr
&{} \oplus \C_{[q+1,p,q+1](j-{1\over 2},j-{1\over 2})}\, , \cr
\A^{2j+p}_{[q,p,q](j,j)}
\simeq {}& \D_{[q,p,q](j,j)} \oplus \D_{[q-1,p+1,q](j-{1\over 2},j)} \oplus
\D_{[q,p+1,q-1](j,j-{1\over 2})} \cr
&{} \oplus \D_{[q-1,p+2,q-1](j-{1\over 2},j-{1\over 2})}\, , \cr
\A^{2j+p}_{[0,p,0](j,j)}
\simeq {}& \D_{[0,p,0](j,j)} \oplus \E_{[0,p+1,0](j-1,j)} \oplus
{\bar \E}_{[0,p+1,0](j,j-1)} \cr
&{} \oplus \C_{[0,p+2,0](j-1,j-1)}\, . \cr }
}
The first case represents the decomposition of a long multiplet into semi-short
multiplets at the unitarity threshold, the second plays a crucial role in chapter 3 
in relating the solution of the superconformal Ward identities to the operator
product expansion.

\vfill
\eject
\appendix{E}{Alternative Derivation of Ward identities for $\N=2$}

The fields we consider have $N$ symmetric fundamental $SU(2)$ indices $\varphi^{(N)}_{i_1\ldots i_N}(x)$.
In the case of even $N=2n$ these can be mapped to $n$ symmetric traceless adjoint $SU(2)$ indices $\varphi^{(2n)}_{r_1\ldots r_n}(x)$.
As in chapter 2 we can contract the $SO(3)$ indices with null vectors $t$. Because of the mapping between $SO(3)$ and $SU(2)$ 
indices we can correspondingly map the $SO(3)$ vector $t$ to a $SU(2)$ spinor $u\tu$ as follows.
In three dimensions a null vector $t$ may be represented \Barg\ in terms of two component
spinors $u=(u^1,u^2)$ by $t_a = u \tau_a \tu$ for $\tu= \ep u^T$ for $\ep$ the
$2\times 2$ antisymmetric matrix. Then $t_1{\cdot t_2} = - 2 (u_2 \tu_1)^2$.
Since $u$ is in the fundamental representation of $SU(2)$ it squares to $0=u^i u^j\epsilon_{ij}$. 
This ensures that the $t_i$ constructed from the $u$'s are in fact null vectors.
Since there is no null condition on $u$ derivatives with respect to $u$ will be ordinary derivatives which is the main reason for the simplicity of this derivation.
In this case the variable $\alpha$ defined in chapter 2 becomes $$\alpha = {u_3 \tu_1 \, u_2 \tu_4 \over u_2 \tu_1 \, u_3 \tu_4}
= 1 - {u_2 \tu_3 \, u_1 \tu_4 \over u_2 \tu_1 \, u_3 \tu_4}.$$ 
Note that also $i\, t_1 \times t_2 \, {\cdot \, t_3} = 
4 \, u_1 \tu_2 \, u_2 \tu_3 \, u_3 \tu_1$. This will allow us to avoid the distinction between two cases as necessary in chapter 2.

So now we can define 
\eqn\ufields{\eqalign{
\varphi^{(N)}(x,u)={}&\varphi_{i_1\ldots i_N}u^{i_1}\ldots u^{i_N}\cr
\psi^{(N-1)}_{\alpha}(x,u)={}&\psi_{\alpha i_1\ldots i_{N-1}}u^{i_1}\ldots u^{i_{N-1}}\cr
\bpsi^{(N-1)}_{\dal}(x,u)={}&\bpsi_{\dal i_1\ldots i_{N-1}}u^{i_1}\ldots u^{i_{N-1}}\cr
J^{(N-2)}_{\alpha\dal}(x,u)={}&J_{\alpha\dal i_1\ldots i_{N-2}}u^{i_1}\ldots u^{i_{N-2}}
}}

In these new variables, the needed superconformal transformations are
\eqn\sct{\eqalign{
\delta\varphi^{(N)}&=\hep_i{}^\alpha u^i \psi^{(N-1)}_{\alpha}+\bpsi{}^{(N-1)}_{\dal} \tu_i \hbep{}^{i\dal},\cr
\delta\psi^{(N-1)}_\alpha&={i\over N}\pr_{\alpha\dal}{\partial\over\partial u^i}\varphi^{(N)}\,\hbep{}^{i\dal}+2{\partial\over\partial u^i}\varphi^{(N)}\,\eta_{\alpha}^i+J^{(N-2)}_{\alpha\dal}\tu_i\hbep{}^{i\dal}.
}}
One can check that these close acting on $\varphi$.
Now we derive the Ward identity by starting from
\eqn\wi{\eqalign{
0={}&\delta\big<\psi^{(N_1-1)}_\alpha(x_1)\varphi^{(N_2)}(x_2)\varphi^{(N_3)}(x_3)\varphi^{(N_4)}(x_4)\big>\cr
={}&{i\over N_1}{\pr\over\pr u_1^i}\pr_{1\alpha\dal}\big<\varphi^{(N_1)}(x_1)\varphi^{(N_2)}(x_2)\varphi^{(N_3)}(x_3)\varphi^{(N_4)}(x_4)\big>\hbep{}^{i\dal}(x_1)\cr
{}&+2{\pr\over\pr u_1^i}\big<\varphi^{(N_1)}(x_1)\varphi^{(N_2)}(x_2)\varphi^{(N_3)}(x_3)\varphi^{(N_4)}(x_4)\big>\eta^{i}_{\alpha}(x_1)\cr
{}&+\big<J_{\alpha\dal}^{(N_1-2)}(x_1)\varphi^{(N_2)}(x_2)\varphi^{(N_3)}(x_3)\varphi^{(N_4)}(x_4)\big>\tu_{1i}\hbep{}^{i\dal}(x_1)\cr
{}&+\big<\psi^{(N_1-1)}_{\alpha}(x_1)\bpsi^{(N_2-1)}_{\dal}(x_2)\varphi^{(N_3)}(x_3)\varphi^{(N_4)}(x_4)\big>\tu_{2i}\hbep{}^{i\dal}(x_2)\cr
{}&+\big<\psi^{(N_1-1)}_{\alpha}(x_1)\varphi^{(N_2)}(x_2)\bpsi^{(N_3-1)}_{\dal}(x_3)\varphi^{(N_4)}(x_4)\big>\tu_{3i}\hbep{}^{i\dal}(x_3)\cr
{}&+\big<\psi^{(N_1-1)}_{\alpha}(x_1)\varphi^{(N_2)}(x_2)\varphi^{(N_3)}(x_3)\bpsi^{(N_4-1)}_{\dal}(x_4)\big>\tu_{4i}\hbep{}^{i\dal}(x_4),}}
where we suppress the $u$ arguments of the fields.
Following the same steps as in chapter 2 equations \deep\ to \hatRJ\ we may decompose this equation into independent ones and separate into $x,{\overline x}$ resulting in
\eqn\ind{\eqalign{
&{1\over N_1}{\pr\over\pr u_1^i}\pr_xF+K\tu_{1i}+{1\over x}T_2\,\tu_{2i}-{1\over 1-x}T_4\,\tu_{4i}=0,\cr
&{\pr\over\pr u_1^i}F=T_2\,{\tu}_{2i}+T_3\,\tu_{3i}+T_4\,\tu_{4i},
}}
which is very similar of the result in chapter 2 in equation \redeq{}.
Now we contract the first equation with $u_1$ to eliminate the derivative in $u_1$. Since $F$ is homogeneous of degree $N_1$ in $u_1$ 
this simply leaves
\eqn\feq{\pr_xF=-{1\over x}T_2\, u_1\tu_2+{1\over 1-x}T_4\, u_1\tu_4.}
The second equation we contract with $u_3$ to project out the term involving $T_3$
\eqn\seq{u_3^i{\pr\over\pr u_1^i}F=T_2\, u_3\tu_2+T_4\, u_3\tu_4.}
We solve this for $T_2$ and insert it into the other equation
\eqn\fineq{\left(u_3^i{\pr\over\pr u_1^i}+{u_3\tu_2\over u_1\tu_2}\,x\pr_x\right)F=T_4\, u_3\tu_4 {1-\alpha x\over 1-x}.}
We now compute the action of the differential operator on the left hand side. Since we know that $F$ is homogeneous 
of degrees $N_1,N_2,N_3,N_4$ in $u_1,u_2,u_3,u_4$ we expand it in the form
\eqn\fexp{F(u,v;u_i)=(u_1\tu_4)^{N_1-E}(u_2\tu_4)^{N_2-E}(u_1\tu_2)^{E}(u_3\tu_4)^{N_3}\F(x,\xb;\alpha).}
where here $2E=N_1+N_2+N_3-N_4$.
The differential operator in \fineq\ acting on this gives
\eqn\dop{\eqalign{\left(u_3^i{\pr\over\pr u_1^i}+{u_3\tu_2\over u_1\tu_2}x\pr_x\right)F(u,v;u_i)
={}&{u_3\tu_2}\,(u_1\tu_4)^{N_1-E}(u_2\tu_4)^{N_2-E}(u_1\tu_2)^{E-1}(u_3\tu_4)^{N_3}\cr{}&\times\left(-\alpha\pr_{\alpha}+x\pr_x+{N_1-E\over 1-\alpha}+E\right)\F(x,\xb;\alpha).}}
For $\alpha={1\over x}$ the right hand side of \fineq\ vanishes.
Therefore the Ward identity we obtain is
\eqn\wi{\pr_x\left(x^{E}(x-1)^{N_1-E}\F(x,\xb;\textstyle{1\over x})\right)=0.}
You can easily check that this is identical to \resF\ we obtained before.

We may also
define $O(3)$ generators by $L_a = {1\over 2}u \tau_a {\pr \over \pr u}$.
Since
$$
(L_1+L_2)^2 (u_1 \tu_4)^a (u_2 \tu_4)^b f(\alpha) =
( - D_\alpha^{(a,b)} + {\ts {1\over 4}}(a+b)(a+b+2) ) f(\alpha) \, ,
$$
the relevant $O(3)$ eigenfunctions for each representation may be directly 
obtained and shown to be identical to the ones obtained before.
\vfill
\eject

\appendix{F}{Mathematica Computation of Conformal Partial Wave Expansions}
In this appendix we will describe how the Mathematica program attached can automate most 
of the steps in the computation of conformal partial wave expansions.
The source code is divided into blocks preceded by ``$In[x]:=$", where we will refer to $x$ as the number of the block of code.

First, to motivate the function of the program, we make the simple observation that the contributions of operators of twist $t$ always have a $u$-leading 
term of the form $c u^{t\over 2}=c (x \xb)^{t\over 2}$. Looking at the expression for $G^{(\ell)}_{\Delta}(x,\xb)$ we see that the contribution of an operator with twist $t$ and spin $\ell$ 
will have leading terms in $x,\xb$ of order ${t\over 2}+\ell+1,{t\over 2}$ respectively.
Therefore by expanding up to a finite order in $x,\xb$, we can simultaneously eliminate all twists and scale dimensions beyond a certain limit. 
We will actually use it to compute the coefficients first twist by twist to then later fit to a general ansatz. 
In our example $p=4$ the overall factor to the amplitude is $u^4$. Thus by only making a power expansion in $\xb$ up to power $4$, we eliminate 
all twist contributions for $t>8$. It is also apparent that no operators with twist $t<8$ can contribute. 
Therefore the leading expansion will automatically reduce to $t=8$ contributions only. Then we can plug the coefficients back into the expansion
which we perform then up to power $5$ in $\xb$ to obtain the coefficients for twist $10$. We keep iterating this until we have enough coefficients to read
off the general twist dependence.

In the following we assume general familiarity with Mathematica. We need the basic commands for simplification of expressions, table management, 
Taylor series and obtaining a list of the coefficients. Also an understanding of the use of rules is required.

Block 1 switches off certain warnings which are not important.

Block 2 sets $p=4$ which is the case for which we will illustrate the use of the program. It can also be used to confirm the 
$p=2,3$ results or to perform the expansion for higher $p$'s. In principle also other functions might be expanded.

Block 3 defines hypergeometric functions of type ${}_2F_1$ as a power series.

Block 5 contains the definition of the basic $G^{(\ell)}_{\Delta}(x,\xb)$ function. Also we set up a function which already contains the universal factor of $u^{t\over 2}$. 
Notice that this function already has a cutoff for powers of $x,\xb$. This will save time since we ultimately only want to compute up to a certain power of $x,\xb$.

Block 8 contains the expressions for the different amplitudes in terms of $\oD$-functions. For different values of $p$ we have to insert the 
appropriate expressions here. At the end of block 8 they are all summed up in one array for convenience. 

Block 15 contains first the choice of amplitude to compute which is just the position of the amplitude in the array. 
Also a variable $even$ is set to $1$=true or $0$=false, signaling if even or odd coefficients are non-zero. This also has to be adjusted for different values of $p$.

Block 16 defines the appropriate expression in the array now as $H0$ which is the function we will try to expand in partial waves. It has parameters
$TT,DD,o$ which set limits for the maximum twist considered, maximum dimension considered and order to which to expand.

Block 17 sets up the expansion coefficients $c[a,\Delta=t+\ell,\ell]$ where $a$ is the index of the amplitude in the array, $\Delta$ is the dimension, $t$ is the twist and $\ell$ the spin. 
Immediately the constraint is implemented that only even or odd twist can contribute depending on which amplitude we deal with. 
This might be too restrictive for other expansions which might have less symmetries.

Block 18 already contains the results computed for the amplitude $A_{22}$ which has index $6$ for the twist range $8-24$. 
Initially this would be empty and is built up successively twist by twist.
Correspondingly Blocks 27,36,45,54,63 contain the already computed values for amplitudes $A_{21},A_{20},A_{11},A_{10},A_{00}$.

Block 72 contains a definition of the $\oD$-functions as a power series in $x,\xb$ \fm.

Block 73 defines the general ansatz we make
\eqn\ans{{\rm HH2}=\sum_{t,\ell}c_{a,\Delta,\ell}u^{t\over 2}G^{(\ell)}_{\Delta},}
where immediately a power expansion up to a fixed order is performed to truncate the series and also limits for $t,\ell$ 
are chosen such that no terms with a too high power of $x,\xb$ are considered. The maximum twist we want to include is specified by the parameter $TT$, 
the maximum dimension by $DD$ and the order to which to expand by $o$.

Block 74 equates the function to expand with the general ansatz and performs a Taylor expansion of the resulting equation. 
The expansion order for $\xb$ is chosen such that only terms at the first twist level not yet computed survive. The expansion order 
in $x$ is chosen high enough to accomodate all the partial waves in the specified dimensional range.

Block 75 computes the resulting equations for the coefficients for a certain twist set by the variable $twist$, here 26, 
which is the first level not yet computed. It produces a set of rules which solve this system of equations.

Block 77 uses the solution of the last step to compute the partial wave coefficients and stores them in a table.

Block 79 strips off universal factors of $\ell$-dependence such that the remainder is polynomial in $\ell$ (schematically)
\eqn\lans{c_{2\tau,\ell}={2^{\ell}(\ell+\tau-1)!(\ell+\tau)!(\tau!)^2\over (2\ell+2\tau-1)!(2\tau)!}c'_{2\tau,\ell}.}
This is an ansatz which needs change if for example other functions are supposed to be expanded. The table contains the coefficients 
for fixed twist $t$, but for a range of values of the spin $\ell$

Block 82 takes these values and tries to fit them against a polynomial of order $4$. This is a crucial point of the procedure. Certainly there exists some sort of expression 
of the table values in terms of $\ell$, but one has to ensure that through the removal of the universal factor only a polynomial dependence is left over.
If the guess in the factorization is roughly correct, but for example one of the factorials has a wrong argument, one can increase the order of the fitting polynomial.
The problem is that eventually this will make the equations used to fit underdetermined and thus requires computation of 
more coefficients to start from. Therefore for the factorization the best known ansatz should be used.

Block 85 uses the polynomial fitted to the numerical values and puts back the universal factor. 
It will generate output which is the final expression of the partial wave expansion coefficients for this particular twist but for general $\ell$. 
Using copy\&paste this now needs to be fed back into one of the Blocks above with the previously computed values and one can return to Block 75 to 
compute for the next higher twist. This is the way the Blocks 18-63 were obtained.

To summarize, so far the program computed the general $\ell$-dependence of the expansion coefficients.
Now the program has a second stage where we will now try to determine the general dependence on twist. The coefficients we determined so far are 
general expressions for $\ell$ but only valid for the specific values of $t$ they were computed for. The final result we want is a general expression
valid for all values of $t,\ell$.

Block 89 contains the definition of an ansatz to take out a universal factor containing factorial and exponential (basically any non-polynomial) 
dependence on $t$ similar to the case of $\ell$ (schematically)
\eqn\tans{c_{t,\ell}={2^{\ell-1}(\ell+\tau+1)!(\ell+2\tau)!(\ell+1)^{1-evenl}(\ell+2\tau+2)^{1-evenl}\over (2\ell+2\tau+1)!}c'_{\tau,\ell}.}
We use the variable $\tau={t\over 2}$ since we already know that only even twists can appear. The variable $evenl$ takes care of the fact that for odd $\ell$ 
certain additional factors appear.
Here the same applies as for block 79/82. This ansatz needs to be close enough to the expected result for the program to work.

Block 90 defines coefficients $d$ which are the expansion coefficients $c$ with the ansatz just made factored off.

Block 91 defines a general ansatz for the $d$ coefficients which is a $4$-th order polynomial in $\ell$ and $8$-th order in $\tau$.
The orders have to be high enough to capture the complexity of the expansion coefficients, but also not too large, since otherwise with a limited number
of coefficients computed in the first part, a system of equations to fit the polynomial to the numerical values would be underdetermined.

Block 92 tries to simplify the $d$-coefficients to speed up the computation.

Block 93 fits the $d$ coefficients twist by twist against the corresponding polynomial ansatzes we just made. The result is then 
returned by the function ploy.

Blocks 94-99 show the output, the function ploy will generate for the $p=4$ case, the individual amplitude expansion coefficients for $A_{00},A_{10},A_{11},A_{20},A_{21},A_{22}$. 
The polynomials obtained still look quite complicated but should describe the general $\tau,\ell$ dependence of the expansion coefficients. 
Further simplifications have to be done by hand.

\vfill\eject
\centerline{\hbox{\epsfxsize=5.5in \epsfbox{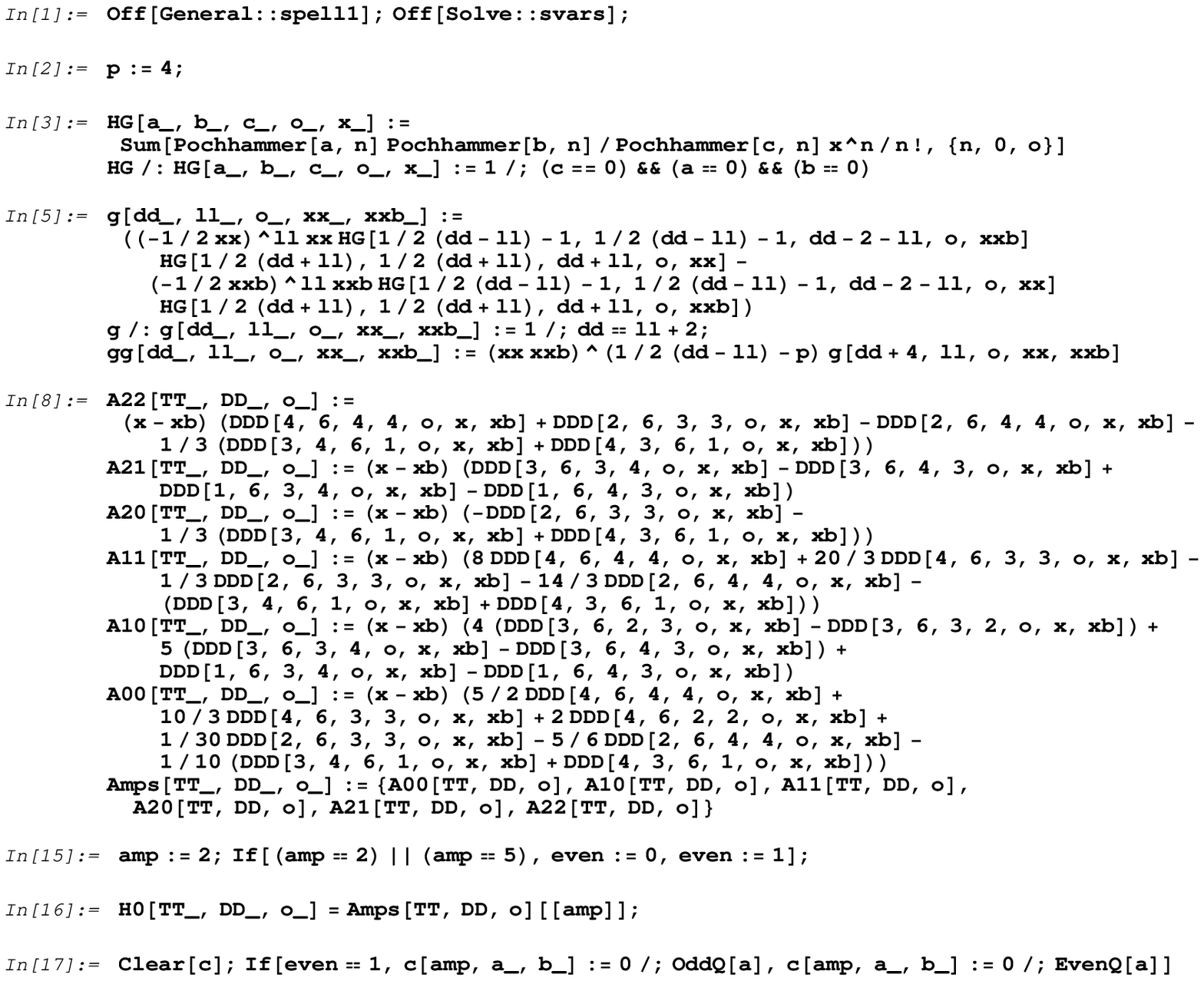}}}
\vfill\eject
\centerline{\hbox{\epsfxsize=5.6in \epsfbox{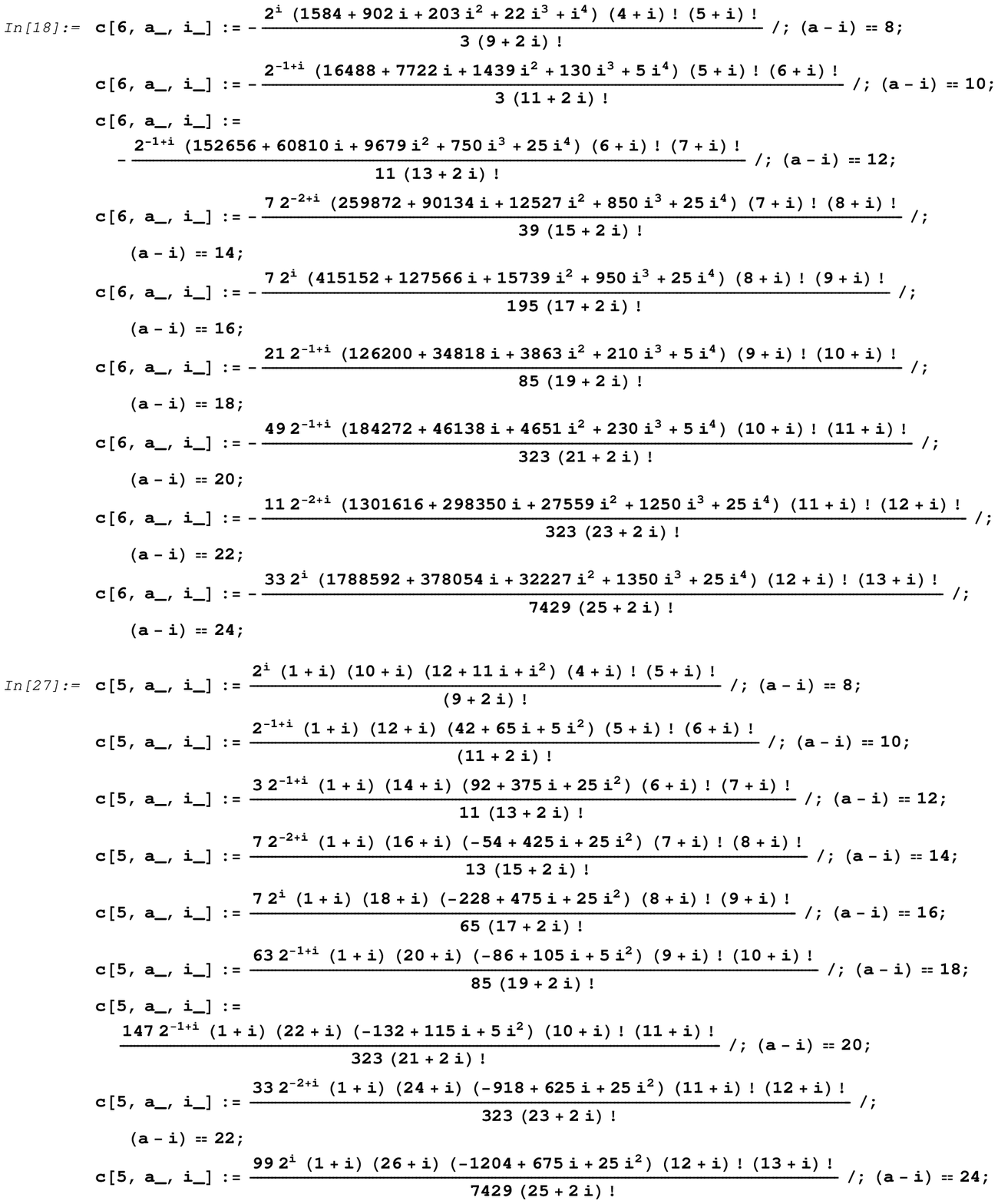}}}
\vfill\eject
\centerline{\hbox{\epsfxsize=5.7in \epsfbox{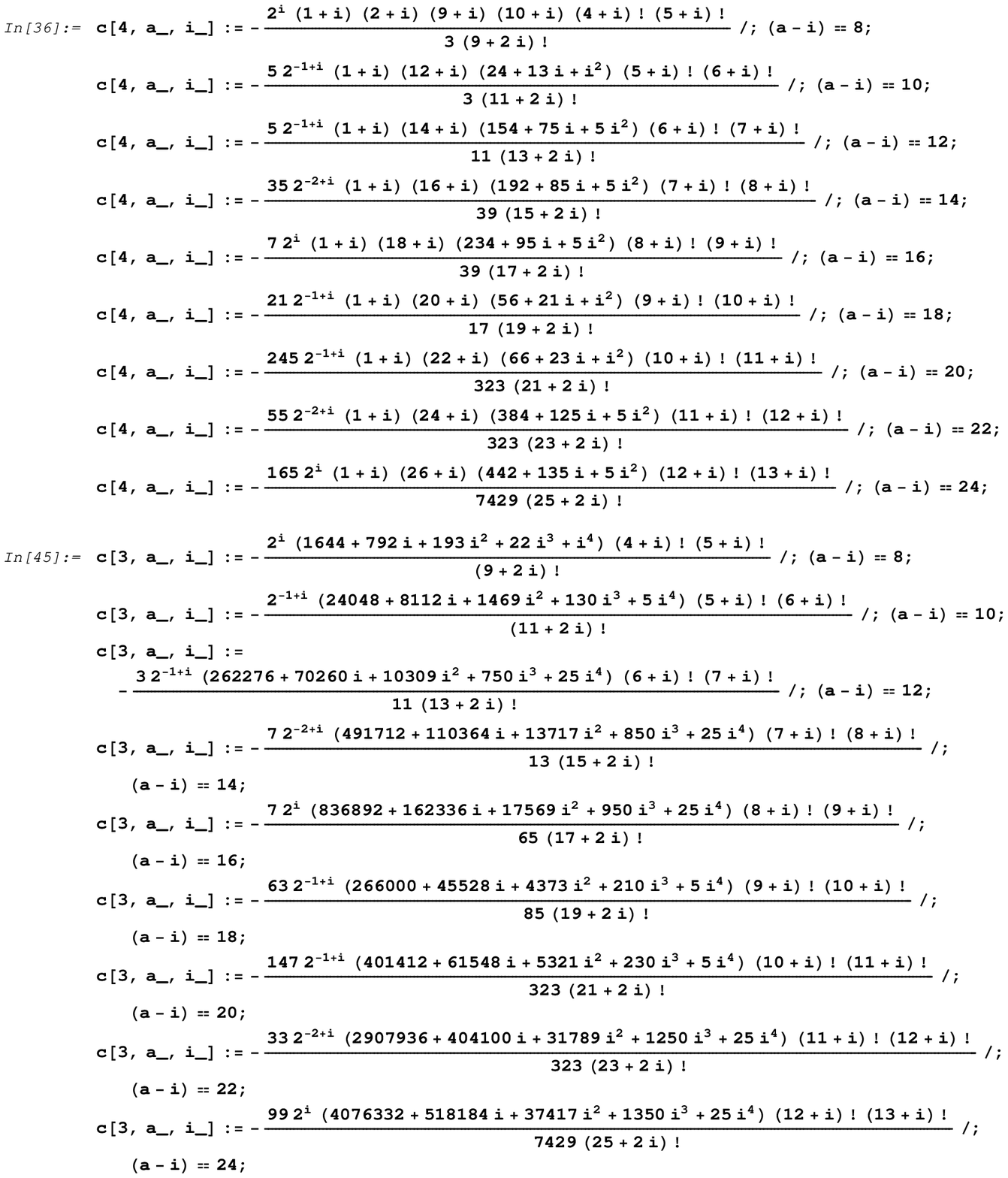}}}
\vfill\eject
\centerline{\hbox{\epsfxsize=5.8in \epsfbox{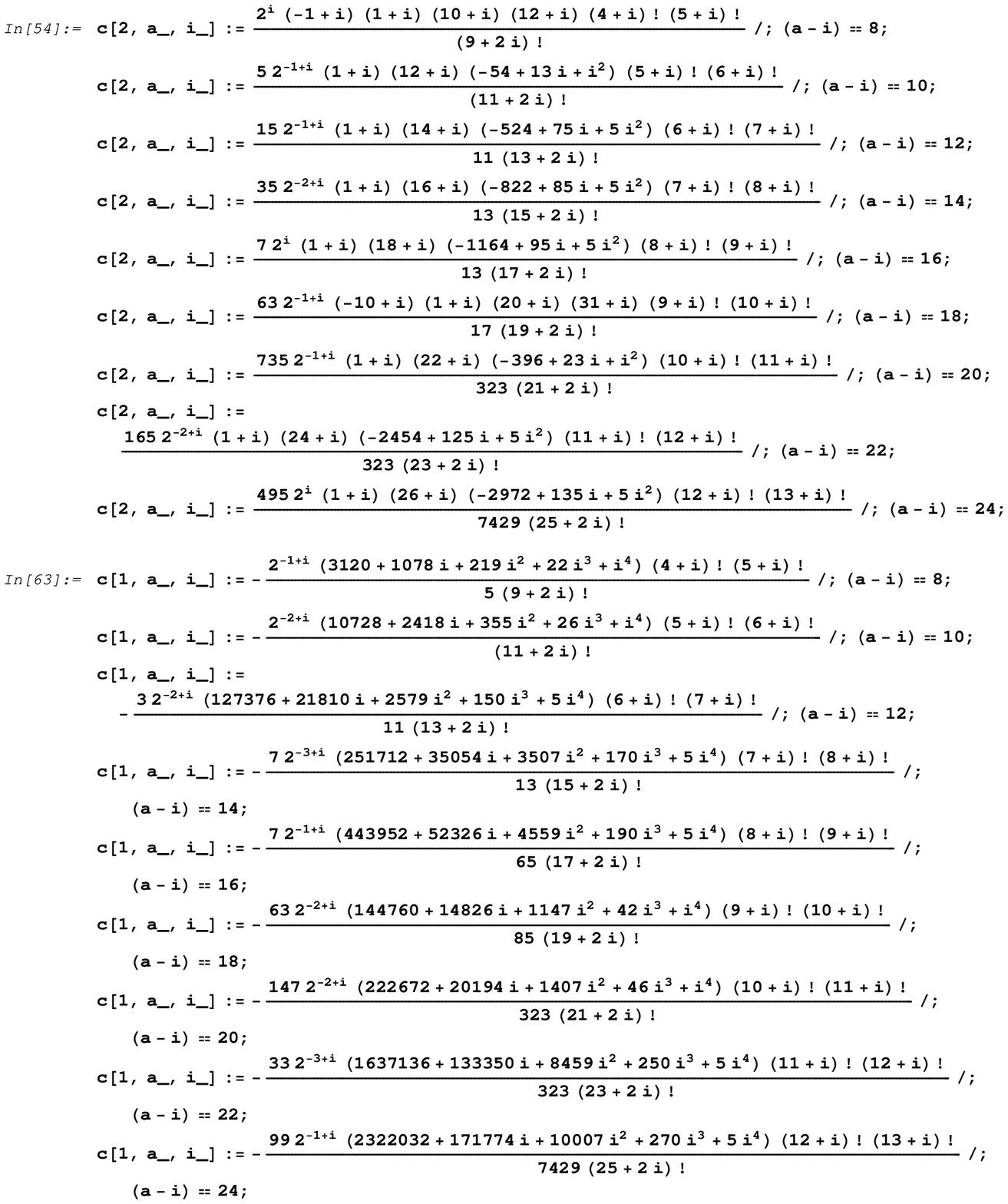}}}
\vfill\eject
\centerline{\hbox{\epsfxsize=5.9in \epsfbox{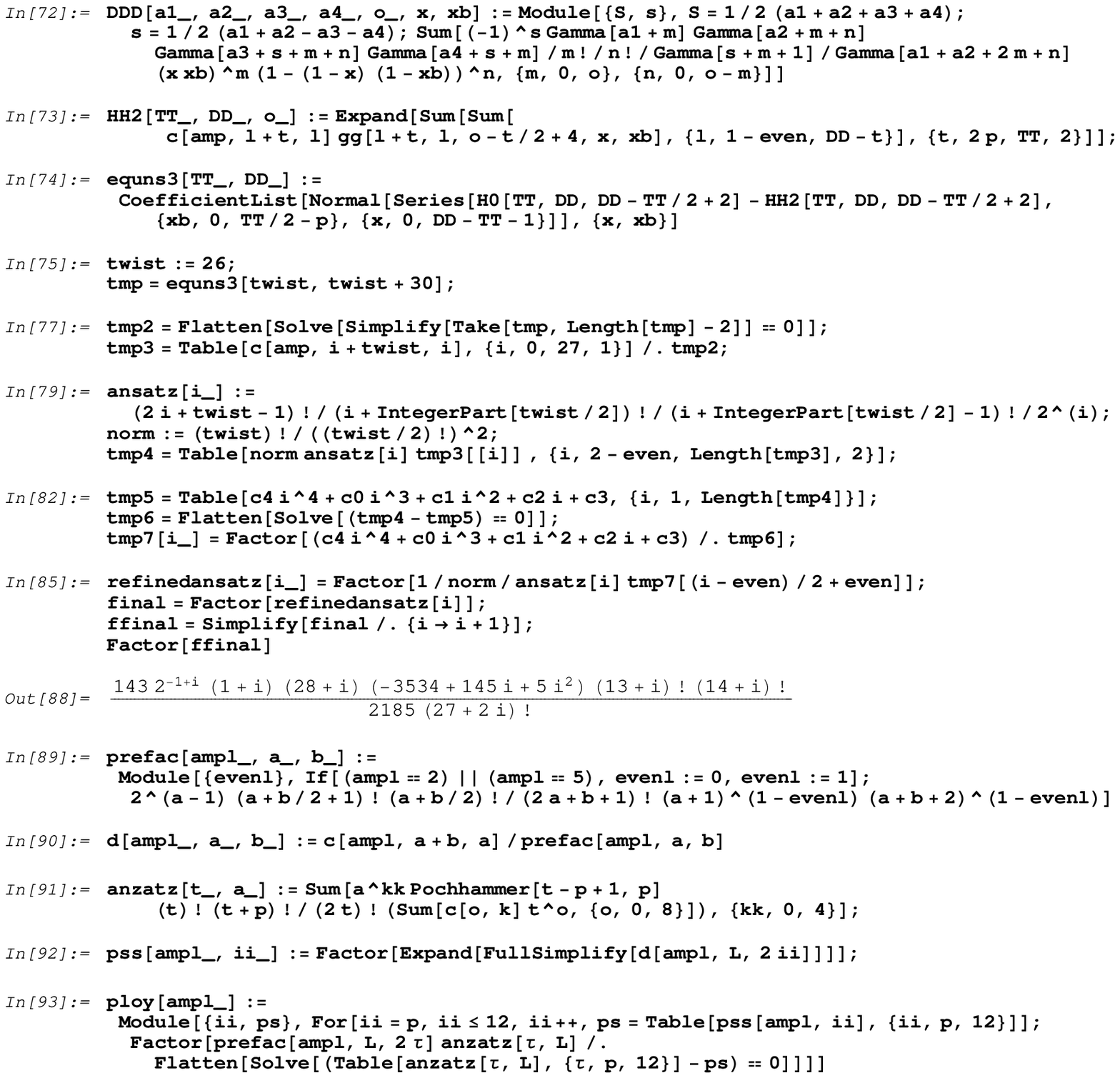}}}
\vfill\eject
\centerline{\hbox{\epsfxsize=6in \epsfbox{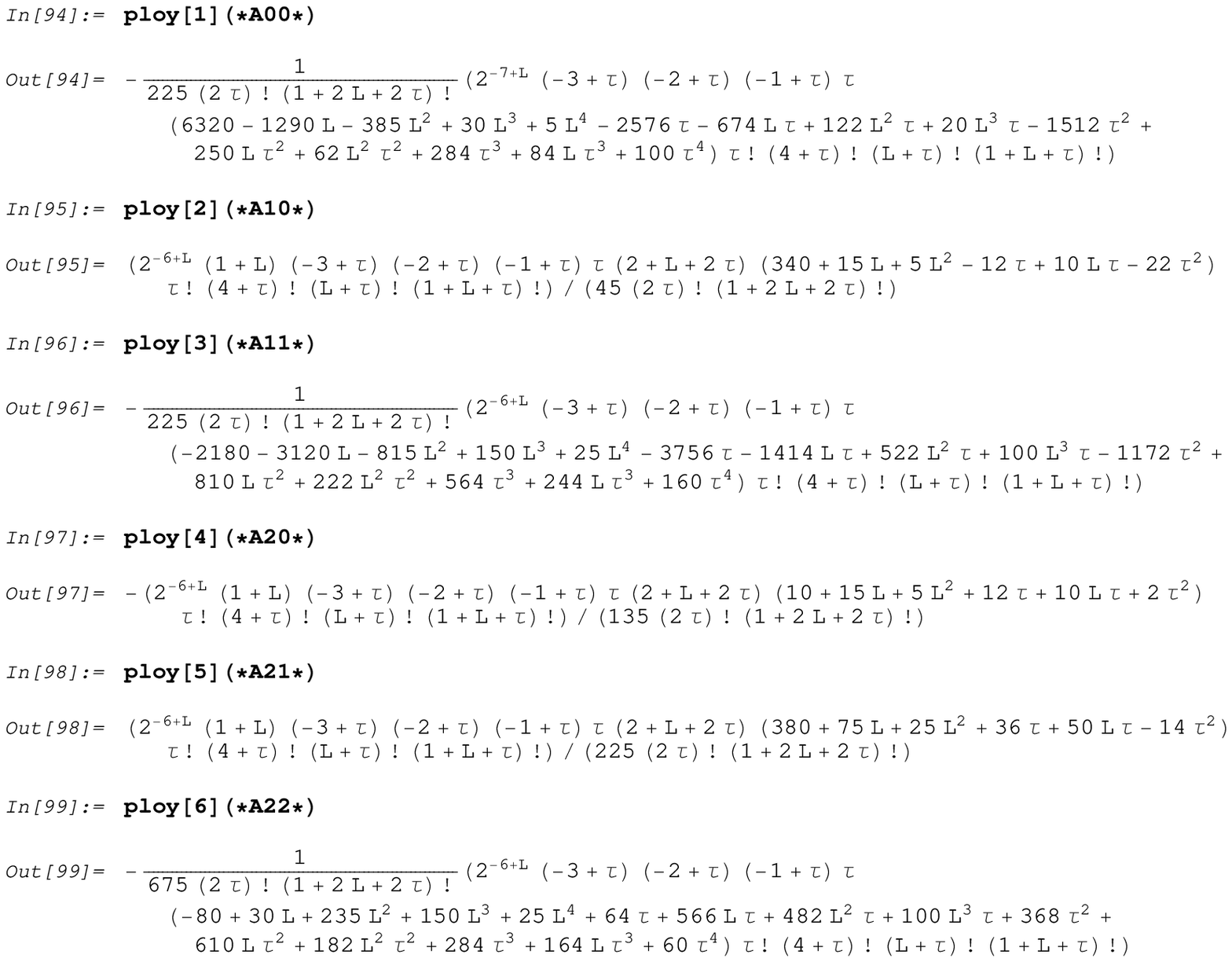}}}
\vfill\eject

\appendix{G}{$\oD$ Identities}
We give a list of standard relations\ADHS\ for $\oD$ functions here since they are used at various points in chapter 5.
First there are identities which relate $\oD$ functions with crossing transformed arguments (in the following we use $\Sigma=\half (\Delta_1+\Delta_2+\Delta_3+\Delta_4)$)
\eqn\didso{\eqalign{
\oD_{\Delta_1\Delta_2\Delta_3\Delta_4}(u,v)={}&v^{-\Delta_2}\oD_{\Delta_1\Delta_2\Delta_4\Delta_3}(u/v,1/v),\cr
={}&v^{\Delta_4-\Sigma}\oD_{\Delta_2\Delta_1\Delta_3\Delta_4}(u/v,1/v),\cr
={}&\oD_{\Delta_3\Delta_2\Delta_1\Delta_4}(v,u),\cr
={}&u^{-\Delta_2}\oD_{\Delta_4\Delta_2\Delta_3\Delta_1}(1/u,v/u).
}}
Then there is a set of relations between individual functions which leave the arguments invariant and permute the indices
\eqna\didst
$$\eqalignno{
\oD_{\Delta_1\Delta_2\Delta_3\Delta_4}(u,v)={}&v^{\Delta_1+\Delta_4-\Sigma}\oD_{\Delta_2\Delta_1\Delta_4\Delta_3}(u,v),&\didst a\cr
\oD_{\Delta_1\Delta_2\Delta_3\Delta_4}(u,v)={}&u^{\Delta_3+\Delta_4-\Sigma}\oD_{\Delta_4\Delta_3\Delta_2\Delta_1}(u,v).&\didst b
}$$
Also there exists a reflection property of the indices
\eqn\didref{\eqalign{
\oD_{\Delta_1\Delta_2\Delta_3\Delta_4}(u,v)=\oD_{\Sigma-\Delta_3\Sigma-\Delta_4\Sigma-\Delta_1\Sigma-\Delta_2}(u,v).
}}
Now we list identities relating $\oD$ functions with different values of $\Sigma$. First there are the two up, two down relations
\eqna\didtutd
$$\eqalignno{
(\Delta_2+\Delta_4-\Sigma)\oD_{\Delta_1\Delta_2\Delta_3\Delta_4}(u,v)={}&\oD_{\Delta_1\Delta_2+1\Delta_3\Delta_4+1}(u,v)-\oD_{\Delta_1+1\Delta_2\Delta_3+1\Delta_4}(u,v),\cr&&\didtutd a\cr
(\Delta_1+\Delta_4-\Sigma)\oD_{\Delta_1\Delta_2\Delta_3\Delta_4}(u,v)={}&\oD_{\Delta_1+1\Delta_2\Delta_3\Delta_4+1}(u,v)-v\oD_{\Delta_1\Delta_2+1\Delta_3+1\Delta_4}(u,v),\cr&&\didtutd b\cr
(\Delta_3+\Delta_4-\Sigma)\oD_{\Delta_1\Delta_2\Delta_3\Delta_4}(u,v)={}&\oD_{\Delta_1\Delta_2\Delta_3+1\Delta_4+1}(u,v)-u\oD_{\Delta_1+1\Delta_2+1\Delta_3\Delta_4}(u,v).\cr&&\didtutd c\cr
}$$
Finally there is a formula for the sum of 3 $\oD$ functions with the same value of $\Sigma$
\eqn\dids{\eqalign{
\Delta_4\oD_{\Delta_1\Delta_2\Delta_3\Delta_4}(u,v)={}&\oD_{\Delta_1+1\Delta_2\Delta_3\Delta_4+1}(u,v)+\oD_{\Delta_1\Delta_2+1\Delta_3\Delta_4+1}(u,v)\cr
&{}+\oD_{\Delta_1\Delta_2\Delta_3+1\Delta_4+1}(u,v).
}}

\vfill\eject
\footatend\vfill\supereject\immediate\closeout\rfile\writestoppt
\baselineskip=14pt\centerline{{\bf References}}\bigskip{\frenchspacing%
\parindent=20pt\escapechar=` \input refs.tmp\vfill\eject}\nonfrenchspacing

\bye

%% file: harvmac2.tex
%
%
%
\def\unredoffs{} 

%
%
%
%
\newbox\leftpage \newdimen\fullhsize \newdimen\hstitle \newdimen\hsbody
\tolerance=1000\hfuzz=2pt
\catcode`\@=11 

\magnification=1200\unredoffs\baselineskip=16pt plus 2pt minus 1pt
\hsbody=\hsize \hstitle=\hsize 
%
%
\newcount\yearltd\yearltd=\year\advance\yearltd by -1900

%
%

\def\draftmode{\message{ DRAFTMODE }\def\draftdate{{\rm preliminary draft:
\number\month/\number\day/\number\yearltd\ \ \hourmin}}%
\headline={\hfil\draftdate}\writelabels\baselineskip=20pt plus 2pt minus 2pt
 {\count255=\time\divide\count255 by 60 \xdef\hourmin{\number\count255}
  \multiply\count255 by-60\advance\count255 by\time
  \xdef\hourmin{\hourmin:\ifnum\count255<10 0\fi\the\count255}}}
\def\nolabels{\def\wrlabeL##1{}\def\eqlabeL##1{}\def\reflabeL##1{}}
\def\writelabels{\def\wrlabeL##1{\leavevmode\vadjust{\rlap{\smash%
{\line{{\escapechar=` \hfill\rlap{\sevenrm\hskip.03in\string##1}}}}}}}%
\def\eqlabeL##1{{\escapechar-1\rlap{\sevenrm\hskip.05in\string##1}}}%
\def\reflabeL##1{\noexpand\llap{\noexpand\sevenrm\string\string\string##1}}}
\nolabels
%
\global\newcount\secno \global\secno=0
\global\newcount\meqno \global\meqno=1
\def\newsec#1{\global\advance\secno by1\message{(\the\secno. #1)}
\global\subsecno=0\eqnres@t\noindent{\bf\the\secno. #1}
\writetoca{{\secsym} {#1}}\par\nobreak\medskip\nobreak}
\def\eqnres@t{\xdef\secsym{\the\secno.}\global\meqno=1\bigbreak\bigskip}
\def\sequentialequations{\def\eqnres@t{\bigbreak}}\xdef\secsym{}
\global\newcount\subsecno \global\subsecno=0
\def\subsec#1{\global\advance\subsecno by1\message{(\secsym\the\subsecno. #1)}
\ifnum\lastpenalty>9000\else\bigbreak\fi
\noindent{\it\secsym\the\subsecno. #1}\writetoca{\string\quad 
{\secsym\the\subsecno.} {#1}}\par\nobreak\medskip\nobreak}
\def\appendix#1#2{\global\meqno=1\global\subsecno=0\xdef\secsym{\hbox{#1.}}
\bigbreak\bigskip\noindent{\bf Appendix #1. #2}\message{(#1. #2)}
\writetoca{Appendix {#1.} {#2}}\par\nobreak\medskip\nobreak}
%
%
\def\eqnn#1{\xdef #1{(\secsym\the\meqno)}\writedef{#1\leftbracket#1}%
\global\advance\meqno by1\wrlabeL#1}
\def\eqna#1{\xdef #1##1{\hbox{$(\secsym\the\meqno##1)$}}
\writedef{#1\numbersign1\leftbracket#1{\numbersign1}}%
\global\advance\meqno by1\wrlabeL{#1$\{\}$}}
\def\eqn#1#2{\xdef #1{(\secsym\the\meqno)}\writedef{#1\leftbracket#1}%
\global\advance\meqno by1$$#2\eqno#1\eqlabeL#1$$}
%
\newskip\footskip\footskip14pt plus 1pt minus 1pt 
\def\footnotefont{\ninepoint}\def\f@t#1{\footnotefont #1\@foot}
\def\f@@t{\baselineskip\footskip\bgroup\footnotefont\aftergroup\@foot\let\next}
\setbox\strutbox=\hbox{\vrule height9.5pt depth4.5pt width0pt}
\global\newcount\ftno \global\ftno=0
\def\foot{\global\advance\ftno by1\footnote{$^{\the\ftno}$}}
%
\newwrite\ftfile   
\def\footend{\def\foot{\global\advance\ftno by1\chardef\wfile=\ftfile
$^{\the\ftno}$\ifnum\ftno=1\immediate\openout\ftfile=foots.tmp\fi%
\immediate\write\ftfile{\noexpand\smallskip%
\noexpand\item{f\the\ftno:\ }\pctsign}\findarg}%
\def\footatend{\vfill\eject\immediate\closeout\ftfile{\parindent=20pt
\centerline{\bf Footnotes}\nobreak\bigskip\input foots.tmp }}}
\def\footatend{}
%
%
\global\newcount\refno \global\refno=1
\newwrite\rfile
\def\ref{[\the\refno]\nref}
\def\nref#1{\xdef#1{[\the\refno]}\writedef{#1\leftbracket#1}%
\ifnum\refno=1\immediate\openout\rfile=refs.tmp\fi
\global\advance\refno by1\chardef\wfile=\rfile\immediate
\write\rfile{\noexpand\item{#1\ }\reflabeL{#1\hskip.31in}\pctsign}\findarg}
\def\findarg#1#{\begingroup\obeylines\newlinechar=`\^^M\pass@rg}
{\obeylines\gdef\pass@rg#1{\writ@line\relax #1^^M\hbox{}^^M}%
\gdef\writ@line#1^^M{\expandafter\toks0\expandafter{\striprel@x #1}%
\edef\next{\the\toks0}\ifx\next\em@rk\let\next=\endgroup\else\ifx\next\empty%
\else\immediate\write\wfile{\the\toks0}\fi\let\next=\writ@line\fi\next\relax}}
\def\striprel@x#1{} \def\em@rk{\hbox{}} 
\def\lref{\begingroup\obeylines\lr@f}
\def\lr@f#1#2{\gdef#1{\ref#1{#2}}\endgroup\unskip}
\def\semi{;\hfil\break}
\def\addref#1{\immediate\write\rfile{\noexpand\item{}#1}} 
\def\startrefs#1{\immediate\openout\rfile=refs.tmp\refno=#1}
\def\xref{\expandafter\xr@f}\def\xr@f[#1]{#1}
\def\refs#1{\count255=1[\r@fs #1{\hbox{}}]}
\def\r@fs#1{\ifx\und@fined#1\message{reflabel \string#1 is undefined.}%
\nref#1{need to supply reference \string#1.}\fi%
\vphantom{\hphantom{#1}}\edef\next{#1}\ifx\next\em@rk\def\next{}%
\else\ifx\next#1\ifodd\count255\relax\xref#1\count255=0\fi%
\else#1\count255=1\fi\let\next=\r@fs\fi\next}
%

%
\newwrite\ffile\global\newcount\figno \global\figno=1
\def\fig{fig.~\the\figno\nfig}
\def\nfig#1{\xdef#1{fig.~\the\figno}%
\writedef{#1\leftbracket fig.\noexpand~\the\figno}%
\ifnum\figno=1\immediate\openout\ffile=figs.tmp\fi\chardef\wfile=\ffile%
\immediate\write\ffile{\noexpand\medskip\noexpand\item{Fig.\ \the\figno. }
\reflabeL{#1\hskip.55in}\pctsign}\global\advance\figno by1\findarg}
\def\xfig{\expandafter\xf@g}\def\xf@g fig.\penalty\@M\ {}
\def\figs#1{figs.~\f@gs #1{\hbox{}}}
\def\f@gs#1{\edef\next{#1}\ifx\next\em@rk\def\next{}\else
\ifx\next#1\xfig #1\else#1\fi\let\next=\f@gs\fi\next}
\newwrite\lfile
{\escapechar-1\xdef\pctsign{\string\%}\xdef\leftbracket{\string\{}
\xdef\rightbracket{\string\}}\xdef\numbersign{\string\#}}
\def\writedefs{\immediate\openout\lfile=labeldefs.tmp \def\writedef##1{%
\immediate\write\lfile{\string\def\string##1\rightbracket}}}
\def\writestop{\def\writestoppt{\immediate\write\lfile{\string\pageno%
\the\pageno\string\startrefs\leftbracket\the\refno\rightbracket%
\string\def\string\secsym\leftbracket\secsym\rightbracket%
\string\secno\the\secno\string\meqno\the\meqno}\immediate\closeout\lfile}}
\def\writestoppt{}\def\writedef#1{}
\def\seclab#1{\xdef #1{\the\secno}\writedef{#1\leftbracket#1}\wrlabeL{#1=#1}}
\def\subseclab#1{\xdef #1{\secsym\the\subsecno}%
\writedef{#1\leftbracket#1}\wrlabeL{#1=#1}}
\newwrite\tfile \def\writetoca#1{}
\def\leaderfill{\leaders\hbox to 1em{\hss.\hss}\hfill}
\def\writetoc{\immediate\openout\tfile=toc.tmp 
   \def\writetoca##1{{\edef\next{\write\tfile{\noindent ##1 
   \string\leaderfill {\noexpand\number\pageno} \par}}\next}}}
\def\listtoc{\centerline{\bf Contents}\nobreak\medskip{\baselineskip=12pt
 \parskip=0pt\catcode`\@=11 \input toc.tex \catcode`\@=12 \bigbreak\bigskip}}
\catcode`\@=12 
%
\edef\tfontsize{\ifx\answ\bigans scaled\magstep3\else scaled\magstep4\fi}
 \tfontsize  \tfontsize
 \tfontsize \font\titlei=cmmi10 \tfontsize
\font\titleis=cmmi7 \tfontsize \font\titleiss=cmmi5 \tfontsize
\font\titlesy=cmsy10 \tfontsize \font\titlesys=cmsy7 \tfontsize
\font\titlesyss=cmsy5 \tfontsize  \tfontsize
\skewchar\titlei='177 \skewchar\titleis='177 \skewchar\titleiss='177
\skewchar\titlesy='60 \skewchar\titlesys='60 \skewchar\titlesyss='60
 \ifx\answ\bigans\else scaled\magstep1\fi
\ifx\answ\bigans\else

 \font\absi=cmmi10 scaled\magstep1
\font\absis=cmmi7 scaled\magstep1 \font\absiss=cmmi5 scaled\magstep1
\font\abssy=cmsy10 scaled\magstep1 \font\abssys=cmsy7 scaled\magstep1
\font\abssyss=cmsy5 scaled\magstep1 
\skewchar\absi='177 \skewchar\absis='177 \skewchar\absiss='177
\skewchar\abssy='60 \skewchar\abssys='60 \skewchar\abssyss='60
\fi
\font\ninerm=cmr9 \font\sixrm=cmr6 \font\ninei=cmmi9 \font\sixi=cmmi6 
\font\ninesy=cmsy9 \font\sixsy=cmsy6 \font\ninebf=cmbx9 
\font\nineit=cmti9 \font\ninesl=cmsl9 \skewchar\ninei='177
\skewchar\sixi='177 \skewchar\ninesy='60 \skewchar\sixsy='60 
\def\ninepoint{\def\rm{\fam0\ninerm}
\textfont0=\ninerm \scriptfont0=\sixrm \scriptscriptfont0=\fiverm
\textfont1=\ninei \scriptfont1=\sixi \scriptscriptfont1=\fivei
\textfont2=\ninesy \scriptfont2=\sixsy \scriptscriptfont2=\fivesy
\textfont\itfam=\ninei \def\it{\fam\itfam\nineit}\def\sl{\fam\slfam\ninesl}%
\textfont\bffam=\ninebf \def\bf{\fam\bffam\ninebf}\rm} 
%
%

\hyphenation{anom-aly anom-alies coun-ter-term coun-ter-terms}
\def\inv{^{\raise.15ex\hbox{${\scriptscriptstyle -}$}\kern-.05em 1}}

\def\Dsl{\,\raise.15ex\hbox{/}\mkern-13.5mu D} 
\def\dsl{\raise.15ex\hbox{/}\kern-.57em\partial}

\def\tr{{\rm tr}} 

\def\lspace{\ifx\answ\bigans{}\else\qquad\fi}
\def\lbspace{\ifx\answ\bigans{}\else\hskip-.2in\fi} 
\def\boxeqn#1{\vcenter{\vbox{\hrule\hbox{\vrule\kern3pt\vbox{\kern3pt
	\hbox{${\displaystyle #1}$}\kern3pt}\kern3pt\vrule}\hrule}}}
\def\mbox#1#2{\vcenter{\hrule \hbox{\vrule height#2in
		\kern#1in \vrule} \hrule}}  
%
 \def\CC{{\cal C}}   
   
\def\CB{{\cal B}}  \def\CD{{\cal D}}

\def\darr#1{\raise1.5ex\hbox{$\leftrightarrow$}\mkern-16.5mu #1}
\def\ha{{1\over2}}
\def\half{{\textstyle{1\over2}}} 
\def\roughly#1{\raise.3ex\hbox{$#1$\kern-.75em\lower1ex\hbox{$\sim$}}}

%% file: toc.tex
\noindent {1.} {Introduction} \leaderfill{3} \par 
\noindent \quad{1.1.} {Motivation} \leaderfill{3} \par 
\noindent \quad{1.2.} {Superconformal Symmetry in $d=4$} \leaderfill{6} \par 
\noindent \quad{1.3.} {Physical Fields in ${\fam \tw@ N}=4$ SYM} \leaderfill{7} \par 
\noindent \quad{1.4.} {Conformal Scalar $2,3,4$-point Functions in $d=4$} \leaderfill{7} \par 
\noindent \quad{1.5.} {Operator Product Expansion} \leaderfill{9} \par 
\noindent \quad{1.6.} {${\fam \tw@ N}=4$ Superconformal Multiplets} \leaderfill{11} \par 
\noindent \quad{1.7.} {${\fam \tw@ N}=2$ Superconformal Multiplets} \leaderfill{15} \par 
\noindent \quad{1.8.} {Large $N$ Amplitudes} \leaderfill{17} \par 
\noindent \quad{1.9.} {Known Results about Superconformal Ward Identities} \leaderfill{18} \par 
\noindent \quad{1.10.} {Known large $N$ results} \leaderfill{20} \par 
\noindent \quad{1.11.} {Outline} \leaderfill{21} \par 
\noindent {2.} {${\fam \tw@ N}=2$} \leaderfill{23} \par 
\noindent \quad{2.1.} {Superconformal Ward Identities} \leaderfill{23} \par 
\noindent \quad{2.2.} {Solution of Identities} \leaderfill{32} \par 
\noindent \quad{2.3.} {OPE Analysis} \leaderfill{33} \par 
\noindent \quad{2.4.} {Long Operators} \leaderfill{35} \par 
\noindent \quad{2.5.} {Semishort Operators} \leaderfill{35} \par 
\noindent \quad{2.6.} {Short Operators} \leaderfill{37} \par 
\noindent \quad{2.7.} {(Next-to-) Extremal Correlators} \leaderfill{38} \par 
\noindent \quad{2.8.} {Summary} \leaderfill{40} \par 
\noindent {3.} {${\fam \tw@ N}=4$} \leaderfill{41} \par 
\noindent \quad{3.1.} {Superconformal Ward Identities} \leaderfill{41} \par 
\noindent \quad{3.2.} {Solution of Identities} \leaderfill{48} \par 
\noindent \quad{3.3.} {OPE Analysis} \leaderfill{49} \par 
\noindent \quad{3.4.} {Long Operators} \leaderfill{52} \par 
\noindent \quad{3.5.} {Semi-Short and Non-Unitary Operators} \leaderfill{54} \par 
\noindent \quad{3.6.} {Short Operators} \leaderfill{58} \par 
\noindent \quad{3.7.} {Identity Operator} \leaderfill{60} \par 
\noindent \quad{3.8.} {(Next-to-) Extremal Case} \leaderfill{60} \par 
\noindent \quad{3.9.} {Summary} \leaderfill{62} \par 
\noindent {4.} {Crossing Symmetry} \leaderfill{64} \par 
\noindent \quad{4.1.} {${\fam \tw@ N}=4$} \leaderfill{64} \par 
\noindent \quad{4.2.} {${\fam \tw@ N}=2$} \leaderfill{70} \par 
\noindent \quad{4.3.} {(Next-to-)Extremal Case} \leaderfill{73} \par 
\noindent {5.} {Large $N$ Results} \leaderfill{75} \par 
\noindent \quad{5.1.} {Simplification of the Amplitude for $p=4$} \leaderfill{75} \par 
\noindent \quad{5.2.} {Computation of First Order Anomalous Dimensions} \leaderfill{77} \par 
\noindent \quad{5.3.} {Conjectures for General Chiral Four Point Functions} \leaderfill{82} \par 
\noindent \quad{5.4.} {Applications for Low $p$} \leaderfill{89} \par 
\noindent \quad{5.5.} {General Solutions} \leaderfill{93} \par 
\noindent {6.} {Conclusion and Future Investigations} \leaderfill{102} \par 
\noindent Appendix {A.} {Results for Null Vectors} \leaderfill{104} \par 
\noindent Appendix {B.} {Two Variable Harmonic Polynomials} \leaderfill{106} \par 
\noindent Appendix {C.} {Calculation of Differential Operators} \leaderfill{114} \par 
\noindent Appendix {D.} {Non Unitary Semi-short Representations} \leaderfill{118} \par 
\noindent Appendix {E.} {Alternative Derivation of Ward identities for ${\fam \tw@ N}=2$} \leaderfill{124} \par 
\noindent Appendix {F.} {Mathematica Computation of Conformal Partial Wave Expansions} \leaderfill{127} \par 
\noindent Appendix {G.} {${\overline D}$ Identities} \leaderfill{138} \par 

%% file: labeldefs.tex
\def\nh{[1]}
\def\adscft{[2]}
\def\DZ{[3]}
\def\dims{(1.1)}
\def\short{[4]}
\def\dims{(1.2)}
\def\defrij{(1.3)}
\def\defdij{(1.4)}
\def\stpf{(1.5)}
\def\sthpf{(1.6)}
\def\sfpf{(1.7)}
\def\defuv{(1.8)}
\def\fgrel{(1.9)}
\def\scft{[5]}
\def\Howe{[6]}
\def\eiguv{(1.10)}
\def\OPE{(1.11)}
\def\opp{(1.12)}
\def\oppp{(1.13)}
\def\sdl{(1.14)}
\def\deft{(1.15)}
\def\gegendef{(1.16)}
\def\Gform{(1.17)}
\def\gegendf{(1.18)}
\def\Gsimp{(1.19)}
\def\Dos{[7]}
\def\ope{(1.20)}
\def\GF{(1.21)}
\def\Gsym{(1.22)}
\def\Gexpansion{(1.23)}
\def\cpwope{(1.24)}
\def\Dob{[8]}
\def\sc{(1.25)}
\def\sto{(1.26)}
\def\coms{(1.27)}
\def\Nfour{(1.28)}
\def\irr{(1.29)}
\def\tphi{(1.30)}
\def\ADHS{[9]}
\def\hpert{[10]}
\def\nul{(1.31)}
\def\Barg{[11]}
\def\Dobb{[12]}
\def\Fourp{(1.32)}
\def\deflu{(1.33)}
\def\Fourpp{(1.34)}
\def\SigE{(1.35)}
\def\Ntwo{(1.36)}
\def\irrr{(1.37)}
\def\Arut{[13]}
\def\Degen{[14]}
\def\dss{(1.38)}
\def\od{[15]}
\def\Df{(1.39)}
\def\fm{(1.40)}
\def\Intr{[16]}
\def\nilpo{[17]}
\def\Ext{[18]}
\def\Sei{[19]}
\def\HFS{[20]}
\def\OST{[21]}
\def\HSW{[22]}
\def\pert{[23]}
\def\Hokone{[24]}
\def\OPEN{[25]}
\def\OPEW{[26]}
\def\Edent{[27]}
\def\Except{[28]}
\def\BKRS{[29]}
\def\HMR{[30]}
\def\ASok{[31]}
\def\Htwo{(1.41)}
\def\Hthree{(1.42)}
\def\ptph{(2.1)}
\def\tpsi{(2.2)}
\def\scep{(2.3)}
\def\ptps{(2.4)}
\def\clos{(2.5)}
\def\varp{(2.6)}
\def\vars{(2.7)}
\def\condit{(2.8)}
\def\ward{(2.9)}
\def\exw{(2.10)}
\def\fourpsi{(2.11)}
\def\fourJ{(2.12)}
\def\defX{(2.13)}
\def\deep{(2.14)}
\def\Wardrel#1{\hbox {$(2.15#1)$}}
\def\part{(2.16)}
\def\hatRJ{(2.17)}
\def\redeq#1{\hbox {$(2.18#1)$}}
\def\cons{(2.19)}
\def\UVS{(2.20)}
\def\VU{(2.21)}
\def\VS{(2.22)}
\def\VSt{(2.23)}
\def\genL{(2.24)}
\def\Fx{(2.25)}
\def\eqJ{(2.26)}
\def\eqJV{(2.27)}
\def\FYn{(2.28)}
\def\Vz{(2.29)}
\def\Sz{(2.30)}
\def\contw{(2.31)}
\def\relSU#1{\hbox {$(2.32#1)$}}
\def\ttT{(2.33)}
\def\LLU{(2.34)}
\def\LUW{(2.35)}
\def\order{(2.36)}
\def\inn{(2.37)}
\def\exF{(2.38)}
\def\conlm{(2.39)}
\def\defa{(2.40)}
\def\FhF{(2.41)}
\def\exF{(2.42)}
\def\WW{(2.43)}
\def\crit{(2.44)}
\def\DDD{(2.45)}
\def\superid#1{\hbox {$(2.46#1)$}}
\def\FDD{(2.47)}
\def\diF{(2.48)}
\def\resF{(2.49)}
\def\inc{(2.50)}
\def\exFc{(2.51)}
\def\idthree{(2.52)}
\def\crittwo{(2.53)}
\def\fourp{(2.54)}
\def\GF{(2.55)}
\def\defy{(2.56)}
\def\defz{(2.57)}
\def\exG{(2.58)}
\def\Gf{(2.59)}
\def\Gsol{(2.60)}
\def\OPEG{(2.61)}
\def\Gexp{(2.62)}
\def\af{(2.63)}
\def\asol{(2.64)}
\def\asoln{(2.65)}
\def\pwave{(2.66)}
\def\recurG{(2.67)}
\def\defst{(2.68)}
\def\azG{(2.69)}
\def\longR{(2.70)}
\def\ineq{(2.71)}
\def\unit{(2.72)}
\def\ttwo{(2.73)}
\def\gell{(2.74)}
\def\recurg{(2.75)}
\def\expaf{(2.76)}
\def\afl{(2.77)}
\def\ACC{(2.78)}
\def\sem{(2.79)}
\def\Gell{(2.80)}
\def\CB{(2.81)}
\def\aBR{(2.82)}
\def\Gzero{(2.83)}
\def\azero{(2.84)}
\def\CBtwo{(2.85)}
\def\Gext{(2.86)}
\def\Gextr{(2.87)}
\def\next{(2.88)}
\def\nextG{(2.89)}
\def\Gextexp{(2.90)}
\def\Solg{(2.91)}
\def\text{(2.92)}
\def\semin{(2.93)}
\def\ss{(2.94)}
\def\ash{(2.95)}
\def\Non{[32]}
\def\bpsN{[33]}
\def\decomp{(2.96)}
\def\susp{(3.1)}
\def\conp{(3.2)}
\def\conJ{(3.3)}
\def\varp{(3.4)}
\def\Class{[34]}
\def\Fourpf{(3.5)}
\def\redeqf#1{\hbox {$(3.6#1)$}}
\def\consa{(3.7)}
\def\TV{(3.8)}
\def\seld{(3.9)}
\def\sdual{(3.10)}
\def\consis{(3.11)}
\def\VUp{(3.12)}
\def\FU{(3.13)}
\def\KV{(3.14)}
\def\KS{(3.15)}
\def\defL{(3.16)}
\def\Xsd{(3.17)}
\def\VT{(3.18)}
\def\LLK{(3.19)}
\def\Wbar{(3.20)}
\def\KW{(3.21)}
\def\LVW#1{\hbox {$(3.22#1)$}}
\def\Ttt{(3.23)}
\def\WUp{(3.24)}
\def\exFp{(3.25)}
\def\ppp{(3.26)}
\def\Oprod{(3.27)}
\def\FUc{(3.28)}
\def\exWp{(3.29)}
\def\ABC{(3.30)}
\def\sumW{(3.31)}
\def\exVp{(3.32)}
\def\OV{(3.33)}
\def\defO{(3.34)}
\def\solIJ{(3.35)}
\def\VABC{(3.36)}
\def\relABC{(3.37)}
\def\FUW{(3.38)}
\def\ABCW{(3.39)}
\def\WW{(3.40)}
\def\defab{(3.41)}
\def\relF{(3.42)}
\def\WarF{(3.43)}
\def\Wwp{(3.44)}
\def\solW{(3.45)}
\def\GFp{(3.46)}
\def\Gyy{(3.47)}
\def\yy{(3.48)}
\def\solid{(3.49)}
\def\real{(3.50)}
\def\Gsolt{(3.51)}
\def\OPEf{(3.52)}
\def\poly{(3.53)}
\def\ff{(3.54)}
\def\Gsoltt{(3.55)}
\def\Kexp{(3.56)}
\def\longr{(3.57)}
\def\defBm{(3.58)}
\def\Azero{(3.59)}
\def\expf{(3.60)}
\def\afour{(3.61)}
\def\FF{(3.62)}
\def\afz{(3.63)}
\def\afone{(3.64)}
\def\AZG{(3.65)}
\def\delL{(3.66)}
\def\Ared{(3.67)}
\def\AA{(3.68)}
\def\expA{(3.69)}
\def\anm{(3.70)}
\def\Azer{(3.71)}
\def\Asym{(3.72)}
\def\unitl{(3.73)}
\def\unit{(3.74)}
\def\tzero{(3.75)}
\def\fg{(3.76)}
\def\decD{(3.77)}
\def\single{(3.78)}
\def\bell{(3.79)}
\def\ttwob{(3.80)}
\def\DD{(3.81)}
\def\CC{(3.82)}
\def\CD{(3.83)}
\def\resD{(3.84)}
\def\relDC{(3.85)}
\def\CDe{(3.86)}
\def\decC{(3.87)}
\def\resC{(3.88)}
\def\CBfour{(3.89)}
\def\decB{(3.90)}
\def\BB{(3.91)}
\def\resB{(3.92)}
\def\resBm{(3.93)}
\def\CBfs{(3.94)}
\def\Bhalf{(3.95)}
\def\CBh{(3.96)}
\def\Id{(3.97)}
\def\azero{(3.98)}
\def\Gident{(3.99)}
\def\extp{(3.100)}
\def\ppp{(3.101)}
\def\extPG{(3.102)}
\def\exts{(3.103)}
\def\Fourext{(3.104)}
\def\next{(3.105)}
\def\deco{(3.106)}
\def\Fn{(3.107)}
\def\azz{(3.108)}
\def\semsh{(3.109)}
\def\bell{(3.110)}
\def\CBB{(3.111)}
\def\Bext{(3.112)}
\def\Bex{(3.113)}
\def\decompB{(3.114)}
\def\cross{(4.1)}
\def\polylm{(4.2)}
\def\countN{(4.3)}
\def\Syy{(4.4)}
\def\lmyy{(4.5)}
\def\repA{(4.6)}
\def\mix{(4.7)}
\def\anti{(4.8)}
\def\mixl{(4.9)}
\def\mixlm{(4.10)}
\def\mult{(4.11)}
\def\scc{(4.12)}
\def\crossf{(4.13)}
\def\crossG{(4.14)}
\def\fsym{(4.15)}
\def\crossff{(4.16)}
\def\Fnew{(4.17)}
\def\Sinvv{(4.18)}
\def\antiyy{(4.19)}
\def\mixzz{(4.20)}
\def\mixyy{(4.21)}
\def\SSinv{(4.22)}
\def\Smix{(4.23)}
\def\relAS{(4.24)}
\def\solss{(4.25)}
\def\gexp{(4.26)}
\def\relb{(4.27)}
\def\factf{(4.28)}
\def\crossK{(4.29)}
\def\Sy{(4.30)}
\def\Soly{(4.31)}
\def\Sinv{(4.32)}
\def\fgen{(4.33)}
\def\antiy{(4.34)}
\def\mixz{(4.35)}
\def\mixy{(4.36)}
\def\crosstwo{(4.37)}
\def\scct{(4.38)}
\def\crossGt{(4.39)}
\def\fSn{(4.40)}
\def\rela{(4.41)}
\def\defpq{(4.42)}
\def\fgr{(4.43)}
\def\factwo{(4.44)}
\def\Gcross{(4.45)}
\def\Gext{(4.46)}
\def\fcross{(4.47)}
\def\fsol{(4.48)}
\def\fgen{(4.49)}
\def\Gwex{(4.50)}
\def\HFtFP{(5.1)}
\def\cs{(5.2)}
\def\csh{(5.3)}
\def\oldab{(5.4)}
\def\rewf{(5.5)}
\def\tfsone{(5.6)}
\def\tfstwo{(5.7)}
\def\tfsthree{(5.8)}
\def\tfsfour{(5.9)}
\def\tfsfive{(5.10)}
\def\tfssix{(5.11)}
\def\tfsseven{(5.12)}
\def\tFf{(5.13)}
\def\Hfour{(5.14)}
\def\GlN{(5.15)}
\def\flN{(5.16)}
\def\Hztwo{(5.17)}
\def\Hzthree{(5.18)}
\def\Hzfour{(5.19)}
\def\Hreps{(5.20)}
\def\did{(5.21)}
\def\spl{(5.22)}
\def\cpwe{(5.23)}
\def\cpweu{(5.24)}
\def\cpwer{(5.25)}
\def\azzero{(5.26)}
\def\azero{(5.27)}
\def\et{(5.28)}
\def\td{(5.29)}
\def\v{(5.30)}
\def\etf{(5.31)}
\def\Gfuniv{(5.32)}
\def\csr{(5.33)}
\def\csb{(5.34)}
\def\ab{(5.35)}
\def\sab{(5.36)}
\def\check{(5.37)}
\def\csh{(5.38)}
\def\lna{(5.39)}
\def\T{(5.40)}
\def\symT{(5.41)}
\def\TT{(5.42)}
\def\exH{(5.43)}
\def\unit{(5.44)}
\def\hyper{(5.45)}
\def\crossf{(5.46)}
\def\ssD{(5.47)}
\def\gh{(5.48)}
\def\ies{(5.49)}
\def\ca{(5.50)}
\def\DDD{(5.51)}
\def\TTT{(5.52)}
\def\ijk{(5.53)}
\def\pp{(5.54)}
\def\num{(5.55)}
\def\Huniv{(5.56)}
\def\expF{(5.57)}
\def\kl{(5.58)}
\def\univ{(5.59)}
\def\Htwo{(5.60)}
\def\Hthree{(5.61)}
\def\cth{(5.62)}
\def\cp{(5.63)}
\def\Hfour{(5.64)}
\def\Hexp{(5.65)}
\def\one{(5.66)}
\def\sig{(5.67)}
\def\cfour{(5.68)}
\def\cfourp{(5.69)}
\def\Hexpt{(5.70)}
\def\cfoura{(5.71)}
\def\Funiv{(5.72)}
\def\Hfive{(5.73)}
\def\resfive{(5.74)}
\def\Ffour{(5.75)}
\def\solfour{(5.76)}
\def\Hsix{(5.77)}
\def\csix{(5.78)}
\def\Ffive{(5.79)}
\def\solfive{(5.80)}
\def\Hexpall{(5.81)}
\def\abz{(5.82)}
\def\solz{(5.83)}
\def\solcz{(5.84)}
\def\Fsa{(5.85)}
\def\recura{(5.86)}
\def\abase{(5.87)}
\def\solca{(5.88)}
\def\relcc{(5.89)}
\def\solcaf{(5.90)}
\def\recurf{(5.91)}
\def\solcat{(5.92)}
\def\recurs{(5.93)}
\def\Sols{(5.94)}
\def\recurp{(5.95)}
\def\symc{(5.96)}
\def\sta{(5.97)}
\def\Fnew{(5.98)}
\def\recurab{(5.99)}
\def\abbase{(5.100)}
\def\solcab{(5.101)}
\def\relcab{(5.102)}
\def\solcabf{(5.103)}
\def\recurt{(5.104)}
\def\solcabt{(5.105)}
\def\symca{(5.106)}
\def\symcb{(5.107)}
\def\solcc{(5.108)}
\def\symcab{(5.109)}
\def\diffo{(\hbox {A.}1)}
\def\diffm{(\hbox {A.}2)}
\def\defR{(\hbox {A.}3)}
\def\tdiff{(\hbox {A.}4)}
\def\Lap{(\hbox {A.}5)}
\def\dr{(\hbox {A.}6)}
\def\div{(\hbox {A.}7)}
\def\gend{(\hbox {A.}8)}
\def\LLe{(\hbox {A.}9)}
\def\eiSO{(\hbox {A.}10)}
\def\Vd{(\hbox {A.}11)}
\def\VD{(\hbox {A.}12)}
\def\Vdiv{(\hbox {A.}13)}
\def\Ls{(\hbox {B.}1)}
\def\gen{(\hbox {B.}2)}
\def\polyY{(\hbox {B.}3)}
\def\eigL{(\hbox {B.}4)}
\def\Llu{(\hbox {B.}5)}
\def\LDd{(\hbox {B.}6)}
\def\Dad{(\hbox {B.}7)}
\def\defw{(\hbox {B.}8)}
\def\eigc{(\hbox {B.}9)}
\def\eig{(\hbox {B.}10)}
\def\Koo{[35]}
\def\ortho{[36]}
\def\defQ{(\hbox {B.}11)}
\def\Qact{(\hbox {B.}12)}
\def\Qeig{(\hbox {B.}13)}
\def\Ynm{(\hbox {B.}14)}
\def\Ynn{(\hbox {B.}15)}
\def\Vretare{[37]}
\def\LD{(\hbox {B.}16)}
\def\Dop{(\hbox {B.}17)}
\def\eigD{(\hbox {B.}18)}
\def\eigfour{(\hbox {B.}19)}
\def\DF{(\hbox {B.}20)}
\def\eigsix{(\hbox {B.}21)}
\def\defpnm{(\hbox {B.}22)}
\def\PFe{(\hbox {B.}23)}
\def\DFe{(\hbox {B.}24)}
\def\FPnm{(\hbox {B.}25)}
\def\recur{(\hbox {B.}26)}
\def\defq{(\hbox {B.}27)}
\def\eigeight{(\hbox {B.}28)}
\def\eigLab{(\hbox {B.}29)}
\def\Lab{(\hbox {B.}30)}
\def\defDab{(\hbox {B.}31)}
\def\LDa{(\hbox {B.}32)}
\def\matM{(\hbox {B.}33)}
\def\eigab{(\hbox {B.}34)}
\def\Yab{(\hbox {B.}35)}
\def\Ynnab{(\hbox {B.}36)}
\def\Dopab{(\hbox {B.}37)}
\def\Dthreeab{(\hbox {B.}38)}
\def\eigsixab{(\hbox {B.}39)}
\def\Lab{(\hbox {B.}40)}
\def\Leig{(\hbox {B.}41)}
\def\eigthreeab{(\hbox {B.}42)}
\def\defT{(\hbox {B.}43)}
\def\Lcross{(\hbox {B.}44)}
\def\TTid{(\hbox {B.}45)}
\def\eigf{(\hbox {B.}46)}
\def\solcr{(\hbox {B.}47)}
\def\eigftwo{(\hbox {B.}48)}
\def\Ld{(\hbox {C.}1)}
\def\defDD{(\hbox {C.}2)}
\def\DD{(\hbox {C.}3)}
\def\DDD{(\hbox {C.}4)}
\def\simD{(\hbox {C.}5)}
\def\Dtwo{(\hbox {C.}6)}
\def\Dlth{(\hbox {C.}7)}
\def\Dtw{(\hbox {C.}8)}
\def\Dinv{(\hbox {C.}9)}
\def\LDD{(\hbox {C.}10)}
\def\Dltf{(\hbox {C.}11)}
\def\TY{(\hbox {C.}12)}
\def\solY{(\hbox {C.}13)}
\def\exYp{(\hbox {C.}14)}
\def\Utwo{(\hbox {C.}15)}
\def\UVW{(\hbox {C.}16)}
\def\UUU{(\hbox {C.}17)}
\def\req{(\hbox {C.}18)}
\def\phd{[38]}
\def\algQS{(\hbox {D.}1)}
\def\MQS{(\hbox {D.}2)}
\def\MJ{(\hbox {D.}3)}
\def\Car{(\hbox {D.}4)}
\def\hw{(\hbox {D.}5)}
\def\DSw{(\hbox {D.}6)}
\def\weight{(\hbox {D.}7)}
\def\Qti{(\hbox {D.}8)}
\def\sQ{(\hbox {D.}9)}
\def\shQ{(\hbox {D.}10)}
\def\sDel{(\hbox {D.}11)}
\def\snew{(\hbox {D.}12)}
\def\sDelid{(\hbox {D.}13)}
\def\dem{(\hbox {D.}14)}
\def\bc{(\hbox {D.}15)}
\def\dcel{(\hbox {D.}16)}
\def\weightL{(\hbox {D.}17)}
\def\semiC{(\hbox {D.}18)}
\def\semiD{(\hbox {D.}19)}
\def\DCrel{(\hbox {D.}20)}
\def\semiE{(\hbox {D.}21)}
\def\deCDE{(\hbox {D.}22)}
\def\ufields{(\hbox {E.}1)}
\def\sct{(\hbox {E.}2)}
\def\wi{(\hbox {E.}3)}
\def\ind{(\hbox {E.}4)}
\def\feq{(\hbox {E.}5)}
\def\seq{(\hbox {E.}6)}
\def\fineq{(\hbox {E.}7)}
\def\fexp{(\hbox {E.}8)}
\def\dop{(\hbox {E.}9)}
\def\wi{(\hbox {E.}10)}
\def\ans{(\hbox {F.}1)}
\def\lans{(\hbox {F.}2)}
\def\tans{(\hbox {F.}3)}
\def\didso{(\hbox {G.}1)}
\def\didst#1{\hbox {$(\hbox {G.}2#1)$}}
\def\didref{(\hbox {G.}3)}
\def\didtutd#1{\hbox {$(\hbox {G.}4#1)$}}
\def\dids{(\hbox {G.}5)}